\documentclass[twocolumn,twocolappendix]{aastex631}

\usepackage{enumerate}
\usepackage{enumitem}
\usepackage{natbib,ifthen}
\usepackage{graphicx}
\usepackage{fancyhdr}
\usepackage{amssymb,amsmath}
\usepackage{multirow,bigdelim}
\usepackage{wasysym}  
\usepackage{amssymb}
\usepackage{xcolor}
\usepackage[version=4]{mhchem}

\usepackage[T1]{fontenc}

\definecolor{blue-violet}{rgb}{0.54, 0.17, 0.89}

\shorttitle{Hot Cores in the LMC}
\shortauthors{Sewi{\l}o et al.}

\begin{document}


\title{ALMA Observations of Molecular Complexity in the Large Magellanic Cloud: The N\,83 Star-Forming Region}


\correspondingauthor{Marta Sewi{\l}o}
\email{marta.m.sewilo@nasa.gov}

\author[0000-0003-2248-6032]{Marta Sewi{\l}o }
\affiliation{Exoplanets and Stellar Astrophysics Laboratory, NASA Goddard Space Flight Center, Greenbelt, MD 20771, USA}
\affiliation{Department of Astronomy, University of Maryland, College Park, MD 20742, USA}
\affiliation{Center for Research and Exploration in Space Science and Technology, NASA Goddard Space Flight Center, Greenbelt, MD 20771} 

\author[0000-0002-1048-9618]{Amanda Broadmeadow}
\altaffiliation{Our colleague, Amanda Broadmeadow, passed away on 2026 May 14.}
\affiliation{Department of Astronomy, University of Maryland, College Park, MD 20742, USA}

\author[0000-0003- 1993-2302]{Roya Hamedani Golshan} 
\affiliation{I. Physikalisches Institut der Universit{\"a}t zu K{\"o}ln, Z{\"u}lpicher Str. 77, 50937, K{\"o}ln, Germany}

\author[0000-0001-6752-5109]{Steven B. Charnley}
\affiliation{Astrochemistry Laboratory, NASA Goddard Space Flight Center, Greenbelt, MD 20771, USA}

\author[0000-0002-8876-0690]{Lee Mundy}
\affiliation{Department of Astronomy, University of Maryland, College Park, MD 20742, USA}

\author[0000-0002-2062-1600]{Kazuki Tokuda}
\affiliation{Faculty of Education, Kagawa University, Saiwai-cho 1-1, Takamatsu, Kagawa 760-8522, Japan}

\author[0000-0002-9277-8025]{Thomas M{\"o}ller}
\affiliation{I. Physikalisches Institut der Universit{\"a}t zu K{\"o}ln, Z{\"u}lpicher Str. 77, 50937, K{\"o}ln, Germany}

\author[0000-0002-1143-6710]{Jennifer Wiseman}
\affiliation{NASA Goddard Space Flight Center, 8800 Greenbelt Rd, Greenbelt, MD 20771, USA}

\author[0000-0002-3925-9365]{C.-H. Rosie Chen}
\affiliation{Max-Planck-Institut f{\"u}r Radioastronomie, Auf dem H{\"u}gel 69 D-53121 Bonn, Germany}

\author[0000-0002-4663-6827]{Remy Indebetouw}
\affiliation{Department of Astronomy, University of Virginia, PO Box 400325, Charlottesville, VA 22904, USA}
\affiliation{National Radio Astronomy Observatory, 520 Edgemont Rd, Charlottesville, VA 22903, USA}

\author[0000-0002-0861-7094]{Joana M. Oliveira}
\affiliation{Lennard-Jones Laboratories, Keele University, ST5 5BG, UK}

\author[0000-0003-2141-5689]{Peter Schilke}
\affiliation{I. Physikalisches Institut der Universit{\"a}t zu K{\"o}ln, Z{\"u}lpicher Str. 77, 50937, K{\"o}ln, Germany}


\begin{abstract}

We report the results of the 1.2 mm continuum and molecular line $\sim$0.1 parsec-scale observations with the Atacama Large Millimeter/submillimeter Array of two fields in the star-forming region N\,83 in the Large Magellanic Cloud (LMC).  This study is third in a series of investigations exploring molecular complexity in a low-metallicity and high UV flux environment of the LMC.  We determine the physical and chemical properties of five sources detected with ALMA based on spectral modeling with XCLASS and the analysis of the continuum emission. The temperatures exceeding 100 K, small sizes, and high H$_2$ volume densities of sources N\,83A--mm A and B indicate that they are hot cores.  N\,83A--mm A/B are associated with complex organic molecules: methanol (\ce{CH3OH}), methyl cyanide (\ce{CH3CN}), and dimethyl ether (\ce{CH3OCH3}).  We also found \ce{CH3OH} in three cold cores, and extended cold \ce{CH3OH} emission in both fields. We detected \ce{SO}, \ce{^33SO}, \ce{SO2}, \ce{OCS}, \ce{H2CS}, \ce{CS}, \ce{C^33S}, \ce{HC^15N}, \ce{H^13CN}, \ce{HNCO}, \ce{HDCO}, \ce{H^13CO+}, \ce{SiO}, and \ce{CH2CO} in at least one core. Molecular abundances with respect to \ce{H2} fit into the trends previously reported for the Magellanic hot cores: the \ce{CH3OH} abundance varies significantly, while the abundances of S-bearing species show small variations.  We found evidence for strong UV radiation and complex kinematics in both ALMA fields in N\,83. The feedback-produced shocks are important drivers of the chemistry of S-bearing species, \ce{SiO}, and \ce{CH3OH} in and outside hot cores. Local effects have a large impact on the chemical composition and abundances of star-forming cores in the LMC.

 \end{abstract}


\section{Introduction}
\label{s:intro}

Complex organic molecules (COMs), defined as carbon-bearing molecules containing six or more atoms, are key molecular species tracing the chemical and physical processes occurring during  the formation of stars and planetary systems (e.g., \citealt{herbst2009,jorgensen2020}).  COMs are thought to be fundamental building blocks of larger molecules, including critical prebiotic species such as amino acids, sugars, and nucleobases (e.g., \citealt{caselli2012,kwon2016}).  Understanding how the formation, evolution, and destruction of COMs depend on their environment is one of the grand challenges in astrochemistry. 

The Large Magellanic Cloud (LMC), the gas-rich dwarf companion of the Milky
Way, has become a benchmark for studies on complex organic chemistry in low-metallicity ($Z_{\rm LMC}$=0.3--0.5$ Z_{\odot}$; e.g.,  \citealt{russell1992,rolleston2002,toribiosancipriano2017}) and high UV radiation (e.g., \citealt{browning2003,welty2006}) systems.  Located at the distance of $\sim$50 kpc (\citealt{pietrzynski2019}), it facilitates studies of resolved stellar populations on sub-parsec scales. Since the physical and chemical properties of the LMC's environment are distinct from both the inner and outer Galaxy, and its even lower-metallicity neighbor, the Small Magellanic Cloud (SMC; $Z_{\rm SMC}$=0.1--0.2$ Z_{\odot}$),  it provides an important piece of the puzzle in elucidating the role of the environment on complex organic chemistry.  The metallicities of the Magellanic Clouds are similar to galaxies at earlier cosmological epochs (z$\sim$1--2; \citealt{pei1999,mehlert2002,madau2014}).  Thus, studying complex chemistry in the Magellanic Clouds helps shed light on metallicity effects on the processes in the past Universe, even though other physical conditions may be quite different. 

We expect that different conditions in the LMC's interstellar medium, compared to the Galaxy, will result in differences in complex organic (including prebiotic) chemistry. Due to the lower elemental abundances of gaseous C, O, and N atoms in the LMC compared to the Galaxy, fewer key atoms are available for chemistry. The lower dust-to-gas ratio in the LMC results in fewer dust grains for surface chemistry and less shielding than in the Galaxy (e.g., \citealt{dufour1975,koornneef1984,duval2014}).  Due to less shielding, UV photons can penetrate deeper into the dense gas; this effect combined with the higher intensity of the UV radiation (e.g., \citealt{browning2003,welty2006}) leads to higher dust temperature (e.g., \citealt{vanloon2010lmc}), which in turn, results in the less efficient formation of COMs (e.g., \citealt{acharyya2015,shimonishi2016a}). Gamma-ray observations indicate that the cosmic-ray density in the LMC is a factor of 4 lower than that measured in the solar neighborhood, resulting in cosmic-ray-induced UV radiation less effective in driving ice-mantle chemistry (e.g., \citealt{abdo2010,knodlseder2013}). The consequences of all these effects are more pronounced in the SMC which, compared to the LMC, has lower metallicity and dust-to-gas ratio, more intense UV radiation, and lower cosmic-ray density.  

In the protostellar systems, the formation of COMs is mainly a result of the chemical processes taking place on the surfaces and within the icy grain mantles in cold protostellar envelopes (e.g., \citealt{herbst2009,vandishoeck2014,oberg2016}). The ice chemistry products become observable after icy grain mantles are sublimated close to the protostar (in a so called `hot core' in massive protostars) or sublimated and sputtered in shocks in the protostellar jets/outflows (low- to high-velocity shocks) or at the envelope-disk interface (low-velocity shocks; e.g., \citealt{arce2008,aota2015,miura2017}). Some COMs may be the products of the gas-phase chemistry through ion-molecule reactions or neutral-neutral reactions following ice sublimation (e.g., \citealt{taquet2016}).   Hot cores along with their low-mass counterparts known as `hot corinos' are among the most COM-rich sources in the Galaxy.  A typical Galactic hot core is compact ($D\lesssim0.1$ pc),  dense ($n_{\rm H}\gtrsim10^{6-7}$ cm$^{-3}$), and hot ($T_{\rm kin}\gtrsim100$ K; e.g., \citealt{garay1999}; \citealt{kurtz2000}; \citealt{cesaroni2005}; \citealt{palau2011}). 

\begin{figure*}
\centering
\includegraphics[width=\textwidth]{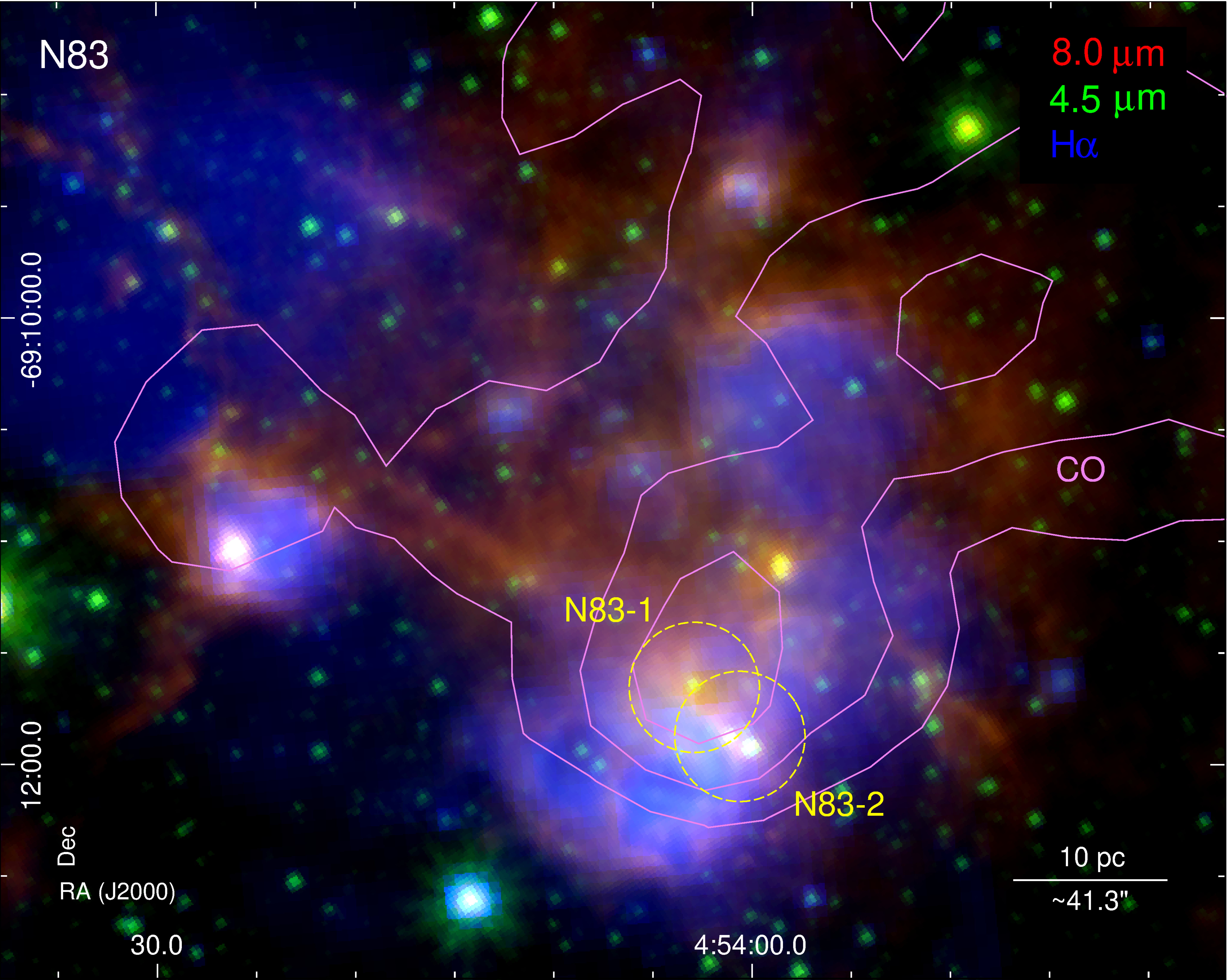}
\caption{Three-color composite image of the N\,83 star-forming region combining the {\it Spitzer}/SAGE IRAC 8.0 $\mu$m (red), 4.5 $\mu$m (green; \citealt{meixner2006}), and MCELS H$\alpha$ (blue; \citealt{smith1998}) images. The magenta contours correspond to the single-dish $^{12}$CO ($J$ = 1--0) emission from the Magellanic Mopra Assessment (MAGMA) survey with an effective angular resolution of $\sim$45$''$ (for comparison, an angular resolution of the {\it Spitzer} images is $\sim$2$''$), with contour levels of (20, 40, 70)\% of the $^{12}$CO integrated intensity peak of 14.8 K km s$^{-1}$ (\citealt{wong2011}). The ALMA fields (true to size) are indicated with dashed yellow circles and labeled; they are located in the H$\alpha$ nebula N\,83A  (\citealt{henize1956}).  The bright compact object in the east is N\,83B--1 (in the H$\alpha$ nebula N\,83B), a member of  a distinct class of H\,{\sc ii} regions in the Magellanic Clouds called `high-excitation blobs' (HEBs; \citealt{heydari1990,heydari2001}).  The bright star in the south is Sk~$-$69$^{\circ}$30, the visually brightest star in the OB association LH\,5 (e.g., \citealt{lucke1970}; see Section~\ref{s:intro}).  \label{f:n83}}
\end{figure*}

 To date, only eight LMC hot cores with COMs have been reported in the literature, with methanol (\ce{CH3OH}) being the only COM identified toward all of them
(\citealt{sewilo2018,sewilo2019,shimonishi2020,sewilo2022n105,hamedanigolshan2024,broadmeadow2025,shimonishi2026}).  Methyl cyanide (\ce{CH3CN}) is detected when covered by the observations (in 5 out of 8 hot cores). Only a handful of larger gas-phase COMs have been found in the LMC: dimethyl ether (\ce{CH3OCH3}; in 6/8), methyl formate (\ce{HCOOCH3}; in 1/8), ethanol (\ce{CH3CH2OH}; in 1/8), and tentatively, formamide (\ce{NH2CHO}; in 1/8).  Hot \ce{SO2} emission but no COMs have been detected toward two LMC sources  (\citealt{shimonishi2016b,shimonishi2026}). 

The early observational searches for gas-phase COMs and theoretical studies indicated a deficiency of \ce{CH3OH} and other COMs in the LMC (e.g., \citealt{shimonishi2016a,acharyya2015}). However, there is growing evidence that \ce{CH3OH} may be widespread in the LMC. Both compact and extended \ce{CH3OH} emission has been found in many locations in the LMC outside hot cores (`cold \ce{CH3OH}'; \citealt{sewilo2022n105,hamedanigolshan2024,broadmeadow2025,mondal2026}).  

The ALMA gas-phase observations of the LMC hot cores reveal a large diversity in their chemical makeup, despite a small sample size.  The particularly large variations are found for gas-phase \ce{CH3OH} abundance with respect to \ce{H2}. The \ce{CH3OH} abundance ranges from the metallicity-scaled values similar to Galactic hot cores, to nondetections with upper limits two orders of magnitude below Galactic hot core abundances. In contrast, the abundances of S-bearing species such as \ce{SO2} and \ce{SO} do not change significantly.  These trends extend to the lower-metallicity environment of the SMC where only two hot cores have been identified to date (\citealt{shimonishi2023}).  In these two SMC hot cores, however, \ce{CH3OH} likely originates in the outflows rather than the hot core region, as indicated by its morphology and low temperature.  

The origin of the molecular abundance patterns observed in the LMC and SMC is still under investigation. The physical and chemical conditions in star-forming regions hosting hot cores during the collapse phase (e.g., the strength of the UV radiation, metallicity, and cosmic-ray ionization rate) and source properties (e.g., luminosity, protostellar feedback) may all contribute to driving the chemical diversity in the Magellanic hot cores (e.g., \citealt{sewilo2022n105,shimonishi2026}). 

COMs larger than \ce{CH3OH} have only recently been found in ices in the LMC (for the first time outside the Galaxy) with the James Webb Space Telescope (JWST).  Based on observations with the Medium Resolution Spectrograph (MRS) of the Mid-Infrared Instrument (MIRI; 4.9--27.9 $\mu$m), \citet{sewilo2025} reported the detection of  methanol  (\ce{CH3OH}), acetaldehyde (\ce{CH3CHO}), ethanol (\ce{CH3CH2OH}), methyl formate (\ce{HCOOCH3}), and acetic acid (\ce{CH3COOH}) ices toward the protostar ST6 in the LMC.  \ce{CH3CHO}, \ce{CH3CH2OH}, and \ce{HCOOCH3} ices were observed outside the Galaxy for the first time, while the detection of \ce{CH3COOH} ice was the first conclusive detection in an astrophysical context.  The detection of COM ices in ST6 confirms that COMs are the products of grain-surface chemistry, supporting the results of chemical modeling and laboratory experiments (e.g., \citealt{garrod2022}). It also provides evidence that COMs can form efficiently in the extragalactic environment that is much harsher than in the Galaxy.  Modeling of the MIRI MRS spectrum of ST6 revealed differences in ice abundances with respect to \ce{H2O} ice between ST6 and four Galactic protostars with measured COM ice abundances. All simple and COM ice abundances with respect to \ce{H2O} ice for overlapping species (\ce{CH4}, \ce{SO2}, \ce{H2CO}, \ce{HCOOH}, \ce{OCN-}, \ce{HCOO-}, \ce{CH3OH}, \ce{CH3CH2OH}, \ce{CH3CHO}, and \ce{HCOOCH3}) are lower in ST6 by a factor ranging from 1.2 (\ce{HCOOH} ice) to $\sim$27 (\ce{H2CO} ice). The \ce{CH3COOH/HCOOCH3} ratio is a factor of 4--46 higher when compared to the predictions of the \citet{garrod2022} {\it Galactic} chemical models. The observed differences could be attributed to the lower metallicity and higher UV flux in the LMC compared to the Galaxy. Chemical model grid for a selection of key parameters appropriate for the LMC (such as elemental abundances, cosmic-ray ionization rate, and UV field strength) is needed to better understand the solid and gas-phase molecular abundances observed in the LMC. 

In this paper, we present the results of the observations of two fields in the star-forming region LHA\,120--N\,83 (hereafter N\,83, \citealt{henize1956}; or DEM\,22, \citealt{davies1976}) with ALMA in four spectral windows covering $\sim$7.5 GHz between $\sim$241.5 and $\sim$260.6 GHz.  This is the third paper in a series exploring the molecular complexity in star-forming regions of the LMC, following \citet{sewilo2022n105} and \citet{broadmeadow2025}, based on the ALMA observations on hot core scales (0$\rlap.{''}$5 or $\sim$1.2 pc) of seven fields (2 in N\,105, 1 in N\,160, 2 in N\,83, and 1 in N\,171). The detection of two hot cores with COMs in N\,105 by \citet{sewilo2022n105} and one hot core in N\,160 by \citet{broadmeadow2025} has demonstrated the effectiveness of the program's target selection criteria (provided in Section~\ref{s:observations} and  in \citealt{sewilo2022n105}) in identifying these objects.  Here, we report the detection of a hot core in each of the two ALMA fields (N\,83--1 and N\,83--2) in the star-forming region N\,83. 

N\,83 is located just off the western edge of the LMC bar.  It is associated with the radio emission revealed in single-dish surveys at centimeter wavelengths: radio source MC\,16 (3.4 and 6 cm, half-power beam width, HPBW of $\sim$2$\rlap.{'}$5 and $\sim$4$'$, respectively; \citealt{mcgee1972,mcgee1978}) or B0454$-$6916 (several bands between 3.5 and 21 cm, HPBW$\sim$2$\rlap.{'}$7--15$\rlap.{'}$2; \citealt{filipovic1998}). \citet{henize1956} identified four optical nebulae in N\,83 (in the order of decreasing H$\alpha$ surface brightness): N\,83A, N\,83B, N\,83C, and N\,83D. The ALMA fields N\,83--1 and N\,83--2 are located in N\,83A. 

The N\,83A nebula lies roughly at the center of the OB association LH\,5 \citep{lucke1970} and is excited by the main sequence star Sk~$-$69$^{\circ}$25 of a spectral type O6  (\citealt{sanduleak1970,sanduleak2008}; \citealt{massey2000}). LH\,5 is $\sim$2.2 Myr old, spans over $\sim$16 squared arcmin and contains 26 blue stars (\citealt{lucke1974}; \citealt{massey2000}).  The visually brightest star in LH\,5 is  Sk~$-$69$^{\circ}$30, an evolved supergiant of a spectral type G5 Ia, coeval with the rest of the massive stars.  LH\,5 also hosts a Wolf-Rayet star of type WN2  (BAT99--5 or Brey\,4; e.g., \citealt{conti1989}; \citealt{breysacher1999}). N\,83A is one of the most reddened star-forming regions in the LMC (e.g., \citealt{lucke1974}; \citealt{ye1998}).

The paper is organized as follows. In Section~\ref{s:observations}, we describe the ALMA observations used in this paper. In Section~\ref{s:analysis}, we present the analysis of the ALMA 1.2 mm continuum and spectral line data, leading to the determination of the chemical and physical properties of the sources. We discuss the results in Section~\ref{s:discussion}, and provide the summary and conclusions in Section~\ref{s:conclusions}.

\begin{figure*}
\centering
\includegraphics[width=0.498\textwidth]{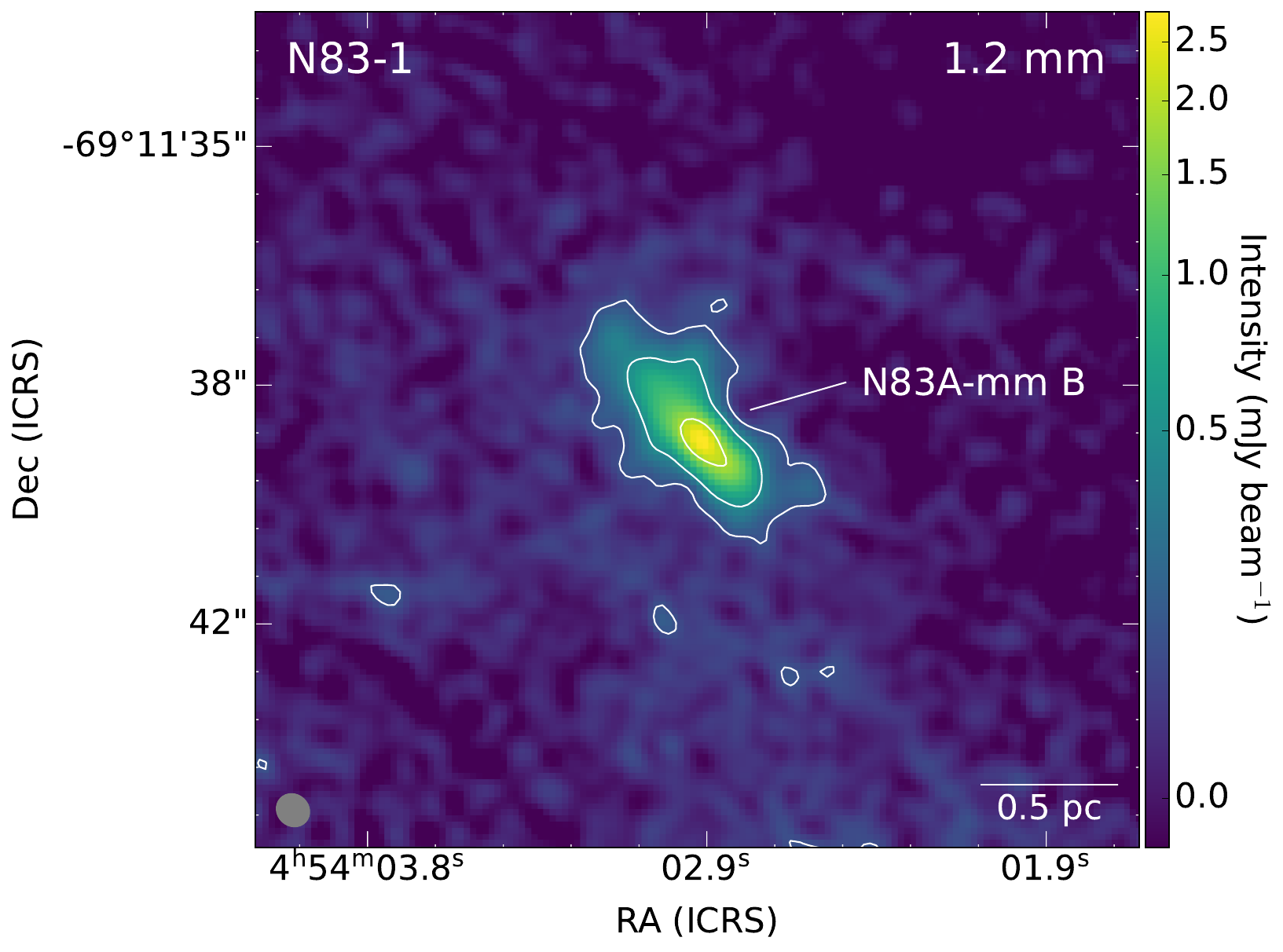}
\includegraphics[width=0.48\textwidth]{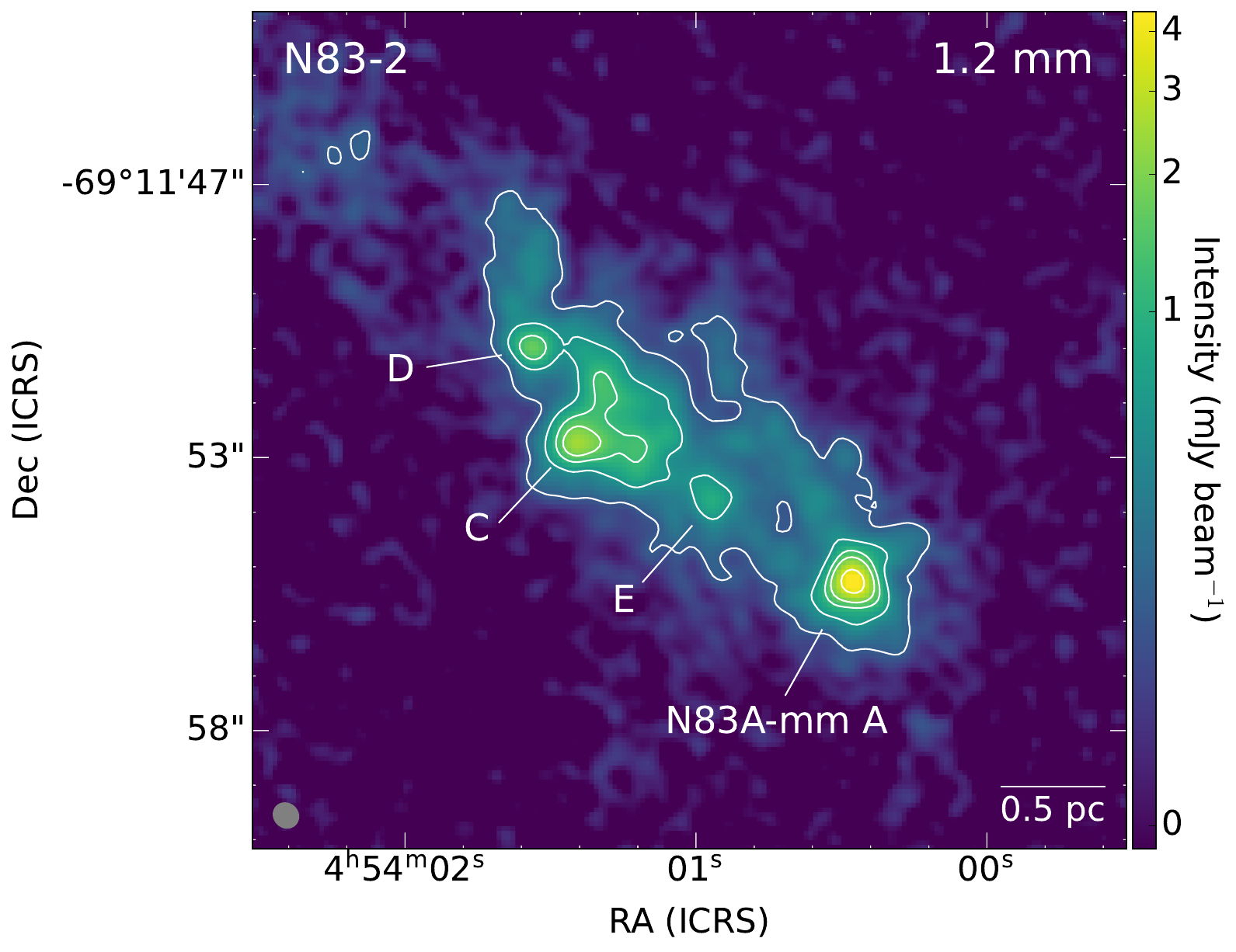}
\caption{The 1.2 mm continuum images of N\,83--1 (left) and N\,83--2 (right).  The contour levels correspond to (3, 10, 40)/(3, 10, 20, 30, 60) $\times$  the 1.2 mm continuum image rms noise ($\sigma$) of 4.7$\times$10$^{-5}$/5.9$\times$10$^{-5}$Jy beam$^{-1}$ for N\,83--1/N\,83--2.  \label{f:N83B6cont}}
\end{figure*}

\section{ALMA Observations}
\label{s:observations}

Two fields in the star-forming region N\,83 (N\,83--1 and N\,83--2) were observed with ALMA 12m Array in Band 6 as part of the Cycle 7 project 2019.1.01720.S (PI M. Sewi{\l}o).   The program consist of the observations of seven fields in the LMC that are associated with {\it Spitzer} YSOs, masers (H$_2$O, OH, or CH$_3$OH), and all but one, with the SO emission (the hot core tracer) -- the characteristics of the fields hosting hot cores N\,113 A1 and B3 that at the time of the target selection, were the only known extragalactic sources exhibiting hot core complex chemistry \citep{sewilo2018,sewilo2019}.  The results for two ALMA fields in the star-forming region N\,105 are presented in \citet{sewilo2022n105}, and for one field in N\,160 in \citet{broadmeadow2025}.  \citet{sewilo2022n105} provide a detailed description of the source selection. 

The single-pointing observations of all targets were executed three times, twice on October 21 and once on October 23, 2019, with 43 antennas and baselines from 15 to 783 m on October 21 and from 15 to 782 m on October 23. The positions of two ALMA pointings in N\,83 are provided in Table~\ref{t:imagesinfo}. The bandpass, flux, and phase calibrators were J0519-4546, J0519-4546, and J0440-6952, respectively, for the first run on October 21 and on October 23. For the second run on October 21, the bandpass, flux, and phase calibrators were J0538-4405, J0538-4405, J0511-6806, respectively.  The total on-source integration time for all seven fields was $\sim$45.9 min. for each execution.  The spectral setup included four 1875 MHz spectral windows centered on frequencies of 242.4, 244.8, 257.85, and 259.7 GHz, each with 3840 channels of 488.3 kHz width, resulting with a spectral resolution of 976.6 kHz ($\sim$1.21--1.13 km~s$^{-1}$; two times the channel width due to the default Hanning smoothing of the data). 

The data were calibrated and imaged with version 5.6.1-8 of the ALMA pipeline in CASA (Common Astronomy Software Applications; \citealt{mcmullin2007}).  The continuum in each spectral window was identified and subtracted before cube imaging. The CASA task \texttt{tclean} was used for imaging using the Hogbom deconvolver, standard gridder, Briggs weighting with a robust parameter of 0.5, and masking using the `auto-multithresh' algorithm.   The spectral cubes have a cell size of  $0\rlap.{''}092 \times 0\rlap.{''}092 \times 0.56$ km s$^{-1}$.  The size of the synthesized beam and achieved sensitivity for each 242.4 GHz (1.2 mm) continuum image and each data cube are provided  in Table~\ref{t:imagesinfo}.  All the images have been corrected for primary beam attenuation. 

In our analysis, we also use the ALMA Band 3 observations of a young stellar object (YSO) 045400.10$-$691155.5 from our Cycle 5 project 2017.1.00093.S (PI T. Onishi, YSO42) that cover both Band 6 fields (N\,83--1 and N\,83--2).  We utilize the Band 3 data from spectral windows centered  on the $^{13}$CO (1--0)  line (110.20135  GHz, an upper state energy, $E_{\rm U}=5.3$ K; combined 12-m and 7-m Array data) and the C$^{18}$O (2--1) line (219.56035 GHz, $E_{\rm U}=15.8$ K; Total Power only). The Band 3 observations and data processing are described in \citet[all target YSOs were observed during each program execution]{sewilo2023}.  For YSO42, the sensitivity of 19.6 mJy per $2\rlap.{''}29\times1\rlap.{''}67$ beam  (0.51 K) was achieved in the $^{13}$CO (1--0) data cube. The sensitivity of 0.45 Jy per $29\rlap.{''}63\times29\rlap.{''}63$ beam  (13 mK) was achieved in the C$^{18}$O (2--1) data cube.  The channel width is 1.0 and 0.17 km s$^{-1}$ for the $^{13}$CO and C$^{18}$O data cubes, respectively.  The sensitivity of $1.5\times10^{-4}$ Jy per $2\rlap.{''}35\times1\rlap.{''}64$ beam (3.95 mK) was achieved in the continuum (105.2 GHz or 2.85 mm) constructed from the line-free channels.

\begin{deluxetable*}{lccccccc}[ht!]
\centering
\tablecaption{ALMA Pointings and 1.2 mm Continuum Image and Spectral Cube Parameters\label{t:imagesinfo}}
\tablewidth{0pt}
\tablehead{
\colhead{Field} &
\colhead{RA} &
\colhead{Decl.} &
\colhead{Spectral} &
\colhead{Frequency Range} & 
\colhead{Synth. Beam: ($\Theta_B$, PA)}  &
\multicolumn{2}{c}{Image rms\tablenotemark{\footnotesize a}} \\
\cline{7-8}
\colhead{} & 
\colhead{($^{\rm h}$ $^{\rm m}$ $^{\rm s}$)} &
\colhead{($^{\rm \circ}$ $'$ $''$)} &
\colhead{window} &
\colhead{(GHz)} &
\colhead{($'' \times ''$, $^{\circ}$)} &
\colhead{(mJy beam$^{-1}$)}  &
\colhead{(K)}
}
\startdata
N\,83--1 & 04:54:02.95 & -69:11:39.4 & 242 GHz & 241.27653--243.14971 & $0.547\times0.494$, 46.6 & 1.84 & 0.14 \\ 
& & & 245 GHz & 243.66769--245.54088 & $0.543\times0.490$, 45.2 & 1.62 & 0.12  \\ 
& & & 258 GHz & 256.71495--258.58814 & $0.520\times0.468$, 47.8 & 1.70 & 0.13 \\ 
& & & 260 GHz & 258.54806--260.42125 & $0.516\times0.466$, 45.8 & 1.96 & 0.15  \\ 
& & & \multicolumn{2}{c}{Continuum (242.4 GHz / 1.2 mm)} & $0.514\times0.467$, 46.9 & 0.047 & 0.004   \\                         N\,83--2 & 04:54:00.64 & -69:11:52.6 & 242 GHz & 241.27653--243.14971 & $0.548\times0.494$, 46.6 & 1.89 & 0.15 \\    
& & & 245 GHz & 243.66769--245.54088 & $0.543\times0.490$, 45.3 & 1.66 &  0.13 \\      
& & & 258 GHz & 256.71495--258.58814 & $0.520\times0.469$, 48.0 & 1.75 & 0.13 \\   
& & & 260 GHz & 258.54806--260.42125 & $0.516\times0.466$, 45.9 & 2.02 & 0.15  \\. 
& & & \multicolumn{2}{c}{Continuum (242.4 GHz / 1.2 mm)} & $0.514\times0.467$, 46.7 & 0.059 & 0.005 \\
\enddata
\tablenotetext{a}{For the data cubes, the `Image rms' is the rms noise per $\sim$0.6 km s$^{-1}$ channel estimated with the CASA task {\sc imstat} in line-free channels.  For the continuum images, the `Image rms' is the rms noise measured in the emission-free regions of the 1.2 mm continuum image.}
\end{deluxetable*}

\section{Data Analysis and Results}
\label{s:analysis}

\subsection{The 1.2 mm Continuum Emission and Source Identification}
\label{s:cont}

Figure~\ref{f:N83B6cont} shows the 1.2 mm continuum images of the ALMA fields N\,83--1 and N\,83--2.  The image of N\,83--1 reveals a single elongated continuum source surrounded by an extended emission. In contrast, multiple continuum peaks are detected in N\,83--2 that similarly to N\,83--1, are associated with an extended emission.  We assign source names to the 1.2 mm continuum peaks with a signal-to-noise ratio larger than 10 in the order of decreasing peak intensity: N\,83A--mm A, N\,83A--mm B, etc. Since the N\,83--1 and N\,83--2 fields are overlapping (Figure~\ref{f:n83}), we consider all the sources from both fields when assigning names.  As a result, N\,83A--mm B lies in N\,83--1, while the remaining sources are located in N\,83--2 (see Figure~\ref{f:N83B6cont}). We refer to individual sources throughout the paper by providing either their full name or only the letter reflecting their relative peak intensity.

\subsection{The Inventory of Detected Molecular Species}
\label{s:specid}

The initial inspection of the data cubes and the spectral line identification were carried out in the CASA task {\tt imview} which uses the NRAO's spectral line database Splatalogue\footnote{\url{http://www.cv.nrao.edu/php/splat}}. The line identification was later verified by detailed spectral modeling described in Section~\ref{s:modeling}. In this section, we describe our detection criteria and present an overview of the molecular species detected in N\,83--1 and N\,83--2. 

Throughout the paper, we use molecular data from the Cologne Database for Molecular Spectroscopy (CDMS\footnote{\url{http://www.astro.uni-koeln.de/cdms}}; \citealt{muller2005}) where available; otherwise, we use the Jet Propulsion Laboratory (JPL) Millimeter and Submillimeter Spectral Line Catalog\footnote{\url{http://spec.jpl.nasa.gov/}} (\citealt{pickett1998}).  We adopt the CDMS quantum number notation. 

For N\,83A--mm A--D, the 1D spectra used in our analysis were extracted from the primary beam-corrected data cubes (listed in Table~\ref{t:imagesinfo}) as the mean within the area enclosed by the contour corresponding to the 50\% of a given source's 1.2 mm continuum intensity peak. For N\,83A--mm E, we used the 60\% of the 1.2 mm continuum intensity peak contour instead since the 50\% contour partly overlaps with nearby cores. 

We adopt the line detection criteria from \citet{broadmeadow2025}, the previous paper in our series on the LMC hot cores. For a line to be considered as detected,  
the emission intensity in three consecutive channels must be at least three times the noise level in the spectrum. This noise level ($\sigma_{\rm s,i}$) is calculated for each source `s' and spectral window `i' using a sigma clipping  method. For {\it tentative} detections, we require the emission intensity in four consecutive channels to be above $2\sigma_{\rm s,i}$ with at least one channel above $3\sigma_{\rm s,i}$.  Since the adjacent channels in the ALMA data cubes are correlated,\footnote{\url{https://help.almascience.org/kb/articles/how-is-noise-correlated-in-casa-when-mstransform-and-tclean-are-used} (2026 October 1)} the selection of four consecutive channels for the identification of tentative detections increases the confidence that these faint features are real spectral lines rather than noise.

A complete inventory of molecular line transitions detected (reliably or tentatively) toward each source is presented in Table~\ref{t:detections} in Appendix~\ref{app:detections}. We distinguish between $(3-5)\sigma_{\rm s,i}$ and $>$$5\sigma_{\rm s,i}$ detections in Table~\ref{t:detections}, allowing for a quick assessment of the relative line strengths. 

We detected 3 COMs in the ALMA fields in N83: CH$_3$OH, CH$_3$CN, and CH$_3$OCH$_3$. Methanol has been detected toward all the sources, and CH$_3$CN toward N\,83A--mm A and B only.  CH$_3$OCH$_3$ is only present in the spectrum N\,83A--mm B, making it the fifth source in the LMC (and the Magellanic Clouds) with the detection of this molecule.  

In addition to COMs, we detected simple 
S-bearing species (SO, $^{33}$SO, SO$_2$, OCS, H$_2$CS, CS, C$^{33}$S),  
N-bearing species (HC$^{15}$N, H$^{13}$CN, HNCO), one  
deuterated molecule (HDCO), 
as well as H$^{13}$CO$^{+}$, SiO, and CH$_2$CO.
We also report a tentative detection of HDO, HDCS, and $^{34}$SO$_2$. N\,83A--mm B is the most chemically-rich source in our ALMA fields in N\,83. 

\begin{figure*}
\centering
\includegraphics[width=\textwidth]{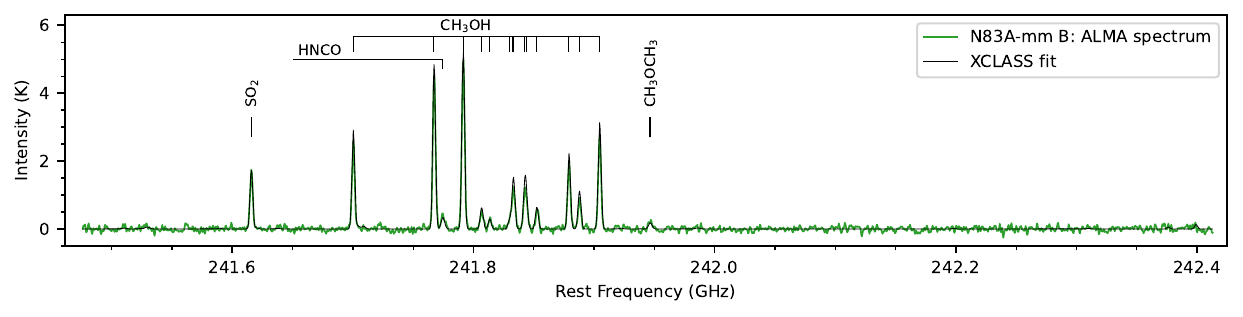}
\includegraphics[width=\textwidth]{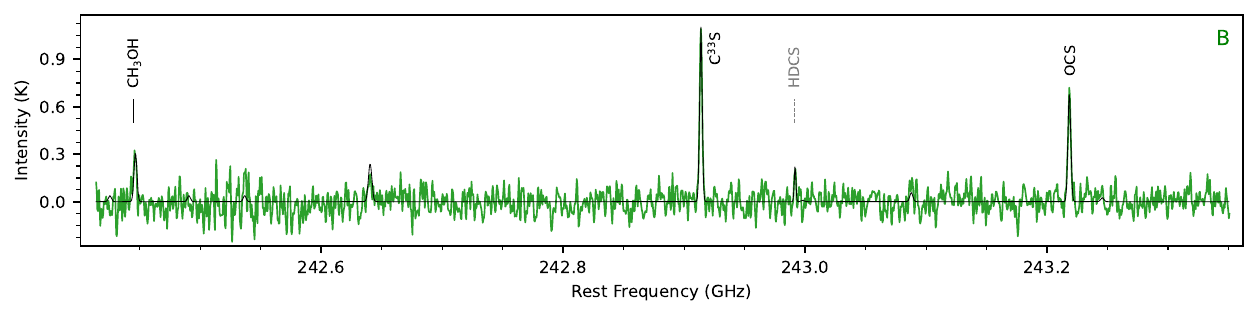}
\includegraphics[width=\textwidth]{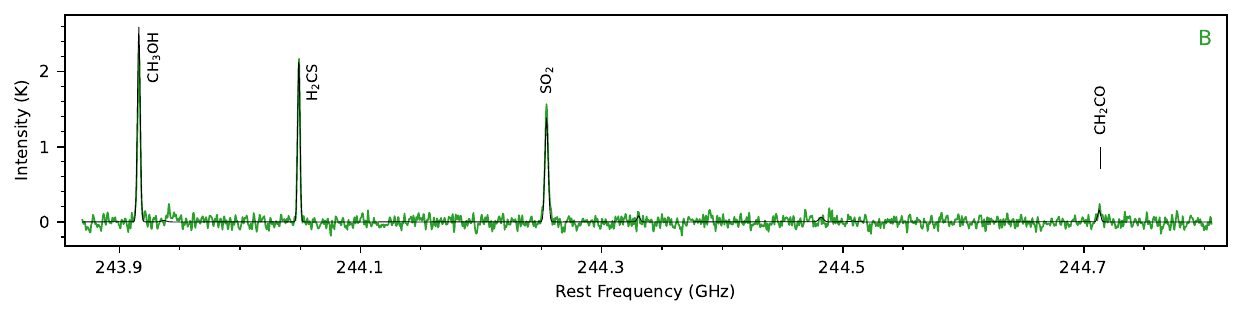}
\includegraphics[width=\textwidth]{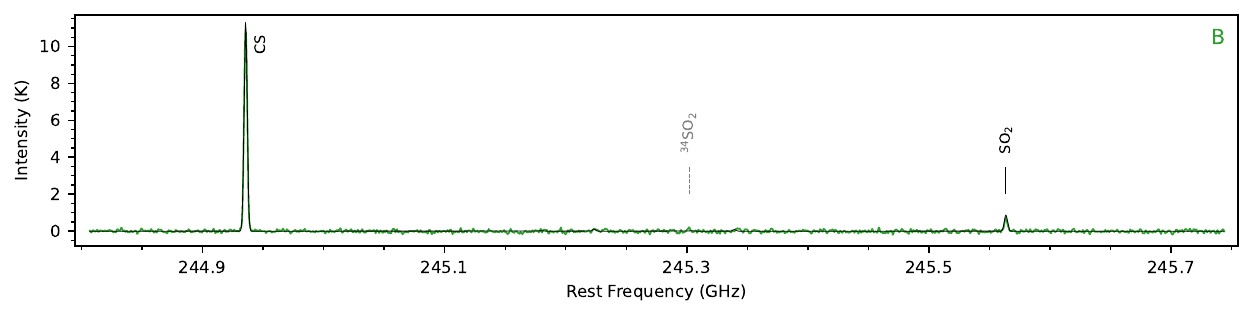}
\caption{ALMA Band 6 spectra of source N\,83A--mm\,B in the $\sim$242 GHz (top two panels) and $\sim$245 GHz (bottom two panels) spectral windows. The spectra were extracted as the mean over the area enclosed by the 50\% of the 1.2 mm continuum peak contour. Firm and tentative detections are indicated in black and gray, respectively. \label{f:specB_1}}
\end{figure*}

\begin{figure*}
\centering
\includegraphics[width=\textwidth]{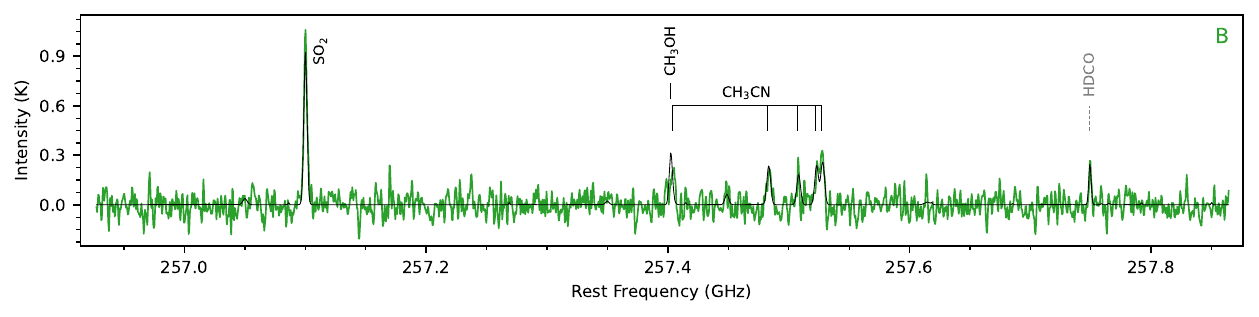}
\includegraphics[width=\textwidth]{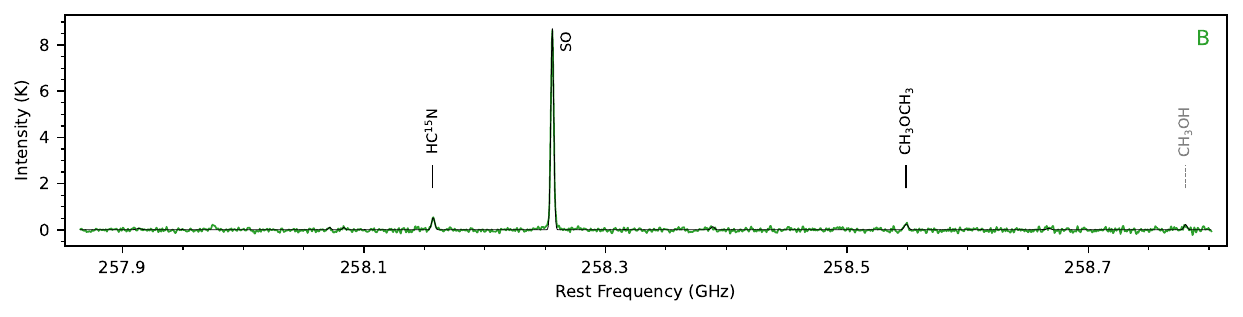}
\includegraphics[width=\textwidth]{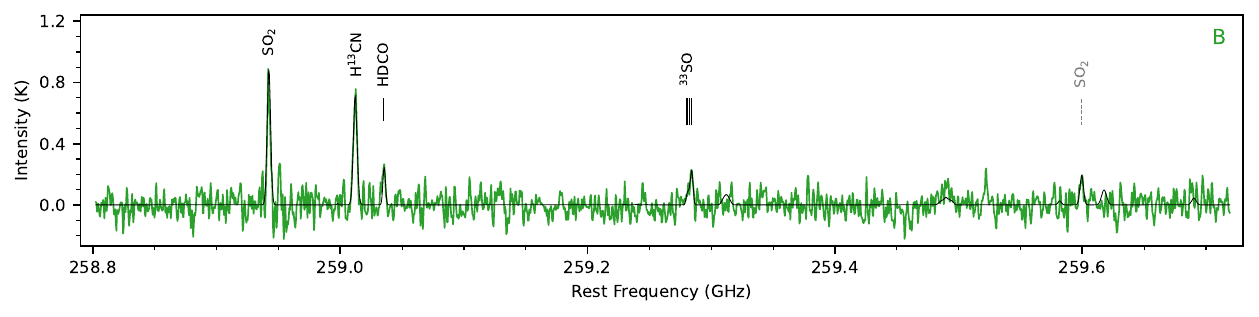}
\includegraphics[width=\textwidth]{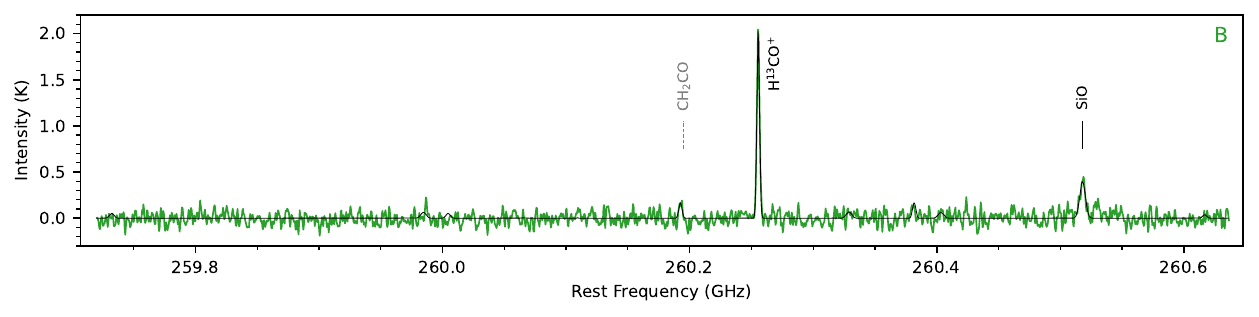}
\caption{The same as Figure~\ref{f:specB_1}, but for the $\sim$258 GHz (top two panels) and $\sim$260 GHz (bottom two panels) spectral windows. \label{f:specB_2}}
\end{figure*}

\subsection{Spatial Distribution of the Molecular Line Emission}
\label{s:mom}

\begin{figure*}
\centering
\includegraphics[width=0.32\textwidth]{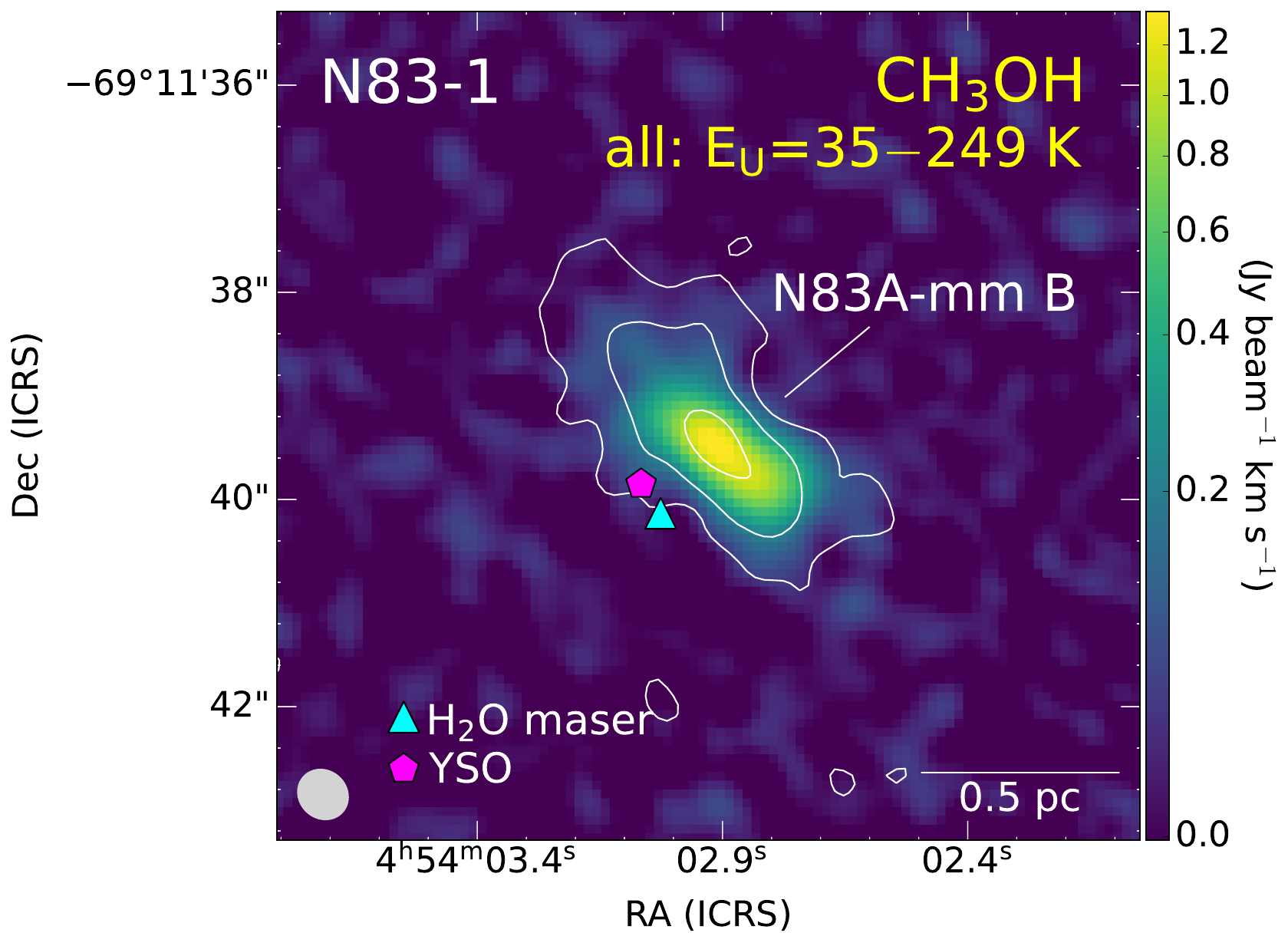}
\includegraphics[width=0.32\textwidth]{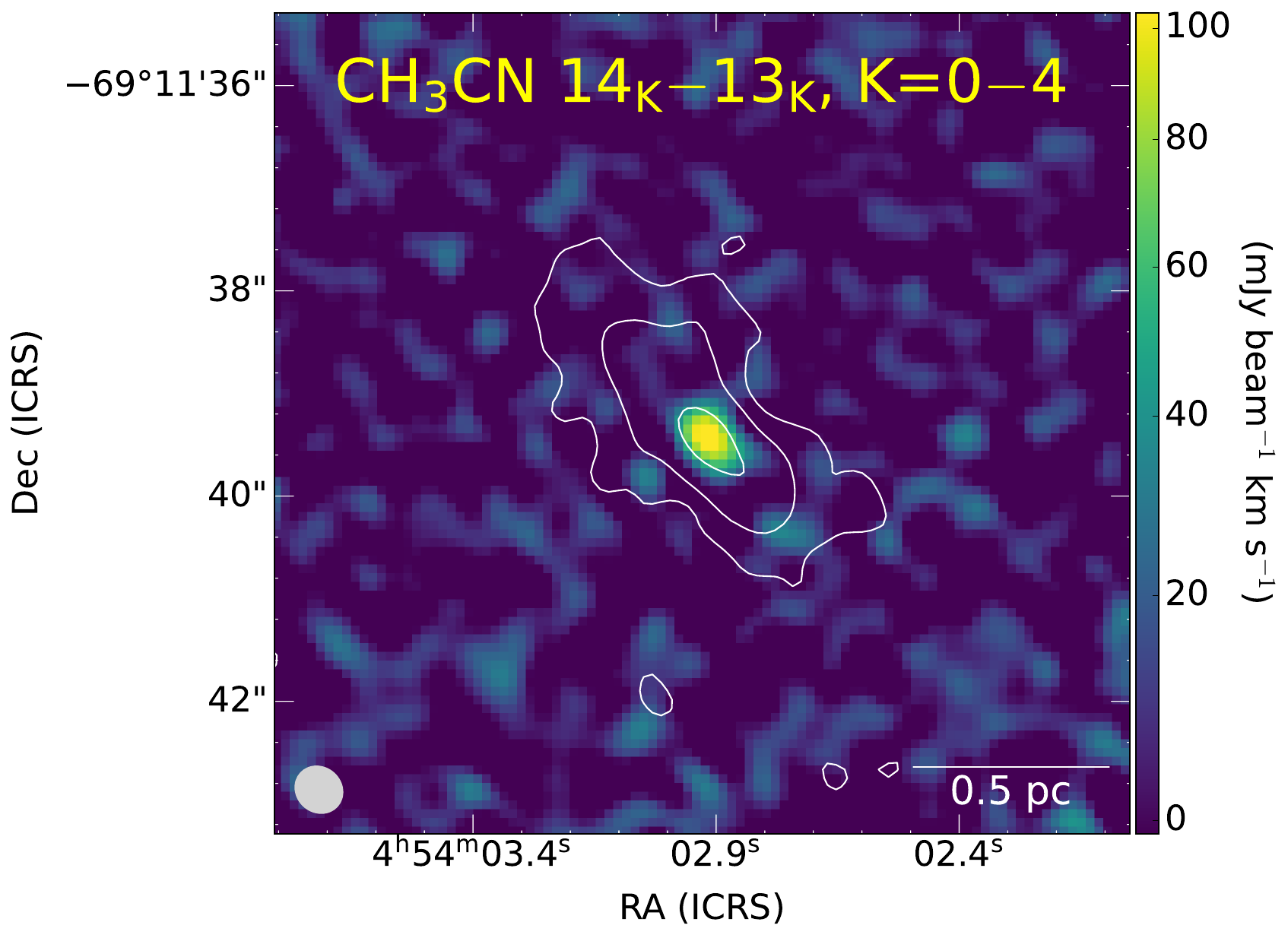}
\includegraphics[width=0.32\textwidth]{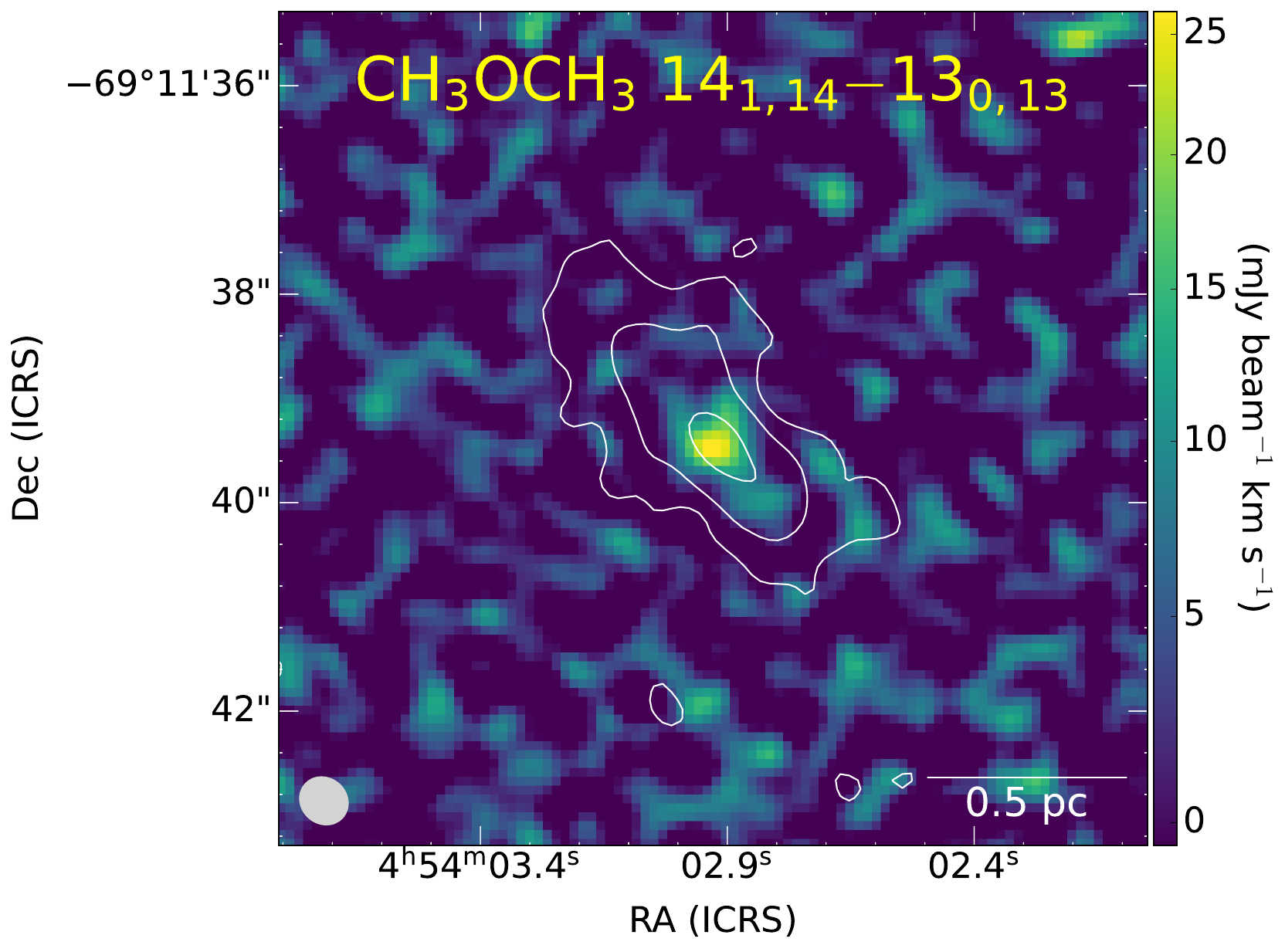}
\includegraphics[width=0.32\textwidth]{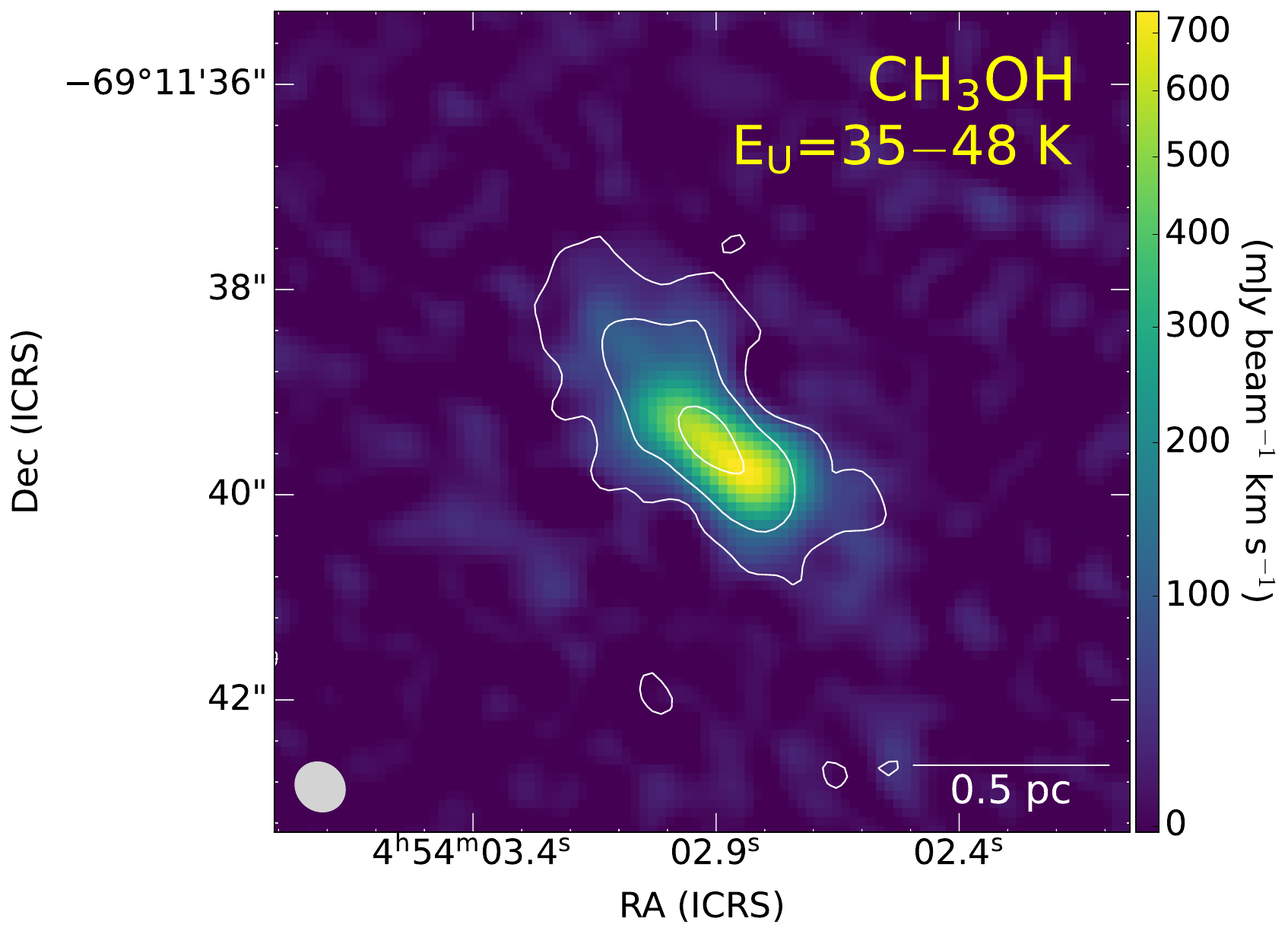}
\includegraphics[width=0.32\textwidth]{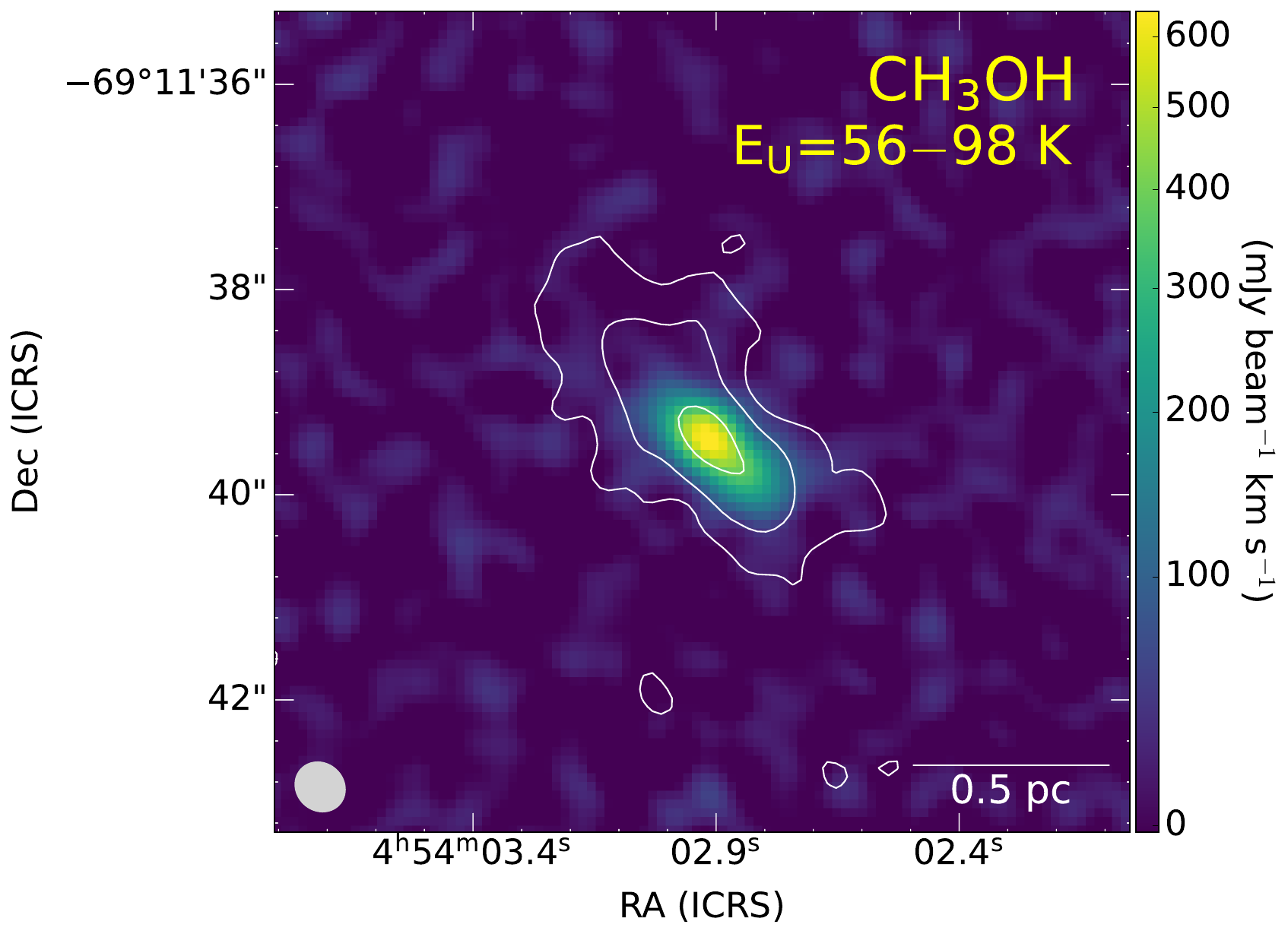}
\includegraphics[width=0.32\textwidth]{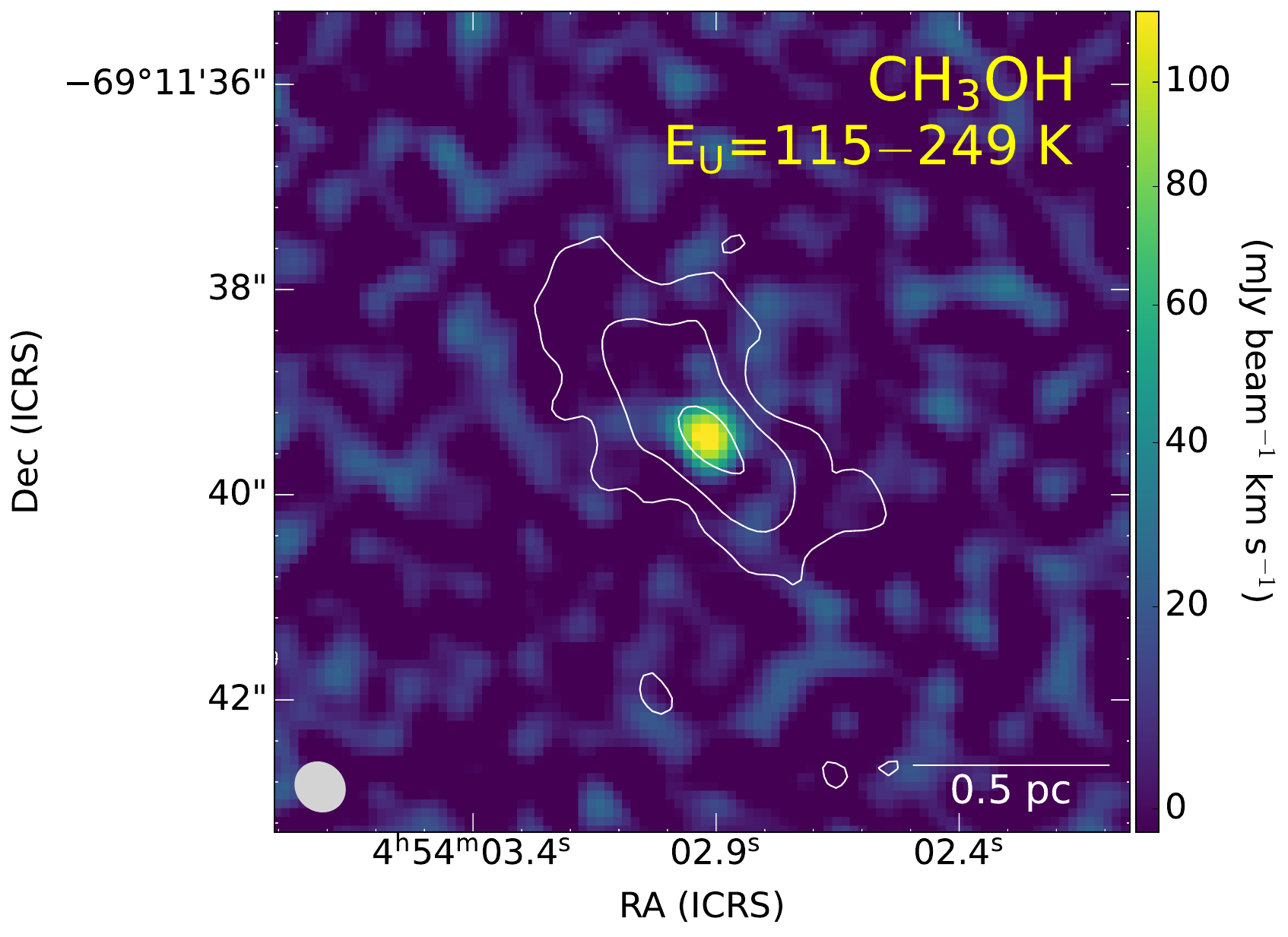}
\caption{The integrated intensity images of COMs detected toward N\,83A--mm\,B in the ALMA field N\,83--1.  The top panel shows the CH$_3$OH image integrated over all CH$_3$OH transitions detected in the $\sim$242~GHz spectral window (left), the CH$_3$CN image integrated over the $K$=0--4 components of the CH$_3$CN 14$_K$--13$_K$ ladder (center), and the CH$_3$OCH$_3$ 14$_{1,14}$--13$_{0,13}$ image.  In the lower panel, we show the CH$_3$OH images integrated over the $\sim$242~GHz transitions with $E_{\rm U}$ in the 35--48 K (left), 56--98 K (center), and  115--249 K (right) range. The 1.2 mm continuum contours with contour levels as in Figure~\ref{f:N83B6cont} are overlaid in each image for reference. The position of the H$_2$O maser and the {\it Spitzer} YSO are shown in the CH$_3$OH image in the upper left panel.  The size of the ALMA synthesized beam is indicated in the lower left corner of each image.
\label{f:N831mom0COMs}}
\end{figure*}

\begin{figure*}
\centering
\includegraphics[width=0.32\textwidth]{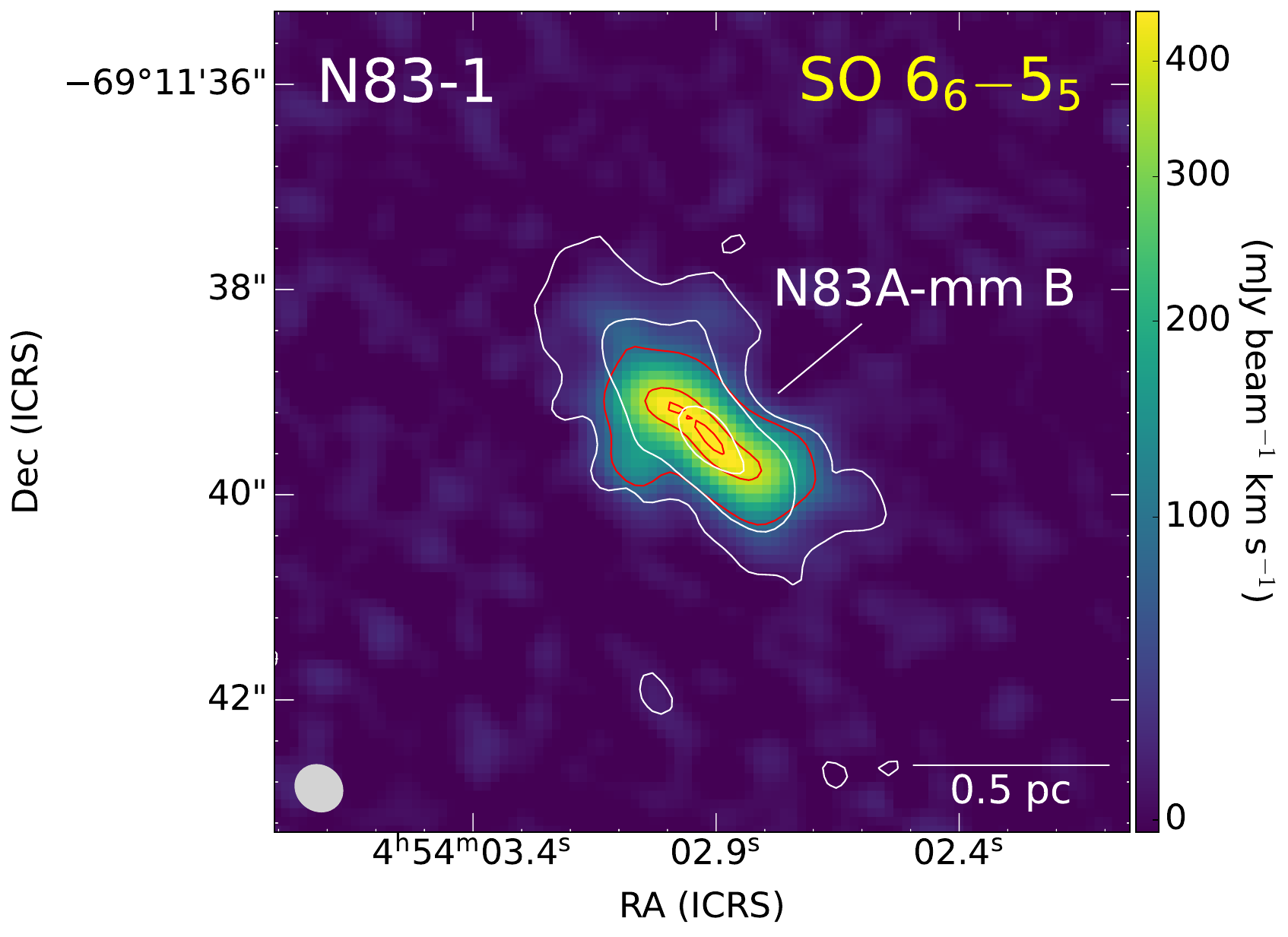}
\includegraphics[width=0.32\textwidth]{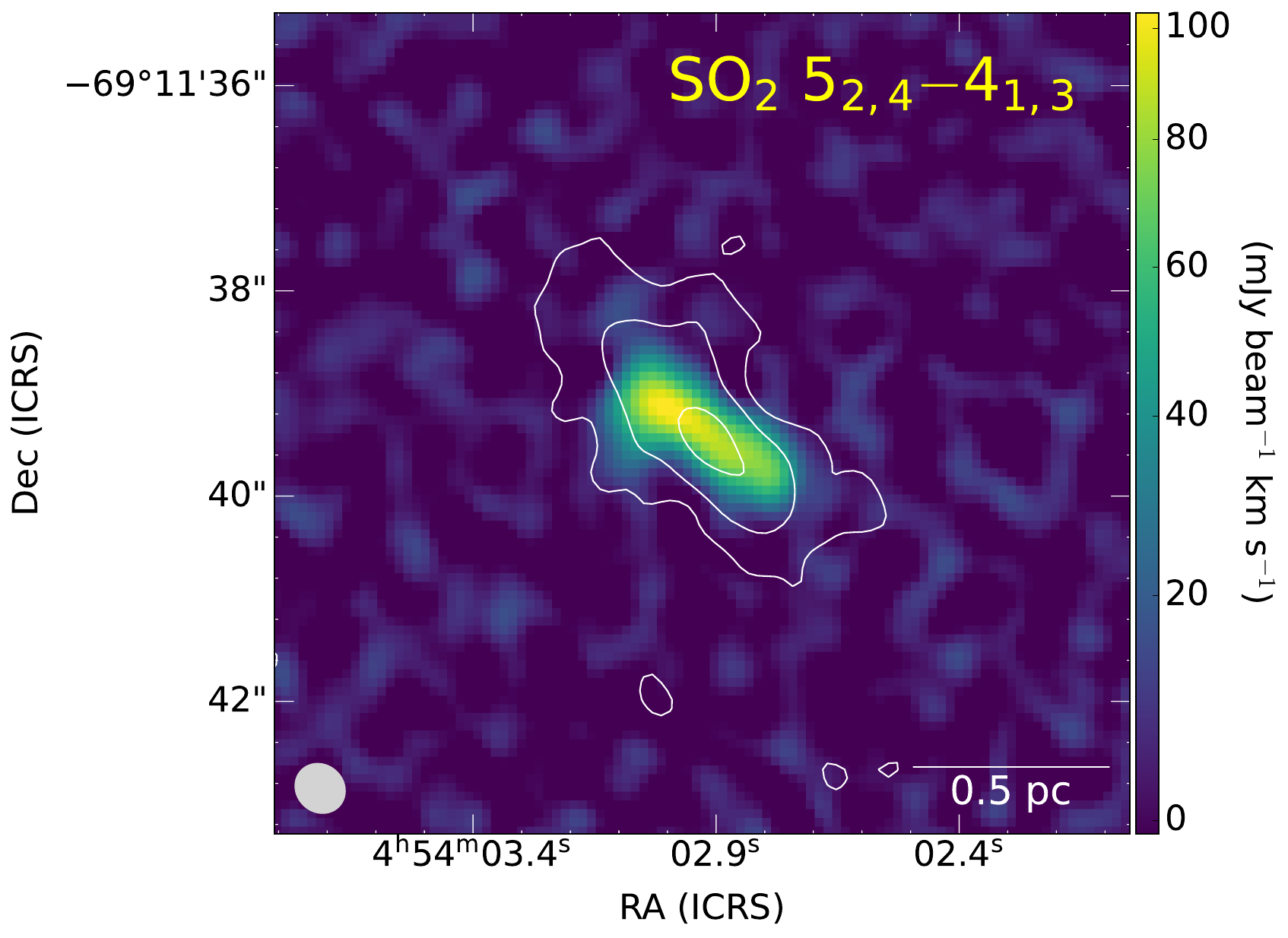}
\includegraphics[width=0.32\textwidth]{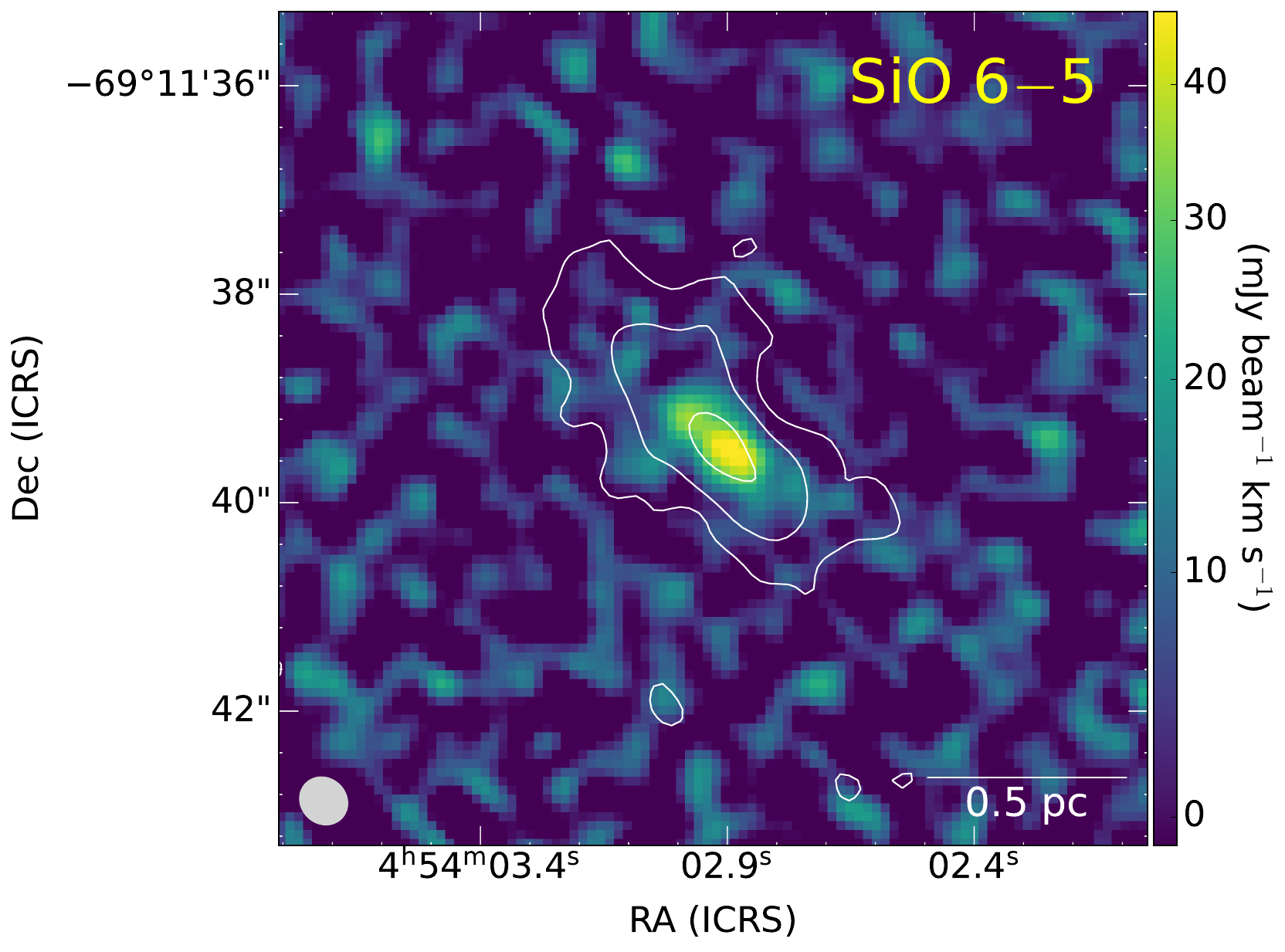}
\includegraphics[width=0.32\textwidth]{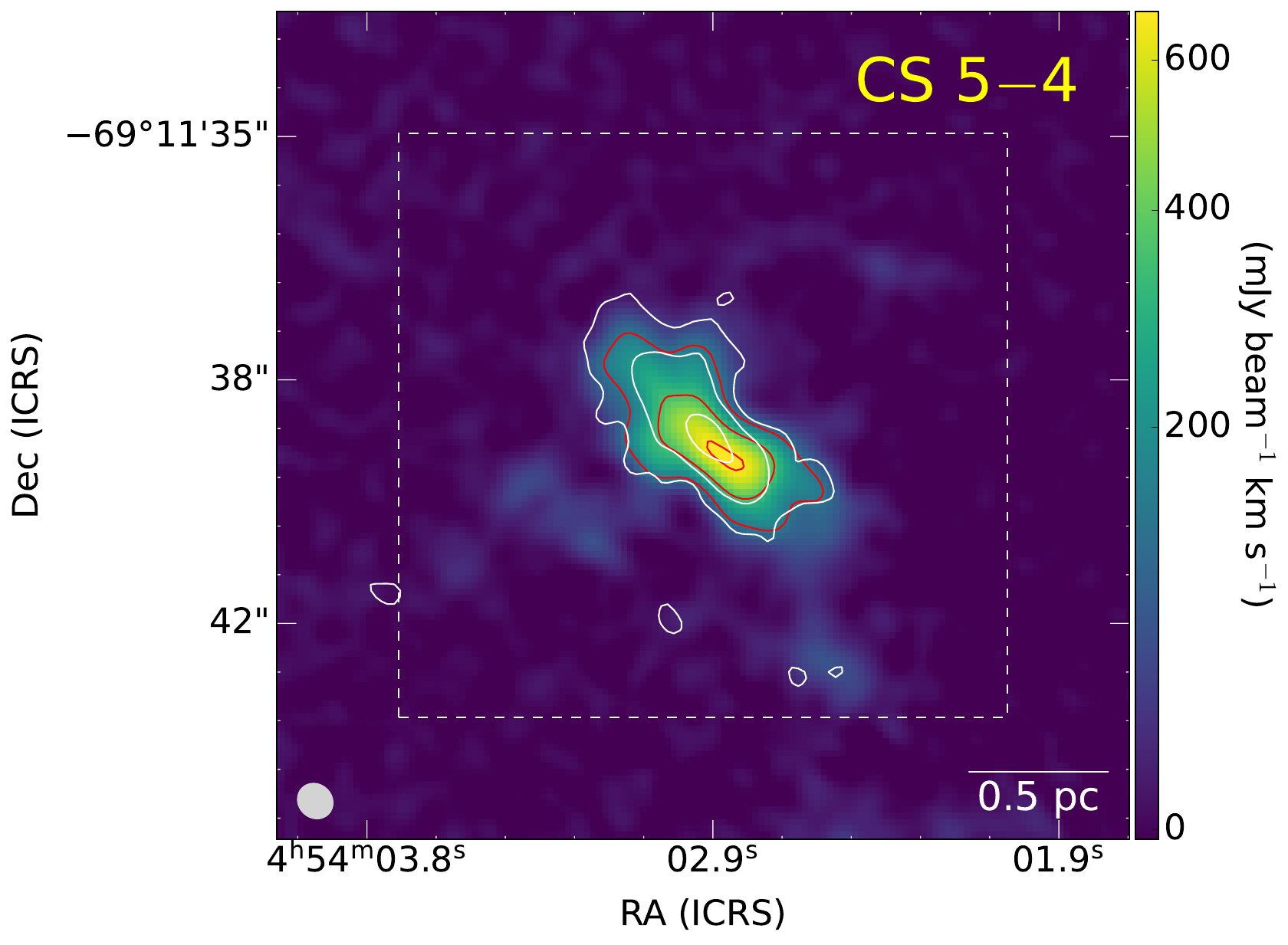}
\includegraphics[width=0.32\textwidth]{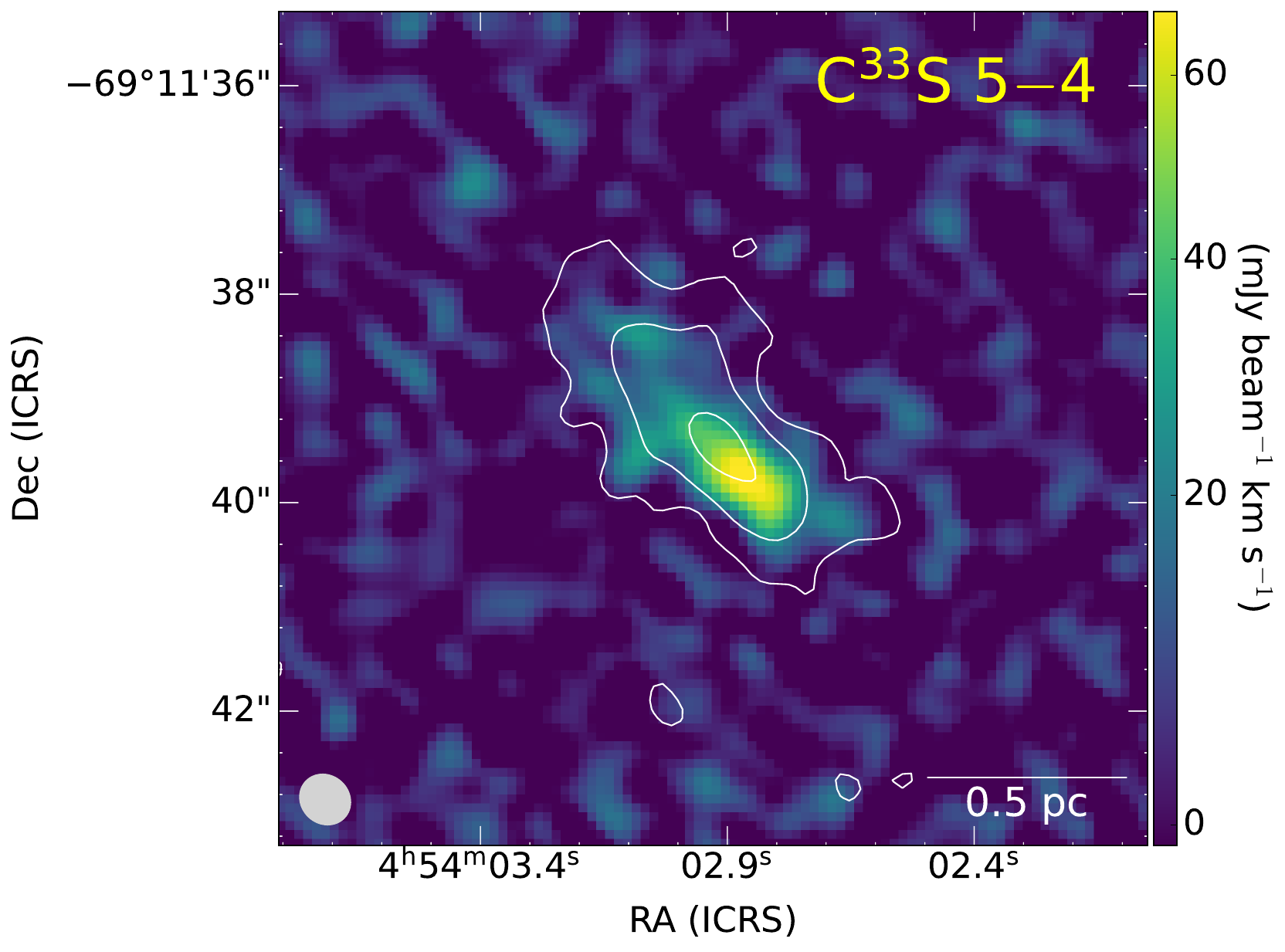}
\includegraphics[width=0.32\textwidth]{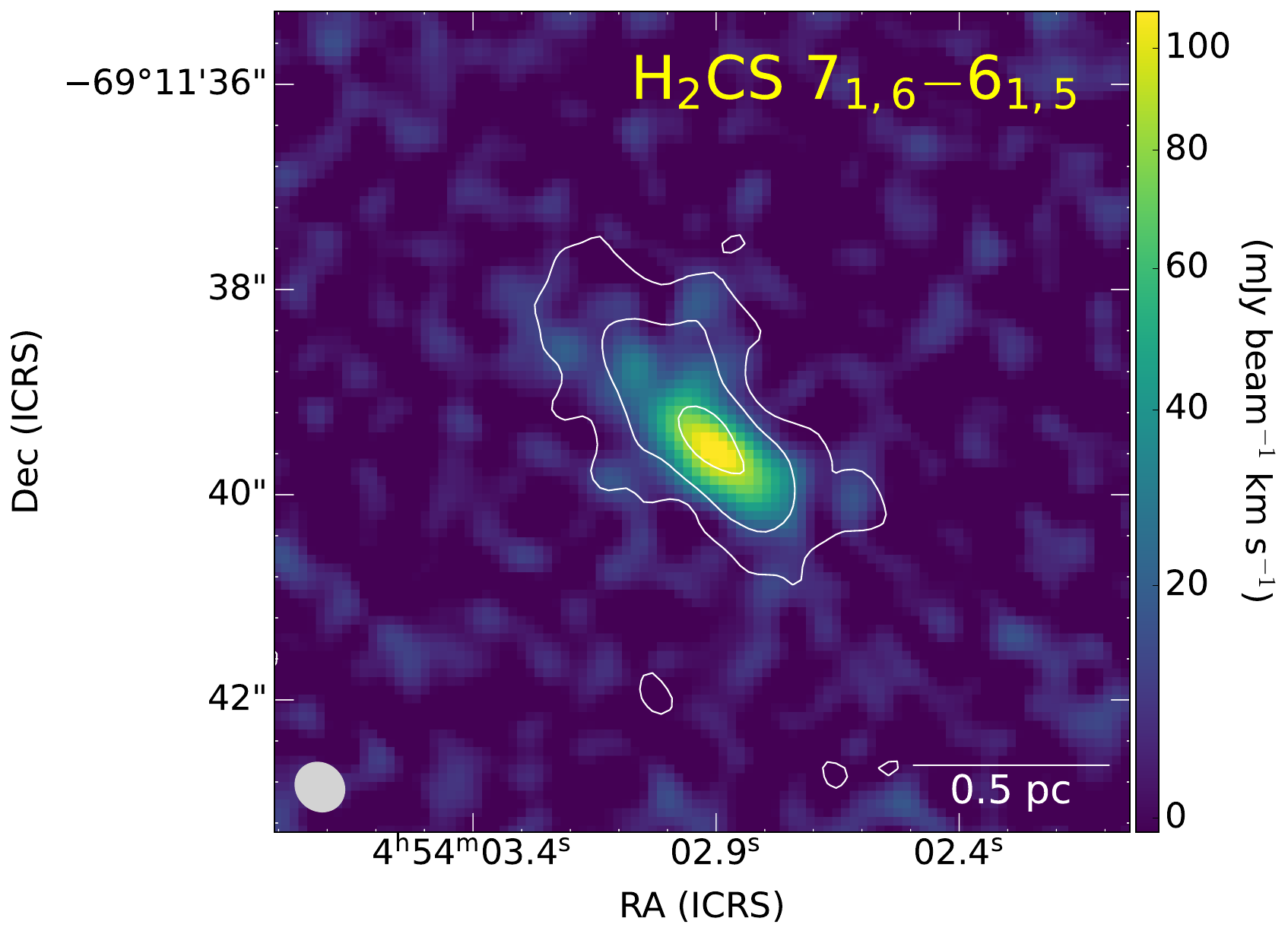}
\includegraphics[width=0.32\textwidth]{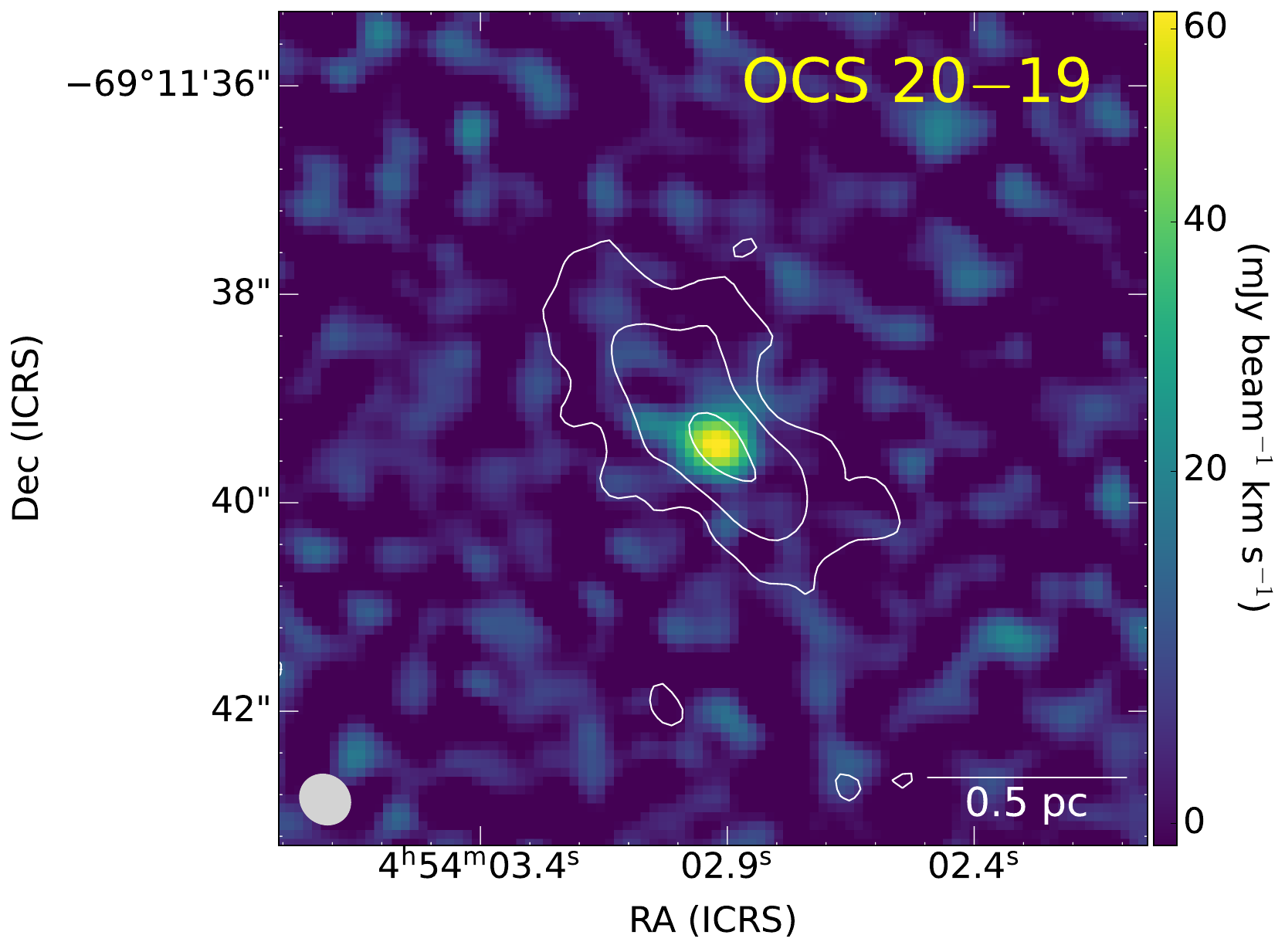}
\includegraphics[width=0.32\textwidth]{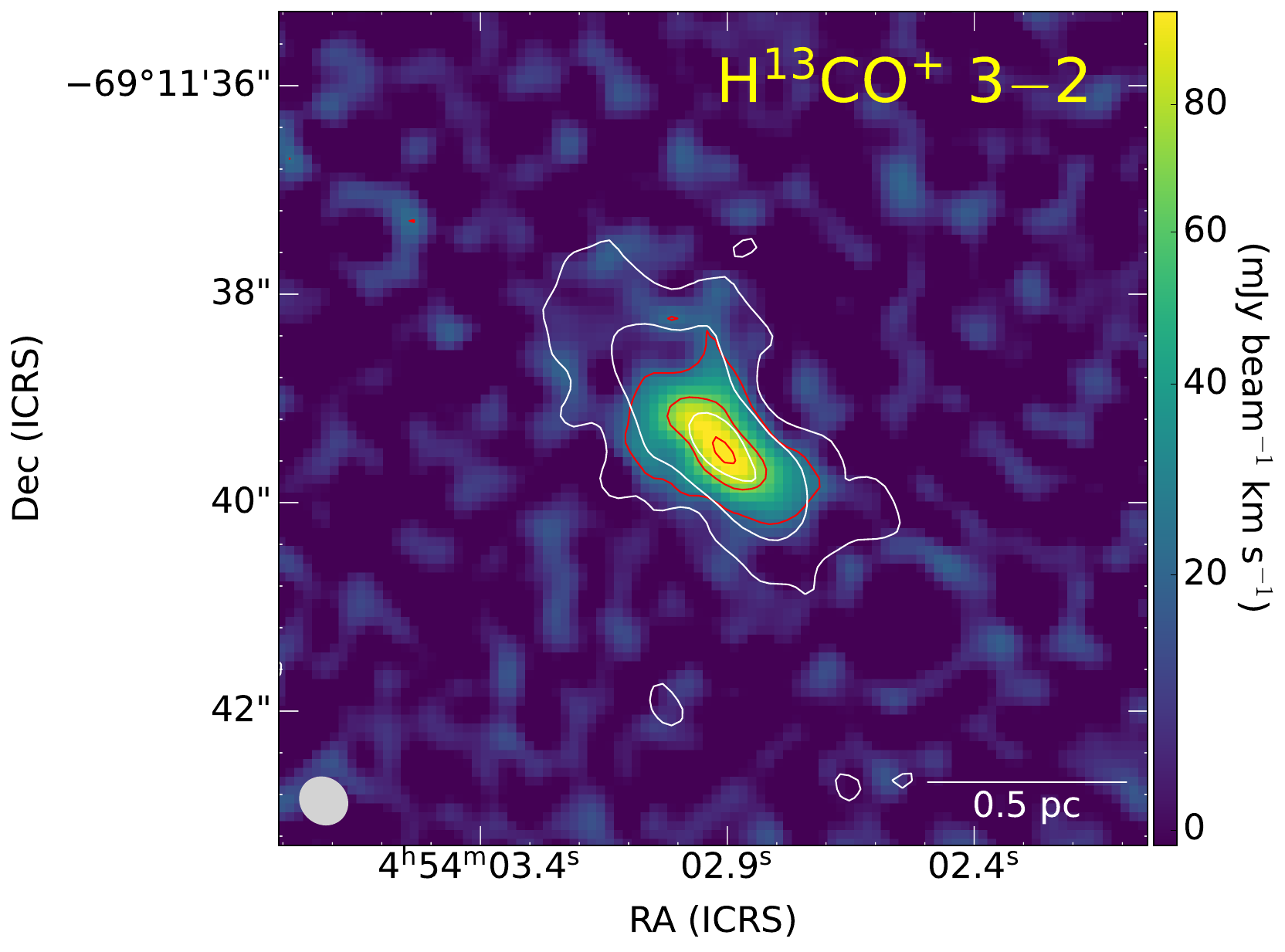}
\includegraphics[width=0.32\textwidth]{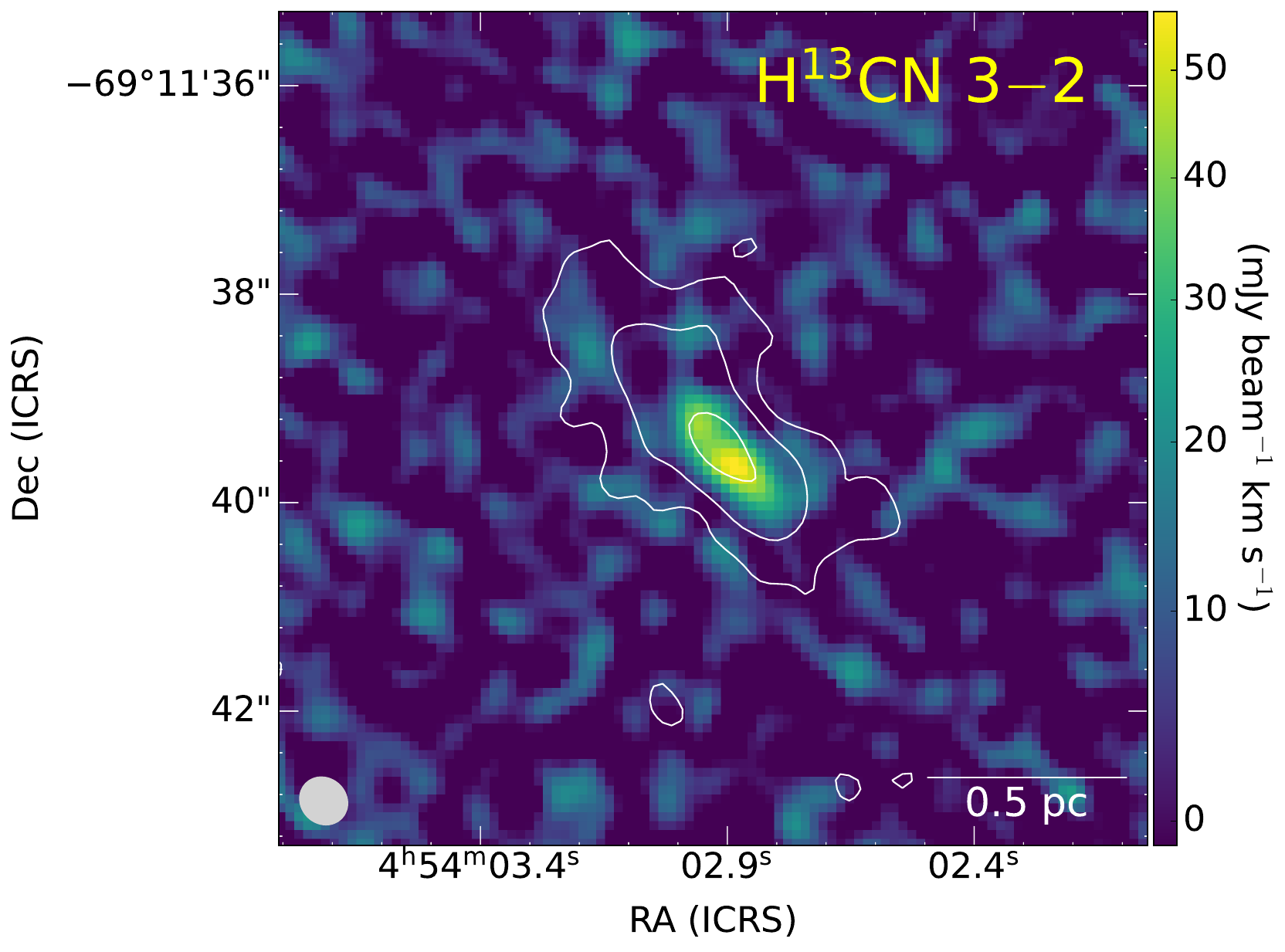}
\includegraphics[width=0.32\textwidth]{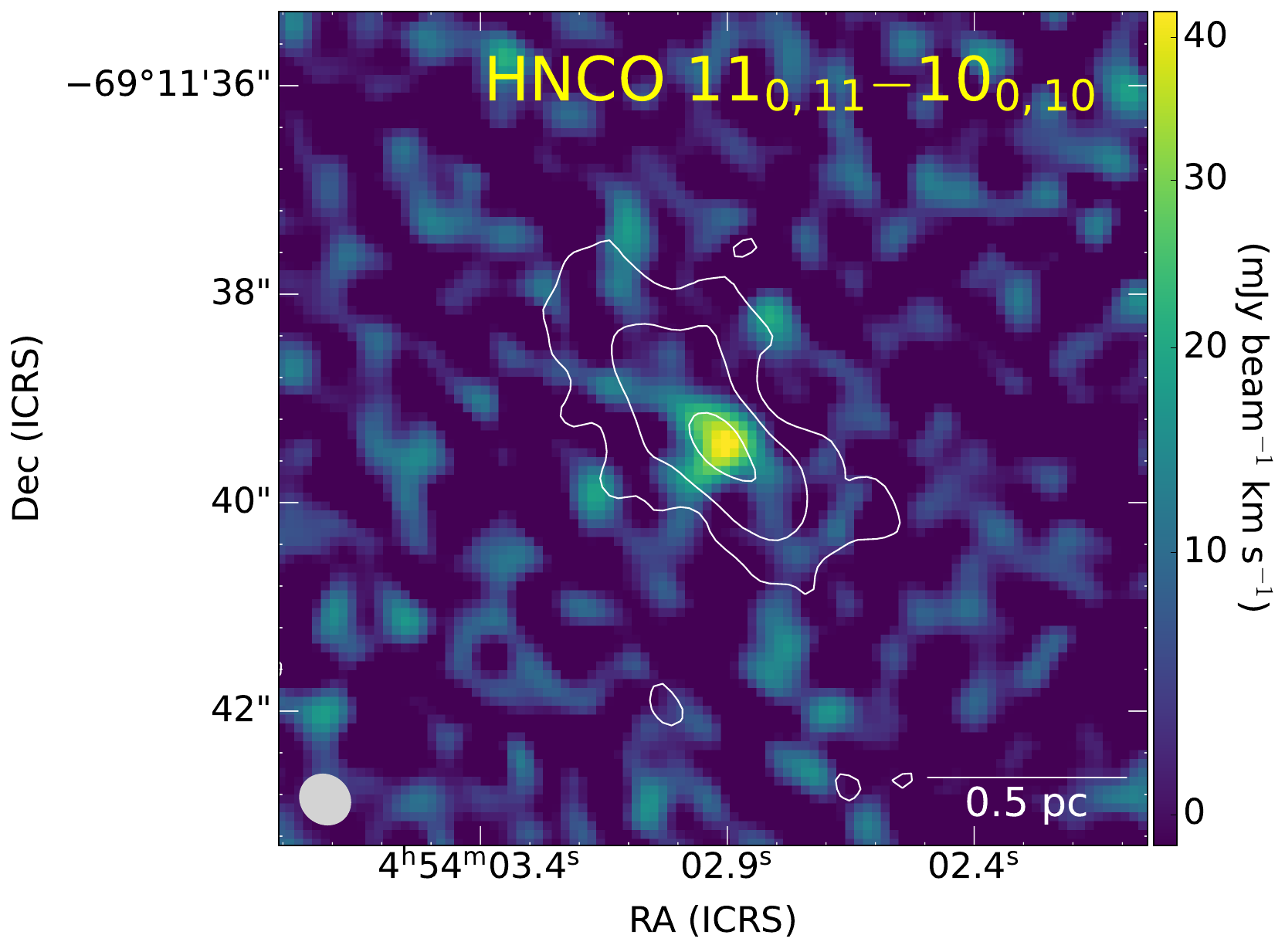}
\includegraphics[width=0.32\textwidth]{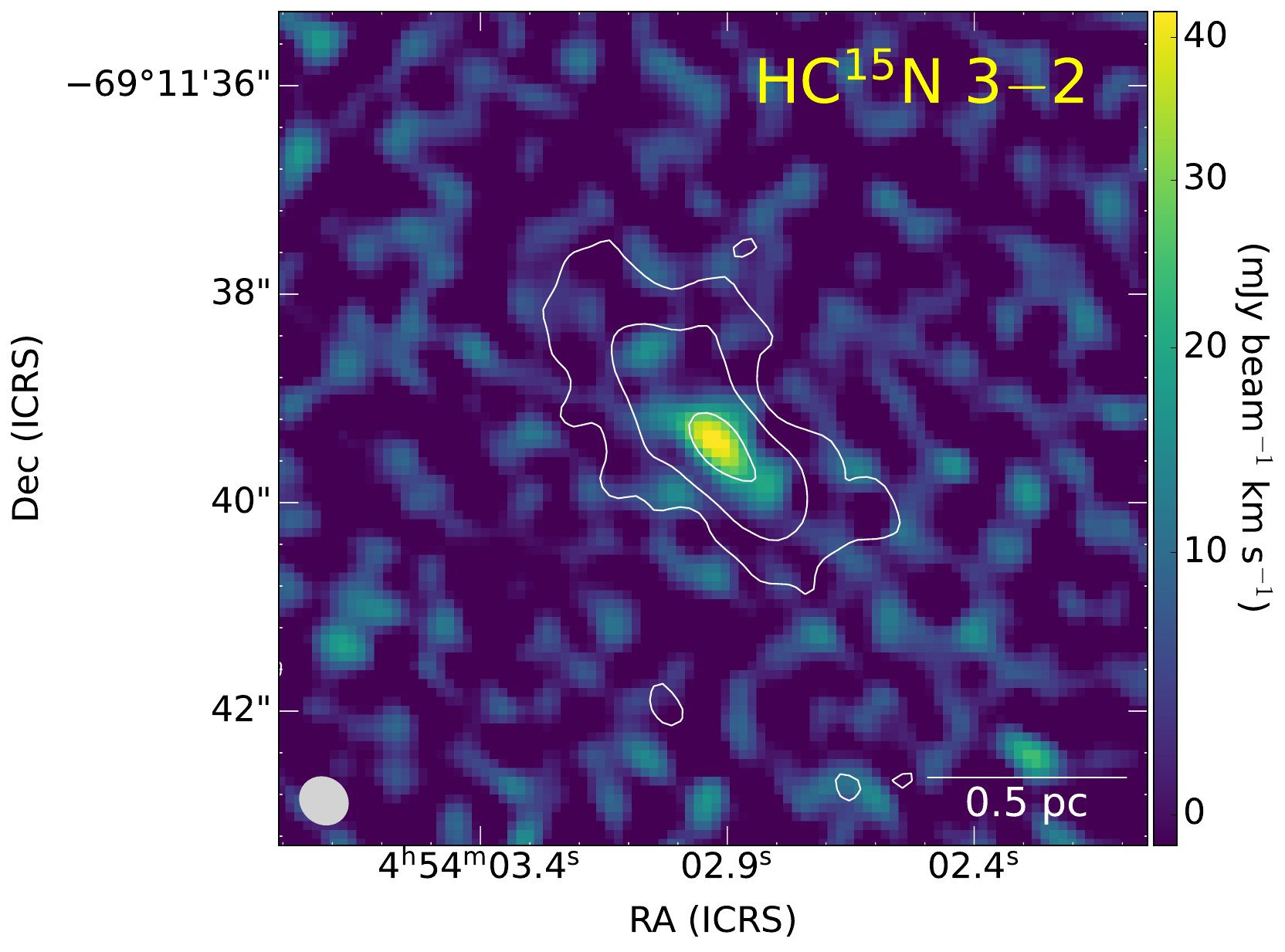}
\caption{Same as Figure~\ref{f:N831mom0COMs}, but for simple molecular species detected toward N\,83A--mm\,B (from left to right, top to bottom): 
SO 6$_6$--5$_5$ with red contours corresponding to (20, 80, 97)\% $\times$ the SO integrated intensity peak of 477 mJy~beam$^{-1}$~km~s$^{-1}$,
SO$_2$ 5$_{2, 4}$--4$_{1, 3}$,
SiO 6--5,
CS 5--4 with red contours corresponding to (20, 50, 95)\% $\times$ the CS integrated intensity peak of 697 mJy~beam$^{-1}$~km~s$^{-1}$,
C$^{33}$S 5--4,
H$_2$CS 7$_{1, 6}$--6$_{1, 5}$,
OCS 20--19,
H$^{13}$CO$^{+}$ 3--2 with red contours corresponding to (20, 60, 95)\% $\times$ the H$^{13}$CO$^{+}$ integrated intensity of 108 mJy~beam$^{-1}$~km~s$^{-1}$,
H$^{13}$CN 3--2,
HNCO 11$_{0, 11}$--10$_{0, 10}$, and
HC$^{15}$N 3--2.
\label{f:N831mom0simple}}
\end{figure*}

\begin{figure*}
\centering
\includegraphics[width=0.32\textwidth]{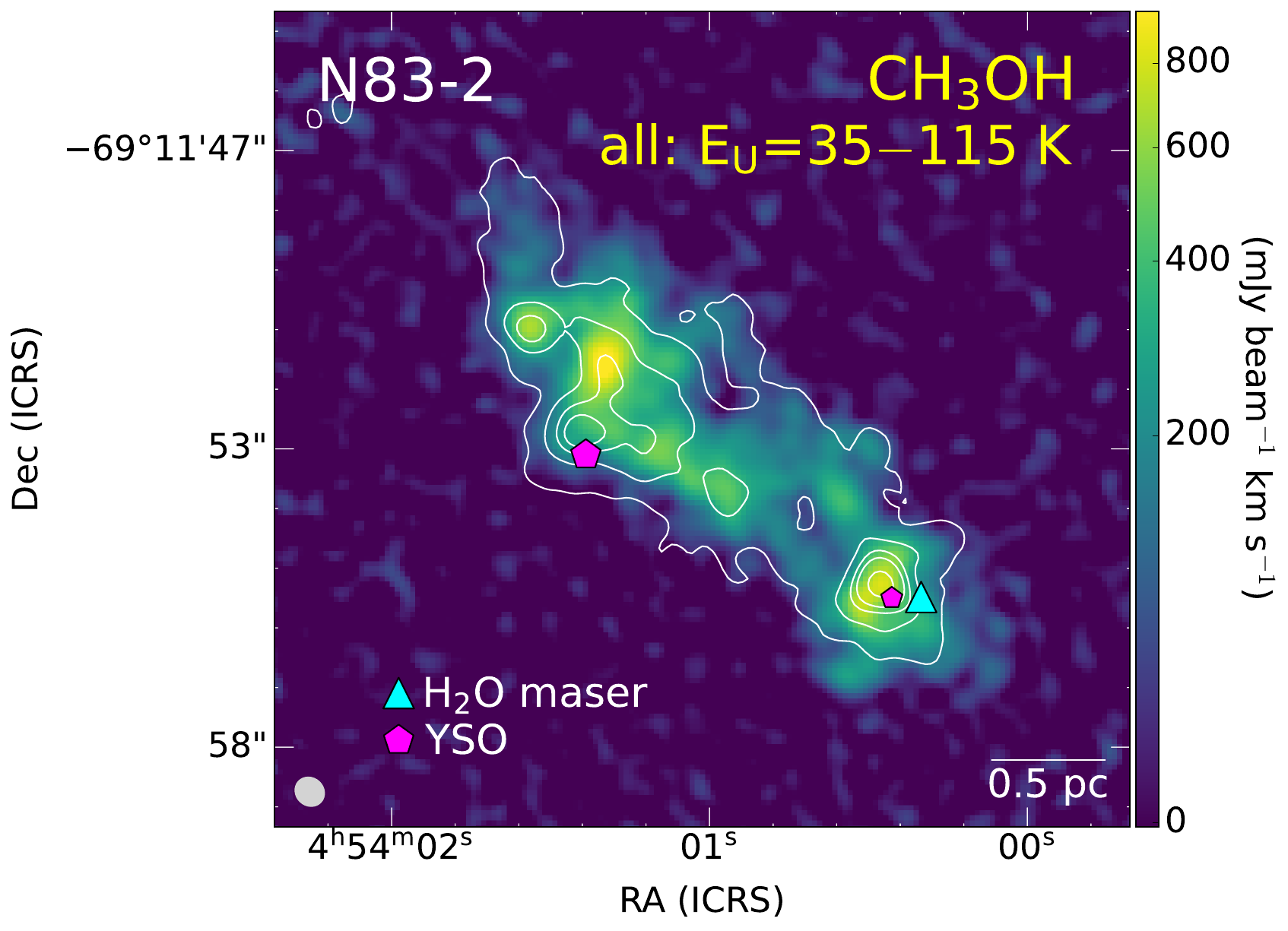}
\includegraphics[width=0.32\textwidth]{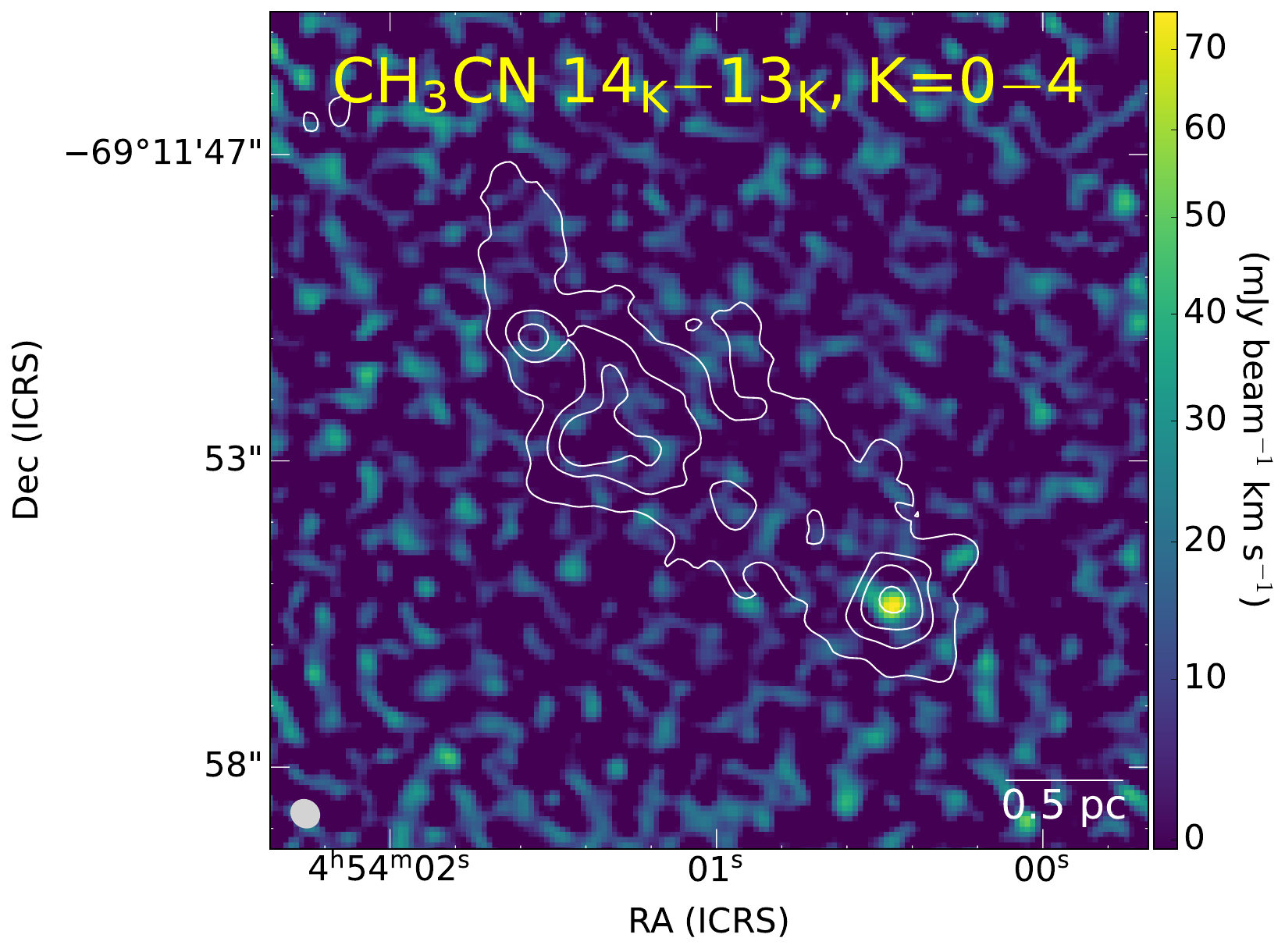}\\
\includegraphics[width=0.32\textwidth]{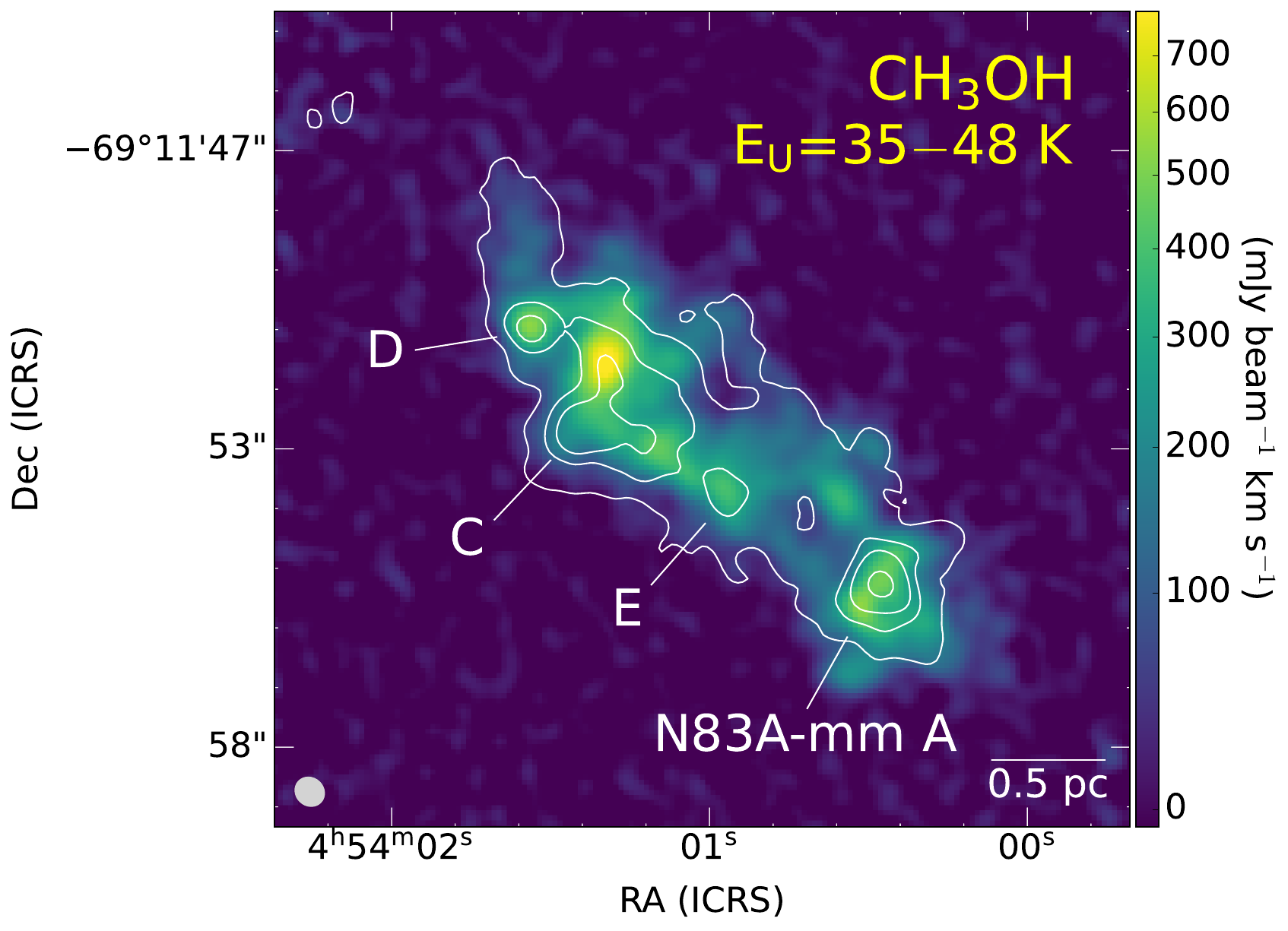}
\includegraphics[width=0.32\textwidth]{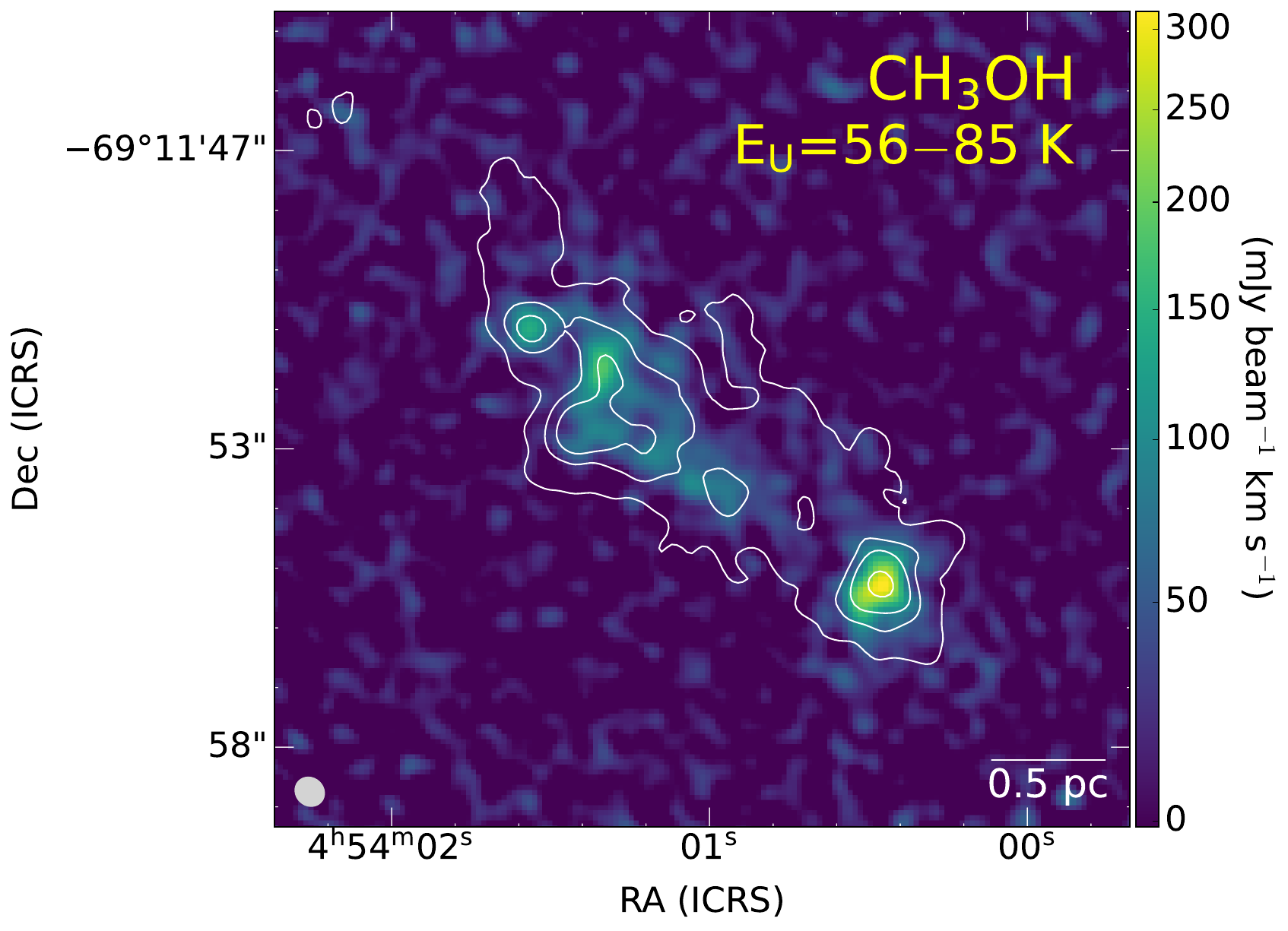}
\includegraphics[width=0.32\textwidth]{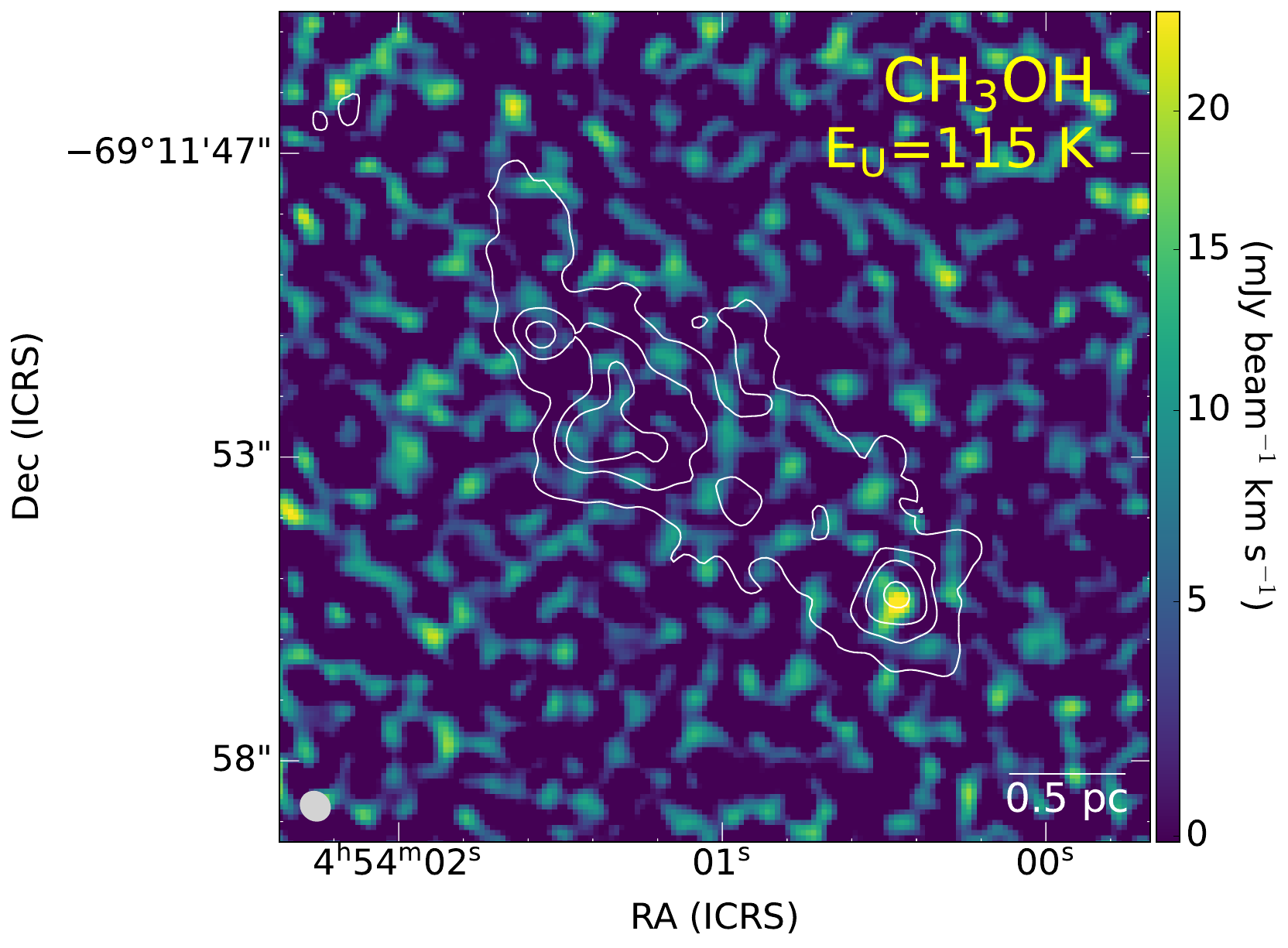}
\caption{The integrated intensity images of COMs detected toward N\,83A--mm\,A, C, D, and E in the ALMA field N\,83--2.  In the top panel, we show the CH$_3$OH image integrated over all CH$_3$OH transitions detected in the $\sim$242~GHz spectral window (left) and the CH$_3$CN image integrated over the $K$=0--4 components of the CH$_3$CN 14$_K$--13$_K$ ladder (right).  The lower panel shows the CH$_3$OH images integrated over the $\sim$242~GHz transitions with $E_{\rm U}$=35--48 K (left),  56--85 K (center), and 115 K (right, see Table~\ref{t:detections}). The 1.2 mm continuum contours with contour levels of 3, 10, 20, 30, and 60$\sigma$ (the upper left panel) or 3, 10, 20, and 60$\sigma$ (the remaining panels) are overlaid in each image for reference (see Figure~\ref{f:N83B6cont}). The position of the H$_2$O maser and the {\it Spitzer} YSOs are shown in the CH$_3$OH image in the upper left panel.  The size of the ALMA synthesized beam is indicated in the lower left corner of each image. 
\label{f:N832mom0COMs}}
\end{figure*}

\begin{figure*}
\centering
\includegraphics[width=0.32\textwidth]{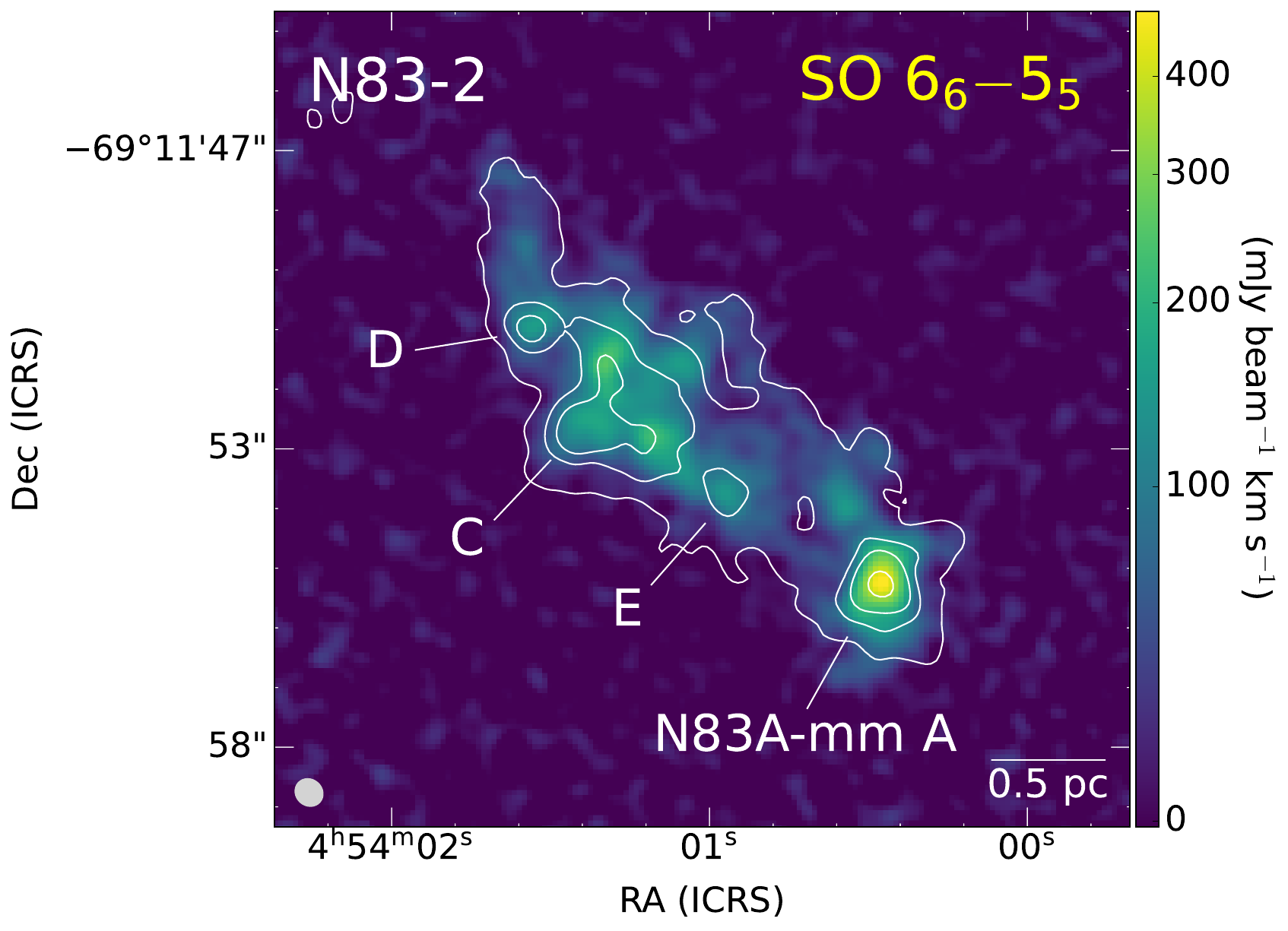}
\includegraphics[width=0.32\textwidth]{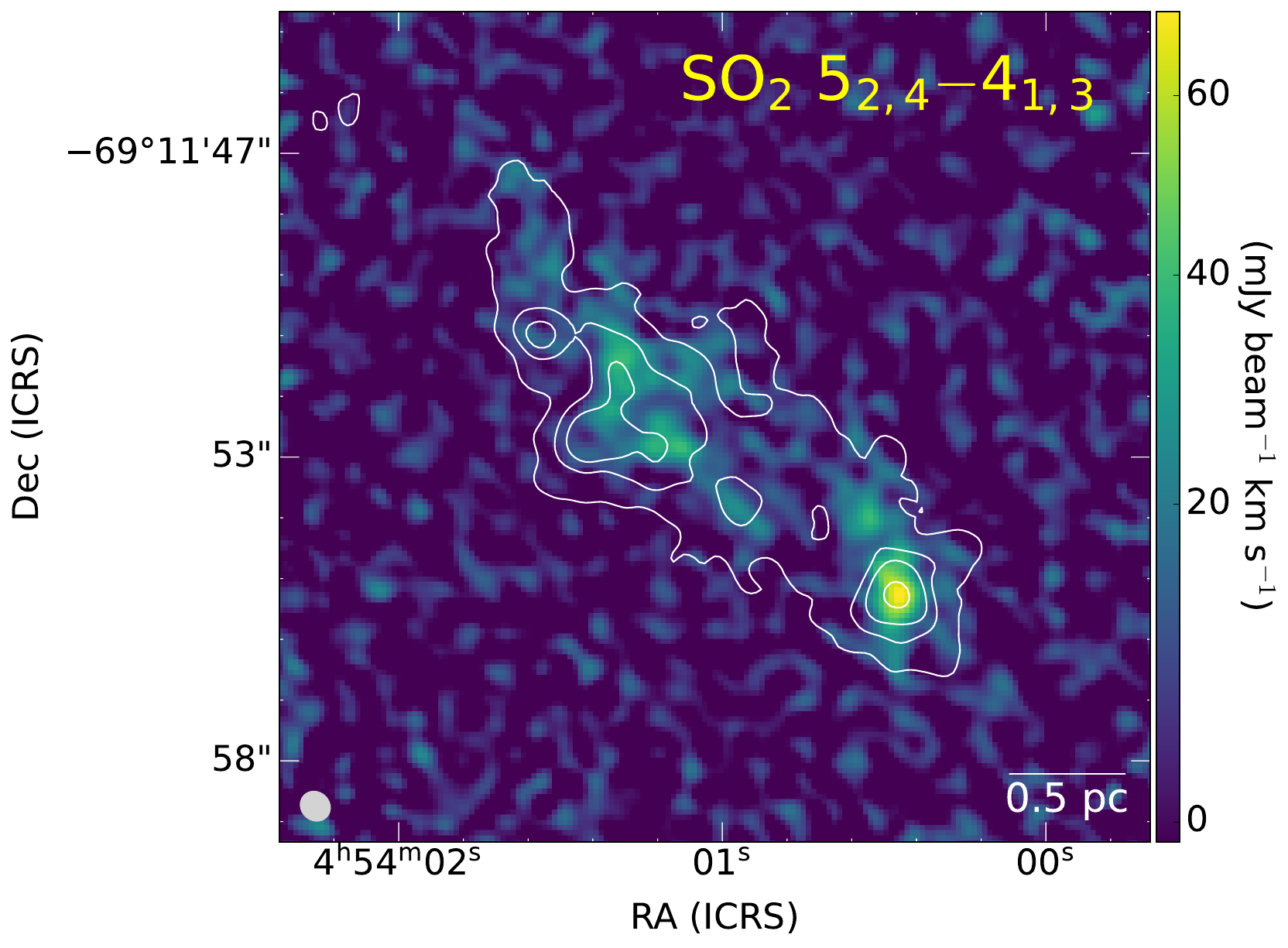}
\includegraphics[width=0.32\textwidth]{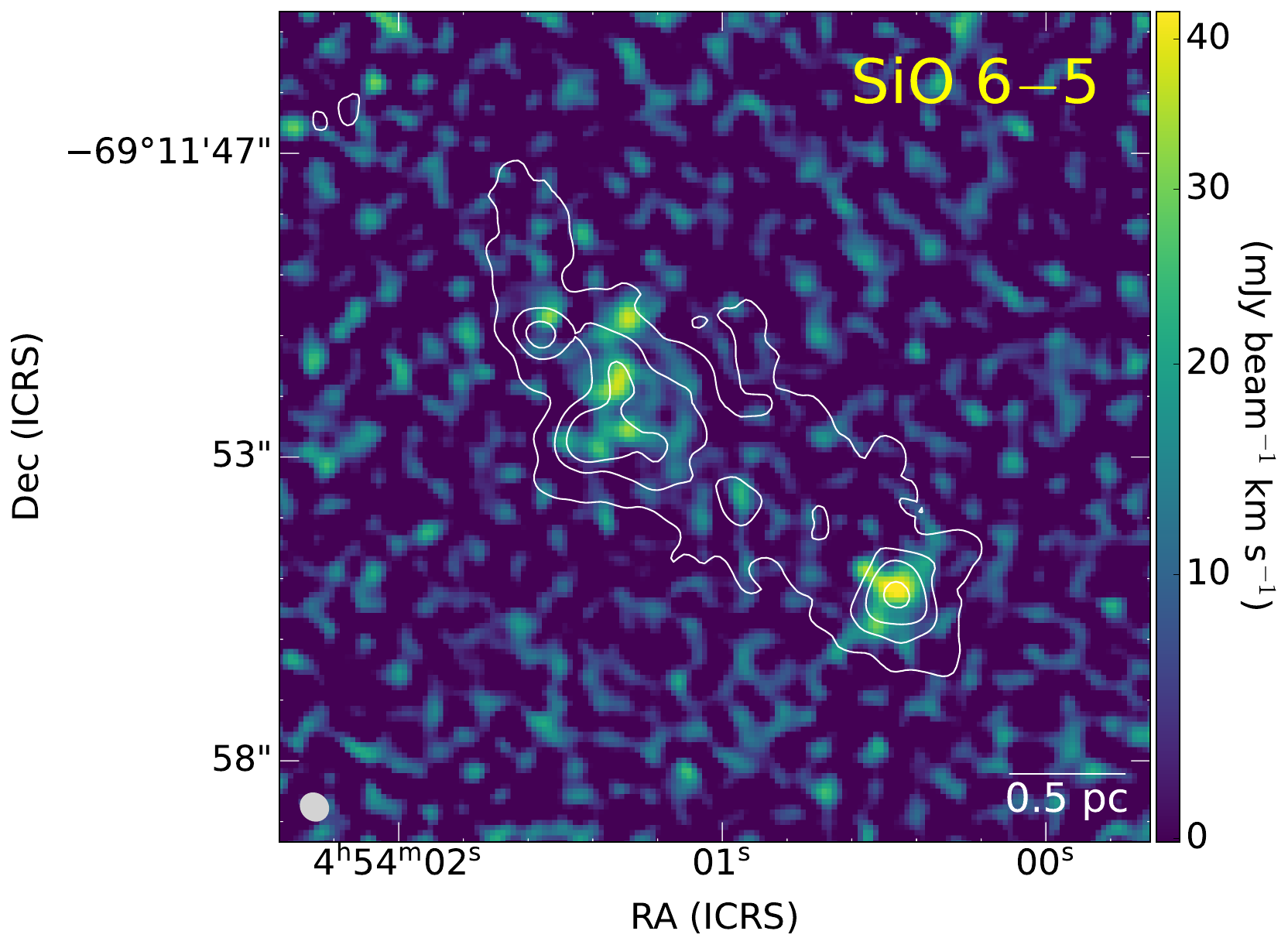}
\includegraphics[width=0.32\textwidth]{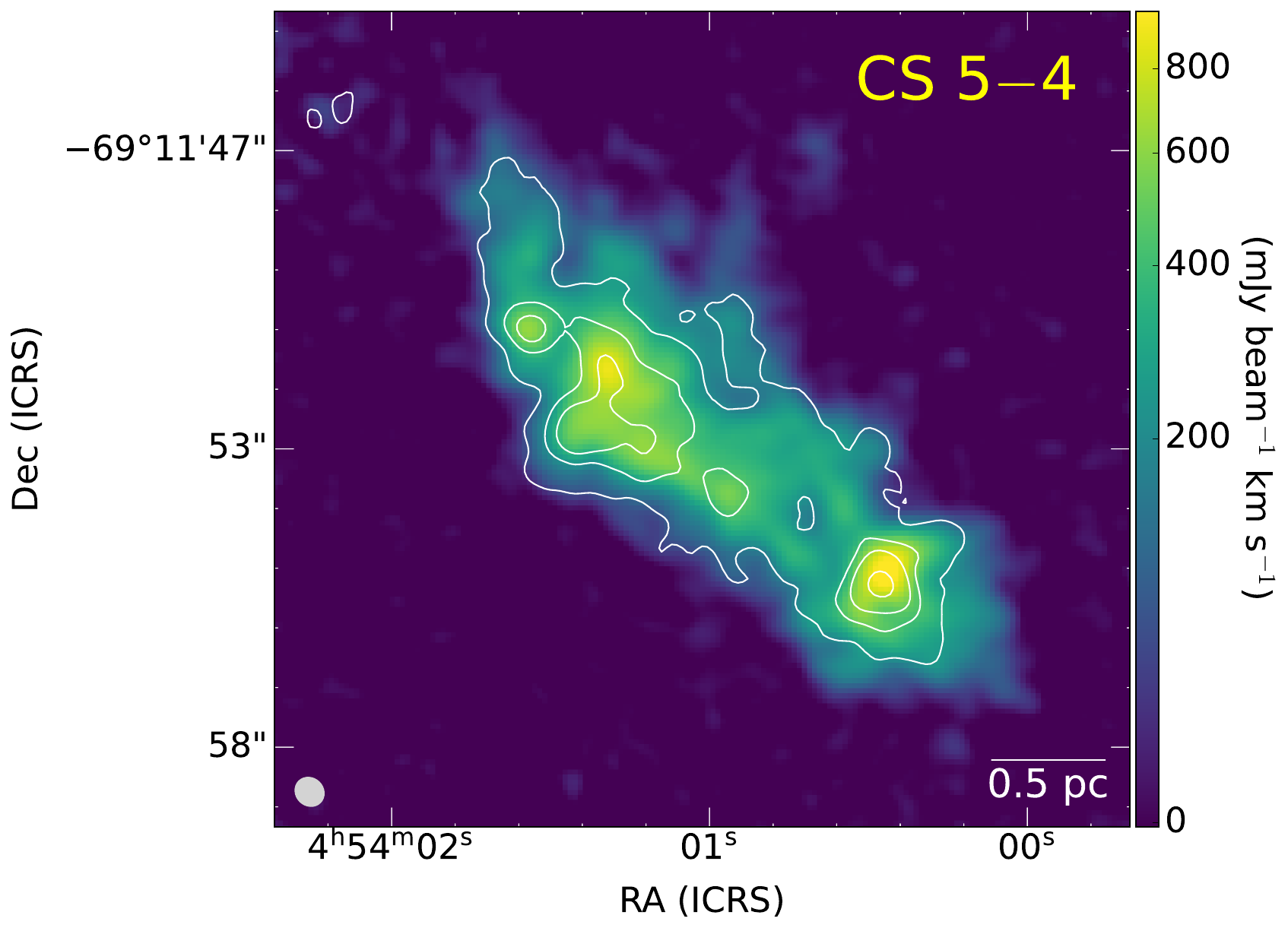}
\includegraphics[width=0.32\textwidth]{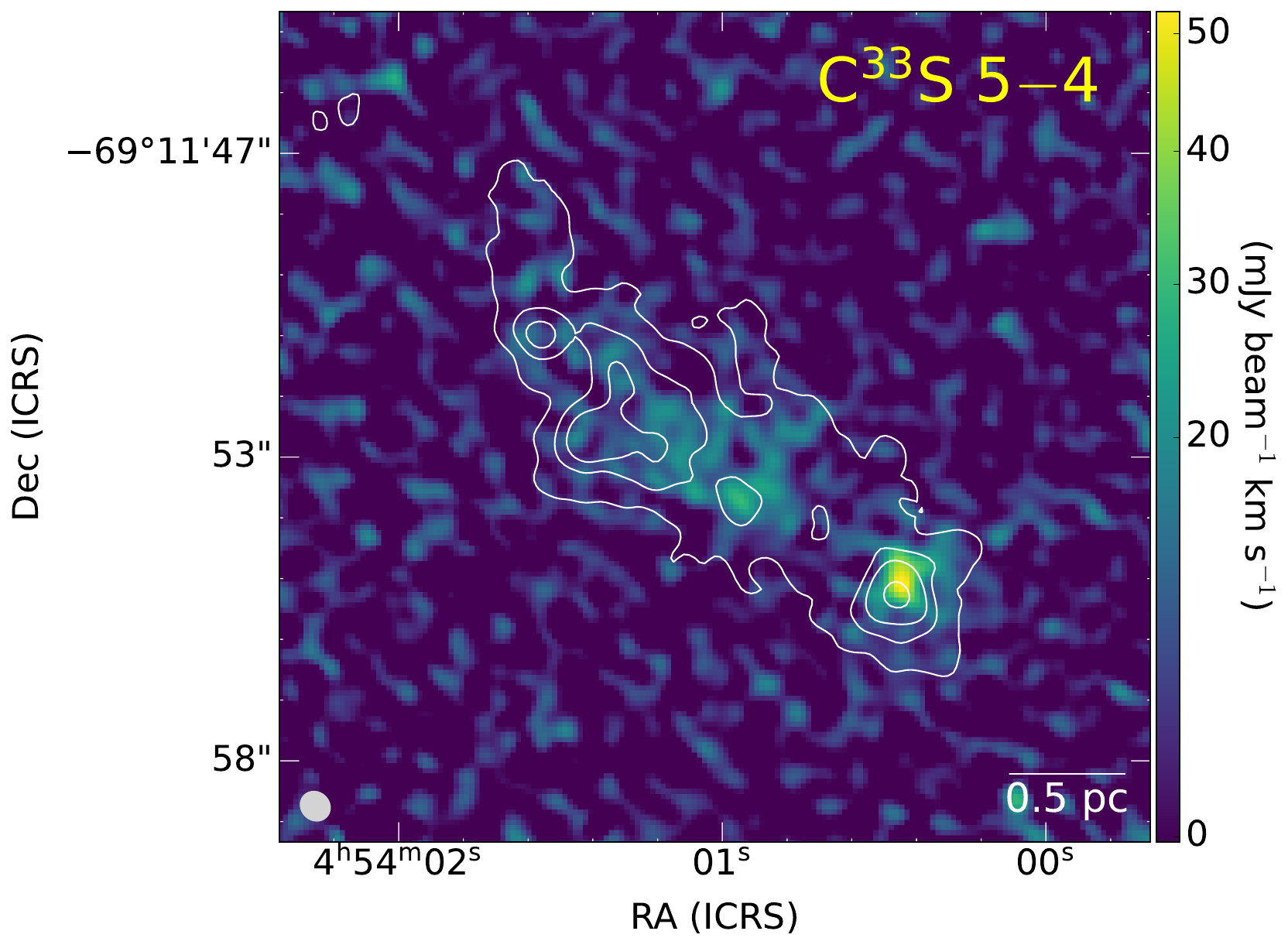}
\includegraphics[width=0.32\textwidth]{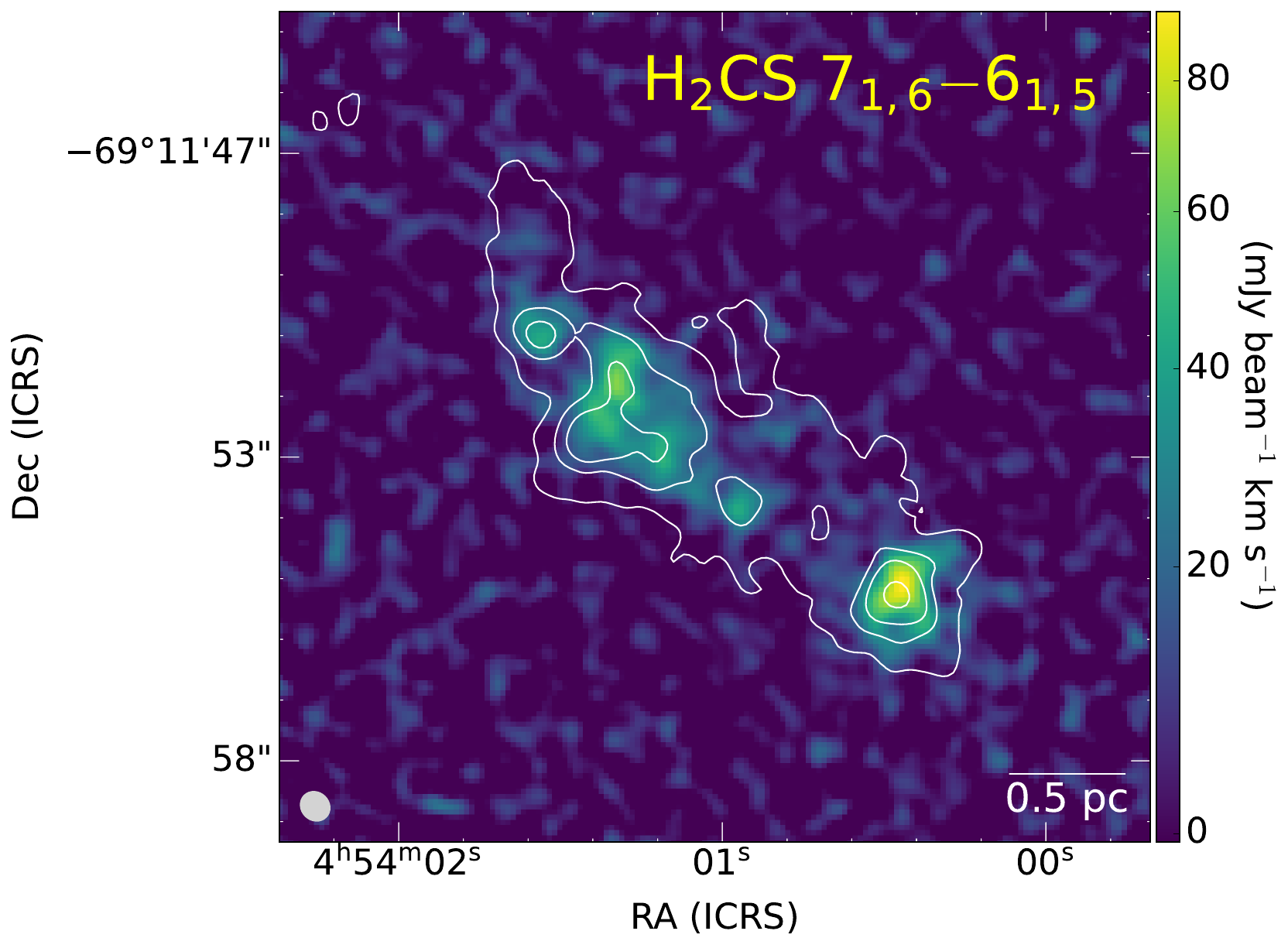}
\includegraphics[width=0.32\textwidth]{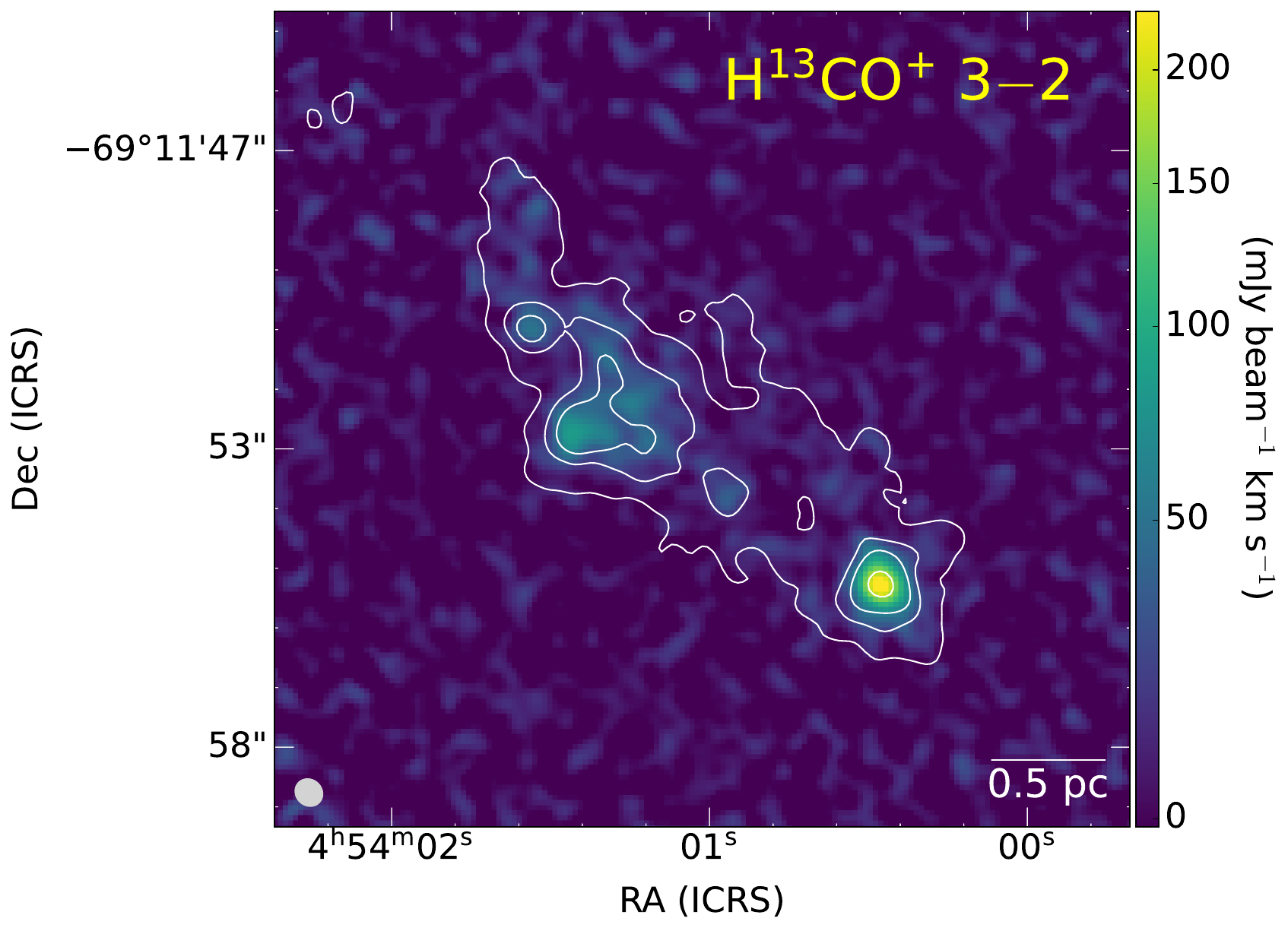}
\includegraphics[width=0.32\textwidth]{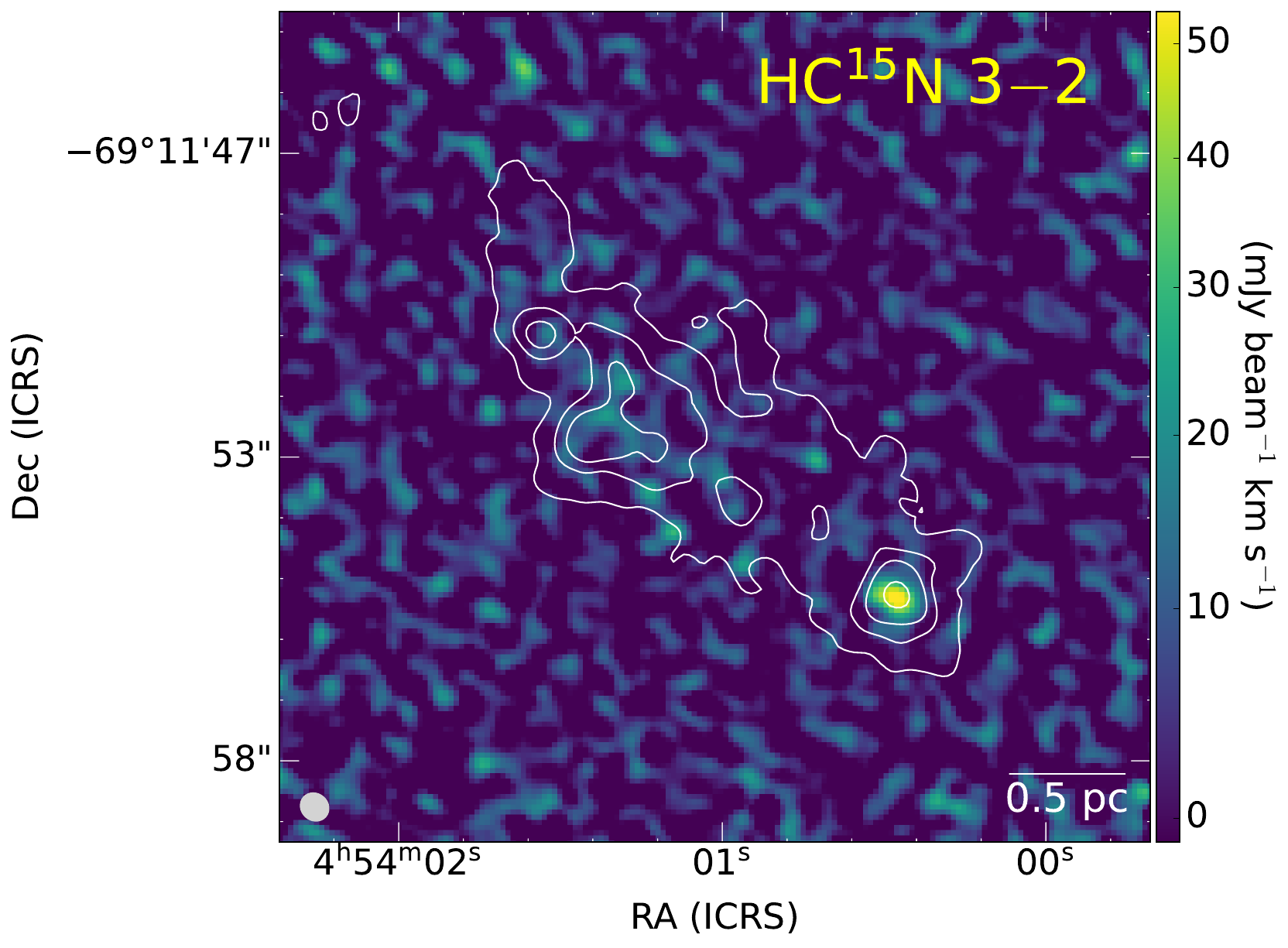}
\includegraphics[width=0.32\textwidth]{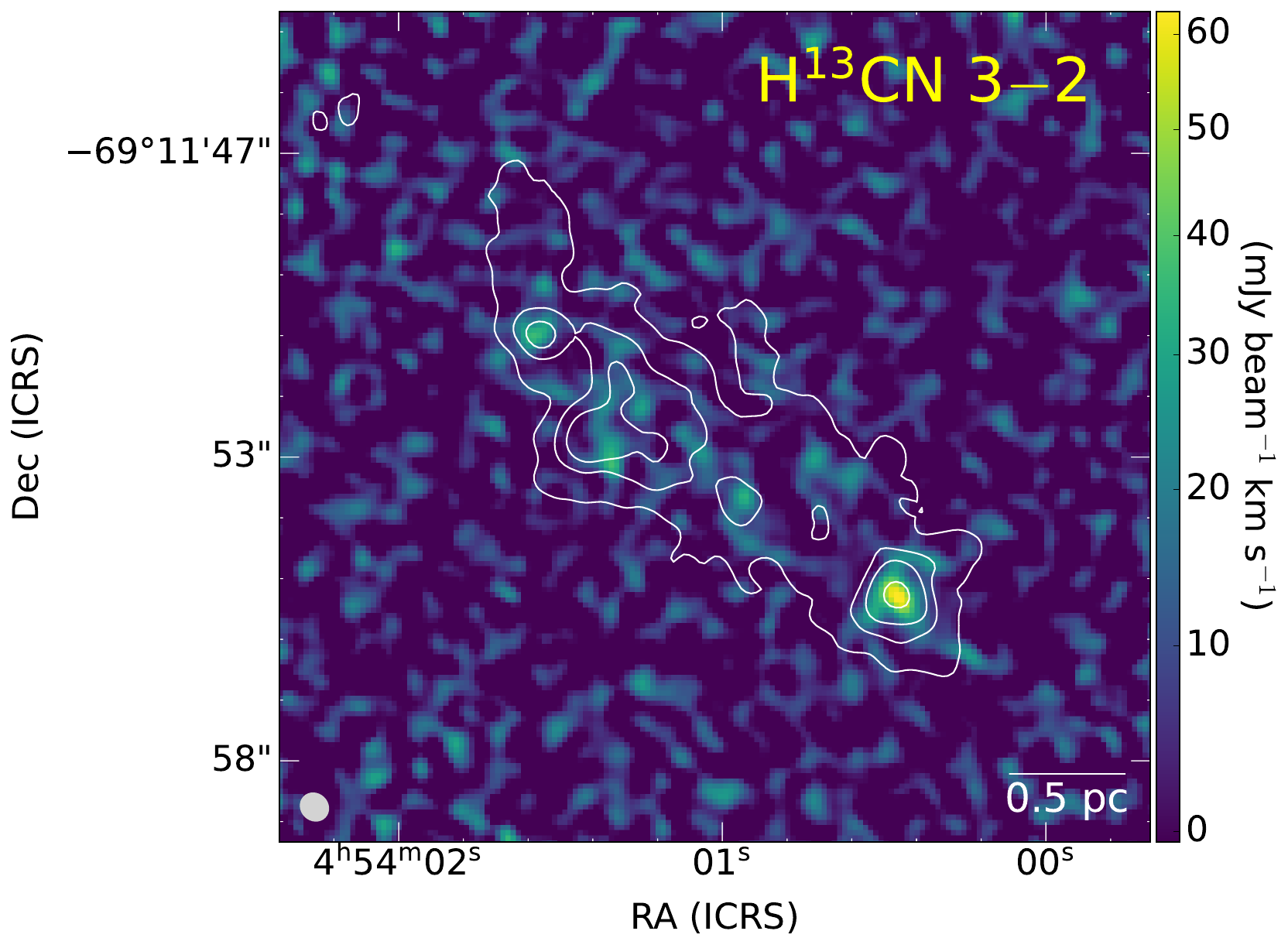}
\caption{Same as Figure~\ref{f:N832mom0COMs}, but for simple molecular species detected toward N\,83A--mm\,A, C, D, and E (from left to right, top to bottom): 
SO 6$_6$--5$_5$,
SO$_2$ 5$_{2, 4}$--4$_{1, 3}$,
SiO 6--5,
CS 5--4,
C$^{33}$S 5--4,
H$_2$CS 7$_{1, 6}$--6$_{1, 5}$,
H$^{13}$CO$^{+}$ 3--2,
H$^{13}$CN 3--2, and 
HC$^{15}$N 3--2. The 1.2 mm continuum contours with contour levels of (3, 10, 20, 60)$\sigma$ are overlaid in each image for reference. 
\label{f:N832mom0simple}}
\end{figure*}

The integrated intensity (moment 0) images are shown in Figures~\ref{f:N831mom0COMs} and \ref{f:N831mom0simple} for the ALMA field N\,83--1, and Figures~\ref{f:N832mom0COMs} and \ref{f:N832mom0simple} for N\,83--2.  For a given species, the emission is integrated over a range of channels covering a single line and/or several transitions if more than one transition is detected, as indicated in the images and/or figure captions. We do not include the integrated intensity images for species with the detection of only one transition that is too faint to produce a reliable image (e.g., HDO). Our observations covering the $\sim$241.3--260.4 GHz sky frequency range are not sensitive to scales larger than 5$\rlap.{''}$6--5$\rlap.{''}$2 (the maximum recoverable scale calculated from the 5$^{\rm th}$ percentile baseline length).

Figures \ref{f:N831mom0COMs} and \ref{f:N832mom0COMs} show the distribution of the COM emission in N\,83--1 (hosting source B) and N\,83--2 (hosting A, C, D, and E), respectively: CH$_3$OH and CH$_3$CN for both fields, and CH$_3$OCH$_3$ for N\,83--1 only.

In N\,83--1, the images reveal a compact hot CH$_3$OH emission ($>$100~K) associated with the 1.2 mm continuum peak (N\,83A--mm\,B) and an extended  component that follows the morphology of the continuum emission. The CH$_3$CN and  CH$_3$OCH$_3$ emission toward N\,83A--mm\,B is very compact with the emission peaks coinciding with the continuum peak.     
  
In N\,83--2, the 1.2 mm continuum emission extends over a $\sim$3 times larger region, elongated roughly in the same northeast-southwest direction as in N\,83--1, and has several peaks.  All the sources we selected for the spectral analysis except N\,83A--mm\,C (the second brightest continuum source in N\,83--2) are associated with the CH$_3$OH emission peaks. The brightest CH$_3$OH emission peak in N\,83--2 is located $\sim$1$\rlap.{''}$3 toward north-northwest from source C. The compact hot CH$_3$OH emission is only associated with source A and coincides with the continuum peak. The distribution of the CH$_3$CN emission is similar to that of hot CH$_3$OH. 

Our observations show that the CH$_3$OH emission is widespread in the LMC, in agreement with the results of our previous ALMA studies (\citealt{sewilo2022n105, broadmeadow2025,hamedanigolshan2024}). In all ALMA fields, both the compact and extended CH$_3$OH emission is present.

Figures \ref{f:N831mom0simple} and \ref{f:N832mom0simple} show the distribution of the emission of simple species in N\,83--1 and N\,83--2, respectively.  In both regions, the CS emission is the brightest and most extended, followed by SO and H$^{13}$CO$^{+}$. CS, SO, H$_2$CS, H$^{13}$CN, and H$^{13}$CO$^{+}$ are the only simple species reliably detected toward all the sources. For all species, the molecular line emission is the brightest toward source A in N\,83--2 and B in N\,83--1, the brightest sources in the 1.2 mm continuum. Interestingly, for source C, most of the molecular emission peaks are offset from the 1.2 mm continuum peak.


\subsection{Spectral Modeling with XCLASS}
\label{s:modeling}

We model the observed ALMA spectra using the eXtended CASA Line Analysis Software Suite (XCLASS\footnote{\url{https://xclass-pip.astro.uni-koeln.de/}}; \citealt{moller2017}) with additional extensions, including non-LTE (Non-Local Thermodynamic Equilibrium) fitting using the \citet{vandertak2007} RADEX routines (XCLASS version 1.4.5, T. M\"{o}ller, priv. comm.; see e.g., \citealt{sewilo2023}; \citealt{hamedanigolshan2024}; \citealt{broadmeadow2025}). 

 XCLASS solves the radiative transfer equation for an isothermal object in one dimension under either LTE or non-LTE conditions. XCLASS can simultaneously model an unlimited number of molecular species, accounting for line blending and opacity effects,  generating synthetic spectra most closely reproducing the observations. In XCLASS, a contribution of each species to a model spectrum is described by a certain number of components, with each component described by the source size, excitation temperature, total column density, line full width at half maximum, velocity offset with respect to the source's local standard of rest (LSR) velocity, and the position along the line of sight. These parameters are either allowed to vary in a provided parameter range or remain fixed. A given component can be described as a core (located in the center of the beam) or a foreground (an envelope around the core or an object along the line-of-sight).    

Spectral fitting within XCLASS is performed by minimizing the $\chi^2$ between modeled and observed spectra, using a suite of optimization algorithms. Our XCLASS fitting procedure employs the Trust Region Reflective (TRF) optimization algorithm, implemented through the \texttt{scipy} optimization package.  The velocity offsets for N\,83A--mm (A, B, C, D, E) are measured with respect to the kinematic LSR velocities of  (247.0, 247.2, 244.2, 245.5, 246.2)~km~s$^{-1}$. 

Both the rotational temperature and total column density can only be reliably determined for species with the detection of more than two transitions. These include CH$_3$OH for all five sources, and CH$_3$CN and SO$_2$ for N\,83A-mm\,A and B. For each source, the initial run of XCLASS only includes these molecules (when detected), assumes LTE conditions, and describes the source as a core, i.e., we assume that all the emission originates in the same region.  To address an issue of a degeneracy between the source size and column density when no optically thick transition is present, we assume that the source is beam-filling. 

For N\,83A-mm\,C, D, and E, the initial LTE modeling of CH$_3$OH resulted in unrealistically low temperatures ($<$10~K) and overall poor fits with some CH$_3$OH transitions being under- or overfitted. We therefore adopted a non-LTE approach, which produced the significantly improved results. The single component non-LTE synthetic spectra of CH$_3$OH for N\,83A-mm\,C, D, and E provide a good match to the observations. 

For N\,83A--mm\,A and B, a single-component LTE synthetic spectrum of CH$_3$OH also provided unsatisfactory fits. For these two sources, a two-component model consisting of a  hot core and an extended cold envelope provide a significantly better fit to the observed spectra. In this setup, the core was modeled with a sub-beam source size (non-beam-filling) and in LTE, while the envelope was beam filling and in non-LTE. 

In the non-LTE framework, two additional parameters are required: the A/E CH$_3$OH ratio and the volume density of the collision partner, H$_2$ ($n_{\rm H_2}$). We fixed the A/E ratio to unity and treated $n_{\rm H_2}$ as a free parameter, finding best-fit values in the range of $\sim$10$^6$–10$^7$~cm$^{-3}$. The resulting rotational temperatures of CH$_3$OH for N\,83A--mm\,A--E range between 17 and 27 K (see Table~\ref{t:XCLASSresults}).

We tested the assumption that the A/E CH$_3$OH ratio is equal to 1 by running XCLASS with the A/E ratio as a free parameter.  The resulting values of the A/E ratio deviated from 1 by only a few percent. This variation is negligible compared to the absolute flux calibration uncertainty ($\sim$10–15\%) and the uncertainties due to line overlaps caused by COMs present simultaneously in our model.  Since for the relatively low signal-to-noise spectra too many free parameters may lead to overfitting, we adopted the A/E \ce{CH3OH} ratio of 1 in our analysis.

Similarly to CH$_3$OH, the SO$_2$ spectra of N\,83A--mm\,A and B require a two-component model to obtain a good fit. In this case, however, both hot compact and cold extended components are in LTE.  In addition to CH$_3$OH and SO$_2$, the detection of multiple transitions of CH$_3$CN allow for an additional temperature determination toward sources A and B. For both sources, a single-component LTE model of CH$_3$CN provides a good fit to the observed spectrum. 

The subsequent XCLASS runs included all the detected and tentatively detected molecules. For all the species with the detection of only one or two transitions other than $^{34}$SO$_2$, SO, and $^{33}$SO, the excitation temperature was fixed to the value derived for CH$_3$OH (a hot component if two are present). Where available, we adopt the SO$_2$ temperature for $^{34}$SO$_2$, SO, and $^{33}$SO. For all the molecules included in our initial XCLASS fit, we fix the XCLASS input parameters to their best-fit values. We utilized the XCLASS’s linking feature to tie the detected isotopologues to their respective main species, sharing the same temperature and velocity structure. This enables the derivation of isotopic ratios based on their fitted column densities.

For each source, we estimate column density upper limits for molecules detected toward any other source in the ALMA fields N\,83--1 and N\,83--2. For N\,83A--mm\,A and B, we also estimate upper limits on column densities of COMs detected (reliably or tentatively) in previously known LMC hot cores, i.e., HCOOCH$_3$, NH$_2$CHO, CH$_3$CHO, and CH$_3$OCH$_3$; the latter for source A only as CH$_3$OCH$_3$ is detected toward source B. To determine upper limits on column density for a given molecule, we only fit its column density with XCLASS; all the other input parameters (the rotational temperature, velocity, and the velocity offset) are fixed to the values we obtained for CH$_3$OH (a hot component if two are present). All the other molecules are included in the XCLASS run with their parameters set to best-fit values.

Following \citet{broadmeadow2025}, we use the Monte Carlo noise resampling to estimate the uncertainties of the parameters resulting from the XCLASS modeling of the ALMA spectra. For each source, we randomly injected noise into the best-fit XCLASS model spectra and fit the resulting spectra with XCLASS. The random noise is drawn from a normal distribution multiplied by the rms noise in a given spectrum ($\sigma_{\rm s,i}$, see Section~\ref{s:specid}). We repeat this procedure 300 times to obtain a well-sampled Gaussian distribution for each XCLASS output parameter, and adopt its standard deviation as a given parameter's uncertainty.  

We provide the XCLASS fitting results in Table~\ref{t:XCLASSresults}: the rotational temperatures ($T_{\rm rot}$), total column densities ($N$), the kinematic LSR velocities ($v_{\rm LSR}$), and line full widths at half maximum ($\Delta v_{\rm FWHM}$). Table~\ref{t:XCLASSresults} also lists fractional molecular abundances with respect to H$_2$ for each detected species (${\rm X}$), $N({\rm X})/N({\rm H_2})$. We describe the H$_2$ column density determination in Section~\ref{s:NH2}. As a result of extracting the spectra as mean intensities (see Section~\ref{s:specid}), the physical parameters of N\,83A--mm A--E determined based on spectral modeling provide averages over the spectral extraction areas. 

We present the ALMA spectra of N\,83A--mm A--E with the XCLASS model fits overlaid in Figures~\ref{f:specA_1}--\ref{f:specE_2} in Appendix~\ref{app:specobsfits}.

\subsection{Physical Properties of N\,83A-mm A--E}
\label{s:phys}

To determine the physical properties of N\,83A-mm\,A--E based on the 1.2 mm continuum emission (H$_2$ column and number densities, sizes, and masses; Sections~\ref{s:NH2}--\ref{s:nH2}), we follow the analysis presented in \citet{sewilo2022n105} for sources in the star-forming region N\,105 (including two hot cores, N\,105--2\,A and 2\,B).

\subsubsection{H$_2$ Column Densities}
\label{s:NH2} 
 
Assuming that the thermal equilibrium between the gas and dust ($T_{\rm dust} \sim T_{\rm gas} \sim T$) holds in high density regions ($n_{\rm H_2}\gtrsim10^5$ cm$^{-3}$; e.g.,  \citealt{goldsmith1978}; \citealt{ceccarelli1996}; \citealt{kaufman1998}), we calculate H$_2$ column density, $N({\rm H_2})$, using the formula from \citet{kauffmann2008}:
\begin{eqnarray}
\label{e:NH2}
N({\rm H_2}) = 2.02\cdot10^{20} (e^{1.439(\lambda/{\rm mm})^{-1}(T/{\rm 10\,K})^{-1}}-1) \\ \nonumber
\cdot \left(\frac{\kappa_{{\rm \nu, d}}/R_{\rm gd}}{0.01\,{\rm cm^2\,g^{-1}}}\right)^{-1}\,\left(\frac{I_{\rm \nu}^{\rm beam}}{{\rm mJy\,beam^{-1}}}\right) \\ 
\cdot \left(\frac{\theta_{\rm HPBW}}{{\rm 10\,arcsec}}\right)^{-2}\,\left(\frac{\lambda}{{\rm mm}}\right)^3 {\rm cm^{-2}}, \nonumber 
\end{eqnarray} 
where $\kappa_{\rm {\nu, d}}$ is the dust opacity per unit mass at a frequency $\nu$ (e.g., \citealt{hildebrand1983}; \citealt{shirley2000}); 
$R_{\rm gd}$ is the gas-to-dust mass ratio;
$I_{\rm \nu}^{\rm beam}$  is the dust continuum intensity;
$\theta_{\rm HPBW}= \sqrt{\theta_{\rm maj}\cdot\theta_{\rm min}}$, where $\theta_{\rm maj}$ and $\theta_{\rm min}$ are the major and minor axes of the synthesized beam, respectively; and $\lambda$ is the wavelength of the dust continuum emission.  Since no radio continuum emission or hydrogen recombination lines have been detected toward N\,83A-mm\,A--E (\citealt{indebetouw2004} and this work), we assume that all the 1.2 mm continuum emission is the dust thermal emission, i.e., the contribution from the free-free emission is negligible.

In Equation~\ref{e:NH2}, we adopt the rotational temperatures determined based on CH$_3$OH, $T = T({\rm CH_3OH})$.  For hot cores N\,83A--mm\,A and B where both the cold and hot CH$_3$OH components are present, we use the temperature of the hot component. If a reliable CH$_3$CN and/or SO$_2$  temperature is available, we also estimate $N({\rm H_2})$  for $T=T({\rm CH_3CN})$ and/or $T=T({\rm SO_2})$ for comparison. 

Following \citet{sewilo2022n105}, we adopt a Galactic dust opacity from \citet{ossenkopf1994} for the model with the initial MRN distribution \citep{mathis1977} with thin ice mantles after 10$^{5}$ years of coagulation at a hydrogen gas density of 10$^{6}$ cm$^{-3}$.  For $\nu=242.4$ GHz ($\lambda=1.24$ mm), the opacity per unit dust mass, $\kappa_{\rm 242.4\,GHz, d}$, is 0.993 cm$^{2}$ g$^{-1}$ (Table 1 in \citealt{ossenkopf1994}).  We  assume $R_{\rm gd}$ of 316 for the LMC, estimated in \citet{sewilo2022n105}.  Finally, $I_{\rm \nu}^{\rm beam}$ was measured as a mean intensity within the region used to extract spectra ($I_{\rm mean}$), i.e., the area enclosed by the contour corresponding to 50\% of the 1.2 mm continuum peak of a given source.  

The resulting values of $N({\rm H_2})$ based on the temperature of \ce{CH3OH} (all sources), \ce{SO2} and \ce{CH3CN} (hot cores N\,83A--mm A and B) are provided in Table~\ref{t:datanh2}. Table~\ref{t:datanh2} also lists $I_{\rm mean}$, along with the 1.2 mm continuum peak positions (RA and Decl.) and intensities ($I_{\rm peak}$). 

We take advantage of the availability of the $^{13}$CO (1--0) data (see Section~\ref{s:observations}) to determine $N(\ce{H2})$ using the method based on the  \ce{^13CO} (1--0) emission and check for consistency with $N(\ce{H2})$ measured from the 1.2 mm continuum. The \ce{^13CO} (1--0) data allow us to construct the $N(\ce{H2})$ map (covering both N\,83--1 and N\,83--2 ALMA fields) by calculating $N(\ce{H2})$ on the pixel-by-pixel basis. We assume LTE, a uniform excitation temperature ($T_{\rm ex}$) of 20 K across the region, and the H$_2$ to  \ce{^13CO} abundance ratio of $3.1\times10^6$.  The adopted H$_2$ to  \ce{^13CO} abundance ratio accounts for the reduced CO abundance in the LMC compared to the solar neighborhood (see \citealt{mizuno2010,fujii2014}), and is consistent with the results from other independent tracers in the ALMA studies of the LMC, which have consistently reproduced H$_2$ column densities (e.g., \citealt{naslim2018}; \citealt{tokuda2019}; \citealt{wong2022}).  The assumption of $T_{\rm ex}$ of 20 K is based on the empirical expectation from multi-line CO analyses that molecular gas in the LMC is somewhat warmer than typical Galactic molecular clouds for which a temperature of $\sim$10 K is often adopted (e.g., \citealt{minamidani2011,finn2021}). The resulting $N(\ce{H2})$ map with the resolution of $\sim$2$''$ ($\sim$0.5 pc) is shown in Figure~\ref{f:N83B3B6_13CO}.  The map reveals that the two $N(\ce{H2})$ determination methods (based on the $^{13}$CO 1--0 and 1.2 mm continuum emission) provide consistent results. The $N(\ce{H2})$ map also shows that ALMA fields N\,83--1 (hosting N\,83A--mm B) and N\,83--2 (hosting N\,83A--mm A, C--E) coincide with the two $N(\ce{H2})$ peaks in the region. N\,83--2 is associated with the higher \ce{H2} column density gas (see Section~\ref{s:multiwav} for the discussion on the large-scale physical setting of N\,83A--mm A--E). 

\begin{deluxetable*}{ccccccccccccc}
\centering
\rotate
\tablecaption{Physical Properties of N\,83A-mm A--E Based on the 1.2 Continuum Emission \label{t:datanh2}}
\tablewidth{0pt}
\tablehead{
\multicolumn{1}{c}{Source} &
\colhead{RA (J2000)\tablenotemark{\footnotesize a}} &
\colhead{Decl. (J2000)\tablenotemark{\footnotesize a}} &
\colhead{$I_{\rm peak}$\tablenotemark{\footnotesize b}} &
\colhead{$I_{\rm mean}$\tablenotemark{\footnotesize c}} &
\colhead{$A$\tablenotemark{\footnotesize c}}  &
\colhead{FWHM$_{\rm eff}$}  &
\colhead{FWHM$_{\rm eff, deconv}$}  &
\colhead{$F_{\rm 50}$\tablenotemark{\footnotesize d}}  &
\colhead{$M_{\rm 50, gas}$\tablenotemark{\footnotesize d}}  &
\colhead{$N(\ce{H2})$\tablenotemark{\footnotesize e}} &
\colhead{$N(\ce{H2})$\tablenotemark{\footnotesize f}} &
\colhead{$N(\ce{H2})$\tablenotemark{\footnotesize g}} \\
\cline{11-13}
\colhead{N\,83A--mm} &
\colhead{($^{\rm h}$~$^{\rm m}$~$^{\rm s}$)} &
\colhead{($^{\rm \circ}$ $'$ $''$)} &
\colhead{(mJy beam$^{-1}$)} &
\colhead{(mJy beam$^{-1}$)} &
\colhead{(arcsec$^{2}$)} &
\colhead{($''$/pc)} &
\colhead{($''$/pc)} &
\colhead{(mJy)} &
\colhead{($M_{\odot}$)} &
\multicolumn{3}{c}{(10$^{22}$ cm$^{-2}$)} 
}
\startdata
A & 4:54:00.141 & $-$69:11:55.24 & 4.968 & 3.646 & 0.330 & 0.65/0.16 & 0.42/0.10 & $4.4\pm0.4$ & $63\pm7$ & $14.8\pm1.7$ & $9.7\pm2.3$ & $11.0\pm1.3$\\
B & 4:54:02.902 & $-$69:11:39.22 & 2.859 & 2.008 & 0.525 & 0.82/0.20 & 0.65/0.16 & $3.9\pm0.4$ & $65\pm7$ & $9.6\pm1.0$ & $14.6\pm3.1$ & $9.3\pm1.0$ \\
C & 4:54:01.161 & $-$69:11:52.52 & 2.511 & 1.703 & 0.855 & 1.04/0.25 & 0.92/0.22 & $5.4\pm0.5$ & $518\pm78$ & $47.0\pm7.3$ & \nodata & \nodata \\
D & 4:54:01.332 & $-$69:11:50.64 & 1.706 & 1.207 & 0.398 & 0.71/0.17 & 0.52/0.12 & $1.8\pm0.2$ & $207\pm30$ & $40.4\pm6.1$ & \nodata & \nodata \\
E & 4:54:00.658 & $-$69:11:53.68 & 0.922 & 0.702 & 0.576 & 0.86/0.21 & 0.70/0.17 & $>$1.5 & $>$147 & $19.8\pm3.4$ & \nodata & \nodata \\
\enddata
\tablenotetext{a}{The 1.2 mm continuum peak positions.} 
\tablenotetext{b}{$I_{\rm peak}$ is the observed 1.2 mm continuum intensity peak.}
\tablenotetext{c}{$I_{\rm mean}$ is the 1.2 mm continuum intensity averaged over the area ($A$) within the contour corresponding to the 50\% (A--D) or 60\% (E) of the 1.2  mm continuum peak.  This is the same area used to extract spectra for the analysis (see Section~\ref{s:modeling}). The beam area for both the N\,83--1 and N\,83--2 observations is 0.272 arcsec$^{2}$.}
\tablenotetext{d}{$F_{\rm 50}$ and $M_{\rm 50, gas}$ are flux densities and masses, respectively, calculated for the area above the 50\% of the peak intensity.}
\tablenotetext{e}{$N(\ce{H2})$  calculated assuming $T = T_{\rm rot} ({\rm CH_{3}OH})$ (see Section~\ref{s:phys} for details).}
\tablenotetext{f}{$N(\ce{H2})$  calculated assuming $T = T_{\rm rot} ({\rm CH_{3}CN})$.}
\tablenotetext{g}{$N(\ce{H2})$  calculated assuming $T = T_{\rm rot} ({\rm SO_{2}})$.}
\end{deluxetable*}

\subsubsection{Masses}
\label{s:mass} 

We estimate gas masses of N\,83A--mm A--E using the formula \citep{kauffmann2008}: 
\begin{eqnarray}
\label{e:mass}
M = 0.12\, M_{\odot} \cdot (e^{1.439(\lambda/{\rm mm})^{-1}(T/{\rm 10\,K})^{-1}}-1) \\
\cdot \left(\frac{\kappa_{{\rm \nu, d}}/R_{\rm gd}}{0.01\,{\rm cm^2\,g^{-1}}}\right)^{-1}\,\left(\frac{F_{\rm \nu}}{{\rm Jy}}\right) \nonumber 
\cdot \left(\frac{D}{{\rm 100\,pc}}\right)^{2} \,\left(\frac{\lambda}{{\rm mm}}\right)^3,  \nonumber
\end{eqnarray}
where $F_{\nu}$ is the flux density, $D$ is the distance to the LMC, and other parameters are the same as in Equation~\ref{e:NH2}. 

For $F_{\nu}$, we adopt the 1.2 mm continuum flux density for the area above 50\% of the peak intensity ($F_{\rm 50}$), i.e., the area used for the spectral extraction for N\,83A--mm\,A--D.  $F_{\rm 50}$ and the corresponding mass, $M_{\rm 50, gas}$, are listed in Table~\ref{t:datanh2}.  Since in our analysis of N\,83A--mm\,E, we use the 60\% of the 1.2 mm continuum peak intensity contour instead (see Section~\ref{s:cont}), we report $F_{\rm 50}$ and $M_{\rm 50, gas}$ for this source as lower limits.  Source areas ($A$) used for the determination of $F_{\rm 50}$ are also provided in Table~\ref{t:datanh2}.

\subsubsection{Source Sizes}
\label{s:size} 

We define the observed FWHM source size as FWHM$_{\rm eff} = 2 \cdot R_{\rm eff}$, where $R_{\rm eff}$ is the ``effective radius''  measured from the source area $A$, $R_{\rm eff} = \sqrt{A/\pi}$.  Assuming the sources can be described by Gaussian profiles, we also calculate the deconvolved FWHM sizes: FWHM$_{\rm {eff, deconv}} = \sqrt{{\rm FWHM}_{\rm eff}^2 - \theta_{\rm HPBW}^2}$, where $\theta_{\rm HPBW}$ is the geometric mean of the minor and major axes of the synthesized beam of the 1.2 mm continuum observations (as in Equation~\ref{e:NH2}).  The observed and deconvolved source sizes (FWHM$_{\rm eff}$ and FWHM$_{\rm {eff, deconv}}$, respectively) are provided in Table~\ref{t:datanh2}.

\subsubsection{H$_2$ Number Densities}
\label{s:nH2} 

Assuming that the extent of the source along the line of sight is the same as in the plane of the sky (${\rm FWHM}_{\rm eff, deconv}$), we estimate the H$_2$ number density ($n_{\rm H_2}$) using the relation $n_{\rm H_2} = N({\rm H_2})/{\rm FWHM}_{\rm eff, deconv}$.  The resulting $n_{\rm H_2}$ of the cores N\,83A--mm A--E ranges from $\sim$$2.0\times10^5$ to $\sim$$1.1\times10^6$ cm$^{-3}$.

\section{Discussion} 
\label{s:discussion}

\subsection{Identification of Hot Cores in N\,83} 
\label{s:HCident}

Based on the physical and chemical characteristics of sources N\,83A--mm A and B, we identify them as bona fide hot cores.  For both sources, the rotational temperatures derived based on multiple tracers exceed 100~K.  The deconvolved effective FWHM sizes of sources A and B are 0.1 and 0.16 pc, respectively.  The size of source A is consistent with the Galactic hot core sizes (e.g., \citealt{kurtz2000}), while the larger size of source B is similar to some of the other LMC hot cores (e.g., \citealt{sewilo2018,sewilo2022n105}).  The H$_2$ number densities for both sources are a few times 10$^{5}$ cm$^{-3}$, and both are associated with COMs. 

\begin{figure}[ht!]
\centering
\includegraphics[width=0.494\textwidth]{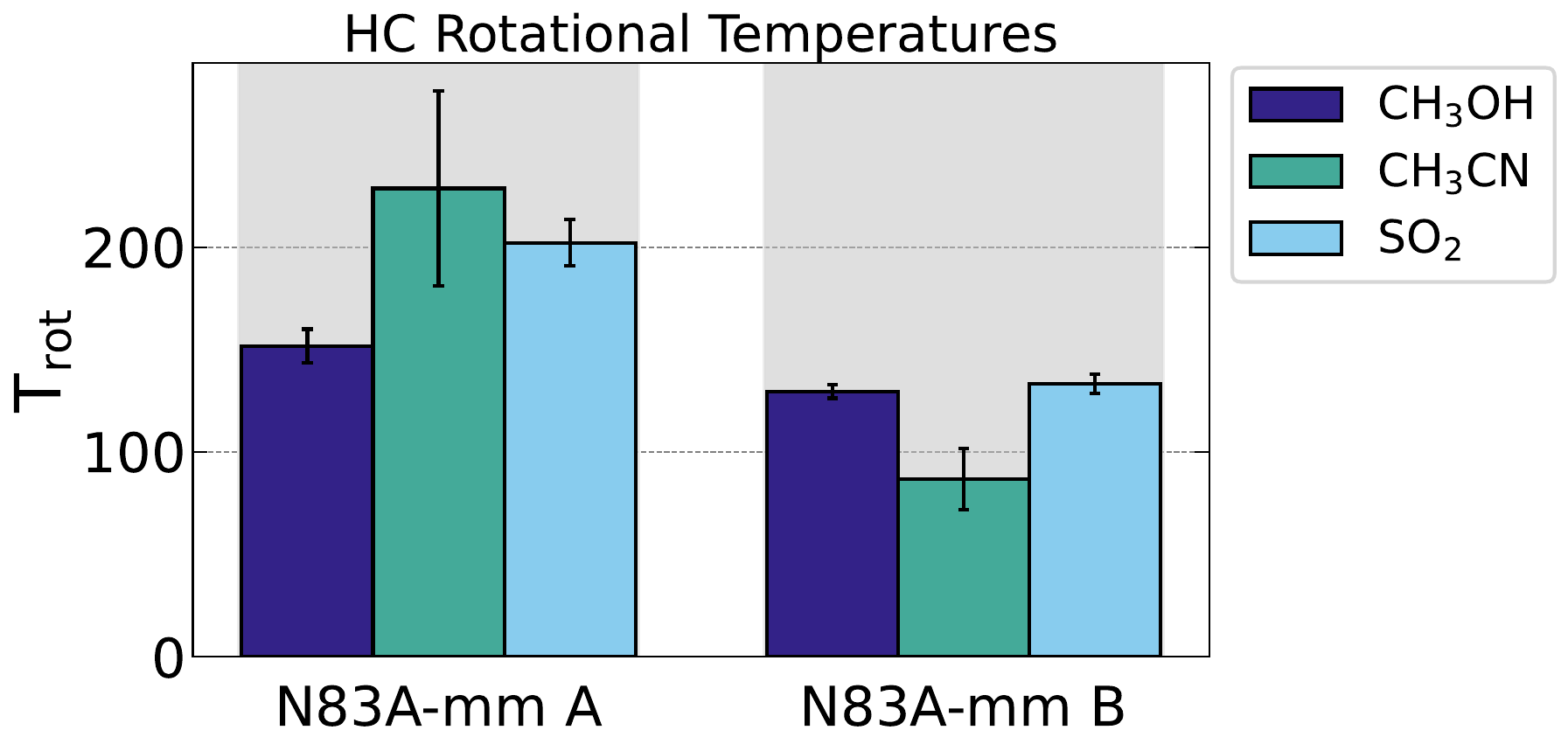}
\caption{Rotational temperatures of hot cores N\,83A--mm A and B determined by XCLASS spectral modeling for molecular species with the detection of multiple transitions -- CH$_3$OH, CH$_3$CN, and SO$_2$. \label{f:hotcoresTN}}
\end{figure}

In Figure~\ref{f:hotcoresTN}, we show the bar plot of rotational temperatures for hot cores N\,83A--mm A and B for species with the detection of multiple transitions: \ce{CH3OH}, \ce{CH3CN}, and \ce{SO2} (hot components for \ce{CH3OH} and  \ce{SO2}).  For all species, $T$ for source A is higher than for source B. With the exception of  $T(\rm \ce{SO2})$ for source B, the temperatures are well above 100 K. In terms of abundances of these key species with respect to \ce{H2} (provided in  Table~\ref{t:XCLASSresults} and shown graphically in Figure~\ref{f:abundN83}), $N({\rm \ce{CH3OH}})/N({\rm \ce{H2}})$ is a factor of $\sim$2.1 higher in source B than in source A. $N({\rm \ce{SO2}})/N({\rm \ce{H2}})$ is the same in both sources within the uncertainties. $N({\rm \ce{CH3CN}})/N({\rm \ce{H2}})$ is a factor of 1.3 higher in source A, but it is associated with much larger uncertainties.  In general, for both sources,  the temperature and column density determination based on \ce{CH3CN} is the least reliable due to the lines being much fainter than most of the \ce{CH3OH} and \ce{SO2} lines. 

\subsubsection{Confirmed YSOs and Water Masers in the ALMA Fields N\,83--1 and N\,83--2}
\label{s:ysosmas}

\begin{figure*}
\centering
\includegraphics[width=0.8\textwidth]{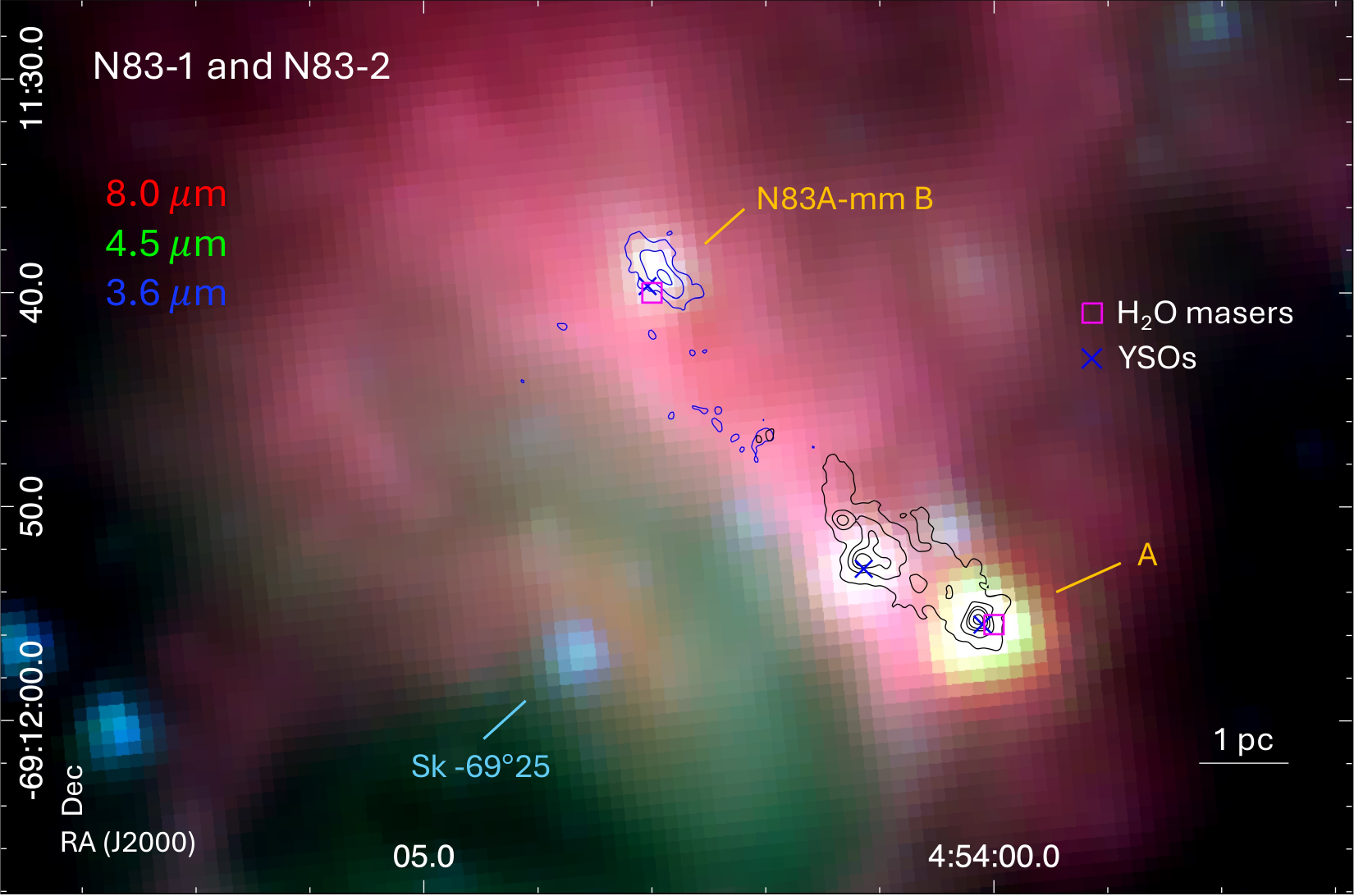}
\caption{Three-color mosaic of N\,83--1 and N\,83--2, combining the {\it Spitzer}/SAGE IRAC 8.0 $\mu$m (red), 4.5 $\mu$m (green), and 3.6 $\mu$m (blue) images \citep{meixner2006}. The catalog positions of the YSOs and H$_2$O masers are indicated with blue `$\times$' symbols and magenta squares, respectively (see Section~\ref{s:ysosmas}).  The 1.2 mm continuum contour levels are (3, 10, 40)$\sigma$ for N\,83--1 (blue) and (3, 10, 20, 30, 60)$\sigma$ for N\,83--2 (black; see also Figure~\ref{f:N83B6cont}).  Hot cores N\,83A--mm A and B are labeled; they are separated by $\sim$21$\rlap.{''}$8 or $\sim$5.2 pc. The blue star east of N\,83A--mm A is Sk~$-$69$^{\circ}$25, the exciting star of the N\,83A nebula (see also Figure~\ref{f:N83AspitzerFOV}). \label{f:N83spitzer}}
\end{figure*}

\begin{figure*}
\centering
\includegraphics[width=0.49\textwidth]{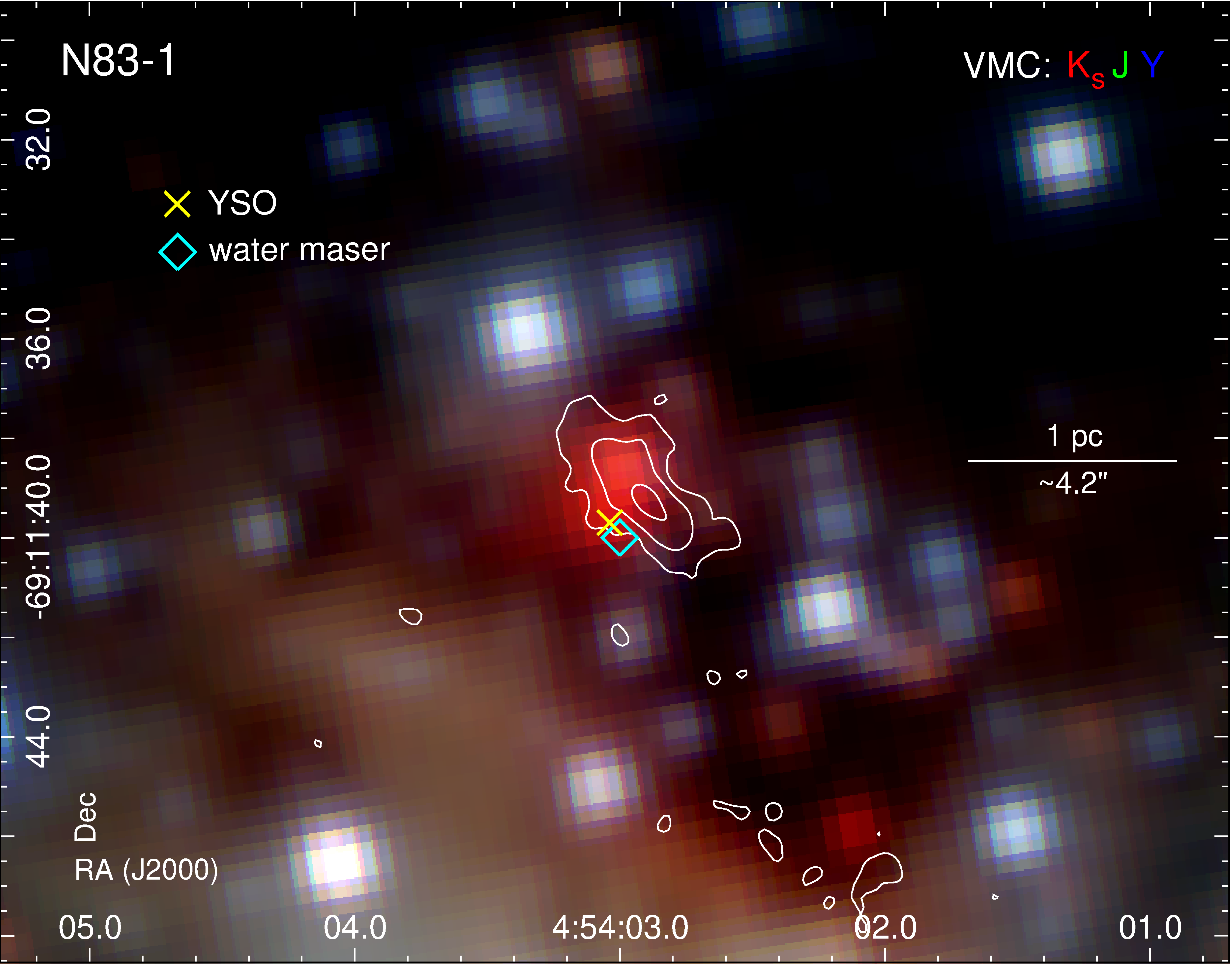}
\includegraphics[width=0.493\textwidth]{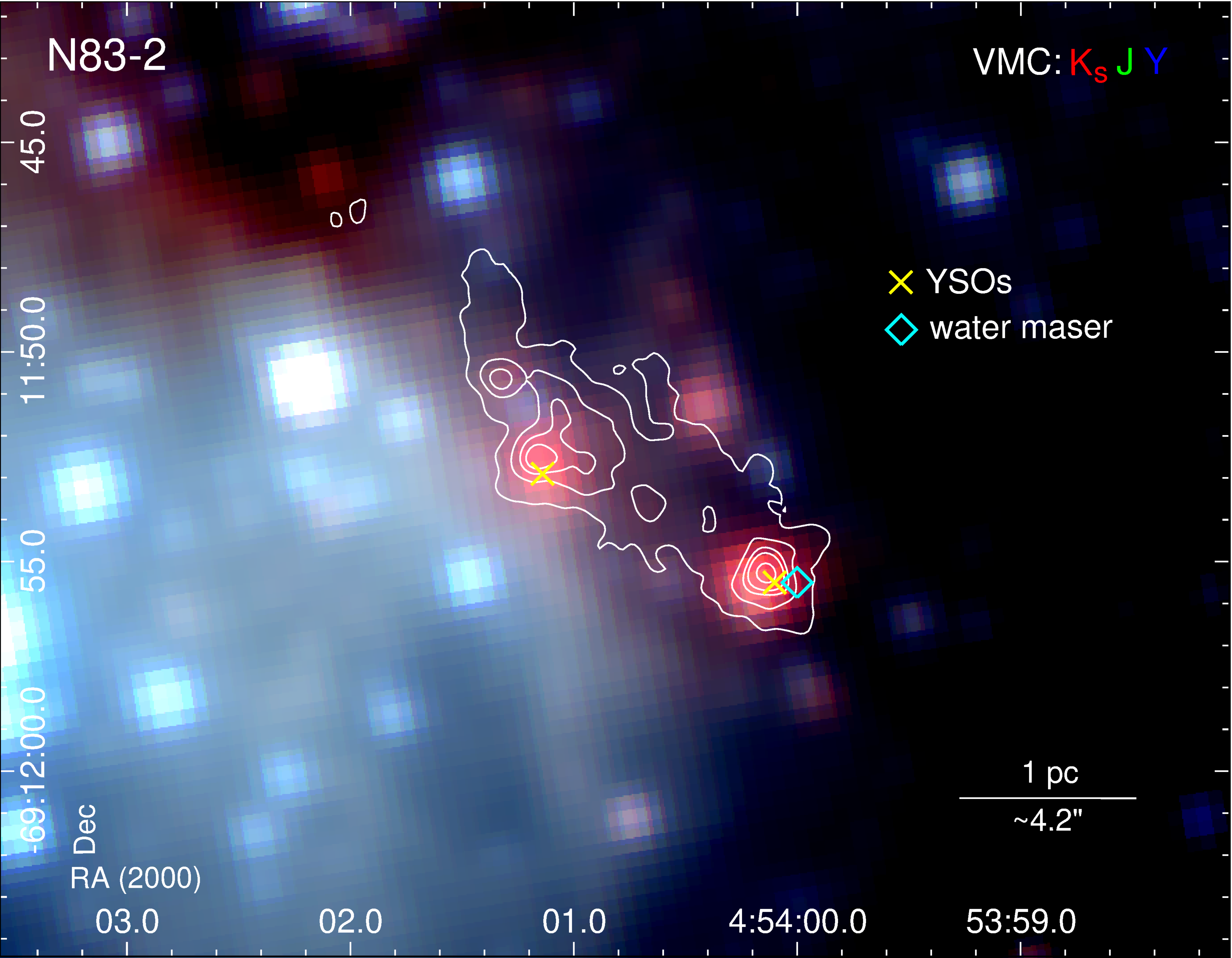}
\caption{Three-color mosaic of N\,83--1 (left) and N\,83--2 (right) combining the VMC $K_{\rm s}$ (red), $J$ (green), and $Y$ (blue) images \citep{cioni2011}.  The 1.2 mm continuum contours are overlaid for reference with contour levels the same as in Figure~\ref{f:N83B6cont}.  The catalog positions of YSOs and H$_2$O masers are indicated (see Section~\ref{s:ysosmas}).  \label{f:N83vmc}} 
\end{figure*}

We required the LMC fields selected for our observations to be associated with {\it Spitzer} YSOs and masers (see Section~\ref{s:observations}). Three YSOs are located in our ALMA fields in N\,83, one in N\,83--1 and two in N\,83--2 (see Figs.~\ref{f:N83spitzer} and \ref{f:N83vmc}).   In N\,83--1, YSO 045403.04$-$691139.7 is associated with the hot core N\,83A--mm B, although its position is offset  from the 1.2 mm continuum peak (\citealt{gruendl2009}; see Section~\ref{s:multiwav}). In N\,83--2, YSOs 045400.10$-$691155.5 (SSTISAGE1C\,J045400.14$-$691155.4 or \#90 in \citealt{whitney2008}) and 045401.14$-$691152.9 are associated with the hot core N\,83A--mm A and the cold core N\,83A--mm C (identified as cold, $T_{\rm rot}$$\sim$27 K,  based on the analysis of the \ce{CH3OH} lines), respectively. All 3 YSOs were first identified as YSO candidates based on the {\it Spitzer} photometric data (3.6--24 $\mu$m; \citealt{gruendl2009,whitney2008}).

The classification of 045400.10$-$691155.5 (A), 045403.04$-$691139.7 (B), and 045401.14$-$691152.9 (C) as YSOs was confirmed spectroscopically based on the {\it Spitzer} Infrared Spectrograph (IRS) observations (5--37 $\mu$m; first reported in \citealt{seale2009}, reanalyzed in \citealt{jones2017}).  The presence of the CO$_2$ ice absorption band at 15.2 $\mu$m in their {\it Spitzer}/IRS spectra indicates that the sources are at the early (embedded) stage of the YSO evolution.

N\,83A--mm A and B are spatially coincident with H$_2$O masers (J. Ott, priv. comm. and \citealt{schwarz2012};  detected with the Australia Telescope Compact Array, ATCA, observations with a pointing accuracy of $\sim$2$''$) which is consistent with their classification as embedded YSOs since H$_2$O masers are often found at this early stage of the protostellar evolution (e.g., \citealt{dejong1973}; \citealt{meehan1998}; \citealt{gray2022}).

\subsubsection{The Large-scale Physical Setting of the ALMA Hot and Cold Cores in N\,83A}
\label{s:multiwav}

We review the data from the previous multi-wavelength observations (from the optical to radio) of the N\,83A nebula to put our ALMA results into the broader physical context. 

In Figure~\ref{f:N83spitzer}, we show the {\it Spitzer} three-color mosaic of the field centered on N\,83--1 and N\,83--2, combining the IRAC 8.0, 4.5, and 3.6 $\mu$m images (0$\rlap.{''}$6 pixel tiles) from the LMC-wide Spitzer Surveying the Agents of Galaxy Evolution (SAGE) survey (\citealt{meixner2006,sage}). The three sources brightest at 8.0 $\mu$m are the YSOs associated with N\,83A--mm A, C, and B (in the order from the brightest to faintest) discussed in Section~\ref{s:ysosmas}.  No clear {\it Spitzer} sources coincide with the 1.2 mm continuum cores N\,83A--mm D and E.  The near-IR images from the Visible and Infrared Survey Telescope for Astronomy (VISTA) $YJK_{\rm s}$ survey of the Magellanic Clouds system (VMC,  \citealt{cioni2011}; the seeing of $\sim$0$\rlap.{''}$7) shown in Figure~\ref{f:N83vmc} reveal the central source of N\,83A--mm D, but no near-IR source coincides with source E. All the remaining cores are associated with the VMC sources. 

The relatively narrow 8.0 $\mu$m emission feature elongated in the northeast-southwest direction seen in Figure~\ref{f:N83spitzer} coincides with the 1.2 mm continuum emission from both (overlapping) ALMA fields. The 1.2 mm continuum 3$\sigma$ emission contours in the region between N\,83A--mm B (in N\,83--1) and N\,83A--mm A/C-E (in N\,83--2) indicate the presence of an extended emission connecting these sources, likely filtered out by our interferometric observations.   Both compact and larger-scale extended continuum emission has been detected with the lower-resolution ($\sim$2$''$) ALMA observations at 3 mm covering both of our Band 6 fields (see Appendix~\ref{app:ancillary}).  Hot cores N\,83A--mm A and B, and source C are associated with the 3 mm continuum peaks (Figure~\ref{f:N83B3B6cont}).

To get the far-IR view of the star-forming region N\,83, we use the data from the {\it Herschel} Space Observatory. In Figure~\ref{f:N83AspitzerFOV}, we show the 100 and 160 $\mu$m images from the {\it Herschel} Inventory of the Agents of Galaxy Evolution (HERITAGE) survey (\citealt{meixner2013,heritage}). The images with FWHM of 8$\rlap.{''}$6/$\sim$1.2 pc and 12$\rlap.{''}$6/$\sim$3.1 pc, respectively, reveal that each ALMA field (N\,83--1 and N\,83--2) coincides with a single bright {\it Herschel} source, indicating the presence of the cold dust. These are the two brightest  {\it Herschel} sources in N\,83A.

The multi-wavelength data reveal signatures of triggered star formation in the N\,83A nebula, including the presence of a ``{\it Spitzer} bubble.'' 
The {\it Spitzer} mosaic with a field of view covering the bubble in its entirety is shown in Figure~\ref{f:N83AspitzerFOV} in the 24, 8.0, and 4.5 $\mu$m color combination typically used to highlight the characteristic morphology of {\it Spitzer} bubbles: the 8.0 $\mu$m shell (green) filled with the 24 $\mu$m emission (red; e.g., \citealt{churchwell2006,deharveng2010}).  The 8.0 $\mu$m emission primarily traces polycyclic aromatic hydrocarbons (PAHs) in the photodissociation regions (PDRs) at the rims of the bubbles, where the ionized gas interacts with the surrounding molecular gas.  The 24 $\mu$m emission is a combination of the thermal emission from warm dust and ionized gas surrounding the central exciting star.  The position of the main sequence star Sk~$-$69$^{\circ}$25 of a spectral type O6 (64 $M_{\odot}$; \citealt{massey2000}) indicates that the bubble was formed by its radiation and winds (Figures~\ref{f:N83spitzer} and \ref{f:N83AspitzerFOV}). Sk~$-$69$^{\circ}$25 is the main ionizing source of the N\,83A nebula. 

The 4.5 $\mu$m band is often used to trace shocked emission and ionized gas as it includes contributions from the nebular continuum and lines such as Br$\gamma$ at 4.05 $\mu$m or the H$_2$ v=0--0 S(9) line at 4.695 $\mu$m. Figure~\ref{f:N83AspitzerFOV} shows that the extended 4.5 $\mu$m nebular emission in N\,83A highlights the ionized gas within the bubble outlined by the 8.0 $\mu$m emission. The interior of the bubble is also traced by the [O\,{\sc III}] 88.36 $\mu$m line emission indicating that the gas is highly ionized  (Figure~\ref{f:herspec}), as well as the H$\alpha$ (Figure~\ref{f:n83}) and 3 mm continuum emission (Figure~\ref{f:N83B3B6cont}; see Appendix~\ref{app:ancillary} for  details).  The ALMA cores N\,83A--mm A--E are embedded in the gas and dust at the rim of the bubble. 

We conclude that the star formation in the ALMA fields N\,83--1 and N\,83--2 has likely been triggered by the expansion of the ionized gas bubble. The expanding bubble compresses the dense molecular gas at its boundary, leading to the fragmentation and the formation of the new generation of stars.  In Figure~\ref{f:N83B3B6_13CO}, we show the ALMA $^{13}$CO (1--0) image with the resolution of  $\sim$2$''$ tracing the dense gas in N\,83A.  The morphology of the  $^{13}$CO emission is similar to that of the 1.2 mm continuum and the 8.0 $\mu$m emission (it appears as a relatively narrow structure at the rim of the bubble), supporting the triggered star formation scenario. Two $^{13}$CO emission peaks are associated with the ALMA fields N\,83--1 and N\,83--2; the $^{13}$CO emission is brighter toward N\,83--2, indicating that the gas is denser (see Figure~\ref{f:N83B3B6_13CO} and Section~\ref{s:NH2}).  This is consistent with the distribution of the C$^{18}$O (2--1) emission detected with the ALMA Total Power Array (HPBW$\sim$30$''$);  the C$^{18}$O emission peak coincides with N\,83--2 (Figure~\ref{f:N83B3B6_13CO}).  N\,83--1 and N\,83--2 are associated with the densest gas in the star-forming region N\,83. 

We did not detect any hydrogen recombination lines covered by our Band 6 and the archival Band 3 observations (e.g., H40$\alpha$, H36$\beta$, H41$\gamma$; see \citealt{sewilo2023}) toward hot cores N\,83A--mm A and B, indicating that the protostars have not started ionizing their immediate environment yet.  No compact cm-wave radio emission was detected toward our ALMA fields by \citet{indebetouw2004} in their high-resolution ATCA observations at 4.8 GHz (6.2 cm) and 8.6 GHz (3.5 cm) with the resolution of $2\rlap.{''}19\times1\rlap.{''}70$ and $1\rlap.{''}82\times1\rlap.{''}24$, respectively.

\begin{figure*}[ht!]
\centering
\includegraphics[width=\textwidth]{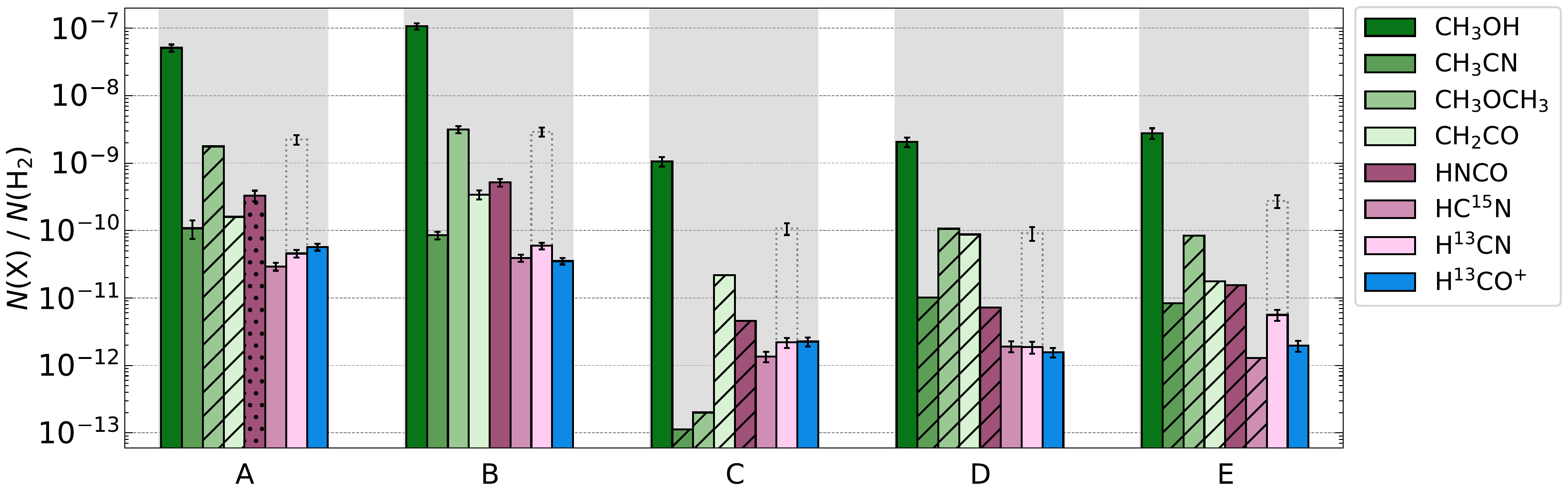}
\includegraphics[width=\textwidth]{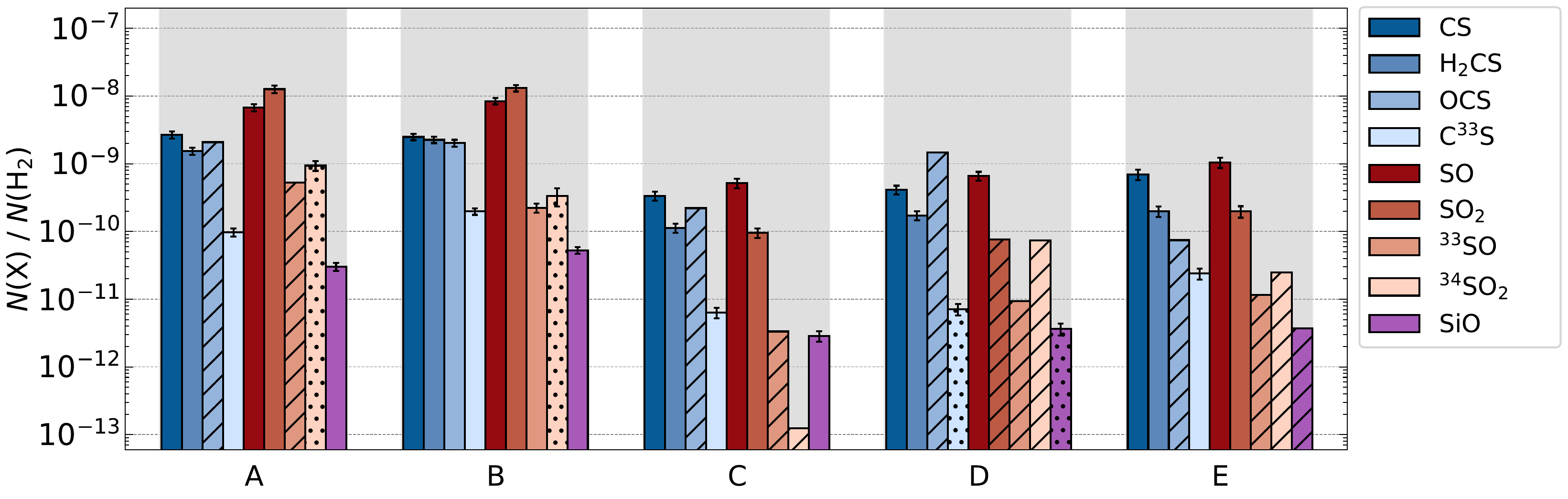}
\caption{Comparison of molecular abundances between all the cores detected in the ALMA fields N\,83--1 (B) and N\,83--2 (A, C--E).  In the upper panel, we show the abundances of COMs, N-bearing species, and \ce{H^13CO+}.  The open bars overlapping with  color bars for \ce{H^13CN} indicate $N(\ce{HCN})/N(\ce{H2})$ derived from $N(\ce{H^13CN})/N(\ce{H2})$ using the LMC \ce{^12C}/\ce{^13C} isotope ratio of $49\pm5$ (\citealt{wang2009}; see the text for details). In the lower panel, we compare the abundances of S-bearing species and SiO.  Upper limits are indicated with diagonal hatch marks and abundances of the tentatively detected species are labeled with dots.  \label{f:abundN83}}
\end{figure*}

\begin{figure*}
\centering
\includegraphics[width=0.49\textwidth]{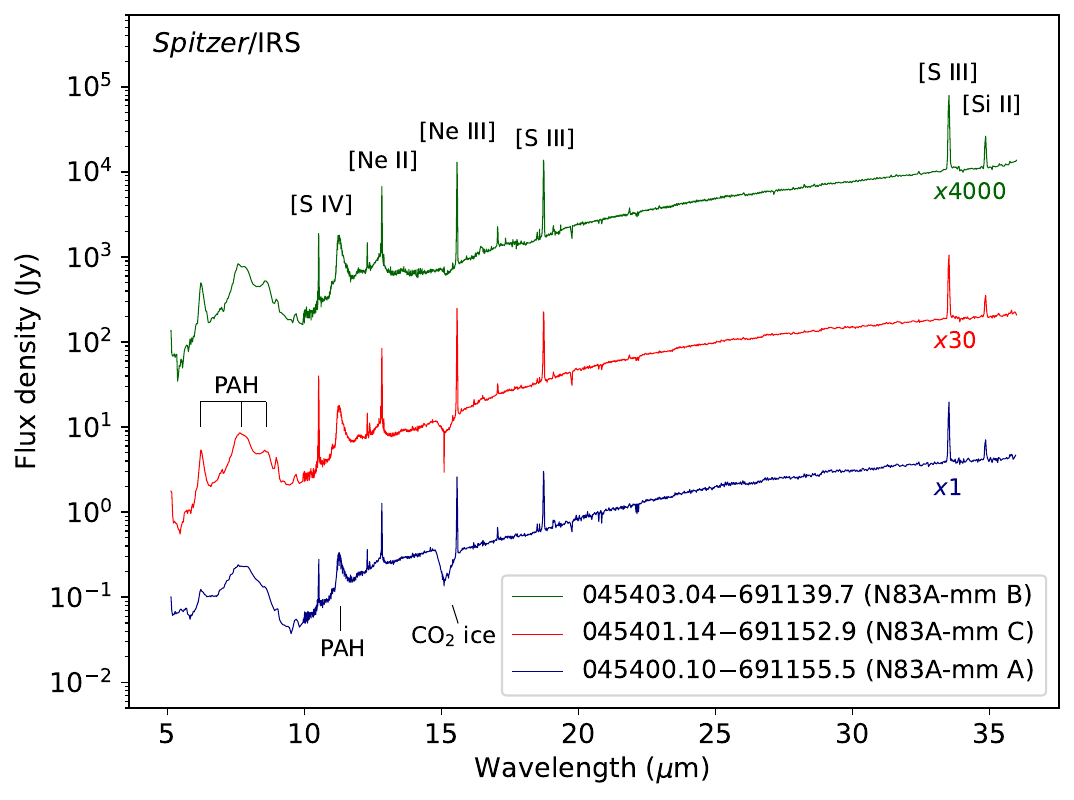}
\includegraphics[width=0.49\textwidth]{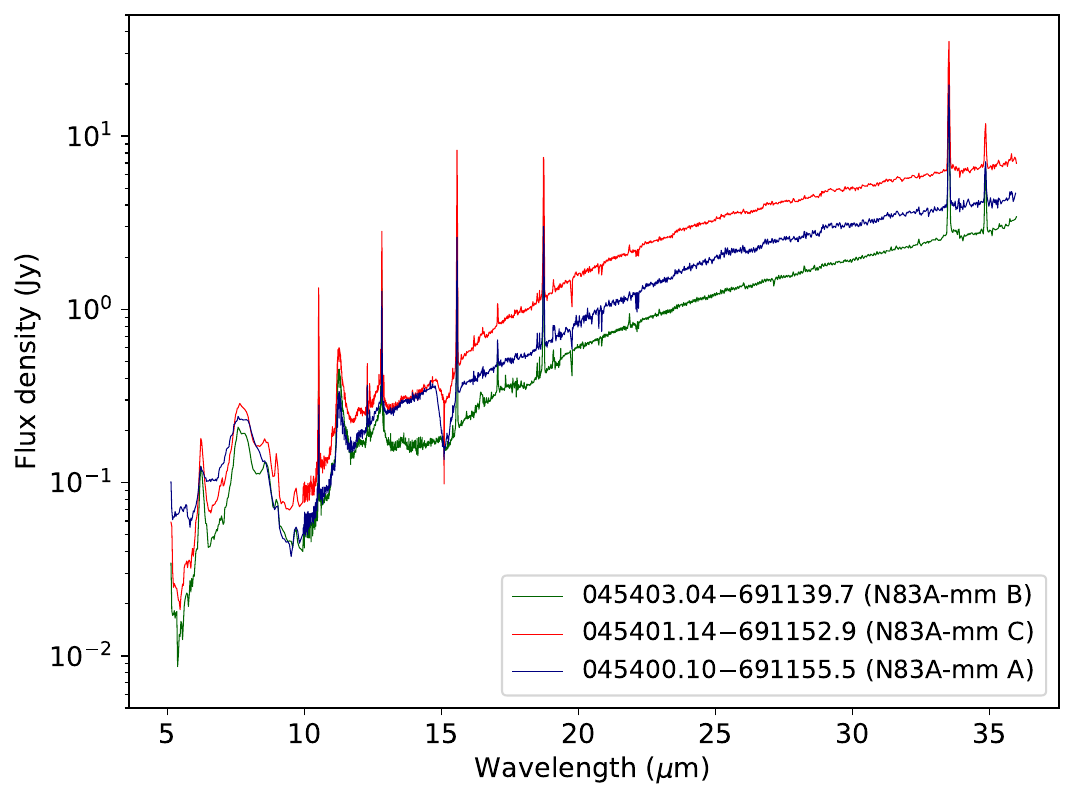}
\caption{The {\it Spitzer}/IRS spectra of embedded YSOs associated with ALMA continuum sources N\,83A--mm\,B (green), N\,83A--mm\,C (red), and N\,83A--mm\,A (navy). PAH and CO$_2$ ice features and fine-structure lines are labeled. The spectra in the left panel have been scaled by the following multiplicative factors for clarity (from bottom to top): 1, 30, and 4000. The narrow absorption feature overlapping with the CO$_2$ ice absorption band in the spectrum of 045401.14$-$691152.9 is an artifact. All spectra were analyzed and classified in \citet{seale2009} and \citet{jones2017}; see Section~\ref{s:ysosmas}.  \label{f:N83spitzerIRS}} 
\end{figure*}


\subsection{Molecular Abundances: N\,83}
\label{s:abunN83}

In the bar plots in Figure~\ref{f:abundN83}, we compare the fractional molecular abundances with respect to H$_2$, $N({\rm X})$/$N({\rm H_2})$, for all the cores analyzed in this paper: N\,83A--mm A--E.  The values of $N({\rm X})$/$N({\rm H_2})$, calculated as described in Sections~\ref{s:modeling} and \ref{s:NH2}, are provided in Table~\ref{t:XCLASSresults}. 

The plots demonstrate that in terms of the number of securely detected species, N\,83A--mm\,B is the most chemically-rich source in our study. It is the only source with the detection of a COM with more than six atoms, \ce{CH3OCH3}.  \ce{CH3OH} is detected toward all five sources with the highest abundance toward source B, followed by source A.  For each source, the \ce{CH3OH} abundance with respect to \ce{H2} is the highest among all the detected species.  \ce{CH3CN} is detected with a comparable abundance (within the uncertainties) only in the two hot cores (see also Section~\ref{s:HCident}). For any species, the abundances in hot cores are higher than in other sources. 

Our results include abundances of two isotopologues of \ce{HCN}: \ce{H^13CN} and \ce{HC^15N}.  We utilize $N(\ce{H^13CN})/N(\ce{H2})$ to determine $N(\ce{HCN})/N(\ce{H2})$ for all the cores by adopting the \ce{^12C}/\ce{^13C} ratio of $49\pm5$.  We selected \ce{H^13CN} since the \ce{H^13CN} line is brighter than the \ce{HC^15N} line for N\,83A--mm A--D, and \ce{HC^15N} was not detected toward source E.  The \ce{^12C}/\ce{^13C} isotope ratio was derived from optically thin HCN lines for the star-forming region N\,113 in the LMC by  \citet{wang2009}.  $N(\ce{HCN})/N(\ce{H2})$ is shown in the bar plot in Figure~\ref{f:abundN83} as open bars overlapping with the $N(\ce{H^13CN})/N(\ce{H2})$ color bars.

The comparison of the main isotopologues of the N- and S-bearing species shows that for hot cores, the abundances of \ce{CH3CN}, \ce{HCN}, and \ce{HNCO} with respect to \ce{H2} are lower than the abundances of \ce{SO} and hot \ce{SO2}.  The abundances of \ce{CH3CN} and \ce{HNCO}, the molecules only detected toward hot cores A and B, are also lower than the abundances of all the C\ce{-}S bond-bearing species species (\ce{CS}, \ce{H2CS}, and in case of source B, \ce{OCS}).  

As in hot cores, the abundance of \ce{HCN} with respect to \ce{H2} in cold cores N\,83A--mm C-E is lower than the abundance of \ce{SO}.  However, the abundance of \ce{HCN} is higher than the abundance of \ce{SO2} in two cold cores where \ce{SO2} is detected (C and E); this is also the case for the cold \ce{SO2} component in hot cores A and B. The \ce{SO} to \ce{SO2} abundance ratio is consistently found to be higher than one in the cold environments; it typically becomes lower than one in hot cores, as observed in N\,83 (e.g., \citealt{charnley1997,fontani2023,martinez2024}; \citealt{ghosh2024} and references therein).  

In the hot core N\,83A--mm\,B separated from A by 21$\rlap.{''}$8 arcsec or 5.2 pc, the abundances of most of the molecular species are higher than in A.  For example, the hot and cold \ce{CH3OH} abundance is a factor of 2.1 and 1.6, respectively, higher in source B than A. The abundances of N-bearing species \ce{H^13CN} and \ce{HC^15N} are a factor of 1.3 higher.  The abundances of shock tracers SO and SiO are, respectively, a factor of 1.2 and 1.8 higher in source B.  The abundance of \ce{CS} is comparable between the two hot cores, but \ce{H2CS} is a factor of $\sim$1.5 higher in B. The species with the higher abundance in source A than in B is \ce{H^13CO+} (by a factor of 1.6) that is sensitive to the cosmic rate ionization rate.  

Since the two hot cores seem to be forming in the same filamentary molecular cloud minimizes the chance of significant differences in the initial conditions (e.g., the temperature at the ice formation stage or metallicity). Thus, these chemical differences between sources A and B are likely regulated locally. We reported similar differences (up to a factor of 4 for S-bearing species) between the two hot cores in the star-forming region N\,105 (N\,105--2\,A and N\,105--2\,B) that are separated by only $\sim$3$\rlap.{''}$2 ($\sim$0.78 pc). In case of the hot cores in N\,105, the much more complex kinematics in 2\,B has the largest contribution to the observed chemical differences (see the discussion in \citealt{sewilo2022n105}).

\subsubsection{The {\it Spitzer}/IRS Spectra: YSOs and Their Environments}
\label{s:N83spitzerIRS}

In Figure~\ref{f:N83spitzerIRS}, we compare the {\it Spitzer}/IRS spectra of the three {\it Spitzer} YSOs in our ALMA fields (corresponding to sources A, B, and C) to check for any correlations between their chemical abundances and spectral characteristics. All spectra show the PAH features, fine-structure lines from the ionized gas, and the \ce{CO2} 15.2 $\mu$m ice band with varying intensities.  The interpretation of the {\it Spitzer}/IRS spectra is often challenging since the point-source spectra can be contaminated by the emission from the wider H\,{\sc ii} region given the large slit width of the IRS that varies with wavelength (3$\rlap.{''}$6/$\sim$0.9 pc at shorter to 11$\rlap.{''}$1/$\sim$2.7 pc at longer wavelengths). There is no indication in the ALMA data that the {\it Spitzer} YSOs in N\,83--1 and N\,83--2 have started ionizing their surroundings (no hydrogen recombination lines have been detected in the Band 6 and Band 3 observations), thus the environmental contribution to the spectra is very likely.  This contribution is confirmed for the hot core N83A--mm\,A that was observed with JWST MIRI MRS (M. Sewi{\l}o et al., in prep.). 

%
%
The MIRI MRS spectrum of N83A--mm\,A is of a deeply embedded protostar with silicate features at 10 and 18 $\mu$m, simple and COM ice bands, H$_2$ emission lines, and no PAH features. The very faint fine-structure lines are detected with intensities increasing with the increasing spectral extraction aperture, indicating the contribution from the environment.  PAH features also appear with a larger aperture. The highly-ionized region is located to the southeast from N83A--mm\,A in the small field of view of MIRI MRS; it is likely part of the ionized bubble's rim. 

The PAH features and fine-structure lines in the {\it Spitzer}/IRS spectra are the weakest toward N\,83A--mm\,A, consistent with this sources being in the densest and thus most shielded against UV radiation gas in the region. The most prominent \ce{CO2} 15.2 $\mu$m ice band indicates that this source is likely the youngest of the three YSOs. However, there is some uncertainty about whether the  {\it Spitzer} YSO in N\,83--1 is the same source as the ALMA source B, and therefore the {\it Spitzer}/IRS spectrum may not be reflecting the evolutionary stage of the hot core N83A--mm\,B.  

 We explore the positional offset between the infrared sources and the 1.2 mm continuum source N\,83A--mm\,B in the three-color mosaic in Figure~\ref{f:Bmultiple}, combining the {\it Spitzer}/IRAC 8.0 $\mu$m, VMC $K_{\rm s}$, and 1.2 mm continuum images.  The morphology of the {\it Spitzer} and VMC emission indicates that two unresolved sources may be present at this location, both offset from source B.  The {\it Spitzer} and VMC sources coincide, and their peaks are offset from source B by $\sim$0$\rlap.{''}$8. For comparison, the other two  {\it Spitzer} YSOs are offset from the 1.2 mm continuum source positions (A and C) by $\sim$0$\rlap.{''}$3, comparable to the VMC pixel size.    

Source B may not be the same source as the nearest YSO detected in the infrared. Each of the two positions correspond to the SO and \ce{SO2} peaks (Figures~\ref{f:so2int} and \ref{f:alma3color}).  The peak of the lower-excitation \ce{SO2} emission ($E_{\rm U}$$\sim$24 K) coincides with the known {\it Spitzer} YSO, while the peak of the high-excitation \ce{SO2} emission ($E_{\rm U}$$\sim$94 K) is associated with the 1.2 mm continuum intensity peak (see Figure~\ref{f:so2int}). The emission peaks of other species coincide with the hot core or are slightly offset to the southwest (e.g., CS; see Figures~\ref{f:N831mom0simple} and \ref{f:alma3color}). 

In N\,83A--mm\,A, both the low-excitation and high-excitation \ce{SO2} emission, as well as SO peaks coincide with the 1.2 mm continuum peak.  The low-excitation \ce{SO2} emission is elongated in the south-north direction with a fainter peak in the northeast.  Similarly to N\,83A--mm\,B, the CS emission peak is slightly offset from the 1.2 mm continuum peak. 

\subsubsection{Kinematics in N\,83--1 and N\,83--2}
\label{s:N83kinematics}

Chemical abundances can be significantly altered in the presence of large-scale motions producing shocks (e.g., large-scale flows, outflows, turbulence).  In Figures~\ref{f:kinematicsCS}, \ref{f:kinematicsSO}, and \ref{f:kinematicsCH3OH}, we show the line peak intensity (moment 8), intensity-weighted velocity (moment 1), and velocity dispersion (moment 2) images for \ce{CS}, \ce{SO}, and \ce{CH3OH}, respectively, allowing us to get a glimpse at the molecular gas kinematics in N\,83--1 and N\,83--2; a detailed analysis of the kinematics is out of the scope of this paper. The emission of the species shown in Figures~\ref{f:kinematicsCS}--\ref{f:kinematicsCH3OH} has the most extended distribution out of all the detected species.  

The images reveal the complex velocity patterns in both fields with velocity variations of up to 10 km s$^{-1}$.  The velocity also spreads over 8--10 km s$^{-1}$ along the line of sight at most locations along the filament (in both fields), exceeding 10 km s$^{-1}$ at the position of N\,83A--mm\,A.  The multi-directional flows appear to be present, as well as overlapping velocity components (e.g., the high- and low-velocity components in the west in N\,83--2). The overall velocity structure is similar between \ce{CS}, \ce{SO}, and \ce{CH3OH}, although the \ce{SO} and \ce{CH3OH} maps are more similar to each other than to \ce{CS}.  The better match between \ce{SO} and \ce{CH3OH} is also evident in the integrated intensity maps with the \ce{SO} and \ce{CH3OH} emission brighter in source B than A.  In contrast, the \ce{CS} integrated intensity is comparable in both ALMA fields. The \ce{SO} and \ce{CH3OH} velocity dispersion maps also show similar patterns that differ from \ce{CS}.

The broadest \ce{SO} and \ce{CH3OH} lines are observed toward the hot core N\,83A--mm\,A, the filament extending from A toward northeast, and in the area northwest from source C. These locations are associated with the brightest \ce{SO2} and \ce{SiO} emission indicating (together with broad lines) the shock wave origin of the \ce{SO}, \ce{SO2}, \ce{SiO}, and \ce{CH3OH} emission.   

Sputtering or destruction of refractory grain cores in shocks releases the S and Si atoms to the gas. The increased number of S and Si atoms available for chemical reactions enhances the production of the S- and Si-bearing molecules (e.g., \citealt{schilke1997,gusdorf2008,vandishoeck2018} and references therein).  

The sputtering of interstellar dust grains also releases a rich molecular inventory of the grain-surface chemistry products (including \ce{CH3OH}) to the gas phase.  Injected from ices, \ce{H2O} and \ce{H2S} can enhance the production of S-bearing molecules such as \ce{SO} and \ce{SO2} (e.g., \citealt{charnley1997}). Thus, the spatial correspondence of the \ce{CH3OH}, \ce{SO}, \ce{SO2}, and \ce{SiO} emission reveals the presence of shocks.  \ce{CS} is thought to form efficiently in the gas phase, and thus it is expected to be less closely correlated with the shock tracers, as observed in N\,83.

In both ALMA fields, the extended cold \ce{CH3OH} and \ce{SO} emission is also detected in regions where no \ce{SO2} and \ce{SiO} emission is present. In these regions, the production of  \ce{CH3OH} and \ce{SO} may be driven by the same processes, but in weaker shocks where the sputtering yields are lower and the refractory grain cores are not destroyed.  In highly-dynamic environments such as in N\,83, feedback-produced slow shocks (e.g., shocks produced by an expanding bubble, supersonic gas turbulence, protostellar outflows, colliding gas flows) may be present throughout the region. 

In addition to shock sputtering and thermal desorption, other desorption mechanisms have been proposed for \ce{CH3OH} and other ice-mantle molecules. These include photodesorption (e.g., \citealt{bertin2013}), transient heating of dust grains in low-velocity grain–grain collisions (e.g., \citealt{markwick2001,dickens2001,buckle2006,harju2020}), grain heating following cosmic-ray impact (e.g., \citealt{leger1985}), cosmic-ray sputtering (e.g., \citealt{wakelam2021}), reactive desorption (e.g., \citealt{minissale2016,chuang2018}).  The grain-grain collision and reactive desorption could be efficient in UV-shielded (inside molecular cloud peaks) and low-cosmic ray environments. We discuss the origin of the hot and cold \ce{CH3OH} and \ce{SO2}/\ce{SO} emission in more detail in \citet[see also Sections~\ref{s:abunN83other} and \ref{s:outflows}]{sewilo2022n105}.

\begin{figure*}[ht!]
\centering
\includegraphics[width=\textwidth]{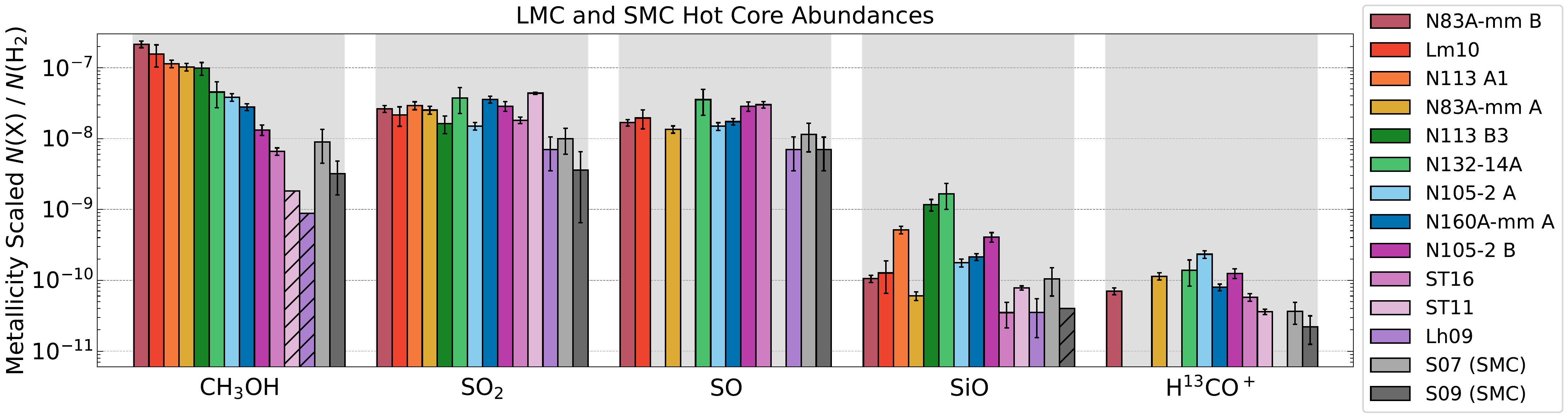}
\includegraphics[width=\textwidth]{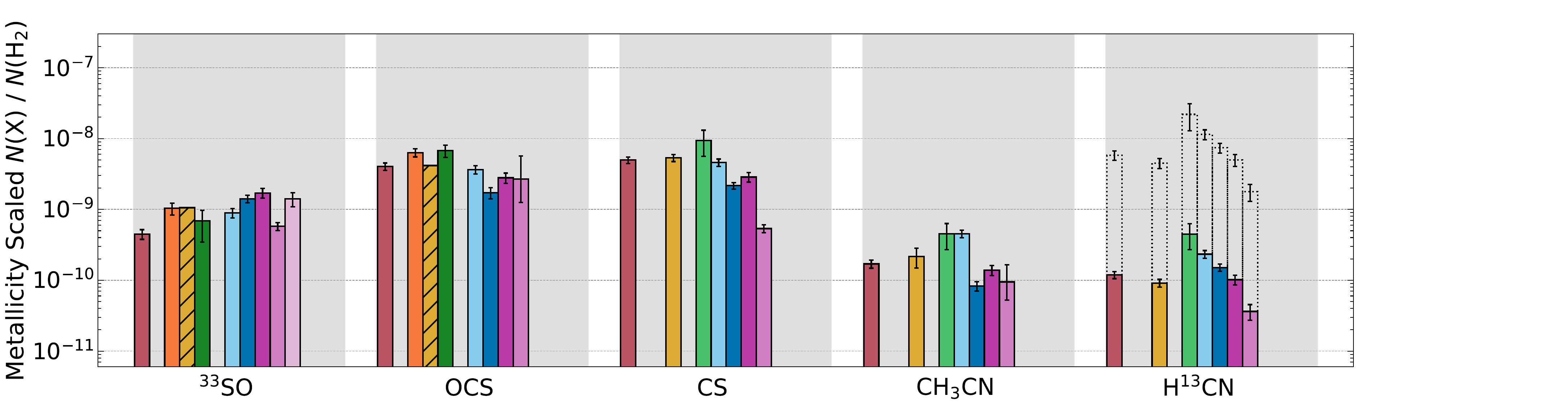}
\caption{Comparison of metallicity-scaled (multiplied by 1/$Z_{\rm LMC}$ or  1/$Z_{\rm SMC}$, i.e., all scaled to 1\,Z$_{\odot}$) abundances with respect to H$_2$ for selected species between the LMC (color bars) and SMC (shades of grey) hot cores. The open bars overlapping with color bars for  \ce{H^13CN}  indicate the metallicity-scaled $N(\ce{HCN})/N(\ce{H2})$ derived from $N(\ce{H^13CN})/N(\ce{H2})$ using the LMC \ce{^12C}/\ce{^13C} isotope ratio of $49\pm5$ (\citealt{wang2009}; see Section~\ref{s:abunN83}).  Upper limits are indicated with diagonal hatch marks.  See Section~\ref{s:abunN83other} for references. \label{f:abund}}
\end{figure*}

\subsection{Molecular Abundances: Newly-identified Hot Cores in N\,83 vs. Other Magellanic Hot Cores}
\label{s:abunN83other}

\begin{figure*}[ht!]
\centering
\includegraphics[width=\textwidth]{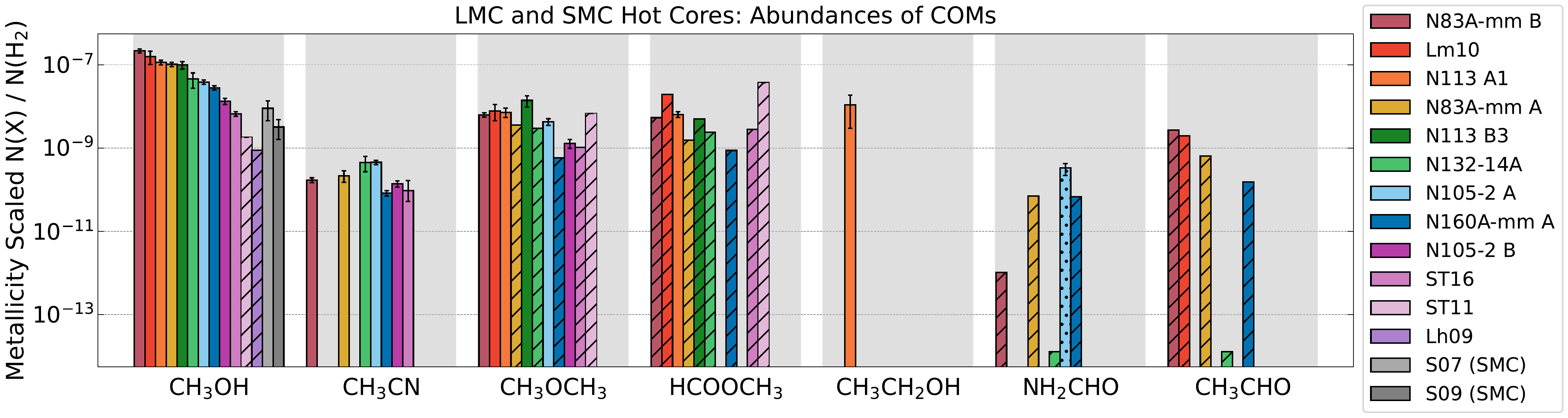}
\caption{Metallicity-scaled COM abundances toward the LMC (color bars) and SMC (shades of grey) hot cores. CH$_3$OH is the only COM detected in the SMC to date. Upper limits are indicated with diagonal hatch marks and the tentative detection is marked with dots. \label{f:abundCOMs}}
\end{figure*}

In Figures~\ref{f:abund} and \ref{f:abundCOMs}, we compare the fractional abundances with respect to H$_2$ of the newly detected hot cores in N\,83 to those of previously known hot cores in both the LMC (10) and SMC (2).  The LMC sample consists of ST11 (\citealt{shimonishi2016b}), N\,113 A1 and B3 (\citealt{sewilo2018}),  ST16 (\citealt{shimonishi2020}), N\,105--2\,A and B (\citealt{sewilo2022n105}), 14A (\citealt{hamedanigolshan2024}), N\,160A--mm A (\citealt{broadmeadow2025}), Lh09 and Lm10 (\citealt{shimonishi2026}).  The SMC hot cores are S07 and S09 (\citealt{shimonishi2023}).  We scale the LMC and SMC abundances to the common (solar) metallicity by multiplying them by a factor of  2 (1/$Z_{\rm LMC}$) and 5 (1/$Z_{\rm SMC}$), respectively. 

The fractional abundances with respect to H$_2$ of all previously known hot cores except Lh09 and Lm10 are as described in \citet{broadmeadow2025}. They are either measured using the same method and the gas-to-dust ratio as for N\,83 (N\,113, N\,105, N\,160) or have been rescaled/recalculated (by revising $N({\rm H_2})$) for consistency (ST11, ST16, and 14A). For the most recently detected hot cores, Lh09 and Lm10, we rescale their abundances to the same value of the gas-to-dust ratio as for the remaining sources.  

The temperature used by \citet{shimonishi2026} to determine $N({\rm H_2})$ for Lm10 is the average of  $T(\ce{CH3OH})$ and $T(\ce{SO2})$ (high-$T$) rather than $T(\ce{CH3OH})$ as for other hot cores in our sample. Since the two temperatures are similar, we do not recalculate $N({\rm H_2})$ for Lm10 using $T(\ce{CH3OH})$. For Lh09, \citet{shimonishi2026} did not detect \ce{CH3OH}, thus $T(\ce{SO2})$ (high-$T$) was used to determine $N({\rm H_2})$.

In the bar plots in Figures~\ref{f:abund} and \ref{f:abundCOMs}, the sources are plotted in the order of decreasing \ce{CH3OH} abundance (from left to right). Figure~\ref{f:abund} compares the abundances of \ce{CH3OH}, \ce{CH3CN}, and simple species detected toward most of the known hot cores, while Figure~\ref{f:abundCOMs} compares abundances of COMs. The abundances of hot cores N\,83A--mm A and B fit into the trends previously reported for the Magellanic hot cores: the large variations in the \ce{CH3OH} abundances and less significant variations for S-bearing species. 

The abundance of SiO shows large variations in the Magellanic hot cores, but there is no correlation with the abundance of \ce{CH3OH}. No clear correlation is also evident between the abundance of \ce{H^13CO+} (sensitive to the radiation fields) and \ce{CH3OH}. In general, \ce{SO2} and \ce{SO} have the highest abundances out of all the species shown in Figure~\ref{f:abund}. Compared to all S-bearing species (including the \ce{SO} isotopologue, \ce{^33SO}), \ce{CH3CN} has the lowest abundances with respect to \ce{H2}.  The abundances of another N-bearing species, \ce{HCN} (derived from \ce{H^13CN} as described in Section~\ref{s:abunN83}), are not as low as for \ce{CH3CN}; for some hot cores they are comparable to the abundances of S-bearing species. 

Figure~\ref{f:abundCOMs} shows that \ce{CH3OH}, \ce{CH3CN}, and \ce{CH3OCH3} are the only COMs detected toward two or more hot cores. N\,83A--mm\,B has the highest abundance of \ce{CH3OH}, followed by Lm10, and N\,113\,A1. \ce{CH3OCH3} has been detected toward all three sources. \ce{CH3CN} has been detected toward N\,83A--mm\,B, but the observations of Lm10 and N\,113\,A1 did not cover this species. N\,113\,A1, one of the first two hot cores with COMs identified in the LMC, has the largest number of COM detections among the Magellanic hot cores:   \ce{CH3OH}, \ce{CH3OCH3}, \ce{HCOOCH3}, and recently detected by \citet{shimonishi2026}, \ce{CH3CH2OH}. \ce{CH3OH} is the only COM detected in the SMC to date. The \ce{CH3OH} abundances of the SMC hot cores are at the lower end of those measured in the LMC, but the \ce{CH3OH} emission appears to originate in the outflows rather than hot cores \citep{shimonishi2023}.

The \ce{CH3OH}, \ce{SO}, and \ce{SO2} abundance patterns observed in the LMC and SMC have been discussed in the literature in the context of differences in their formation pathways (see e.g., \citealt{shimonishi2023}). \ce{CH3OH} is primarily the grain-surface chemistry product (e.g., \citealt{garrod2022}), and thus its abundance at the hot core stage strongly depends on the dust temperature at the earlier, colder phase when CO starts to catastrophically freeze-out onto the dust grains.  For example, astrochemical simulations show that warmer dust grains (as found in the LMC and SMC) result in less efficient production of \ce{CH3OH} in the icy grain mantles (\citealt{acharyya2015,shimonishi2016a}).  The spatial variations of the dust temperature in the cold phase would lead to differences in the \ce{CH3OH} abundance in the hot core stage. 

\ce{SO} and \ce{SO2} form efficiently in hot gas-phase chemistry following ice sublimation (e.g., \citealt{charnley1997,vidal2018}; see also Section~\ref{s:outflows}).  The chemistry of \ce{SO} and \ce{SO2} on icy dust grains, and sulfur chemistry in general, are not well understood. The detections of \ce{SO2} ice toward protostellar objects are scarce, even in the Galaxy (e.g., \citealt{rocha2024,chen2024,vandishoeck2025}). The detection of the \ce{SO2} ice in the protostar ST6 in the LMC has recently been reported by \citet{sewilo2025}.  The observed \ce{SO2} ice may be a product of hot gas chemistry followed by freeze-out or form in grain surface reactions through energetic processes (e.g., \citealt{nguyen2024,santos2024,martindomenech2025}).  \citet{santos2024} argue that the formation of \ce{SO2} ice likely occur during the \ce{H2O} ice formation stage in the low-density, pre-stellar cores. The \ce{SO2} (and \ce{SO}) are released to the gas by ice sublimation or shock sputtering. Both Galactic and LMC observations show that unlike \ce{CH3OH}, the \ce{SO} and \ce{SO2} gas abundances strongly depend on the gas temperature (e.g., \citealt{fontani2023,martinez2024,shimonishi2026}). 

As discussed in Section~\ref{s:N83kinematics}, the chemistry in star-forming regions can be strongly altered by shocks. In Section~\ref{s:outflows}, we discuss in more detail some expected effects of shock chemistry relevant to the observed \ce{CH3OH}, \ce{SO}, and \ce{SO2}  abundance patterns.

\subsection{Molecular Abundances: Magellanic vs. Galactic Hot Cores}
\label{s:abunN83gal}

\begin{figure*}[ht!]
\centering
\includegraphics[width=\textwidth]{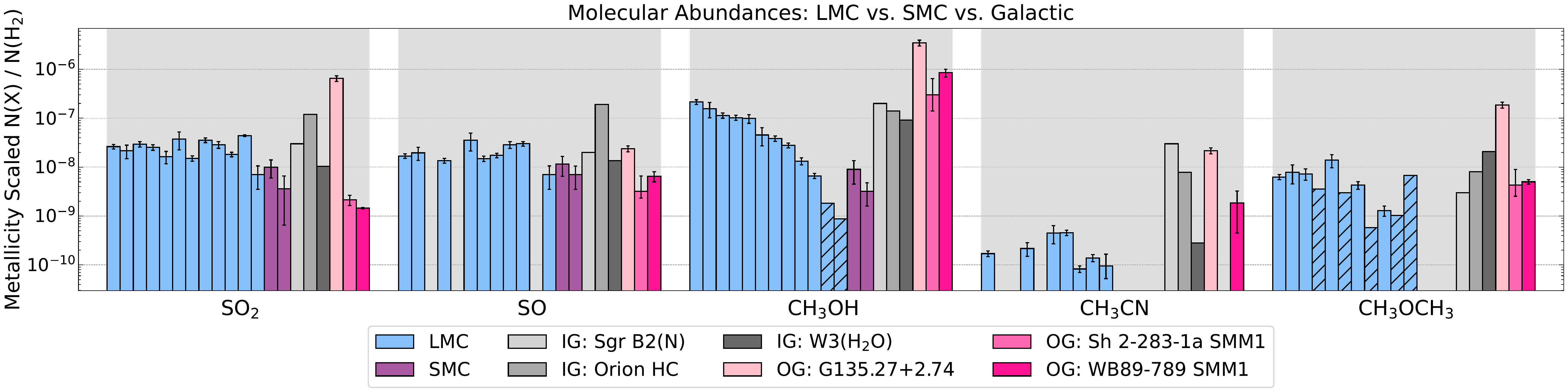}
\caption{Comparison of the metallicity-scaled molecular abundances observed toward the LMC (blue) and SMC (purple) hot cores, selected inner Galaxy hot cores (shades of grey), and all known outer Galaxy hot cores (shades pink). The order of the LMC/SMC hot cores is the same as in Figures~\ref{f:abund}--\ref{f:abundCOMs}. The data for the inner Galaxy hot cores come from \citet{blake1987} and \citet{sutton1995} for Orion Hot Core (HC), from \citet{nummelin2000} for Sgr B2(N), and from \citet{helmich1997} for W3(\ce{H2O}). For the outer Galaxy hot cores WB\,89--789 SMM1, Sh\,2--283--1a SMM1, and G135.27$+$2.79 MM1, the data come from \citet{shimonishi2021}, \citet{ikeda2026}, and \citet{wang2026}, respectively.  Abundance upper limits are indicated with diagonal hatch marks. \label{f:abundGal}}
\end{figure*}

Following \citet{sewilo2022n105} and \citet{broadmeadow2025}, in Figure~\ref{f:abundGal}, we compare the metallicity-scaled abundances of the LMC and SMC hot cores (including the newly detected hot cores in N\,83) to selected inner Galaxy hot cores (i.e., located within the solar circle) with abundances measured based on the single-dish observations on similar physical scales (within a factor of a few) and using similar methods. The sample includes the Orion Hot Core, W3(\ce{H2O}), and Sgr\,B2(N). The fractional abundances for the Orion Hot Core come from \citet{sutton1995} except for \ce{CH3CN} that comes from  \citet{blake1987}.  The data for W3(\ce{H2O}) and Sgr B2(N) come from \citet{helmich1997} and \citet{nummelin2000}, respectively. We include key molecules, namely \ce{SO2}, SO, and COMs securely detected in at least two Magellanic hot cores  (\ce{CH3OH}, \ce{CH3CN}, and \ce{CH3OCH3}). The LMC and SMC hot cores are plotted in the same order as in Figures~\ref{f:abund} and  \ref{f:abundCOMs}. 

 In addition to the inner Galaxy hot cores, we also include the only three known hot cores in the low-metallicity regions beyond the solar circle (i.e., in the outer Galaxy): WB\,89--789 SMM1 (\citealt{shimonishi2021}), Sh\,2--283--1a SMM1 (\citealt{ikeda2026}), and G135.27$+$2.79 MM1 (\citealt{wang2026}). The outer Galaxy represents a distinct environment from the inner Galaxy.  The metallicity,  dust-to-gas mass ratio (\citealt{giannetti2017GDR}),  molecular fraction of the total gas (e.g., \citealt{nakanishi2016}), and the star-formation rate density (e.g., \citealt{elia2022}) decrease with increasing Galactocentric distance, $R_{\rm GC}$.  The mass surface density of molecular clouds and their number density is lower in the outer Galaxy (e.g., \citealt{heyer2015}). With more sparsely distributed molecular clouds, cloud--cloud collisions are less common.  Other external triggers of star formation such as spiral arms or supernova remnants are less significant (\citealt{brand1995}).  The diffuse $\gamma$-ray observations tracing cosmic rays suggest spatial variations of the cosmic ray density throughout the Galaxy (e.g., \citealt{bloemen1987,erlykin1996,abdo2010OG,recchia2016}). The cosmic ray density shows an enhancement in the inner Galaxy (around $\sim$4 kpc), but it drops off in the Galactic center.  It also decreases beyond the solar circle; however, the gradient is significantly flatter than expected from the rapidly declining distribution of the presumed cosmic ray sources (i.e., supernova remnants) in the outer Galaxy.  The interstellar radiation field and the dust temperature decrease with increasing $R_{\rm GC}$ (e.g., \citealt{mathis1983}).  Compared to the outer Galaxy, the UV radiation field is stronger and the dust temperature is higher in the LMC and SMC which are associated with a significantly more active star formation (e.g., \citealt{browning2003,vanloon2010lmc,vanloon2010smc,fukui2008,meixner2006,gordon2011}).  

All the outer Galaxy hot cores reported in the literature have been identified based on the interferometric observations probing the linear scales of $\sim$0.03 pc.  Hot cores (WB\,89--789 SMM1, Sh\,2--283--1a SMM1, G135.27$+$2.79 MM1) are located at $R_{\rm GC}$, of (19.0, 15.7, 13.1) kpc, in regions with the metallicity of about (0.20, 0.30, 0.42) $Z_{\odot}$, as predicted by the negative Galactocentric metallicity gradient (traced by the oxygen abundance) of \citet[assuming 12+{\rm log(O/H)} of 8.69 in the solar neighborhood; \citealt{asplund2009}]{fernandezmartin2017}. The \citet{fernandezmartin2017} analysis is based on the optical spectroscopy data for a sample of 23 H\,{\sc ii} regions with $R_{\rm GC}$ between 11 and 18 kpc.  We use these metallicities to scale molecular abundances of WB\,89--789 SMM1, Sh\,2--283--1a SMM1, G135.27$+$2.79 MM1 to the solar metallicity in Figure~\ref{f:abundGal}. 

For WB\,89--789 SMM1, \citet{shimonishi2021} provide abundances with respect to \ce{H2} based on the spectra extracted from a 0.026 pc diameter region (matching the linear resolution of their observations), as well as a 0.1 pc diameter region (matching the typical size of a Galactic hot core and the resolution of the ALMA observations of the LMC and SMC).  Abundances of WB\,89--789 SMM1 plotted in Figure~\ref{f:abundGal} are calculated for the 0.1 pc diameter region.  For G135.27$+$2.79 MM1, \citet{wang2026} provide \ce{CH3OH}, \ce{CH3OCH3}, and \ce{CH3CN} column densities for two velocity components; the total column densities used to derive $N(\ce{CH3OH})/N(\ce{H2})$, $N(\ce{CH3OCH3})/N(\ce{H2})$, and $N(\ce{CH3CN})/N(\ce{H2})$ shown in Figure 16 are the sum of both components. 

In general, the molecular complexity and abundances in the LMC and SMC hot cores differ from their counterparts in both the inner and outer Galaxy. Despite the low metallicity at large $R_{\rm GC}$, the outer Galaxy hot cores display a great molecular complexity. Multiple COMs are detected toward WB\,89--789 SMM1, Sh\,2--283--1a SMM1, G135.27$+$2.79 MM1 with metallicity-scaled abundances resembling the inner Galaxy hot cores (those with similar luminosities in particular). Based on astrochemical modeling, \citet{wang2026} noted that the sensitivity on the lower metallicity and dust-to-gas ratio varies between (both simple and complex) molecules. 

Figure~\ref{f:abundGal} shows that in terms of the \ce{CH3OH} abundance, five LMC hot cores with the highest \ce{CH3OH} abundances are most similar to the inner and outer Galaxy hot cores. Their metallicity-scaled \ce{CH3OH} abundances are comparable or larger than the \ce{CH3OH} abundance in W3(\ce{H2O}) which is the lowest value among the Galactic hot cores plotted in Figure~\ref{f:abundGal}. The remaining Magellanic hot cores have significantly lower \ce{CH3OH} abundances than in both the inner and outer Galaxy hot cores. Metallicity-scaled \ce{CH3OCH3} abundances in five out of six LMC hot cores with the \ce{CH3OCH3} detection are comparable to the inner and outer Galaxy hot cores.  \ce{CH3OCH3} has not been detected toward the SMC hot cores and the LMC hot cores with the lowest \ce{CH3OH} abundances. 

The efficient {\it cold} \ce{CH3OH} formation at large Galactocentric distances has  been reported in the literature based on single-dish observations (e.g., \citealt{ruffle2007,bernal2021,fontani2022b}), indicating that \ce{CH3OH} is ubiquitous in the low-metallicity environment of the outer Galaxy.  For the outer Galaxy star-forming regions with $R_{\rm GC}$ ranging from 9 kpc to $\sim$24 kpc, \citet{bernal2021} and \citet{fontani2022b} measured $N(\ce{cold CH3OH})/N(\ce{H2})$ that generally scale with metallicity (see also \citealt{gigli2025}).  The ubiquity of  cold \ce{CH3OH} is also evident in the LMC (see Section~\ref{s:mom} and Figure~\ref{f:distribution}). 

Figure~\ref{f:abundGal} reveals that for S-bearing species (\ce{SO} and \ce{SO2}) in hot cores, the metallicity-scaled abundances with respect to \ce{H2} in the LMC and SMC hot cores are comparable to those in the inner Galaxy hot cores; however, the \ce{SO2} and \ce{SO} abundances of Lh09 (LMC) and S09 (SMC) are below the values observed toward all three inner Galaxy hot cores in our comparative sample.  For the two outer Galaxy hot cores furthest away from the Galactic Center, WB\,89--789 SMM1 and Sh\,2--283--1a SMM1, the \ce{SO} and \ce{SO2} abundances are lower than for any other hot core shown in Figure~\ref{f:abundGal}.  It is particularly striking for \ce{SO2}. \citet{shimonishi2021} report that the a significant abundance decrease is not observed for the \ce{C-S} bond-bearing species \ce{CS}, \ce{H2CS}, and \ce{OCS} in WB\,89--789 SMM1. 

\citet{shimonishi2021} and \citet{ikeda2026} suggest that the reduced \ce{SO2} abundances in WB\,89--789 SMM1 and Sh\,2--283--1a SMM1 compared to the inner Galaxy and the LMC/SMC may be due to lower cosmic-ray intensity in the outer Galaxy.  The origin of cosmic rays is still not well understood, particularly for low-energy cosmic rays ($\ll$1 GeV) relevant to astrochemistry (e.g., \citealt{indriolo2013,padovani2020,gabici2022}).  \citet{tang2026} show that low-energy cosmic rays may be generated in situ by the star formation activities rather than being part of the general population of Galactic cosmic rays (described above).  Since the outer Galaxy is a significantly more quiescent environment than the LMC and SMC, the low-energy cosmic ray ionization rate is expected to be lower. \citet{ikeda2026} argue that the weak cosmic-ray environment  results in insufficient production of the O and OH radicals in the icy grain mantles, leading to the lower \ce{SO2} abundance. The weaker cosmic-ray induced UV radiation would also make the formation of \ce{SO} and \ce{SO2} in the hot gas less efficient (e.g., \citealt{vangelder2021}).  

Based on the analysis of the single-dish data for 25 molecular clumps with the detection of \ce{SO}, \citet{gigli2025} found that the $N(\ce{SO})/N(\ce{H2})$ abundance gradient is steeper than that of the elemental abundance of S ([S/H]) between $\sim$12 and $\sim$24 kpc, indicating a lower \ce{SO} formation efficiency in the outer Galaxy (the study does not include \ce{SO2}).  Since \ce{SO} can be a shock tracer, the authors argue that the lower \ce{SO} formation efficiency may be due to the reduced effects of outflow shocks on the gas composition at large $R_{\rm GC}$.  This scenario is supported by the fact that line widths for all species are decreasing with $R_{\rm GC}$, indicating a more quiescent environment in the outer Galaxy. The lower \ce{SO} efficiency leads to a lower \ce{SO2} efficiency as \ce{SO} is an immediate chemical precursor to \ce{SO2} in both hot ($>$100 K) and cold ($\sim$10 K) regions (e.g., \citealt{millar1990,charnley1997,chen2015chem,mifsud2021}).

Finally, Figure~\ref{f:abundGal} shows that the \ce{CH3CN} abundance differs the most between the LMC and Galactic hot cores (no \ce{CH3CN} has been detected in the SMC to date), although the \ce{CH3CN} abundances in 14A and N\,105--2\,A are comparable to that in W3(\ce{H2O}). It should be noted that the molecular abundances reported in the literature for a given Galactic hot core (including W3(\ce{H2O})) can vary significantly depending on the spatial resolution of the observations and the method used to derive molecular column densities (see e.g., \citealt{gerner2014,helmich1997,bisschop2007}). 

The more reliable comparison between hot cores in the LMC, SMC, and the outer Galaxy (and within each of these environments) requires a better characterization of the local physical and chemical conditions in star-forming regions hosting them. The local UV radiation field strengths, cosmic-ray ionization rates, or metallicities have not been quantified. The observations show that large variations in all these properties are expected within each system. For example, the large metallicity variations have been reported in both the LMC and SMC (see Section~\ref{s:inhomogeneities}).  In the outer Galaxy, there are significant differences in the metallicity at a given $R_{\rm GC}$ predicted by different metallicity gradients even for those based on the same type of objects, i.e., Cepheid variables or H\,{\sc ii} regions (e.g., \citealt{balser2011,wenger2019}). The azimuthal variations of the oxygen abundance have also been reported in the literature, contributing to the uncertainties in the metallicity predictions at a given location in the Galaxy (e.g., \citealt{wenger2019,kovtyukh2022}). 
   
\subsection{The Integrated High-excitation SO Line Intensity as a Hot Core Tracer}

In a survey of 30 sources in the LMC with luminosities ranging from 10$^4$ to 10$^{5.5}$  $L_{\odot}$, \citet{shimonishi2026} found 9 hot cores (1 new) and 1 hot core candidate.  The spectra of all 9 hot cores analyzed in the \citet{shimonishi2026} show a strong SO $8_8$--$7_7$ line ($E_{\rm U}=87.5$ K).  The authors demonstrate that the SO integrated intensity clearly distinguishes between hot and cold cores detected in their observations. All the hot cores have the SO integrated intensity above 20 K km s$^{-1}$; they are also associated with high-excitation \ce{SO2} and/or \ce{CH3OH} lines that are not detected toward cold cores.   \citet{shimonishi2026} also found a good correlation between  the SO integrated intensity and the 850 $\mu$m continuum intensity.  The authors suggest that the integrated high-excitation SO line intensity would be useful for identifying hot core candidates. 

In N\,83, we determine the integrated intensity of the SO $6_6$--$5_5$ line ($E_{\rm U}=56.5$ K) of ($28.19\pm0.32$, $30.01\pm0.25$, $9.85\pm0.13$, $9.14\pm0.12$, $8.08\pm0.09$) K km s$^{-1}$ for N\,83A--mm (A, B, C, D, E).  Both hot cores (sources A and B) have the SO integrated intensities $>$20 K km s$^{-1}$, a factor of 3--4 higher than those of cold cores, and they are also the brightest 1.2 mm continuum sources in their fields.  This is consistent with \citet{shimonishi2026}'s results.

\begin{figure*}[ht!]
\centering
\includegraphics[width=\textwidth]{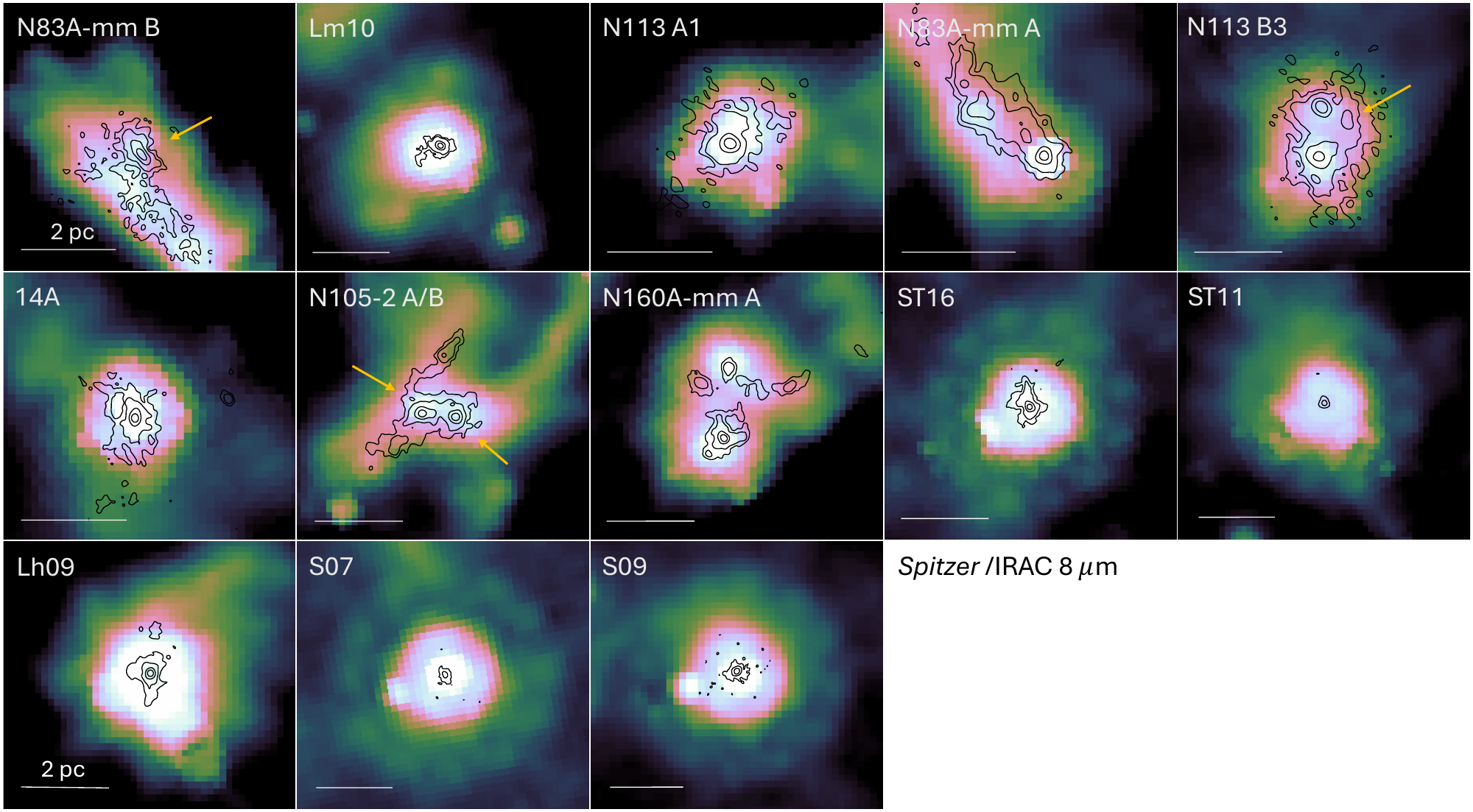}
\caption{The {\it Spitzer}/IRAC 8.0 $\mu$m images (0$\rlap.{''}$6 pixel tiles) of all the ALMA fields hosting hot cores. The hot cores are ordered as in Figure~\ref{f:abund}, i.e., by decreasing hot core's \ce{CH3OH} abundance within each galaxy (left to right, top to bottom; see Figure~\ref{f:abund}), .  Black contours correspond to the (231.9, 231.9, 242.4, 242.4, 242.4, 242.4, 241.8, 351.9, 358.6. 350, 350, 344.7, 344.7) GHz continuum emission detected toward (N\,113\,A1, N\,113\,B3, N\,83A--mm\,B, N\,83A--mm\,A, N\,105--2\,A/B, N\,160A--mm\,A, 14A, ST16, ST11, Lh09, Lm10, S07, S09).  The contour levels are (2, 5, 20, 50)\% for N\,113\,A1, N\,113\,B3, N\,83A--mm\,B, N\,83A--mm\,A, N\,105--2\,A/B, N\,160A--mm\,A, 14A, ST16, and Lh09; (3, 5, 20, 50)\% for Lm10; (10, 50)\% for ST11 and S07; and (5, 20, 50)\% for S09 of the corresponding continuum emission peak.  The ALMA beam sizes range from $\sim$0$\rlap.{''}$4 to 0$\rlap.{''}$8.  The position of a hot core is indicated with an arrow if it is not the brightest {\it Spitzer} source in the ALMA field or multiple hot cores are present. The scale bar in the lower left corner of each image corresponds to 2 pc. \label{f:spitzergrid}}
\end{figure*}

\subsection{The Dependence of the Magellanic Hot Cores' Chemical Makeup on their Environment}
\label{s:environment}

Using the data from our ALMA program, we are consistently finding bona fide hot cores with COMs in similar environments (N\,105--2\,A and 2\,B, \citealt{sewilo2022n105}; N\,160A--mm\,A, \citealt{broadmeadow2025}; N\,83A--mm\,A and B, this paper).  The fields hosting hot cores with COMs were selected for the observations based on their association with the dense gas (traced by the \ce{^13CO}, \ce{C^18O}, and \ce{CS}  emission),  \ce{H2O} masers, {\it Spitzer} YSOs, and the SO  3$_2$--2$_1$ emission at 99.2999 GHz (as a hot core tracer; see Section~\ref{s:observations} and \citealt{sewilo2022n105}). These were the characteristics of the fields where the first (and at the time, the only) extragalactic hot cores exhibiting the hot core chemistry were identified (N\,113\,A1 and B3; \citealt{sewilo2018}). N\,113\,A1 and B3 were found serendipitously in the program that imaged the entire star-forming region N\,113.   

One of the seven fields in our program (N\,105--3) was associated with \ce{CH3OH} rather than \ce{H2O} masers (6.67 GHz and the only 12.2 GHz \ce{CH3OH} maser in the LMC; \citealt{sinclair1992}), and two had no reported maser detections in the literature.  We detected a hot core candidate in the field with \ce{CH3OH} masers \citep{sewilo2022n105} and an emerging ultracompact H\,{\sc ii} region in one of the fields with no masers (N\,105--1; \citealt{sewilo2023}). The second field with no masers hosts an embedded protostar in a pre-hot core stage. Our bona fide hot core detection rate is 100\% (4 out of 4) toward fields hosting H$_2$O masers. 

\citet{shimonishi2026} did not detect \ce{CH3OH} toward N\,160A--mm A (Lh08) in their Band 7 observations. The N\,160A--mm A's \ce{CH3OH} synthetic spectrum from \citet{broadmeadow2025} with $T_{\rm ex}$ = 188.4 K and $N_{\rm tot} = 2.5\times10^{15}$ cm$^{-2}$ predicts line intensities of 0.56 and 0.33 K for the two transitions detected by \citet{shimonishi2026} toward other sources in their hot core sample: 4$_{-1,3}$--3$_{-0,3}$ E at 358.6058 GHz ($E_{\rm U}=44.3$ K)
16$_{1,15}$--15$_{2,14}$ A at 345.9039 GHz ($E_{\rm U}=332.6$ K).  These intensities are below the 3$\sigma$ detection limit of \citet{shimonishi2026}'s observations (the typical rms of 0.24--0.30 K). 

In all the fields from our program, we detected multiple 1.2 mm continuum cores and an extended continuum emission (see Figure~\ref{f:spitzergrid} for the fields with hot cores). If a hot core is present, it is associated with the brightest continuum source in the field (or, as in N\,105--2, the two brightest sources).  All but one cold cores are associated with the compact \ce{CH3OH} emission.  As in N\,83--1 and N\,83--2, the extended (cold) \ce{CH3OH}, SO, and CS emission is present in all the other fields; it is brighter in the fields with hot cores.  

In Figure~\ref{f:spitzergrid}, we show the {\it Spitzer}/IRAC 8.0 $\mu$m images with the ALMA continuum contours overlaid for all known hot cores in the LMC and SMC. For each galaxy, the fields are ordered by the decreasing \ce{CH3OH} abundance in the hosted hot core.  It is striking that the LMC hot cores with no detection or a low abundance of \ce{CH3OH} (ST16, ST11, and Lh09) and the SMC (S07 and S09) hot cores are associated with the brightest {\it Spitzer} YSOs. The mm-wave continuum emission appears to be more compact in the fields hosting them (the observations have comparable maximum recoverable scales) and there are fewer sources than in other fields.  The inspection of the ancillary images indicates that ST11, ST16, S07, and S09 are located away from the molecular cloud peaks and bright H$\alpha$ emission. Lh09 is an exception; however, as a potential super star cluster in the making, this region represents a more extreme environment (e.g., \citealt{ochsendorf2017,nayak2019}). 14A (with the detection of \ce{CH3OH} and \ce{CH3CN}) is another source located away from the densest gas in the hosting star-forming region, but it is much less luminous than ST11, ST16, S07, and S09.  

Even though Lh09 is associated with hot \ce{SO2}, it may be in the post-hot core stage.  Lh09 coincides with the {\it Spitzer} YSO 045153.29$-$692328.6  (the  {\it Herschel} source HSOBMHERICC\,J72.971176$-$69.391112; \citealt{seale2014}), one of the most luminous sources in the LMC ($\sim$$2\times10^6$ $L_{\odot}$; \citealt{ochsendorf2017}) that was proposed to be a precursor to a super star cluster (H72.97$-$69.39; \citealt{nayak2019}).  Its properties resemble those of N\,105--1\,A, an UC H\,{\sc ii} region with strong millimeter hydrogen recombination lines, deeply embedded in its natal molecular cloud \citep{sewilo2022n105,sewilo2023}.  The inspection of the archival Band 3 data for Lh09 reveal the detection of the H40$\alpha$ (99.02295 GHz), H50$\beta$ (99.22521 GHz), and H57$\gamma$ (98.67189 GHz) recombination lines.  \citet{mondal2026} report the detection of H36$\beta$ (260.03278 GHz) and H41$\gamma$ (257.63549 GHz) recombination lines, and \citet{nayak2019} detected the H30$\alpha$ (231.90093 GHz) line. 

In \citet{sewilo2023}, we suggest that the star formation in the region hosting N\,105--1\,A was triggered by the cloud-cloud collision, the scenario proposed for Lh09 by \citet{nayak2019}.  In \citet{sewilo2022n105}, we found the hot \ce{SO2} emission ($\sim$96 K) offset from the 1.2 mm continuum peak, as is observed in Lh09 (\citealt{shimonishi2026}).  We found filaments and streamers in N\,105--1\,A that indicate ongoing accretion onto the clump hosting the UC H\,{\sc ii} region, with the \ce{SO2} and \ce{SO} emission tracing accretion shocks.  Unlike in Lh09, both compact and extended \ce{CH3OH} emission has been detected in N\,105--1\,A, but no \ce{CH3OH} peak coincides with the 1.2 mm continuum peak.  The \ce{CH3OH} emission is present in the field hosting Lh09 though; it has been detected toward two cores north of Lh09 (\citealt{mondal2026}). 

In recent JWST MIRI MRS observations centered on the {\it Spitzer} YSO 045153.29$-$692328.6,  \citet{nayak2024} detected four sources (Y4--Y7), with the brightest source (Y4) coinciding with the ALMA source Lh09. Sources Y5--Y7 are located within the lowest 850 $\mu$m continuum contour shown in Figure~\ref{f:spitzergrid}.  The MIRI MRS spectrum of Y4/Lh09 shows the H\,{\sc i}, H$_2$, and fine-structure emission lines, but no PAH emission or ice absorption features. 

\subsubsection{Insight from the Spitzer/IRS}

\begin{figure*}[ht!]
\centering
\includegraphics[width=0.9\textwidth]{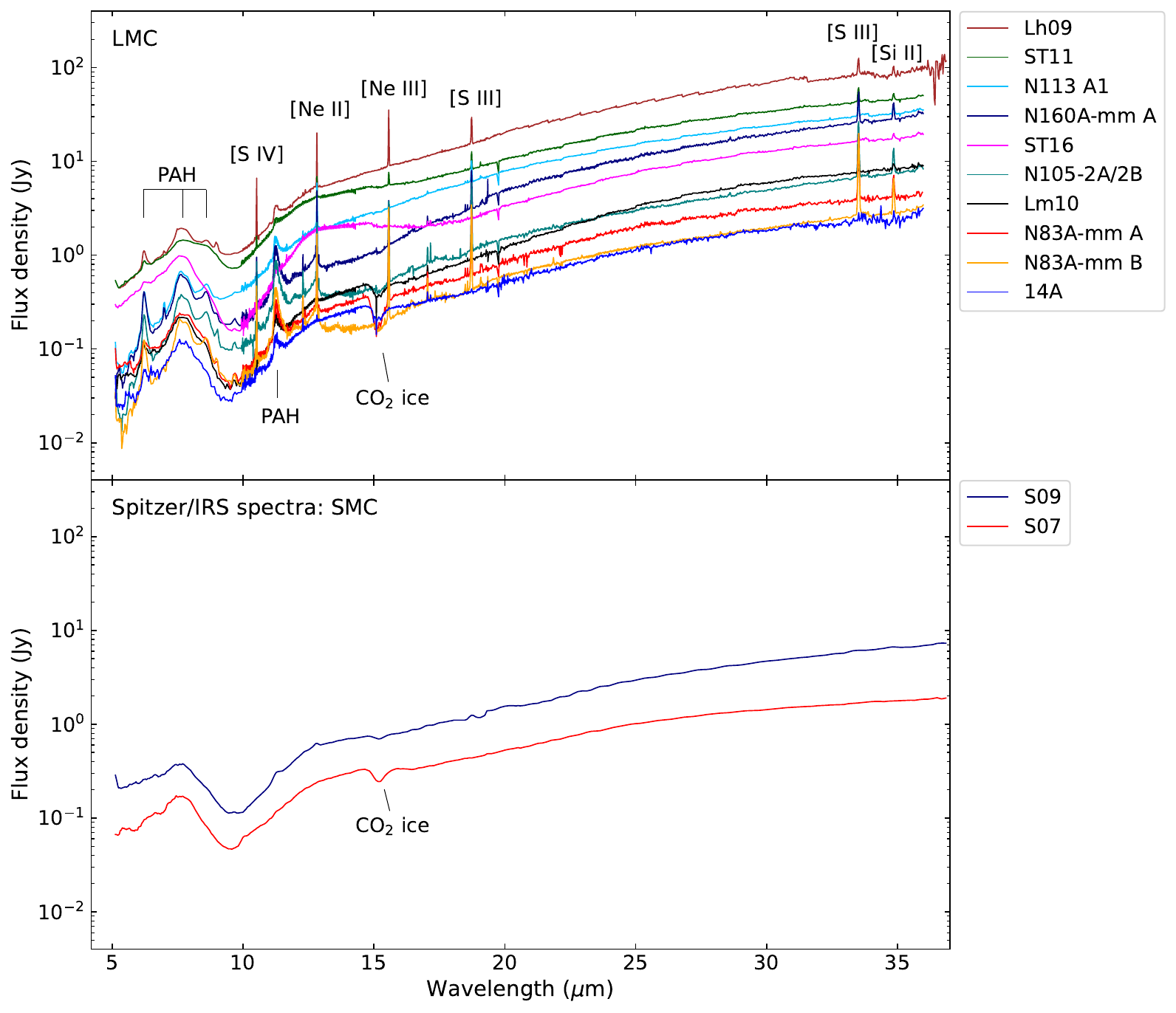}
\caption{The {\it Spitzer}/IRS spectra of the YSOs associated with known hot cores in the LMC (top) and SMC (bottom), color-coded as indicated in the legend.  Due to the size of the {\it Spitzer}/IRS slit, the emission from both the YSOs and their environments contribute to the spectra. The CO$_2$ ice and PAH features, and fine-structure lines are labeled. The LMC spectra were analyzed and classified in \citet{seale2009} and \citet{jones2017}, and the SMC spectra were analyzed and classified in \citet{oliveira2009} and \citet{ruffle2015}. \label{f:spitzerirs}}
\end{figure*}

We extend the discussion on the {\it Spitzer}/IRS spectra of YSOs in the ALMA fields in N\,83 (see Section~\ref{s:N83spitzerIRS}) to the entire sample of known hot cores by comparing a full set of spectra in Figure~\ref{f:spitzerirs} (\citealt{seale2009,oliveira2009,ruffle2015,jones2017}). The spectra are plotted without scaling for a direct comparison; the individual spectra are displayed separately  in Figure~\ref{f:SpitzerIRSseparate} in Appendix~\ref{app:ancillary}. 

The association between {\it Spitzer} YSOs and hot cores identified with ALMA is uncertain not only for N\,83A--mm\,B (Section~\ref{s:N83spitzerIRS}), but also for other hot cores with COMs: N\,105--2\,B and N\,160A--mm\,A (see the discussion in \citealt{sewilo2022n105} and \citealt{broadmeadow2025}, respectively), and N\,113\,A1. Similarly to N\,83A--mm\,B, there is an offset between the ALMA mm-wave continuum peak and the known near- and mid-IR sources.  For N\,113\,A1, the high-resolution near-IR observations with the Very Large Telescope (VLT) SINFONI integral field spectrograph (0$\rlap.{''}$1 pixel scale) revealed 3 sources and none coincides with the 1.3 mm continuum source A1 (\citealt{ward2016}). The morphology of the {\it Spitzer}/IRAC emission hints at a possibility that there is a second source at the position of A1, much fainter and unresolved from the brighter YSO.  Thus, as for N\,83A--mm\,B, the {\it Spitzer}/IRS spectrum likely does not reflect the true nature of the source associated with the hot core.  Another hot core in N\,113, N\,113\,B3, is not associated with any known mid- or near-IR sources; the nearest {\it Spitzer} YSO is $\sim$4$\rlap.{''}$5 away and it coincides with another 1.2 mm continuum source.  The high-resolution and high-sensitivity mid-IR observations will help pinpoining the protostars responsible for the hot core emission.  In a meantime, the {\it Spitzer}/IRAC spectra provide some useful information about the environment of these sources and the star formation history of the regions, e.g., the initial conditions were likely colder in regions located away from the bright H$\alpha$ emission than in those presently associated with the strong UV radiation.

The {\it Spitzer}/IRS spectra of bright YSOs that correspond to hot cores with no COMs (ST11) or low \ce{CH3OH} abundance (ST16, S07, and S09) have no or weak PAH emission features and fine-structure lines. No contamination in the spectra from the surrounding ionized gas is consistent with the sources' location in the LMC away from the bright H$\alpha$ emission (traced by the MCELS H$\alpha$ image).  ST11, S07, and S09 are associated with outflows traced by the CO emission (\citealt{shimonishi2016b,tokuda2022outflow,shimonishi2023}). The presence of the outflow in ST16 is inferred based on the CCH and CN emission tracing outflow cavities (\citealt{shimonishi2020}).  Ice features have been detected toward all four sources; in the near-IR for ST11 and ST16 (\citealt{shimonishi2010,shimonishi2016a}), in the {\it Spitzer} spectrum, there is only a hint of the presence of the \ce{CO2} ice band at 15.2 $\mu$m. 

For hot cores with COMs, there are two types of spectra: one exhibiting strong PAHs and fine-structure lines (N\,83A--mm\,A and B. N\,105--2\,A and B, N\,160A--mm\,A, and N\,113 A1) and the other showing weak PAHs and no or very weak fine-structure lines (14A and Lm10).  

Similarly to the sources with no or low-abundance of \ce{CH3OH},  14A and Lm10 are located away from the bright H$\alpha$ emission. 14A was identified by \citet{hamedanigolshan2024} who specifically selected their ALMA targets with no strong fine-structure lines in their {\it Spitzer}/IRS spectra, i.e., away from the major star-forming regions. In the same program, \ce{CH3OH} was found in 19 out of 20 observed fields, only the cold \ce{CH3OH} emission in 16 fields.  PAHs around 14A and Lm10 have likely been excited by the radiation field dominated by softer, less energetic (non-ionizing) photons in their environment. 

Lm10 corresponds to the {\it Spitzer} YSO 045111.39$-$692646.7 which was the target of the JWST MIRI MRS observations. \citet{nayak2024} detected three mid-IR sources in the MIRI MRS field: Y9, Y10, and Y11.  The ALMA source Lm10 coincides with the brightest mid-IR source, Y11. Y9 and Y10 are located at opposite sides of the elongated 850 $\mu$m continuum emission feature extending toward southeast from Y11 (see Figure~\ref{f:spitzergrid}). The MIRI MRS archival spectrum of Y11 shows that it is an embedded protostar with the silicate and 15.2 $\mu$m CO$_2$ ice absorption features, H$_2$ emission lines, but no PAH or fine-structure emission lines. Lm10 has the second highest \ce{CH3OH} abundance and is one of only a handful of hot cores with the detection of \ce{CH3OCH3}.

The second group of spectra that represent regions hosting hot cores with COMs (N\,83A--mm\,A and B. N\,105--2\,A and B, N\,160A--mm\,A) show prominent PAH features and strong fine-structure lines. In N\,113, the fine-structure lines and PAH features are weaker, but significantly stronger than in 14A and Lm10. Hot cores N\,83A--mm\,A and B, N\,105--2\,A and B, N\,160A--mm\,A, and N\,113\,A1 (all associated with larger COMs) lie in the prominent star-forming regions, close to the bright H$\alpha$ emission. However, since they are close to the densest gas in the hosting regions, they are likely shielded from the most intense radiation field; in such dense regions, the cosmic rays play a central role in initiating chemistry. In all four regions (N\,83, N\,105, N\,113, and N\,160), there are signature of feedback from massive stars (e.g., expanding ionized bubbles) and large scale motions (accretion from filaments, likely filament collisions).  The feedback effects likely contribute to the formation of the extended cold \ce{CH3OH} emission (see Section~\ref{s:N83kinematics}).

\begin{figure*}
\centering
\includegraphics[width=0.6\textwidth]{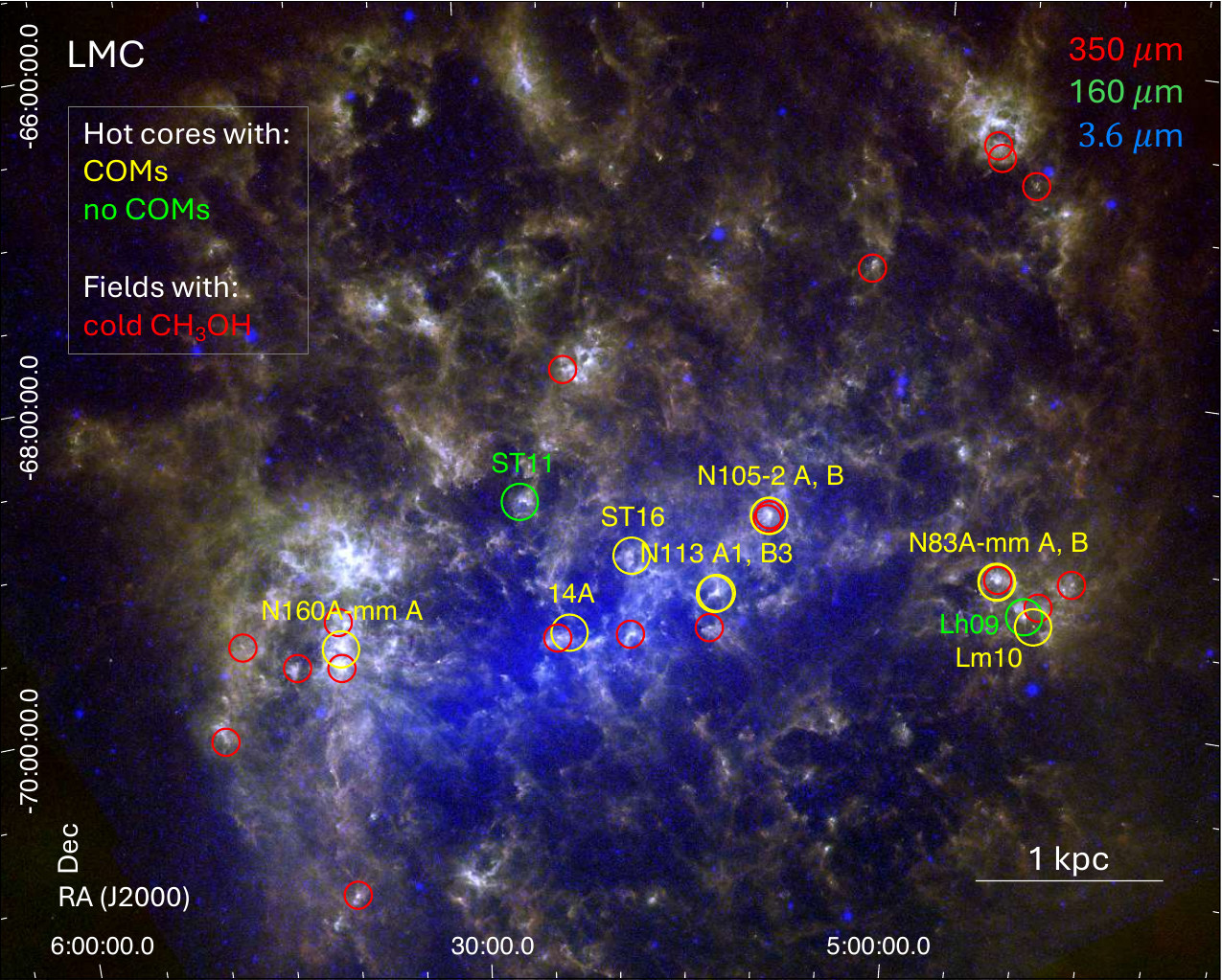} 
\caption{The positions of known hot cores (with and without COMs) and the fields with the cold \ce{CH3OH} detections (outside hot cores; see references in the text) overlaid on the three-color mosaic of the LMC. The mosaic combines the {\it Herschel}/SPIRE 350 $\mu$m,  {\it Herschel}/PACS 160 $\mu$m, and the {\it Spitzer}/IRAC 3.6 $\mu$m images (\citealt{meixner2006,meixner2013}). Multiple cold cores with \ce{CH3OH} have also been detected in the fields hosting hot cores  N\,113\,B3, N\,105--2\,A and B, N\,160A--mm A, Lh09, and N\,83A--mm\,A and B (\citealt{sewilo2019,sewilo2022n105,broadmeadow2025,mondal2026}; this paper). The size of the circles is much larger than the ALMA field of view, thus the ALMA fields located in a single star-forming region seem to overlap. The stellar over-density in the central part of the LMC in the 3.6 $\mu$m image highlights the LMC bar in blue.  \label{f:distribution}}
\end{figure*}

 \subsubsection{Chemical Diversity of Hot Cores vs. Metallicity Inhomogeneities in the LMC Disk}
 \label{s:inhomogeneities}
 
If the significant metallicity inhomogeneities exist in the LMC's disk, they may contribute to the observed diversity in the chemical makeup of hot cores.  The distribution of the known hot cores and regions with the detection of cold \ce{CH3OH} are indicated in Figure~\ref{f:distribution}.  With the increasing sample, we are finding hot cores with COMs outside the LMC's bar. However, the locations of these hot cores are still related to the dynamic features of the LMC.  They lie at the apparent tips of the opposite tidal arms identified in the LMC based on the H\,{\sc i} emission (e.g., \citealt{bekki2007a}). 

The metallicity variations across the LMC disk have been reported in the literature, although the variation pattern is not consistent between different studies. There is evidence for the metallicity radial gradient in the LMC; however, the slope is uncertain as the reported values seem to strongly depend on the method (photometric or spectroscopic) and the type of objects used (e.g., RR Lyrae, asymptotic giant branch, or red giant branch stars).  Yet, all the studies agree that the radial metallicity gradient is shallow (from $-$0.01 to $-$0.05 dex kpc$^{-1}$; see e.g., \citealt{povick2025lmc} and references therein). 

In the most recent work that utilizes the high-resolution {\it H}-band spectroscopic data for 6130 LMC field red giant branch stars, \citet{povick2025lmc} found the metallicity gradient (traced by the iron abundance) of $-0.0296\pm0.0017$ dex kpc$^{-1}$.  Seven of the hot cores with a wide range of \ce{CH3OH} abundance lie within 1 kpc from the dynamic center of the LMC (i.e., (RA, Decl.) = (80$\rlap.^{\circ}$30, $-$69$\rlap.^{\circ}$27); \citealt{vijayasree2025}), and the farthest ones (Lh09 and Lm10) are at 2.3 kpc.  If the metallicity in the center of the LMC is 0.5 $Z_{\odot}$ ($-$0.3 dex), the \citet{povick2025lmc} metallicity gradient predicts the metallicity of $\sim$0.47 $Z_{\odot}$ at 1 kpc  ($\sim$7\% lower than the adopted metallicity of 0.5 $Z_{\odot}$ in the center of the LMC) and $\sim$0.43 $Z_{\odot}$ at 2.3 kpc ($\sim$15\% lower).  For the steepest metallicity gradient reported in the literature ($-$0.05  dex kpc$^{-1}$; \citealt{feast2010}), the predicted metallicity is $\sim$0.45 $Z_{\odot}$ at 1 kpc ($\sim$11\% lower) and  $\sim$0.38 $Z_{\odot}$ at 2.3 kpc  (23\% lower).  Considering the fact that most of the metallicity gradients reported in the literature are very shallow and there are studies that found no evidence for the metallicity gradient in the LMC (e.g., \citealt{pena1987,geisler2003}), the metallicity differences resulting from the metallicity gradient are not expected to impact the molecular abundances in hot cores.

However, \citet{choudhury2016} revealed areas in the LMC significantly deviating from the mean $Z$ after removing a shallow metallicity gradient.  Using the photometric (optical) data for red giant branch stars, they found regions with $Z>0.7$ $Z_{\odot}$ (mostly in the LMC stellar bar) and $Z<0.2$~$Z_{\odot}$ (in the outer regions).  In another study, \citet{romanduval2021} used the interstellar depletions derived from the near- to far-ultraviolet spectra for different elements toward 33 massive stars in the LMC to infer variations in metallicity of up to 0.8 dex across the LMC. The recent study by \citet{lah2024lmc} based on the bright optical nebular lines for 59 sight lines revealed that metallicity (traced by the O abundance) varies by $\sim$0.5 dex across the LMC.  

These second-order variations (relative to the radial gradients) found in the LMC are large in comparison to those observed in other nearby galaxies (e.g., \citealt{groves2023}), and likely reflect a complex star formation history of the LMC. In our previous work (\citealt{sewilo2022n105,hamedanigolshan2024,broadmeadow2025}), we provide a detailed discussion on the possibility that such inhomogeneities may be the result of a close encounter between the LMC and the SMC about 0.2 Gyr ago (e.g., \citealt{fujimoto1990,luks1992,bekki2007a,yozin2014}).  

The numerical simulations show that as the result of the LMC--SMC interaction, the gas stripped from both galaxies by the tidal force  (i.e., containing a large amount of metal-poor gas from the SMC) has been falling back on the LMC disk (\citealt{bekki2007a,yozin2014}). The presence of two H\,{\sc i} velocity components (L- and D-component; H\,{\sc i} flows) in the disk supports this scenario (e.g., \citealt{luks1992}). It was proposed that the collision of these H\,{\sc i} flows driven by the LMC--SMC tidal interaction triggered massive star formation in 30 Dor, N159 (south of 30 Dor), and N44 (north from the center of the LMC; \citealt{fukui2019,tokuda2019,tsuge2019,tokuda2022}). In \citet{tsuge2024}, the authors argue that the gas compression driven by the H\,{\sc i} collisions may be an important high-mass star formation triggering mechanism over the entire galaxy.  The metallicity variation patterns across the LMC disk resulting from the LMC and SMC gas mixing are hard to predict, but the differences would likely be sufficiently large to contribute to observed differences in hot core chemistry.  However, our analysis suggests that local effects other than metallicity (e.g., the UV radiation field strength, the degree of shielding, large-scale kinematics) may have a large impact on shaping the chemical composition of star-forming cores.

\subsection{Outflows vs. Hot and Cold Methanol} 
\label{s:outflows}

In \citet{sewilo2022n105}, we discuss several chemical processes which could in principle account for the presence of \ce{CH3OH}, \ce{SO2}, and \ce{SO} in hot and cold gas, and in compact and extended molecular emission (see also \citealt{mondal2026}). As noted in Section~\ref{s:N83kinematics}, the detection of these molecules and \ce{SiO}, along with \ce{H2O} (masers), indicates that shock waves generally play a key role in the chemistry of N\,83, and LMC hot cores. 

As shown in Figure~\ref{f:abund}, one challenge is to explain the drop-off in the detected \ce{CH3OH} abundance in LMC hot cores of more than an order of magnitude, and overall down by a more than a factor of a hundred when accounting for non-detections in so-called COM-poor sources (e.g., in ST11, see \citealt{shimonishi2016b}). A related issue is that \ce{SO} and \ce{SO2} are detected in all LMC hot cores (including their cold envelopes) with abundances that show relatively little variation (Figure~\ref{f:abund}). Here we consider some expected effects of shock chemistry relevant to these observations. 

In star-forming molecular clouds, shock-induced sputtering of silicate grain cores and their icy mantles is accompanied by a high-temperature chemistry mediated by the evolving physical conditions in the gas, i.e., depending on whether a J-type or a C-type shock is present  (e.g., \citealt{draine1993}).  Assuming non-dissociative shocks (e.g., \citealt{roberge1982}), the chemistry is largely controlled by the gas density and temperature evolution, as well as ambipolar diffusion in the case of multi-fluid C-shocks.  Of the many endoergic chemical reactions that can occur, the most important involve H atom reactions that abstract a H from molecules, and H additions through \ce{H2} reactions with atoms and molecular radicals. Hence, the H/\ce{H2} ratio in the hot gas is a key parameter. 

Shock waves can destroy \ce{CH3OH} molecules in the sequence: 
\begin{equation}
{\rm CH_3OH \buildrel H \over \longrightarrow CH_3O, CH_2OH \buildrel H \over \longrightarrow H_2CO \buildrel H \over \longrightarrow HCO \buildrel H \over \longrightarrow CO   }
\label{eqn:ch3oh+h}
\end{equation} 
Here, the H atom reactions with the  methoxy (\ce{CH3O}), hydroxymethyl (\ce{CH2OH}), and formyl (\ce{HCO}) radicals have no energy barrier. The first steps in the sequence to form the \ce{CH3O} and \ce{CH2OH} radicals have activation energies, $E_{\rm A} \approx 2000-3000$ K  (\citealt{jodkowski1999}), whereas the reverse reactions with \ce{H2} have substantially higher barriers (\citealt{wakelam2024}), e.g., 
\begin{equation}
{\rm CH_2OH  +  H_2} \longrightarrow {\rm CH_3OH + H - 6720\,K} 
\label{eqn:pmeoh}
\end{equation}

As the reverse reaction of \ce{HCO} with \ce{H2} has $E_{\rm A} \approx 9000$~K, the sequence can be suppressed in the hottest  regions of  J-shocks ($T \sim 4000\rm$~K) and CH$_3$OH may be expected to transiently survive until the gas has cooled sufficiently. Consequently, CH$_3$OH destruction should be  more efficient  at the lower peak neutral temperatures produced in C-shocks,  $T\sim1000-3000\rm$ K. 

In both cases, when the gas has cooled to $T \lesssim 300$~K, the sequence in Equation~\ref{eqn:ch3oh+h} ends, the methanol abundance becomes essentially frozen in, and this may explain the observed range both in compact hot cores and in cold gas. However, as the survival of methanol depends on the cooling physics, these estimates should await confirmation by detailed shock modeling. 

Sputtering or thermal sublimation of ices both easily release \ce{H2S} molecules into the gas phase where they can react with atomic H (\citealt{charnley1997}):
\begin{equation}
{\rm H_2S \buildrel H \over \longrightarrow SH \buildrel H \over \longrightarrow S }
\label{eqn:h2s+h}
\end{equation} 

The sequence in Equation~\ref{eqn:h2s+h} produces \ce{SH} and S in dense gas for $T \gtrsim 100$~K and so will  also occur in postshock gas. The barriers for the \ce{H2} reverse reactions are in the range $\approx$8,000--10,000~K (\citealt{millar2024}) and so \ce{H2S} destruction should exhibit a similar  temperature dependence as that outlined above for \ce{CH3OH}. 

Reactions of \ce{SH} and S with O and \ce{OH} (presumably released from water) subsequently produce \ce{SO} and \ce{SO2}:  
\begin{equation}
{\rm SH \buildrel O \over \longrightarrow SO \buildrel OH \over \longrightarrow SO_2   }
\label{eqn:so+o}
\end{equation} 
\begin{equation}
{\rm S \buildrel OH \over \longrightarrow SO \buildrel OH \over \longrightarrow SO_2   }
\label{eqn:s+oh}
\end{equation} 

The detected \ce{H2O} masers are produced in shocks (\citealt{elitzur1989,hollenbach2013}). The large  $E_{\rm A} $ values for H atom abstraction from \ce{H2O} and \ce{OH} may mean that \ce{H2} reactions with O and \ce{OH} may keep these oxygen nuclei largely locked in \ce{H2O}. In this event, it may be necessary to also include the effects of UV photochemistry for water in the postshock gas (\citealt{vangelder2021}). Reactions of H atoms with \ce{SO2} and \ce{SO} have enormous $E_{\rm A}$  values (\citealt{millar2024}) and can be neglected. Hence, once S and \ce{SH} are produced in the  gas,  \ce{SO2} and \ce{SO} are unaffected by H-\ce{H2} reactions and so the weaker abundance variation relative to that of \ce{CH3OH} could be explained. 

\section{Summary and Conclusions}
\label{s:conclusions}

Using the ALMA 1.2 mm continuum and molecular line data covering a broad frequency range ($\sim$2 GHz between $\sim$241.5 and $\sim$260.6 GHz), we explored the physical and chemical characteristics of five sources in the star-forming region N\,83 (dubbed N\,83A--mm A--E) in the low-metallicity LMC.  We determined their physical parameters based on the 1.2 mm continuum emission: \ce{H2} column and number densities, masses, and sizes.  We performed the spectral modeling of the ALMA spectra with the XCLASS software to determine the rotational temperature and column density for species with the detection of multiple transitions, and the column density only for the remaining species. We used the ancillary multi-wavelength photometric and spectroscopic data to investigate the nature of N\,83A--mm A--E and their environments.  The main results are as follows.

\begin{enumerate}

\item We identified sources N\,83A--mm\,A and B as bona fide hot cores. Both sources have high temperatures ($>$100 K), small sizes (0.1 and 0.16 pc for A and B, respectively), and high \ce{H2} number densities (a few times 10$^{5}$ cm$^{-3}$), and are associated with COMs. The remaining three sources (N\,83A--mm\,C, D, and E) are cold cores ($T$$\sim$23--27 K). 

\item In terms of the number of detected species and molecular abundances, the hot core N\,83A--mm\,B is the most chemically-rich source in our ALMA fields in N\,83.  Additionally, it has the highest \ce{CH3OH} abundance with respect to \ce{H2} among all known Magellanic hot cores.

\item We detected three COMs in N\,83: \ce{CH3OH} (in hot cores N\,83A--mm A and B), \ce{CH3CN} (A and B), and \ce{CH3OCH3} (B). In addition to the compact \ce{CH3OH} emission in all the detected cores (cold and hot), an extended cold \ce{CH3OH} emission is present in both ALMA fields. This is consistent with our findings in four other ALMA fields from our program (\citealt{sewilo2022n105,broadmeadow2025}). 
 
\item In addition to COMs, we detected N-bearing species: \ce{HC^15N}, \ce{H^13CN}, \ce{HNCO}; S-bearing species: \ce{SO}, \ce{^33SO}, \ce{SO2}, \ce{OCS}, \ce{H2CS}, \ce{CS}, \ce{C^33S}; deuterated species: \ce{HDCO}; as well as \ce{H^13CO+}, \ce{SiO}, \ce{CH2CO}. We also report a tentative detection of \ce{HDO}, \ce{HDCS}, and \ce{^34SO2}. Only a handful of simple species have been detected toward all sources: \ce{SO}, \ce{CS}, \ce{H2CS}, \ce{H^13CN}, and \ce{H^13CO+}. For each source in N\,83, the \ce{CH3OH} abundance is the highest among all the detected species.

\item The abundances of hot cores N\,83A--mm A and B fit into the trends previously reported for the Magellanic hot cores: the large variations in the \ce{CH3OH} abundances and less significant variations for the sulfur oxide species (\ce{SO} and \ce{SO2}). These trends are discussed in the context of differences in their formation pathways. We also  discuss some expected effects of shock chemistry relevant to these observations.

\item  The velocity patterns in both fields are complex. The multi-directional flows appear to be present, as well as overlapping velocity components. The feedback-produced high- and low-velocity shocks in the region are likely important drivers of the chemistry of S-bearing species, \ce{SiO}, and \ce{CH3OH} in and outside hot cores. 

\item The integrated intensity of the high-excitation SO line for hot cores N\,83A--mm\,A and B is $>$20 K km s$^{-1}$, in agreement with the cutoff value found by \citet{shimonishi2026} for a sample of 9 LMC hot cores. Our results therefore confirm that the integrated intensity of the high-excitation SO line may be a useful hot core candidate selection criterion.

\item N\,83A--mm A--E are embedded in the gas and dust at the rim of the ionized gas bubble formed by the radiation and winds from an O6-type star. The {\it Spitzer}/IRS spectra of the three {\it Spitzer} YSOs in the ALMA fields N\,83--1 and N\,83--2 (associated with sources A, B, and C) reveal that they are embedded YSOs in the environment characterized by the strong UV radiation field and abundant in PAHs, consistent with their position near PDRs.  
 
\item Our analysis of the fields hosting COM-rich hot cores (including those in N\,83) reveals that COMs may form in significantly different environments, but the abundances (low/high) do not correlate with the type of the environment.  

\item Hot cores with no COMs (ST11, Lh09) or low \ce{CH3OH} abundance (ST16, S07, and S09) are associated with the most luminous {\it Spitzer} YSOs in our sample. They are embedded protostars with the detection of outflows, and (with the exception of Lh09), they are located away from the molecular cloud peaks and bright H$\alpha$ emission; they are also the only continuum sources in their fields. As more luminous YSOs drive more powerful and energetic outflows, the effects of shocks on the hot core chemistry (e.g., \ce{CH3OH} destruction) is likely more significant for these sources.  

\item If the significant metallicity inhomogeneities exist in the LMC's disk (e.g., due to the interaction with the SMC), they may contribute to the observed diversity in the chemical makeup of hot cores. With the increasing sample, we are finding hot cores with COMs outside the LMC's bar. However, their locations are still related to the dynamic features of the LMC. Our analysis demonstrates that in addition to metallicity variations, the other local effects (e.g., the UV radiation field strength, the degree of shielding, and large-scale kinematics) and source properties have a large impact on shaping the chemical composition of star-forming cores in the LMC.  Their combined effects may impact different molecular species in a distinct way depending on their formation paths. 

\item The above conclusion is supported by the comparison between the Magellanic and outer Galaxy hot cores, the latter representing a significantly more quiescent environment in terms of the star formation activity. Despite a comparable metallicity range, the Magellanic hot cores are dissimilar to the outer Galaxy hot cores. Compared to the outer Galaxy hot cores, the COM formation efficiency is lower and the formation efficiency of S-bearing species (such as \ce{SO} and \ce{SO2}) is higher in the LMC and SMC hot cores. 

\end{enumerate}

The larger samples of hot cores in the Magellanic Clouds and the outer Galaxy with well-characterized environments (e.g., the measured local metallicity,  the cosmic-ray ionization rate, and the UV radiation field strength) are needed to disentangle the effects of different factors responsible for shaping the chemical composition and abundances of both cold and hot cores.  

\clearpage

\begin{acknowledgements}
We thank the anonymous reviewer for their constructive feedback that helped us improve the manuscript. The material is based upon work supported by NASA under award number 80GSFC24M0006 (M. S.). R.HG. gratefully acknowledges Dr. Anna F. McLeod for her support during the research stay at the Centre for Extragalactic Astronomy, Durham University. We thank Dr. J{\"u}rgen Ott for making an unpublished list of water masers in the LMC available to us. The National Radio Astronomy Observatory and Green Bank Observatory are facilities of the U.S. National Science Foundation operated under cooperative agreement by Associated Universities, Inc. This paper makes use of the following ALMA data: ADS/JAO.ALMA\#2019.1.01720.S and ADS/JAO.ALMA\#2017.1.00093.S. ALMA is a partnership of ESO (representing its member states), NSF (USA) and NINS (Japan), together with NRC (Canada), MOST and ASIAA (Taiwan), and KASI (Republic of Korea), in cooperation with the Republic of Chile. The Joint ALMA Observatory is operated by ESO, AUI/NRAO and NAOJ.
This research made use of APLpy, an open-source plotting package for Python \citep{robitaille2012,aplpy2019}.  
\end{acknowledgements}

\facilities{ALMA, Spitzer, Herschel}

\bibliographystyle{aasjournal}
\bibliography{refs_all.bib}

\clearpage

\appendix 
\counterwithin{figure}{section}

\clearpage 

\section{Spectral Lines Detected Toward N\,83A--mm\,A--E }
\label{app:detections}

In Table~\ref{t:detections}, we list all the molecular lines detected toward sources N\,83A--mm\,A--E in two ALMA fields in the star-forming region N\,83.  The rest frequencies and upper state energies ($E_{\rm U}$) for each transition were taken from the CDMS catalog for all species except HDO for which the data were taken from the JPL database (see Sections~\ref{s:specid} and \ref{s:modeling}). We adopt the CDMS quantum number notation in Table~\ref{t:detections} and throughout the paper.

\startlongtable
\begin{deluxetable*}{lcccccccc}
\centering
\tablecaption{Spectral Lines Detected Toward Continuum Sources in N\,83A--mm\tablenotemark{\footnotesize a} \label{t:detections}}
\tablewidth{0pt}
\tablehead{
\multicolumn{1}{c}{Species} &
\multicolumn{1}{c}{Transition} &
\colhead{Frequency} &
\colhead{$E_{\rm u}$}&
\multicolumn{5}{c}{N\,83A--mm} \\
\cline{5-9}
\colhead{} &
\colhead{} & 
\colhead{(MHz)} &
\colhead{(K)} &
\colhead{A} &
\colhead{B} &
\colhead{C} &
\colhead{D} &
\colhead{E}
}
\startdata
\multicolumn{9}{c}{COMs} \\
\hline
CH$_3$OH & 5$_{ 0, 5}$--4$_{ 0, 4}$ E, v$_t$=0 & 241,700.159 & 47.94 & \checkmark\checkmark & \checkmark\checkmark & \checkmark\checkmark & \checkmark\checkmark & \checkmark\checkmark  \\
CH$_3$OH & 5$_{1, 5}$--4$_{1, 4}$ E, v$_t$=0 & 241,767.234 & 40.39 & \checkmark\checkmark & \checkmark\checkmark & \checkmark\checkmark & \checkmark\checkmark & \checkmark\checkmark  \\
CH$_3$OH & 5$_{0, 5}$--4$_{0, 4}$ A, v$_t$=0 & 241,791.352 & 34.82 & \checkmark\checkmark & \checkmark\checkmark & \checkmark\checkmark & \checkmark\checkmark & \checkmark\checkmark  \\
CH$_3$OH & 5$_{4, 2}$--4$_{4, 1}$ A, v$_t$=0 & 241,806.524 & 115.17 & \checkmark? & \checkmark\checkmark & - & - & -  \\
CH$_3$OH & 5$_{4, 1}$--4$_{4, 0}$ A, v$_t$=0 & 241,806.525 & 115.17 & \checkmark? & \checkmark\checkmark & - & - & -  \\
CH$_3$OH & 5$_{4, 2}$--4$_{4, 1}$ E, v$_t$=0 & 241,813.255 & 122.73 & - & \checkmark\checkmark & - & - & -  \\
CH$_3$OH & 5$_{ 4, 1}$--4$_{ 4, 0}$ E, v$_t$=0 & 241,829.629 & 130.82 & \checkmark & \checkmark & - & - & -  \\
CH$_3$OH & 5$_{3, 3}$--4$_{3, 2}$ A, v$_t$=0 & 241,832.718 & 84.62 & \checkmark\checkmark & \checkmark\checkmark & - & - & -  \\
CH$_3$OH & 5$_{3, 2}$--4$_{3, 1}$ A, v$_t$=0 & 241,833.106 & 84.62 & \checkmark\checkmark & \checkmark\checkmark & - & - & -  \\
CH$_3$OH & 5$_{2, 4}$--4$_{2, 3}$ A, v$_t$=0 & 241,842.284 & 72.53 & \checkmark\checkmark & \checkmark\checkmark & - & - & -  \\
CH$_3$OH & 5$_{ 3, 3}$--4$_{ 3, 2}$ E, v$_t$=0 & 241,843.604 & 82.53 & \checkmark\checkmark & \checkmark\checkmark & - & - & -  \\
CH$_3$OH & 5$_{3, 2}$--4$_{3, 1}$ E, v$_t$=0 & 241,852.299 & 97.53 & \checkmark?\tablenotemark{\footnotesize c} & \checkmark\checkmark & - & - & -  \\
CH$_3$OH & 5$_{ 1, 4}$--4$_{ 1, 3}$ E, v$_t$=0 & 241,879.025 & 55.87 & \checkmark\checkmark & \checkmark\checkmark & \checkmark\checkmark & \checkmark\checkmark & \checkmark\checkmark  \\
CH$_3$OH & 5$_{2, 3}$--4$_{2, 2}$ A, v$_t$=0 & 241,887.674 & 72.54 & \checkmark & \checkmark\checkmark & - & - & -  \\
CH$_3$OH & 5$_{2, 3}$--4$_{2, 2}$ E, v$_t$=0 & 241,904.147 & 60.73 & \checkmark\checkmark & \checkmark\checkmark & \checkmark\checkmark & \checkmark\checkmark & \checkmark\checkmark  \\
CH$_3$OH & 5$_{ 2, 4}$--4$_{ 2, 3}$ E, v$_t$=0 & 241,904.643 & 57.07 & \checkmark\checkmark & \checkmark\checkmark & \checkmark\checkmark & \checkmark\checkmark & \checkmark\checkmark  \\
CH$_3$OH & 14$_{1, 14}$--13$_{2, 11}$ E, v$_t$=0 & 242,446.084 & 248.94 & - & \checkmark & - & - & -  \\
CH$_3$OH & 5$_{1, 4}$--4$_{1, 3}$ A, v$_t$=0 & 243,915.788 & 49.66 & \checkmark\checkmark & \checkmark\checkmark & \checkmark\checkmark & \checkmark\checkmark & \checkmark\checkmark  \\
CH$_3$OH & 18$_{3, 16}$--18$_{2, 17}$ A, v$_t$=0 & 257,402.086 & 446.55 & \checkmark?\tablenotemark{\footnotesize b} & \checkmark & - & - & -  \\
CH$_3$OH & 19$_{3, 17}$--19$_{2, 18}$ A, v$_t$=0 & 258,780.248 & 490.58 & - & \checkmark? & - & - & -  \\
CH$_3$OCH$_3$ & 13$_{1, 13}$--12$_{0, 12}$ EA & 241,946.249 & 81.13 & - & \checkmark & - & - & -  \\
CH$_3$OCH$_3$ & 13$_{1, 13}$--12$_{0, 12}$ AE & 241,946.249 & 81.13 & - & \checkmark & - & - & -  \\
CH$_3$OCH$_3$ & 13$_{1, 13}$--12$_{0, 12}$ EE & 241,946.542 & 81.13 & - & \checkmark & - & - & -  \\
CH$_3$OCH$_3$ & 13$_{1, 13}$--12$_{0, 12}$ AA & 241,946.835 & 81.13 & - & \checkmark & - & - & -  \\
CH$_3$OCH$_3$ & 14$_{1, 14}$--13$_{0, 13}$ EA & 258,548.819 & 93.33 & - & \checkmark\checkmark & - & - & -  \\
CH$_3$OCH$_3$ & 14$_{1, 14}$--13$_{0, 13}$ AE & 258,548.819 & 93.33 & - & \checkmark\checkmark & - & - & -  \\
CH$_3$OCH$_3$ & 14$_{1, 14}$--13$_{0, 13}$ EE & 258,549.063 & 93.33 & - & \checkmark\checkmark & - & - & -  \\
CH$_3$OCH$_3$ & 14$_{1, 14}$--13$_{0, 13}$ AA & 258,549.308 & 93.33 & - & \checkmark\checkmark & - & - & -  \\
CH$_3$CN & 14$_5$--13$_5$ & 257,403.585 & 271.23 & \checkmark\tablenotemark{\footnotesize b} & \checkmark & - & - & -  \\
CH$_3$CN & 14$_3$--13$_3$ & 257,482.792 & 156.99 & \checkmark\tablenotemark{\footnotesize b} & \checkmark & - & - & -  \\
CH$_3$CN & 14$_2$--13$_2$ & 257,507.562 & 121.28 & \checkmark\tablenotemark{\footnotesize b} & \checkmark\tablenotemark{\footnotesize d} & - & - & -  \\
CH$_3$CN & 14$_1$--13$_1$ & 257,522.428 & 99.85 & \checkmark\tablenotemark{\footnotesize b} & \checkmark & - & - & -  \\
CH$_3$CN & 14$_0$--13$_0$ & 257,527.384 & 92.71 & \checkmark\tablenotemark{\footnotesize b} & \checkmark\checkmark & - & - & -  \\
\hline
\multicolumn{9}{c}{Other Molecules} \\
\hline
HNCO & 11$_{0, 11}$--10$_{0, 10}$ & 241,774.032 & 69.63 & \checkmark? & \checkmark\checkmark & - & - & -  \\
HC$^{15}$N & 3--2 & 258,156.996 & 24.78 & \checkmark\checkmark & \checkmark\checkmark & \checkmark & \checkmark & -  \\
H$^{13}$CN & 3--2 & 259,011.798 & 24.86 & \checkmark\checkmark & \checkmark\checkmark & \checkmark\checkmark & \checkmark\checkmark & \checkmark\checkmark  \\
H$^{13}$CO$^{+}$ & 3--2 & 260,255.339 & 24.98 & \checkmark\checkmark &\checkmark\checkmark  & \checkmark\checkmark & \checkmark\checkmark & \checkmark\checkmark  \\
CH$_2$CO & 12$_{1, 11}$--11$_{1, 10}$ & 244,712.269 & 89.40 & - & \checkmark & - & - & -  \\
CH$_2$CO & 13$_{1, 13}$--12$_{1, 12}$ & 260,191.982 & 100.47 & - & \checkmark?\tablenotemark{\footnotesize d} & - & - & -  \\
SO $^{3}\Sigma$ & 6$_6$--5$_5$ & 258,255.826 & 56.50 & \checkmark\checkmark & \checkmark\checkmark & \checkmark\checkmark & \checkmark\checkmark & \checkmark\checkmark  \\
OCS & 20--19 & 243,218.036 & 122.58 & - & \checkmark\checkmark & - & - & -  \\
H$_2$CS & 7$_{1, 6}$--6$_{1, 5}$ & 244,048.504 & 60.03 & \checkmark\checkmark & \checkmark\checkmark & \checkmark\checkmark & \checkmark\checkmark & \checkmark\checkmark  \\
CS & 5--4 & 244,935.557 & 35.27 & \checkmark\checkmark & \checkmark\checkmark & \checkmark\checkmark & \checkmark\checkmark & \checkmark\checkmark  \\
C$^{33}$S & 5--4 & 242,913.610 & 34.98 & \checkmark\checkmark & \checkmark\checkmark & \checkmark\checkmark & \checkmark? & \checkmark\checkmark  \\
SO$_2$ & 5$_{2, 4}$--4$_{1, 3}$ & 241,615.797 & 23.59 & \checkmark\checkmark & \checkmark\checkmark & \checkmark\checkmark & - & \checkmark\checkmark  \\
SO$_2$ & 14$_{0, 14}$--13$_{1, 13}$ & 244,254.218 & 93.90 & \checkmark\checkmark & \checkmark\checkmark & \checkmark & - & \checkmark?  \\ 
SO$_2$ & 26$_{3, 23}$--25$_{4, 22}$ & 245,339.233 & 350.79 & \checkmark? & - & - & - & -  \\
SO$_2$ & 10$_{3, 7}$--10$_{2, 8}$ & 245,563.422 & 72.72 & \checkmark\checkmark & \checkmark\checkmark & \checkmark? & - & -  \\
SO$_2$ & 7$_{3, 5}$--7$_{2, 6}$ & 257,099.966 & 47.84 & \checkmark\checkmark & \checkmark\checkmark & \checkmark? & - & \checkmark?  \\
SO$_2$ & 32$_{4, 28}$--32$_{3, 29}$ & 258,388.716 & 531.12 & \checkmark?\tablenotemark{\footnotesize c} & - & - & - & -  \\
SO$_2$ & 9$_{3, 7}$--9$_{2, 8}$ & 258,942.199 & 63.47 & \checkmark\checkmark & \checkmark\checkmark & - & - & -  \\
SO$_2$ & 30$_{4, 26}$--30$_{3, 27}$ & 259,599.448 & 471.52 & \checkmark?\tablenotemark{\footnotesize c} & \checkmark? & - & - & -  \\
$^{34}$SO$_2$ & 16$_{1, 15}$--15$_{2, 14}$ & 241,509.046 & 130.31 & \checkmark?\tablenotemark{\footnotesize c} & - & - & - & -  \\
$^{34}$SO$_2$ & 6$_{3, 3}$--6$_{2, 4}$ & 245,302.239 & 40.66 & - & \checkmark? & - & - & -  \\
$^{33}$SO & 6$_{7, 6}$--5$_{6, 5}$ & 259,280.331 & 47.12 & - & \checkmark & - & - & -  \\
$^{33}$SO & 6$_{7, 7}$--5$_{6, 6}$ & 259,282.276 & 47.12 & - & \checkmark & - & - & -  \\
$^{33}$SO & 6$_{7, 8}$--5$_{6, 7}$ & 259,284.027 & 47.12 & - & \checkmark & - & - & -  \\
$^{33}$SO & 6$_{7, 9}$--5$_{6, 8}$ & 259,284.027 & 47.12 & - & \checkmark & - & - & -  \\
SiO & 6--5 & 260,518.009 & 43.76 & \checkmark\checkmark & \checkmark\checkmark & \checkmark\checkmark & \checkmark? & -  \\
HDO & 2$_{1, 1}$--2$_{1, 2}$ & 241,561.550 & 95.22 & \checkmark?\tablenotemark{\footnotesize c} & - & - & - & -  \\
HDCO & 4$_{2, 3}$--3$_{2, 2}$ & 257,748.701 & 62.78 & - & \checkmark? & - & - & -  \\
HDCO & 4$_{2, 2}$--3$_{2, 1}$ & 259,034.910 & 62.87 & - & \checkmark & - & - & -  \\
HDCS & 8$_{1, 8}$--7$_{1, 7}$ & 242,991.144 &	 61.48 & - & $\checkmark$? & - & - & - \\
\enddata
\tablenotetext{a}{Spectroscopic parameters were taken from the CDMS catalog for all species except HDO for which the data were taken from the JPL database. The symbols `\checkmark', `\checkmark?', and '-' indicate, respectively, a detection, a tentative detection, and a non detection of a given transition. The symbol '\checkmark\checkmark' indicates the detection is above $5\sigma_{\rm s,i}$.}
\tablenotetext{b}{In the spectrum of N\,83A--mm\,A, the CH$_3$CN $K$=14--13 ladder as a whole is securely detected though individual lines do not pass our detection criteria.}
\tablenotetext{c}{Transitions in the spectrum of N\,83A-mm\,A fulfilling our tentative detection criteria after Hanning smoothing: 7 smoothing windows were required for CH$_3$OH at 241.852 GHz, 5 for SO$_2$ at 258.388 and 259.599 GHz, 5 for $^{34}$SO$_2$ at 241.509 GHz, and 9 for HDO at 241.562 GHz.}
\tablenotetext{d}{Transitions in the spectrum of N\,83A-mm\,B fulfilling our tentative detection criteria after Hanning smoothing: 5 smoothing windows were required for CH$_3$CN at 257.508 GHz, and 9 for CH$_2$CO at 260.192 GHz.} 
\end{deluxetable*}

\clearpage 

\section{The XCLASS Fitting Results and Fractional Molecular Abundances with Respect to H$_2$} 
\label{app:XCLASSresults}

Table~\ref{t:XCLASSresults} provides the results of spectral modeling with XCLASS and the fractional molecular abundances with respect to H$_2$ for all the sources analyzed in this paper: N\,83A--mm\,A, C, D, and E in the ALMA field N\,83--1, and N\,83A--mm\,B in the ALMA field N\,83--2. The XCLASS spectral modeling is described in Section~\ref{s:modeling}.

\startlongtable
\begin{deluxetable*}{ccccccc}
\centering
\tablecaption{The Results of Spectral Modeling with XCLASS and Fractional Abundances with Respect to H$_2$ \label{t:XCLASSresults}}
\tablewidth{0pt}
\tablehead{
\colhead{Source} &
\multicolumn{1}{c}{Species, ${\rm X}$} &
\colhead{$T_{\rm rot} ({\rm X})$\tablenotemark{\footnotesize a}} &
\colhead{$N({\rm X})$}&
\colhead{$v_{\rm LSR}$\tablenotemark{\footnotesize a,b}} &    
\colhead{$\Delta v_{\rm FWHM}$\tablenotemark{\footnotesize a}} &
\colhead{$N(\rm X)/N({\rm H_2})$}  \\
\colhead{} &
\colhead{} &
\colhead{(K)} &
\colhead{(cm$^{-2}$)} &
\colhead{(km s$^{-1}$)} &
\colhead{(km s$^{-1}$)} &
\colhead{}
}
\startdata
N\,83A--mm\,A & CH$_3$OH & $151.76 \pm 8.29$ & $(7.57 \pm0.27)\times10^{15}$ & $246.73 \pm0.09$ & $5.04 \pm0.22$ & $(5.12 \pm 0.62) \times 10^{-8}$ \\ 
              & CH$_3$OH\tablenotemark{\footnotesize c} & $17.04 \pm2.57$ & $(8.80 \pm1.68)\times10^{14}$ & $246.43 \pm0.04$ & $4.53 \pm0.10$ & $(5.95 \pm 1.33) \times 10^{-9}$ \\
              & CH$_3$CN & $228.80 \pm47.54$ & $(1.60 \pm0.46)\times10^{13}$ & $245.40 \pm0.28$ & $4.94 \pm0.63$ & $(1.08 \pm 0.34) \times 10^{ -10 }$ \\
              & SO$_2$ (c) & $202.27 \pm11.24$ & $(1.87 \pm0.10)\times10^{15}$ & $248.06 \pm0.15$ & $7.79 \pm0.36$ & $(1.26 \pm 0.16) \times 10^{ -8 }$ \\
              & SO$_2$ (f) & $24.95 \pm2.03$ & $(1.42 \pm0.11)\times10^{14}$ & $246.80 \pm0.09$ & $3.58 \pm0.24$ & $(9.60 \pm 1.34) \times 10^{ -10 }$ \\
              & $^{34}$SO$_2$ & 202 & $(1.39 \pm0.16)\times10^{14}$ & $248.64 \pm0.22$ & $3.81 \pm0.48$ & $(9.40 \pm 1.54) \times 10^{ -10 }$ \\
              & SO & 202 & $(1.00 \pm0.01)\times10^{15}$ & $246.89 \pm0.01$ & $4.45 \pm0.04$ & $(6.76 \pm 0.79) \times 10^{ -9 }$ \\
              & $^{33}$SO & 202 & $<7.82\times10^{13}$ & 248.06 & 7.79 & $<5.29\times 10^{ -10 }$ \\
              & CS & 152 & $(3.96 \pm0.01)\times10^{14}$ & $246.70 \pm0.01$ & $4.71 \pm0.02$ & $(2.68 \pm 0.31) \times 10^{ -9 }$ \\ 
              & C$^{33}$S & 152 & $(1.44 \pm0.10)\times10^{13}$ & $246.98 \pm0.05$ & $3.80 \pm0.11$ & $(9.73 \pm 1.32) \times 10^{ -11 }$ \\
              & OCS & 152 & $<3.08\times10^{14}$ & 246.73 & 5.04 & $<2.08\times 10^{ -9 }$ \\
              & H$_2$CS & 152 & $(2.29 \pm0.07)\times10^{14}$ & $246.89 \pm0.05$ & $3.83 \pm0.14$ & $(1.55 \pm 0.19) \times 10^{ -9 }$ \\
              & H$^{13}$CN & 152 & $(6.76 \pm0.34)\times10^{12}$ & $246.97 \pm0.06$ & $3.29 \pm0.06$ & $(4.57 \pm 0.58) \times 10^{ -11 }$ \\ 
              & HC$^{15}$N & 152 & $(4.34 \pm0.28)\times10^{12}$ & $247.02 \pm0.05$ & $3.53 \pm0.05$ & $(2.93 \pm 0.39) \times 10^{ -11 }$ \\
              & H$^{13}$CO$^+$ & 152 & $(8.44 \pm0.12)\times10^{12}$ & $246.93 \pm0.03$ & $3.91 \pm0.07$ & $(5.71 \pm 0.67) \times 10^{ -11 }$ \\
              & HDO & 152 & $(1.51 \pm0.20)\times10^{14}$ & $247.01 \pm0.03$ & $3.14 \pm0.03$ & $(1.02 \pm 0.18) \times 10^{ -9 }$ \\ 
              & HNCO & 152 & $(4.90 \pm0.66)\times10^{13}$ & $245.61 \pm0.37$ & $5.33 \pm0.92$ & $(3.31 \pm 0.59) \times 10^{ -10 }$ \\
              & HDCO & 152 & $<2.07\times10^{13}$ & 246.73 & 5.04 & $<1.40\times 10^{ -10 }$ \\
              & HDCS & 152 & $<1.11\times10^{14}$ & 246.73 & 5.04 & $<7.50\times 10^{ -10 }$ \\
              & SiO & 152 & $(4.45 \pm0.34)\times10^{12}$ & $247.00 \pm0.05$ & $3.38 \pm0.05$ & $(3.01 \pm 0.42) \times 10^{ -11 }$ \\  
              & HCOOCH$_3$ & 152 & $<1.14\times10^{14}$ & 246.73 & 5.04 &$<7.71\times 10^{ -10 }$ \\ 
              & CH$_3$OCH$_3$ & 152 & $<2.62\times10^{14}$ & 246.73 & 5.04 & $<1.77\times 10^{ -9 }$ \\
              & NH$_2$CHO & 152 & $<5.18\times10^{12}$ & 246.73 & 5.04 & $<3.50\times 10^{ -11 }$ \\ 
              & CH$_2$CO & 152 & $<2.36\times10^{13}$ & 246.73 & 5.04 & $<1.60\times 10^{ -10 }$ \\
              & CH$_3$CHO & 152 & $<4.76\times10^{13}$ & 246.73 & 5.04 & $<3.22\times 10^{ -10 }$ \\ 
              \hline
N\,83A--mm\,B & CH$_3$OH & $129.59 \pm3.24$ & $(1.03 \pm 0.02)\times10^{16}$ & $246.88 \pm0.03$ & $3.50 \pm0.06$ & $(1.07 \pm 0.12) \times 10^{ -7 }$ \\  
              & CH$_3$OH\tablenotemark{\footnotesize c}  & $20.24 \pm1.32$ & $(9.09 \pm 0.54)\times10^{14}$ & $246.77 \pm0.01$ & $2.93 \pm0.04$ & $(9.46 \pm 1.15) \times 10^{ -9 }$ \\  
              & CH$_3$CN & $86.92 \pm14.97$ & $(8.18 \pm 0.63)\times10^{12}$ & $246.44 \pm0.16$ & $4.21 \pm0.31$ & $(8.52 \pm 1.12) \times 10^{ -11 }$ \\    
              & SO$_2$ (c) & $133.47 \pm4.69$ & $(1.26 \pm 0.04)\times10^{15}$ & $247.11 \pm0.03$ & $3.78 \pm0.08$ & $(1.31 \pm 0.15) \times 10^{ -8 }$ \\  
              & SO$_2$ (f) & $10.54 \pm0.57$ & $(2.34 \pm 0.15)\times10^{14}$ & $246.87 \pm0.05$ & $2.86 \pm0.10$ & $(2.44 \pm 0.30) \times 10^{ -9 }$ \\  
              & $^{34}$SO$_2$ & 133 & $(3.21 \pm 0.91)\times10^{13}$ & $246.52 \pm0.88$ & $5.55 \pm1.95$ & $(3.34 \pm 1.01) \times 10^{ -10 }$ \\  
              & SO & 133 & $(8.08 \pm 0.03)\times10^{14}$ & $247.10 \pm0.01$ & $3.16 \pm0.02$ & $(8.41 \pm 0.89) \times 10^{ -9 }$ \\
              & $^{33}$SO & 133 & $(2.15 \pm 0.25)\times10^{13}$ & $246.88 \pm0.04$ & $2.56 \pm0.03$ & $(2.24 \pm 0.35) \times 10^{ -10 }$ \\
              & CS & 130 & $(2.39 \pm 0.01)\times10^{14}$ & $247.05 \pm0.01$ & $3.61 \pm0.01$ & $(2.49 \pm 0.26) \times 10^{ -9 }$ \\ 
              & C$^{33}$S & 130 & $(1.90 \pm 0.06)\times10^{13}$ & $247.01 \pm0.06$ & $3.03 \pm0.07$ & $(1.98 \pm 0.22) \times 10^{ -10 }$ \\
              & OCS & 130 & $(1.94 \pm 0.10)\times10^{14}$ & $246.76 \pm0.09$ & $3.38 \pm0.09$ & $(2.02 \pm 0.24) \times 10^{ -9 }$ \\ 
              & H$_2$CS & 130 & $(2.17 \pm 0.04)\times10^{14}$ & $246.89 \pm0.02$ & $2.74 \pm0.07$ & $(2.26 \pm 0.24) \times 10^{ -9 }$ \\ 
              & H$^{13}$CN & 130 & $(5.70 \pm 0.23)\times10^{12}$ & $246.88 \pm0.08$ & $3.84 \pm0.10$ & $(5.93 \pm 0.67) \times 10^{ -11 }$ \\  
              & HC$^{15}$N & 130 & $(3.74 \pm 0.22)\times10^{12}$ & $246.85 \pm0.07$ & $3.59 \pm0.07$ & $(3.89 \pm 0.47) \times 10^{ -11 }$ \\
              & H$^{13}$CO$^+$ & 130 & $(3.38 \pm 0.07)\times10^{12}$ & $247.02 \pm0.03$ & $2.85 \pm0.07$ & $(3.52 \pm 0.38) \times 10^{ -11 }$ \\
              & HDO & 130 & $<2.45\times10^{14}$ & 246.88 & 3.50 & $<2.55\times 10^{ -9 }$ \\
              & HNCO & 130 & $(4.96 \pm 0.40)\times10^{13}$ & $246.42 \pm0.17$ & $4.15 \pm0.43$ & $(5.16 \pm 0.69) \times 10^{ -10 }$ \\
              & HDCO & 130 & $(2.55 \pm 0.25)\times10^{13}$ & $246.57 \pm0.06$ & $2.64 \pm0.04$ & $(2.65 \pm 0.38) \times 10^{ -10 }$ \\
              & HDCS & 130 & $(2.76 \pm 0.49)\times10^{13}$ & $246.61 \pm0.04$ & $1.85 \pm0.04$ & $(2.87 \pm 0.59) \times 10^{ -10 }$ \\
              & SiO & 130 & $(5.07 \pm 0.23)\times10^{12}$ & $247.14 \pm0.08$ & $6.10 \pm0.09$ & $(5.28 \pm 0.61) \times 10^{ -11 }$ \\
              & HCOOCH$_3$ & 130 & $<2.58\times10^{14}$ & 246.88 & 3.50 & $<2.69\times 10^{ -9 }$ \\
              & CH$_3$OCH$_3$ & 130 & $(3.01 \pm 0.18)\times10^{14}$ & $247.09 \pm0.06$ & $4.18 \pm0.06$ & $(3.13 \pm 0.38) \times 10^{ -9 }$ \\  
              & NH$_2$CHO & 130 & $<5.00\times10^{10}$ & 246.88 & 3.50 & $<5.21\times 10^{ -13 }$ \\
              & CH$_2$CO & 130 & $(3.28 \pm 0.34)\times10^{13}$ & $246.60 \pm0.05$ & $3.42 \pm0.03$ & $(3.41 \pm 0.51) \times 10^{ -10 }$ \\ 
              & CH$_3$CHO & 130 & $<1.30\times10^{14}$ & 246.88 & 3.50 & $<1.35\times 10^{ -9 }$ \\
              \hline
N\,83A--mm\,C & CH$_3$OH\tablenotemark{\footnotesize c}  & $26.84 \pm2.46$ & $(5.00 \pm 0.24)\times10^{14}$ & $244.24 \pm0.01$ & $2.87 \pm0.02$ & $(1.06 \pm 0.17) \times 10^{ -9 }$ \\  
              & CH$_3$CN & 27 & $<5.26\times10^{10}$ & 244.24 & 2.87 & $<1.12\times 10^{ -13 }$ \\
              & SO$_2$ & 27 & $(4.49 \pm 0.26)\times10^{13}$ & $243.98 \pm0.08$ & $3.04 \pm0.03$ & $(9.55 \pm 1.58) \times 10^{ -11 }$ \\ 
              & $^{34}$SO$_2$ & 27 & $<5.83\times10^{10}$ & 244.24 & 2.87 & $<1.24\times 10^{ -13 }$ \\
              & SO & 27 & $(2.43 \pm 0.02)\times10^{14}$ & $243.83 \pm0.01$ & $2.89 \pm0.03$ & $(5.17 \pm 0.80) \times 10^{ -10 }$ \\
              & $^{33}$SO & 27 & $<1.57\times10^{12}$ & 244.24 & 2.87 & $<3.34\times 10^{ -12 }$ \\
              & CS & 27 & $(1.58 \pm 0.004)\times10^{14}$ & $244.08 \pm0.01$ & $3.47 \pm0.01$ & $(3.37 \pm 0.52) \times 10^{ -10 }$ \\
              & C$^{33}$S & 27 & $(2.97 \pm 0.27)\times10^{12}$ & $244.24 \pm0.08$ & $2.98 \pm0.03$ & $(6.31 \pm 1.13) \times 10^{ -12 }$ \\
              & OCS & 27 & $<1.04\times10^{14}$ & 244.24 & 2.87 & $<2.21\times 10^{ -10 }$ \\
              & H$_2$CS & 27 & $(5.29 \pm 0.15)\times10^{13}$ & $243.97 \pm0.05$ & $2.95 \pm0.03$ & $(1.12 \pm 0.18) \times 10^{ -10 }$ \\ 
              & H$^{13}$CN & 27 & $(1.03 \pm 0.07)\times10^{12}$ & $243.75 \pm0.13$ & $2.89 \pm0.04$ & $(2.19 \pm 0.37) \times 10^{ -12 }$ \\
              & HC$^{15}$N & 27 & $(6.37 \pm 0.51)\times10^{11}$ & $244.04 \pm0.08$ & $2.89 \pm0.02$ & $(1.35 \pm 0.24) \times 10^{ -12 }$ \\ 
              & H$^{13}$CO$^+$ & 27 & $(1.06 \pm 0.02)\times10^{12}$ & $243.55 \pm0.03$ & $2.61 \pm0.05$ & $(2.26 \pm 0.35) \times 10^{ -12 }$ \\ 
              & HDO & 27 & $<5.00\times10^{10}$ & 244.24 & 2.87 & $<1.06\times 10^{ -13 }$ \\
              & HNCO & 27 & $<2.15\times10^{12}$ & 244.24 & 2.87 & $<4.57\times 10^{ -12 }$ \\
              & HDCO & 27 & $<3.78\times10^{12}$ & 244.24 & 2.87 & $<8.04\times 10^{ -12 }$ \\
              & HDCS & 27 & $<9.32\times10^{12}$ & 244.24 & 2.87 & $<1.98\times 10^{ -11 }$ \\
              & SiO & 27 & $(1.34 \pm 0.11)\times10^{12}$ & $244.41 \pm0.08$ & $3.19 \pm0.06$ & $(2.85 \pm 0.50) \times 10^{ -12 }$ \\ 
              & CH$_3$OCH$_3$ & 27 & $<9.42\times10^{10}$ & 244.24 & 2.87 & $<2.00\times 10^{ -13 }$ \\
              & CH$_2$CO & 27 & $<1.02\times10^{13}$ & 244.24 & 2.87 & $<2.17\times 10^{ -11 }$ \\
              \hline
N\,83A--mm\,D & CH$_3$OH\tablenotemark{\footnotesize c}  & $23.01 \pm 1.81$ & $(8.30 \pm 0.43)\times10^{14}$ & $245.30 \pm0.01$ & $3.55 \pm 0.03$ & $(2.05 \pm 0.33) \times 10^{ -9 }$ \\
              & CH$_3$CN & 23 & $<4.12\times10^{12}$ & 245.30 & 3.55 & $<1.02\times 10^{ -11 }$ \\
              & SO$_2$ & 23 & $<3.06\times10^{13}$ & 245.30 & 3.55 & $<7.57\times 10^{ -11 }$ \\
              & $^{34}$SO$_2$ & 23 & $<2.99\times10^{13}$ & 245.30 & 3.55 & $<7.40\times 10^{ -11 }$ \\
              & SO & 23 & $(2.67 \pm 0.04)\times10^{14}$ & $245.35 \pm0.03$ & $3.36 \pm0.07$ & $(6.61 \pm 0.99) \times 10^{ -10 }$ \\
              & $^{33}$SO & 23 & $<3.80\times10^{12}$ & 245.30 & 3.55 & $<9.40\times 10^{ -12 }$ \\
              & CS & 23 & $(1.67 \pm 0.01)\times10^{14}$ & $245.31 \pm0.01$ & $3.73 \pm0.02$ & $(4.13 \pm 0.62) \times 10^{ -10 }$ \\
              & C$^{33}$S & 23 & $(2.88 \pm 0.34)\times10^{12}$ & $245.49 \pm0.14$ & $3.50 \pm0.03$ & $(7.12 \pm 1.36) \times 10^{ -12 }$  \\
              & OCS & 23 & $<5.93\times10^{14}$ & 245.30 & 3.55 & $<1.47\times 10^{ -9 }$ \\
              & H$_2$CS & 23 & $(6.95 \pm 0.25)\times10^{13}$ & $245.29 \pm0.08$ & $3.69 \pm0.05$ & $(1.72 \pm 0.26) \times 10^{ -10 }$ \\ 
               & H$^{13}$CN & 23 & $(7.55 \pm 1.05)\times10^{11}$ & $245.30 \pm0.13$ & $1.71 \pm0.03$ & $(1.87 \pm 0.38) \times 10^{ -12 }$ \\
              & HC$^{15}$N & 23 & $(7.72 \pm 0.83)\times10^{11}$ & $245.81 \pm0.14$ & $3.73 \pm0.04$ & $(1.91 \pm 0.35) \times 10^{ -12 }$ \\  
              & H$^{13}$CO$^+$ & 23 & $(6.29 \pm 0.31)\times10^{11}$ & $245.42 \pm0.08$ & $2.90 \pm0.07$ & $(1.56 \pm 0.25) \times 10^{ -12 }$ \\
              & HDO & 23 & $<4.28\times10^{13}$ & 245.30 & 3.55 & $<1.06\times 10^{ -10 }$ \\
              & HNCO & 23 & $<2.91\times10^{12}$ & 245.30 & 3.55 & $<7.20\times 10^{ -12 }$ \\
              & HDCO & 23 & $<1.65\times10^{12}$ & 245.30 & 3.55 & $<4.08\times 10^{ -12 }$ \\
              & HDCS & 23 & $<2.91\times10^{12}$ & 245.30 & 3.55 & $<7.20\times 10^{ -12 }$ \\
              & SiO & 23 & $(1.47 \pm 0.19)\times10^{12}$ & $245.33 \pm0.15$ & $2.59 \pm0.04$ & $(3.64 \pm 0.72) \times 10^{ -12 }$ \\ 
              & CH$_3$OCH$_3$ & 23 & $<4.30\times10^{13}$ & 245.30 & 3.55 & $<1.06\times 10^{ -10 }$ \\
              & CH$_2$CO & 23 & $<3.54\times10^{13}$ & 245.30 & 3.55 & $<8.76\times 10^{ -11 }$ \\
              \hline
N\,83A--mm\,E & CH$_3$OH\tablenotemark{\footnotesize c}  & $26.43 \pm2.46$ & $(5.49 \pm 0.28)\times10^{14}$ & $246.03\pm0.01$ & $2.75 \pm0.03$ & $(2.78 \pm 0.50) \times 10^{ -9 }$ \\  
              & CH$_3$CN & 26 & $<1.64\times10^{12}$ & 246.03 & 2.75 & $<8.30\times 10^{ -12 }$ \\
              & SO$_2$ & 26 & $(3.91 \pm 0.32)\times10^{13}$ & $246.19\pm0.09$ & $2.37 \pm0.02$ & $(1.98 \pm 0.38) \times 10^{ -10 }$ \\ 
              & $^{34}$SO$_2$ & 26 & $<4.89\times10^{12}$ & 246.03 & 2.75 & $<2.47\times 10^{ -11 }$ \\
              & SO & 26 & $(2.07 \pm 0.03)\times10^{14}$ & $245.89\pm0.01$ & $2.61 \pm0.04$ & $(1.05 \pm 0.18) \times 10^{ -9 }$ \\
              & $^{33}$SO & 26 & $<2.30\times10^{12}$ & 246.03 & 2.75 & $<1.16\times 10^{ -11 }$ \\
              & CS & 26 & $(1.37 \pm 0.005)\times10^{14}$ & $245.90\pm0.01$ & $3.21 \pm0.01$ & $(6.93 \pm 1.21) \times 10^{ -10 }$ \\
              & C$^{33}$S & 26 & $(4.73 \pm 0.33)\times10^{12}$ & $246.02\pm0.09$ & $1.96 \pm0.03$ & $(2.39 \pm 0.45) \times 10^{ -11 }$ \\ 
              & OCS & 26 & $<1.47\times10^{13}$ & 246.03 & 2.75 & $<7.44\times 10^{ -11 }$ \\
              & H$_2$CS & 26 & $(3.94 \pm 0.18)\times10^{13}$ & $245.93\pm0.07$ & $2.53 \pm0.03$ & $(1.99 \pm 0.36) \times 10^{ -10 }$ \\ 
              & H$^{13}$CN & 26 & $(1.11 \pm 0.08)\times10^{12}$ & $246.04\pm0.10$ & $4.02 \pm0.02$ & $(5.62 \pm 1.10) \times 10^{ -12 }$ \\ 
              & HC$^{15}$N & 26 & $<2.53\times10^{11}$ & 246.03 & 2.75 & $<1.28\times 10^{ -12 }$ \\
              & H$^{13}$CO$^+$ & 26 & $(3.87 \pm 0.24)\times10^{11}$ & $245.99\pm0.10$ & $2.75 \pm0.04$ & $(1.96 \pm 0.36) \times 10^{ -12 }$ \\ 
              & HDO & 26 & $<2.99\times10^{13}$ & 246.03 & 2.75 & $<1.51\times 10^{ -10 }$ \\
              & HNCO & 26 & $<3.05\times10^{12}$ & 246.03 & 2.75 & $<1.54\times 10^{ -11 }$ \\
              & HDCO & 26 & $<9.06\times10^{11}$ & 246.03 & 2.75 & $<4.59\times 10^{ -12 }$ \\
              & HDCS & 26 & $<8.27\times10^{12}$ & 246.03 & 2.75 & $<4.19\times 10^{ -11 }$ \\
              & SiO & 26 & $<7.33\times10^{11}$ & 246.03 & 2.75 & $<3.71\times 10^{ -12 }$ \\
               & CH$_3$OCH$_3$ & 26 & $<1.66\times10^{13}$ & 246.03 & 2.75 & $<8.40\times 10^{ -11 }$ \\
              & CH$_2$CO & 26 & $<3.49\times10^{12}$ & 246.03 & 2.75 & $<1.77\times 10^{ -11 }$ \\
 \enddata
\tablenotetext{a}{Parameter values without the uncertainties ($T_{\rm rot}$, $v_{\rm LSR}$, and $\Delta v_{\rm FWHM}$) were fixed for the XCLASS fitting as described in Section~\ref{s:modeling}.}
\tablenotetext{b}{The reported kinematic LSR velocities ($v_{\rm LSR}$) are calculated from the velocity offsets returned by XCLASS  ($v_{\rm LSR, offset}$) using the systemic velocities of N\,83A--mm (A, B, C, D, E) of (247.0, 247.2, 244.2, 245.5, 246.2)~km~s$^{-1}$.} 
\tablenotetext{c}{The XCLASS analysis was performed in non-LTE with H$_2$ as the collision partner. }
\end{deluxetable*}

\clearpage 

\section{Observed Spectra and Model Fits}
\label{app:specobsfits}

\begin{figure*}
\centering
\includegraphics[width=\textwidth]{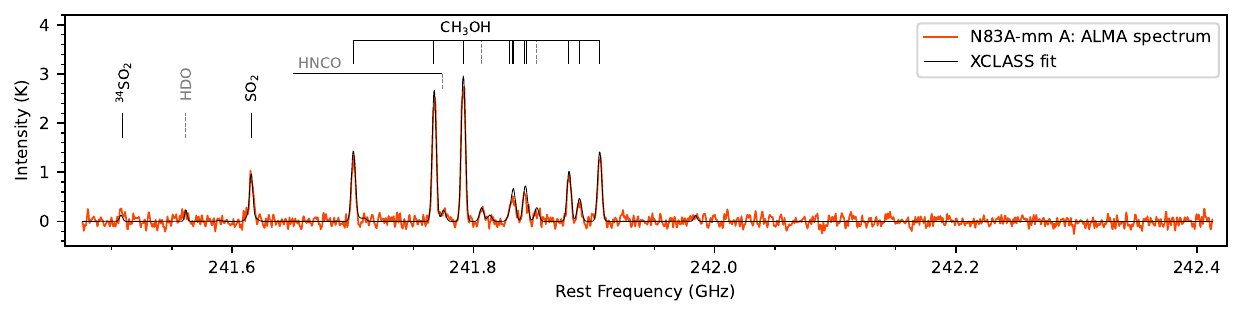}
\includegraphics[width=\textwidth]{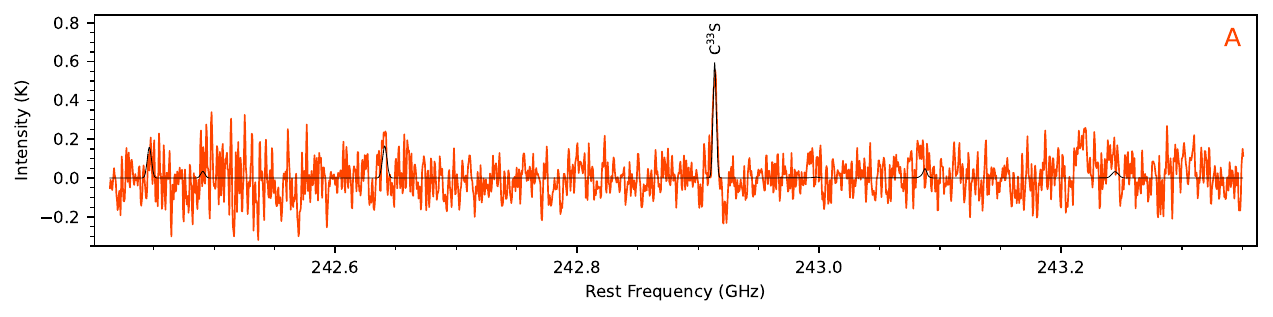}
\includegraphics[width=\textwidth]{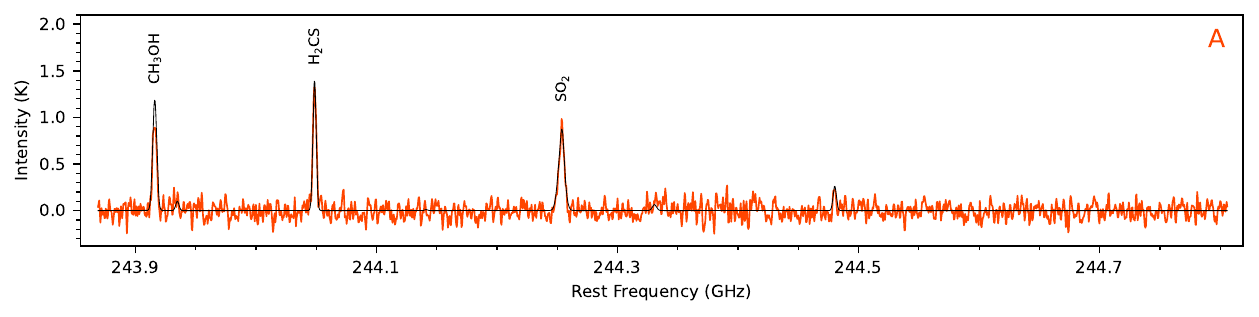}
\includegraphics[width=\textwidth]{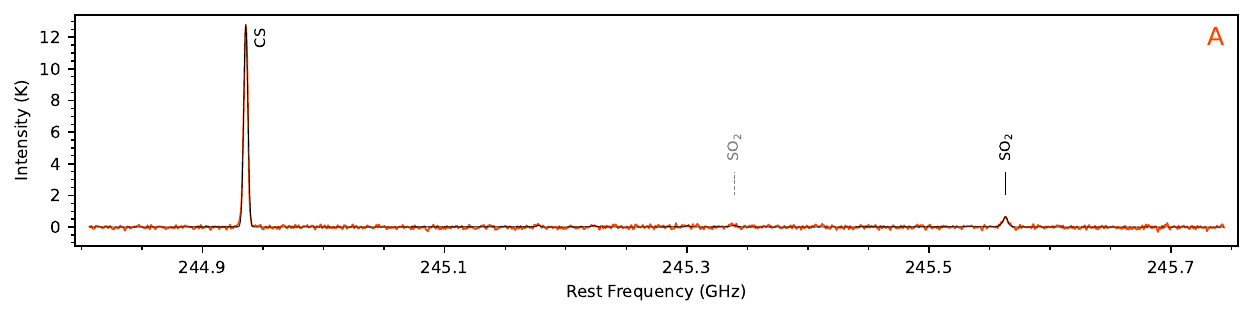}
\caption{ALMA Band 6 spectra of source N\,83A--mm\,A in the $\sim$242 GHz (top two panels) and $\sim$245 GHz (bottom two panels) spectral windows. The spectra were extracted as the mean over the area enclosed by the 50\% of the 1.2 mm continuum peak contour. Firm and tentative detections are indicated in black and gray, respectively.} \label{f:specA_1}
\end{figure*}

\begin{figure*}
\centering
\includegraphics[width=\textwidth]{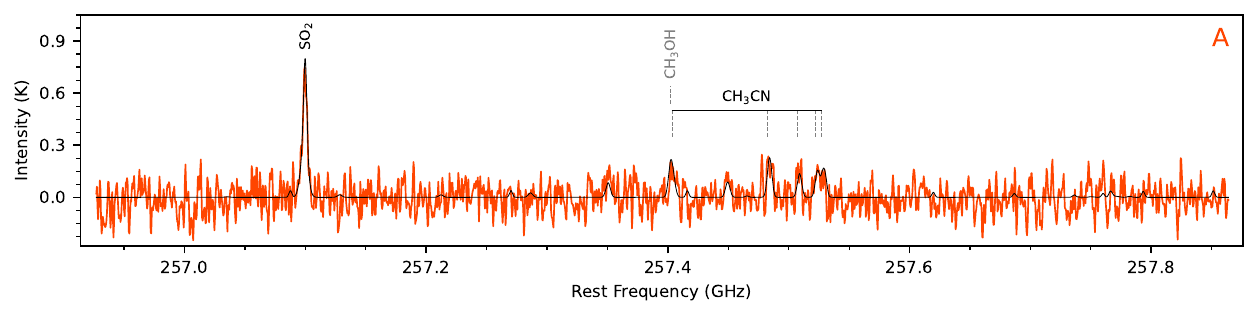}
\includegraphics[width=\textwidth]{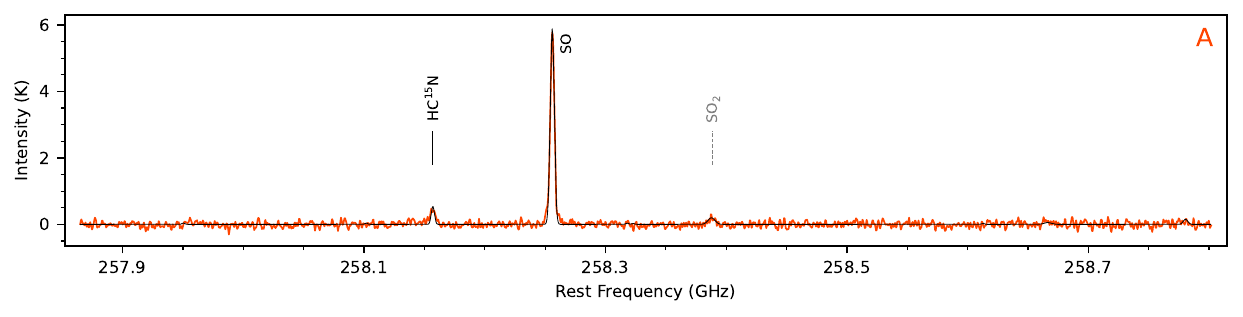}
\includegraphics[width=\textwidth]{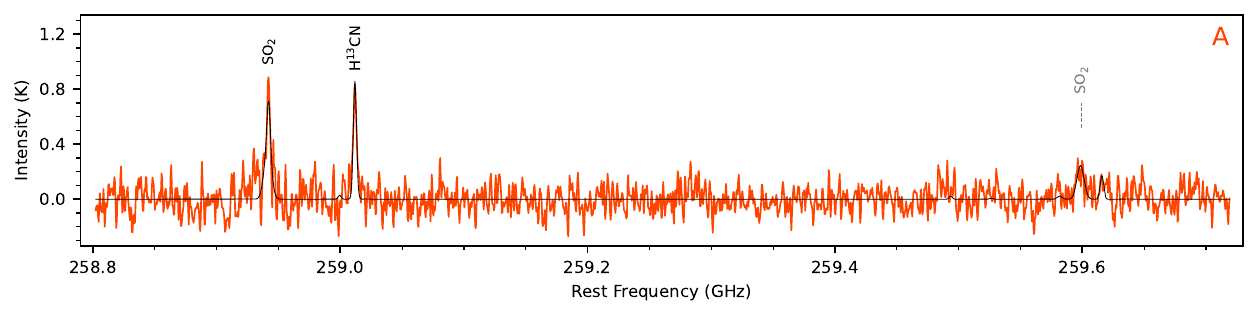}
\includegraphics[width=\textwidth]{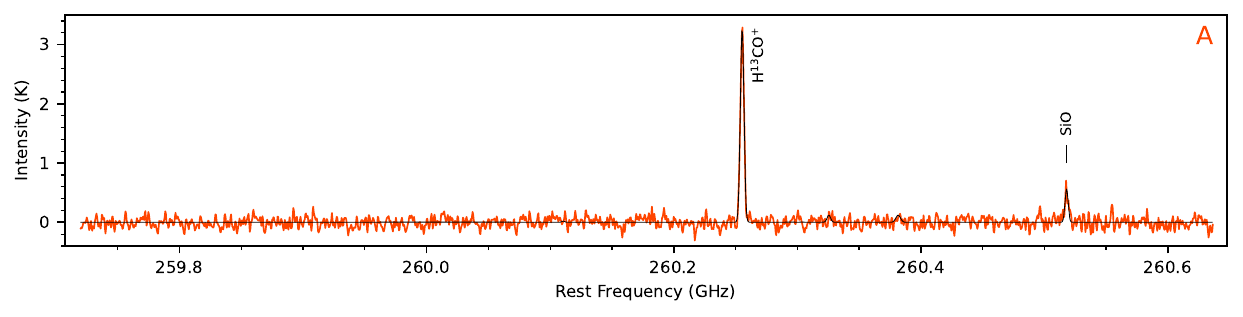}
\caption{The same as Figure~\ref{f:specA_1}, but for the $\sim$258 GHz (top two panels) and $\sim$260 GHz (bottom two panels) spectral windows. \label{f:specA_2}}
\end{figure*}

\begin{figure*}
\centering
\includegraphics[width=\textwidth]{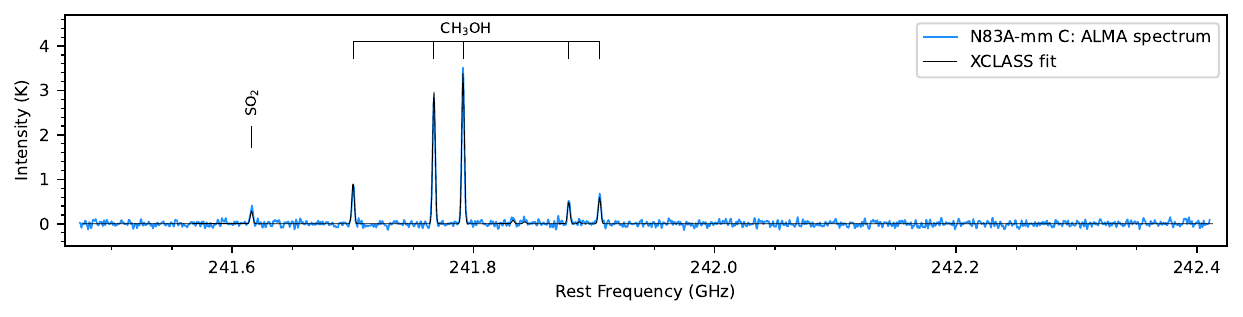}
\includegraphics[width=\textwidth]{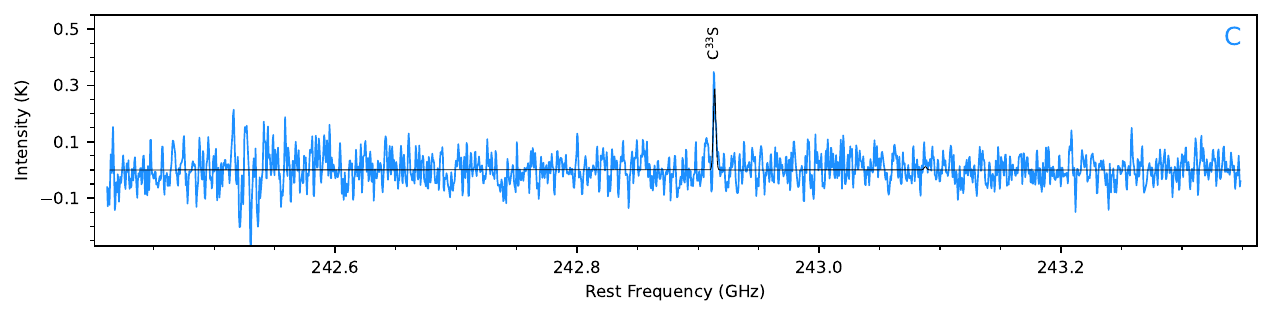}
\includegraphics[width=\textwidth]{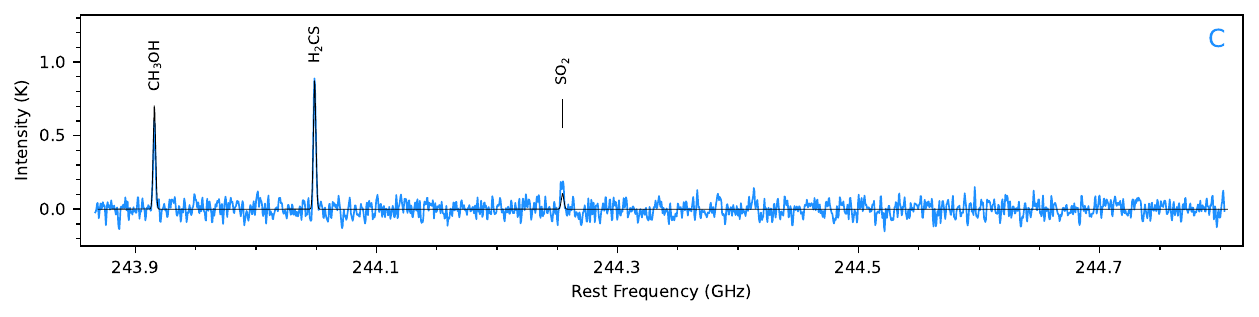}
\includegraphics[width=\textwidth]{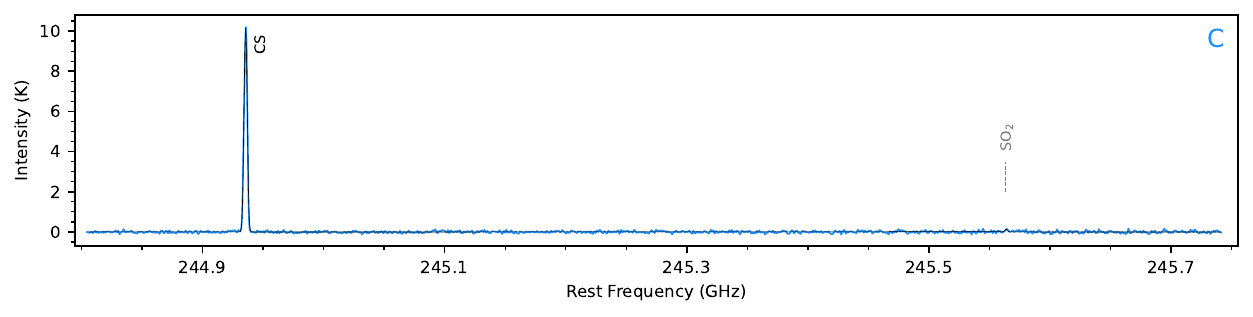}
\caption{The same as Figure~\ref{f:specA_1}, but for source N\,83A--mm C. \label{f:specC_1}}
\end{figure*}

\begin{figure*}
\centering
\includegraphics[width=\textwidth]{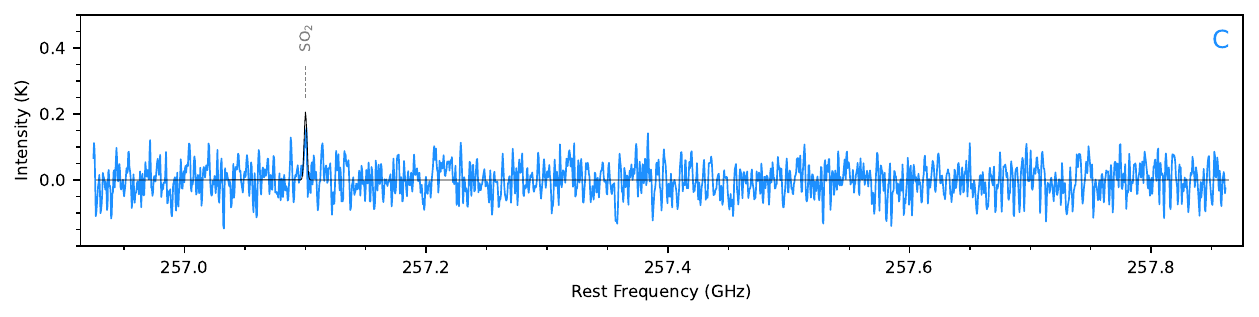}
\includegraphics[width=\textwidth]{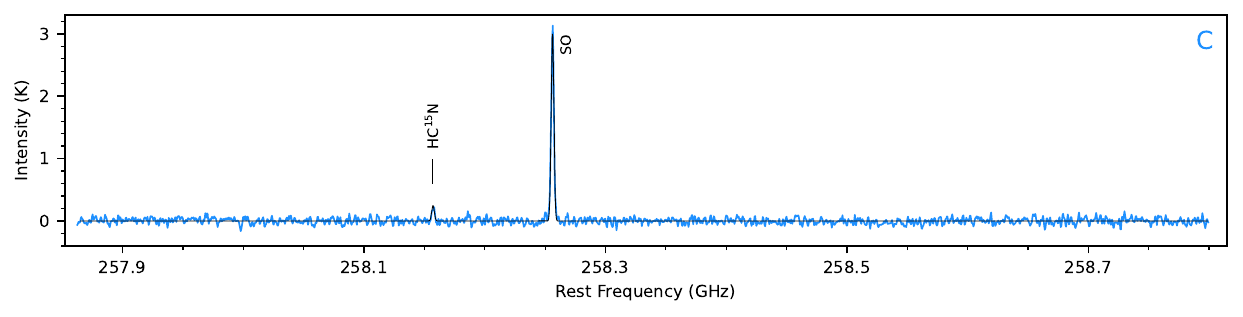}
\includegraphics[width=\textwidth]{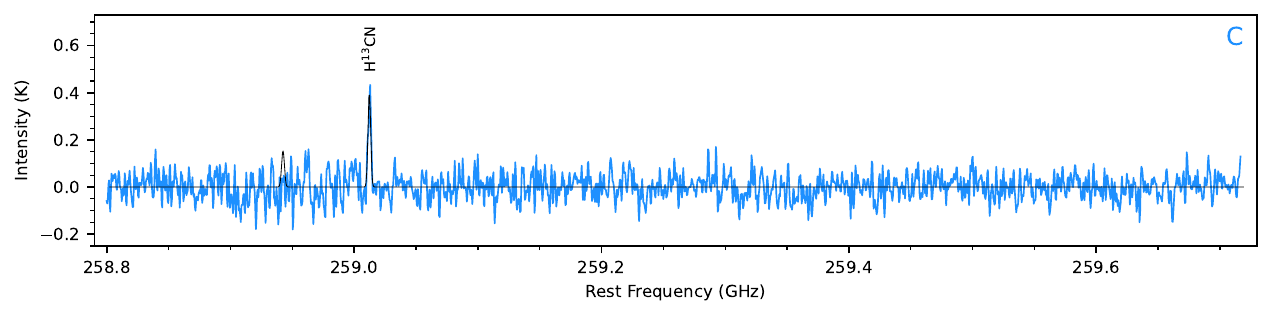}
\includegraphics[width=\textwidth]{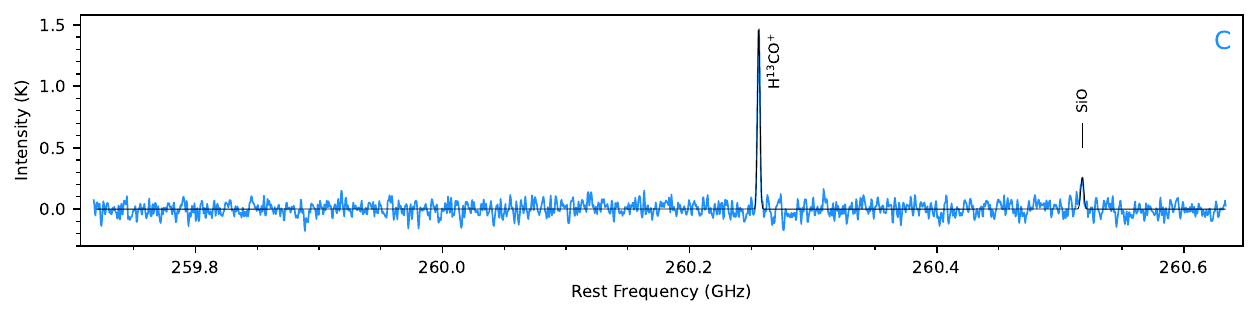}
\caption{The same as Figure~\ref{f:specA_2}, but for source N\,83A--mm C. \label{f:specC_2}}
\end{figure*}

\begin{figure*}
\centering
\includegraphics[width=\textwidth]{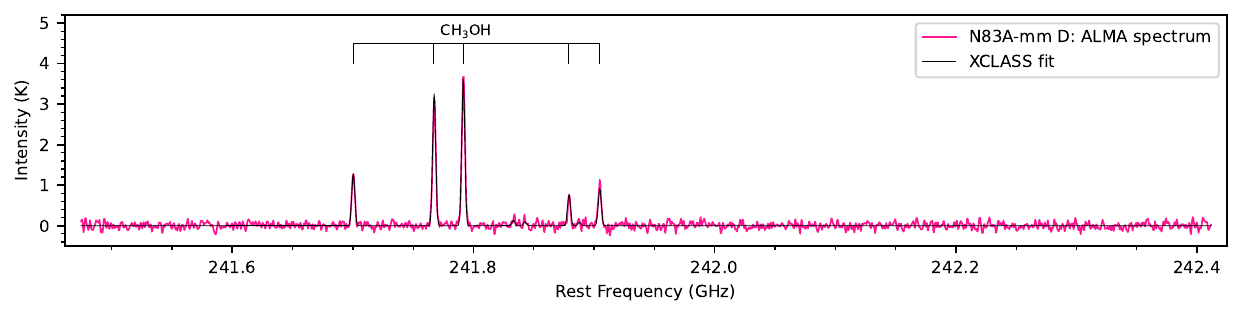}
\includegraphics[width=\textwidth]{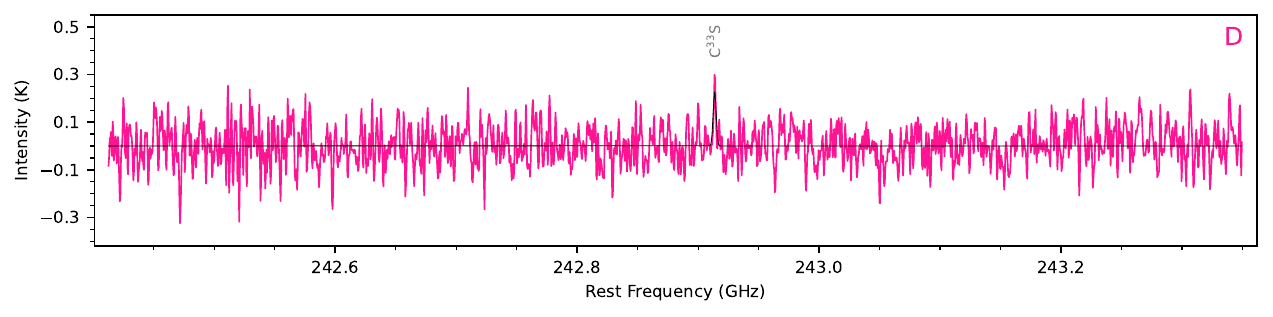}
\includegraphics[width=\textwidth]{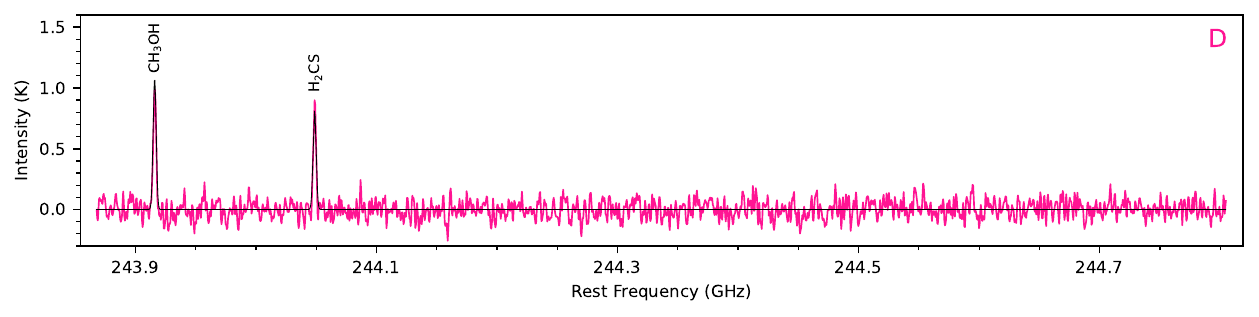}
\includegraphics[width=\textwidth]{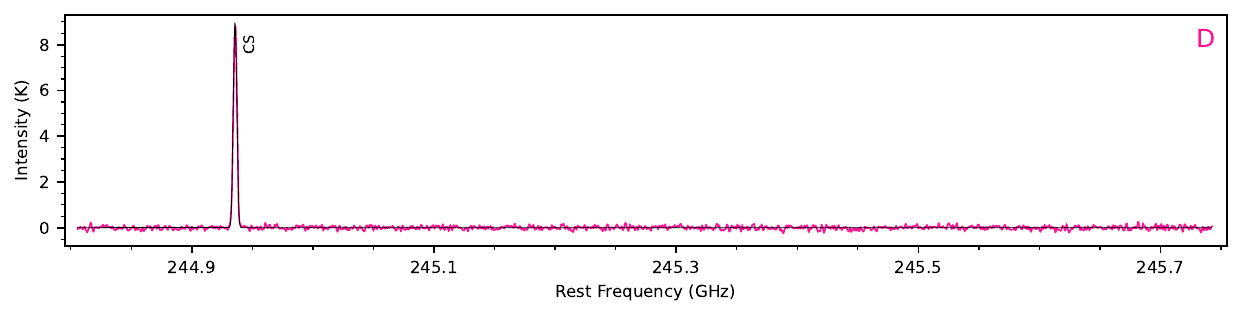}
\caption{The same as Figure~\ref{f:specA_1}, but for source N\,83A--mm D. \label{f:specD_1}}
\end{figure*}

\begin{figure*}
\centering
\includegraphics[width=\textwidth]{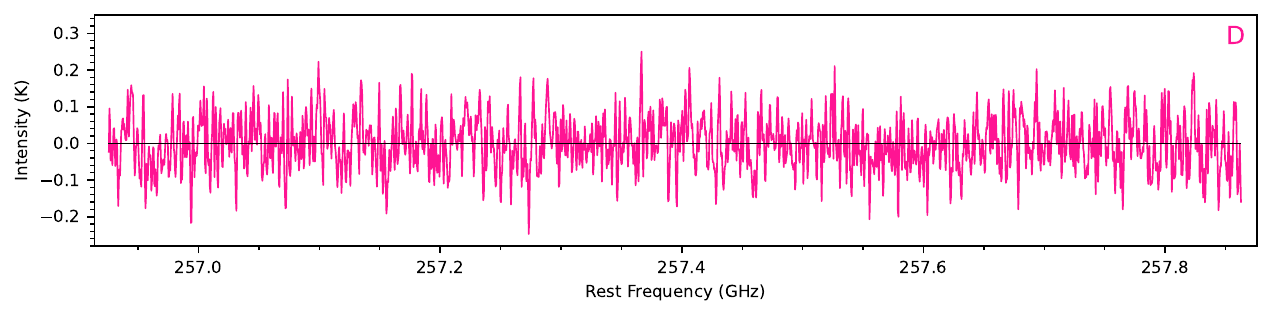}
\includegraphics[width=\textwidth]{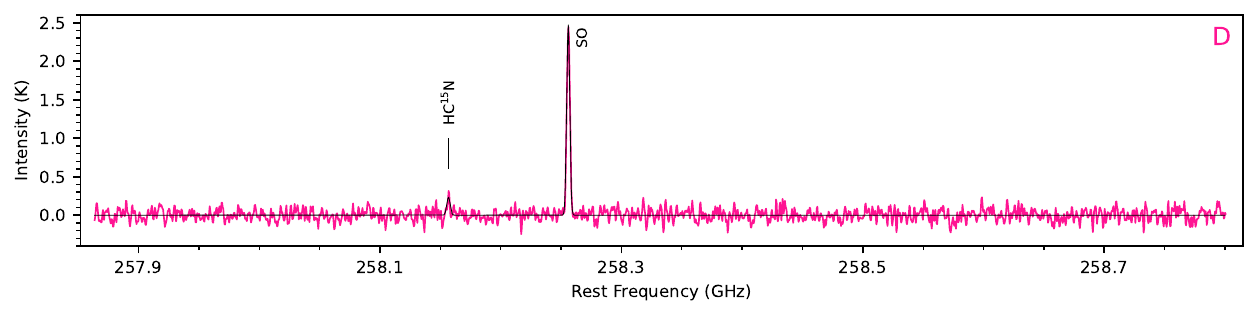}
\includegraphics[width=\textwidth]{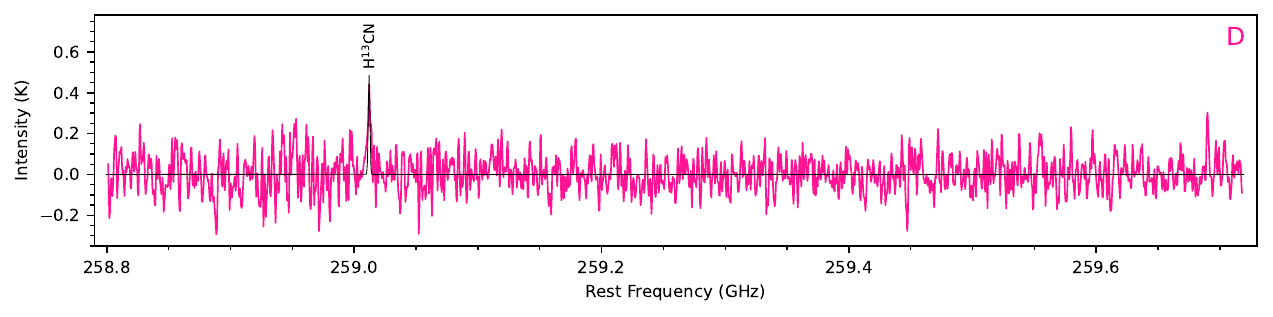}
\includegraphics[width=\textwidth]{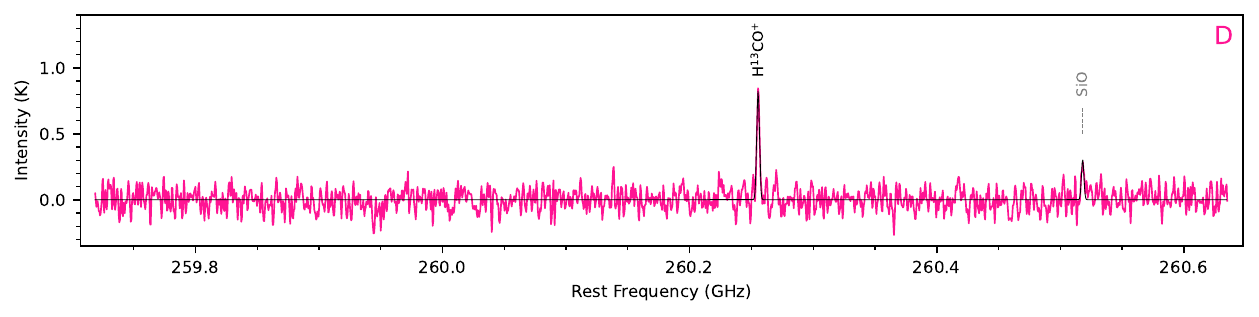}
\caption{The same as Figure~\ref{f:specA_2}, but for source N\,83A--mm D. \label{f:specD_2}}
\end{figure*}

\begin{figure*}
\centering
\includegraphics[width=\textwidth]{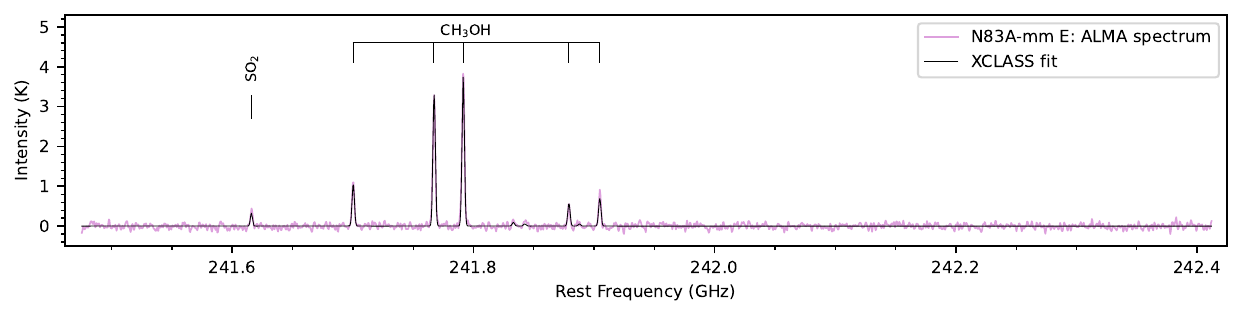}
\includegraphics[width=\textwidth]{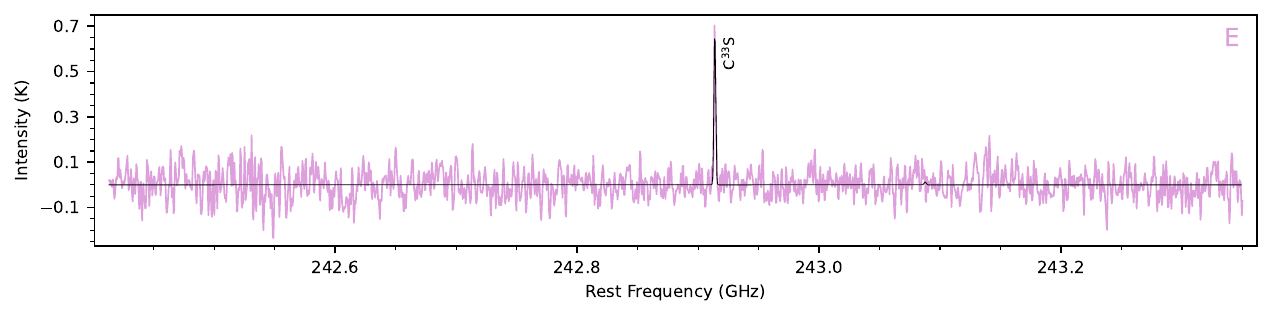}
\includegraphics[width=\textwidth]{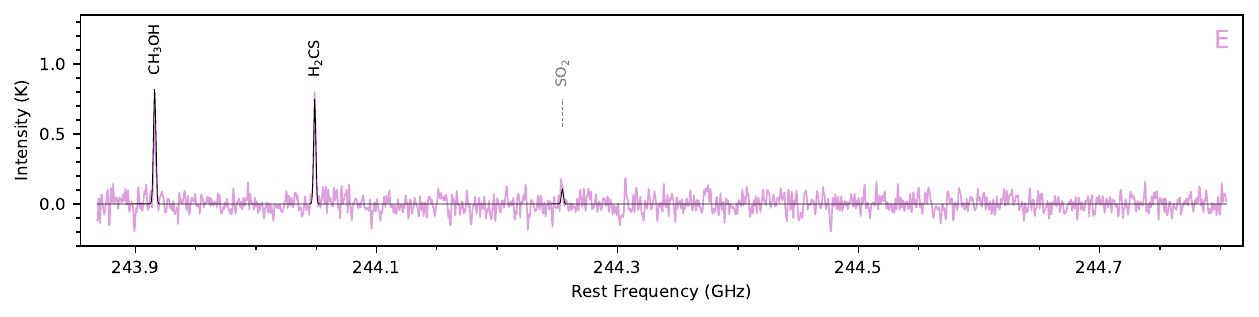}
\includegraphics[width=\textwidth]{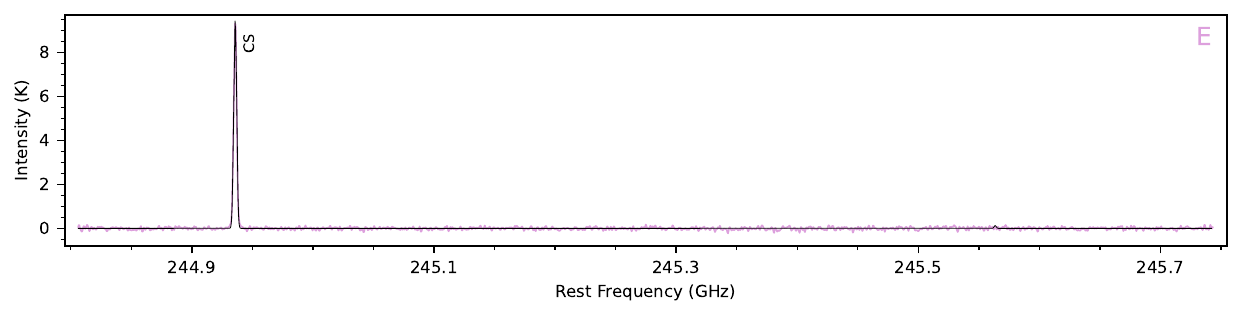}
\caption{The same as Figure~\ref{f:specA_1}, but for source N\,83A--mm E. \label{f:specE_1}}
\end{figure*}

\begin{figure*}
\centering
\includegraphics[width=\textwidth]{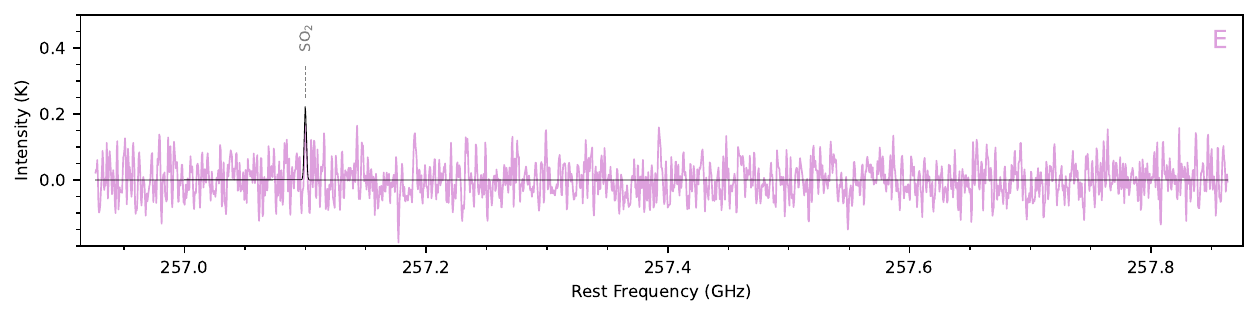}
\includegraphics[width=\textwidth]{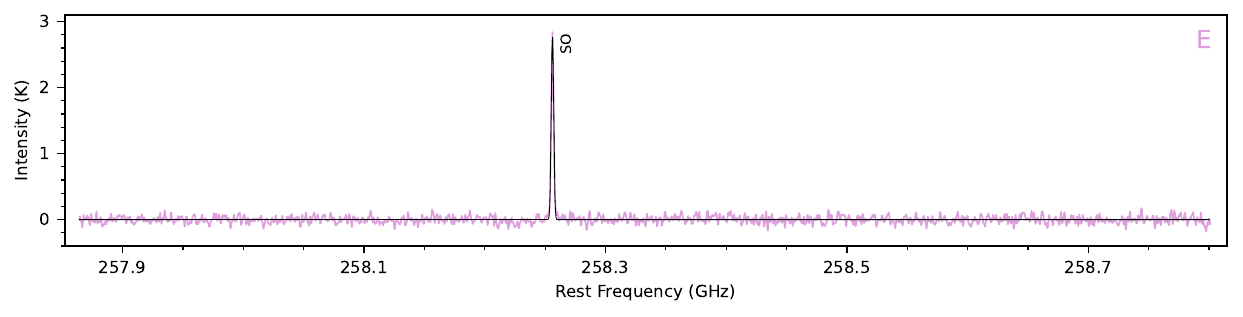}
\includegraphics[width=\textwidth]{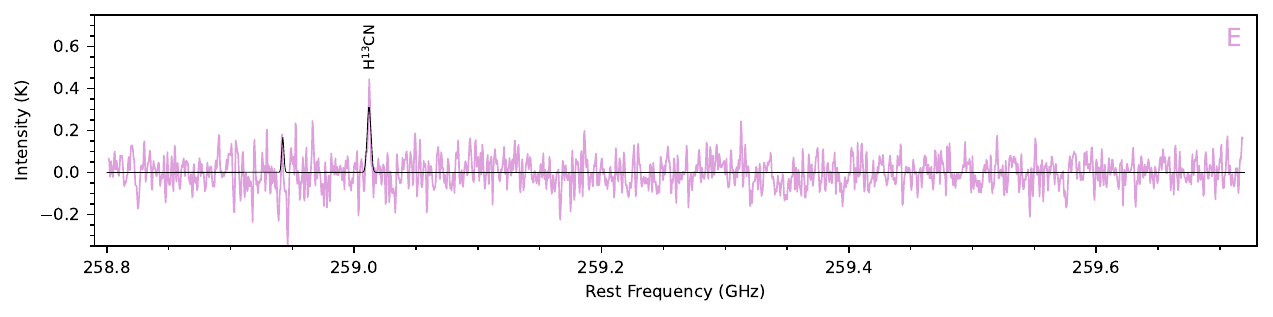}
\includegraphics[width=\textwidth]{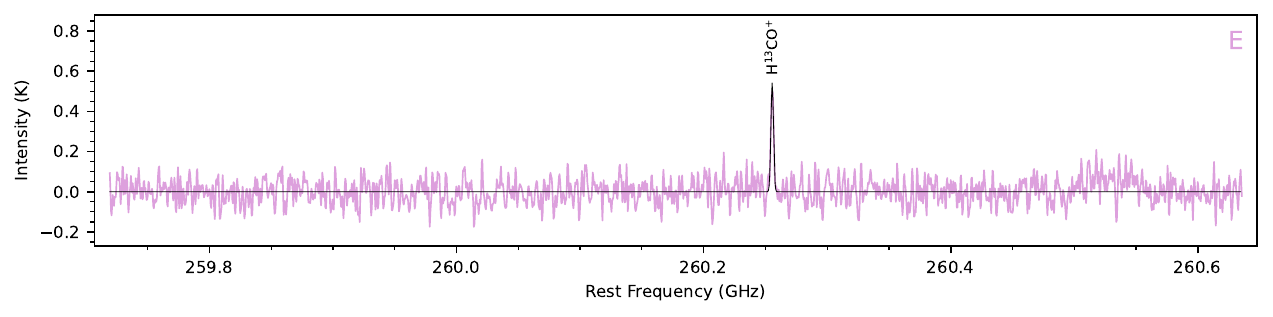}
\caption{The same as Figure~\ref{f:specA_2}, but for source N\,83A--mm E. \label{f:specE_2}}
\end{figure*}

In Figures~\ref{f:specA_1}--\ref{f:specE_2}, we present the ALMA Band 6 spectra of N\,83A--mm A--E with the XCLASS model fits overlaid for all four spectral windows covered by our observations. The XCLASS modeling is described in Section~\ref{s:modeling} and the XCLASS fitting results are provided in Appendix~\ref{app:XCLASSresults}.

\clearpage 

\section{Ancillary Data}
\label{app:ancillary}

\begin{figure*}
\centering
\includegraphics[width=\textwidth]{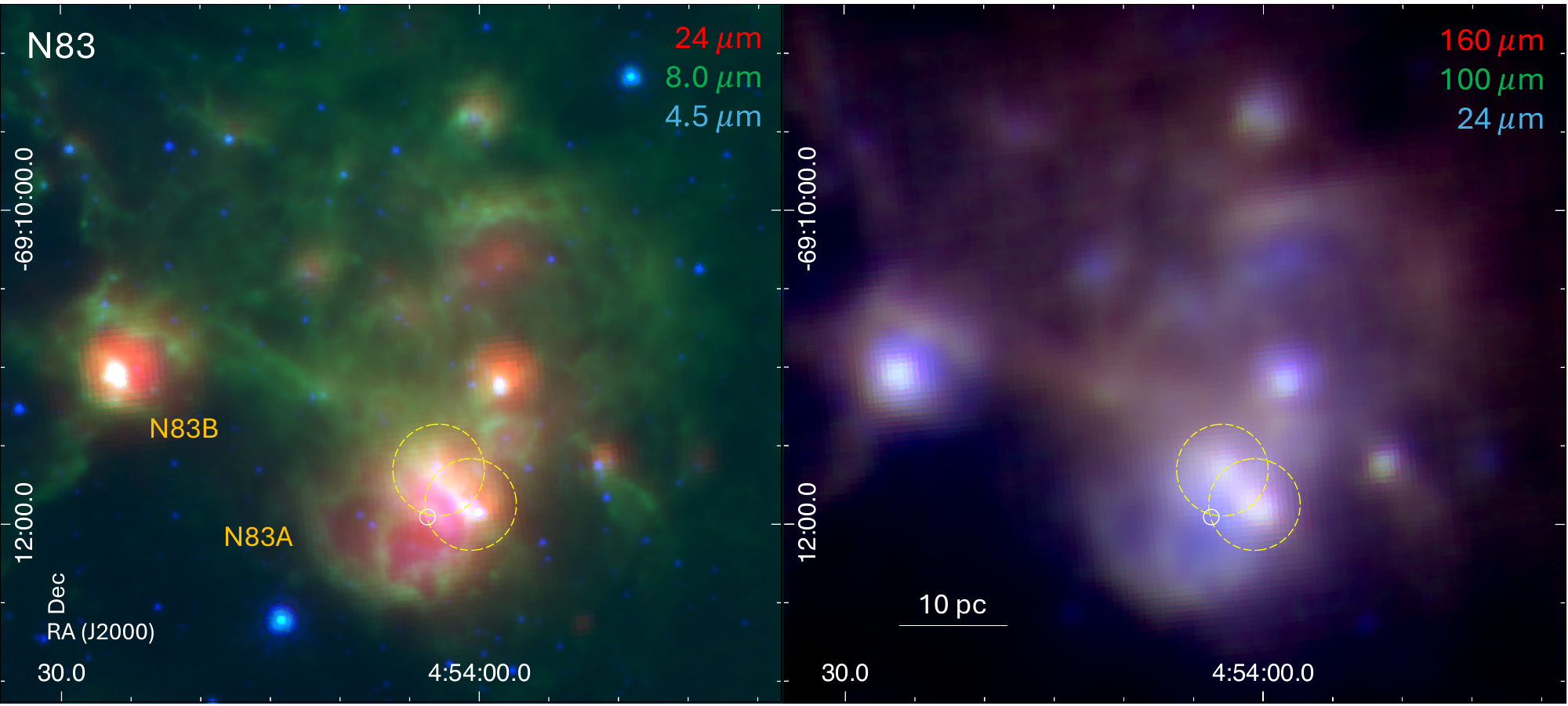}
\caption{Left panel: Three-color mosaic combining the {\it Spitzer} MIPS 24 $\mu$m (red), IRAC 8.0 $\mu$m (green) and 4.5 $\mu$m (blue) images.  The optical nebulae N\,83A and N\,83B are labeled. Right panel: Three-color mosaic combining the {\it Herschel} PACS 160 $\mu$m (red) and 100 $\mu$m (green), and {\it Spitzer} MIPS 24 $\mu$m (blue) images.  In both panels, the white circle indicates the position of the exciting star of the N\,83A nebula, Sk~$-$69$^{\circ}$25. Dashed yellow circles show the areas covered by our ALMA observations: the N\,83--1 (northeast) and N\,83--2 (southwest) fields.  \label{f:N83AspitzerFOV}} 
\end{figure*}

\begin{figure*}[hb!]
\centering
\includegraphics[width=0.8\textwidth]{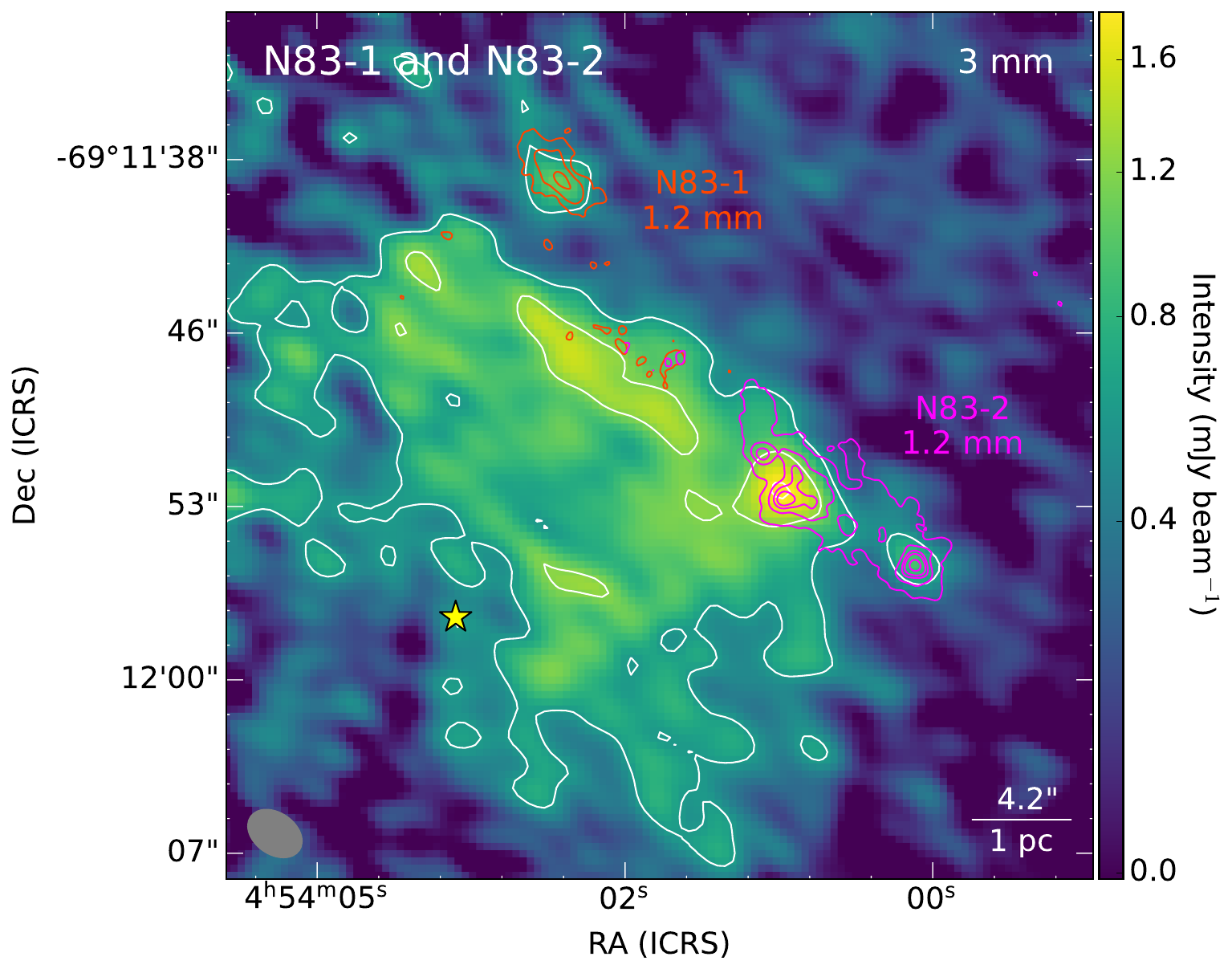}
\caption{The 3 mm continuum image of the N\,83--1 and N\,83--2 fields and the surrounding area covering the brightest part of the N\,83A nebula.  The contour levels correspond to (4, 8, 12) $\times$  the 3 mm continuum image rms noise ($\sigma$) of 1.5$\times$10$^{-4}$ Jy beam$^{-1}$.  The ALMA 3 mm synthesized beam shown in the lower left corner is $2\rlap.{''}44\times1\rlap.{''}71$.  The 1.2 mm contours are overlaid with contour levels the same as in Figure~\ref{f:N83B6cont}.  The yellow star indicates the position of the 64 M$_{\odot}$ star Sk~$-$69$^{\circ}$25 (the spectral type and age of O6~V((f)) and 2.2 Myr, respectively; \citealt{massey2000}), the main ionizing source of the N\,83A nebula. \label{f:N83B3B6cont}}
\end{figure*}

\begin{figure*}
\centering
\includegraphics[width=0.329\textwidth]{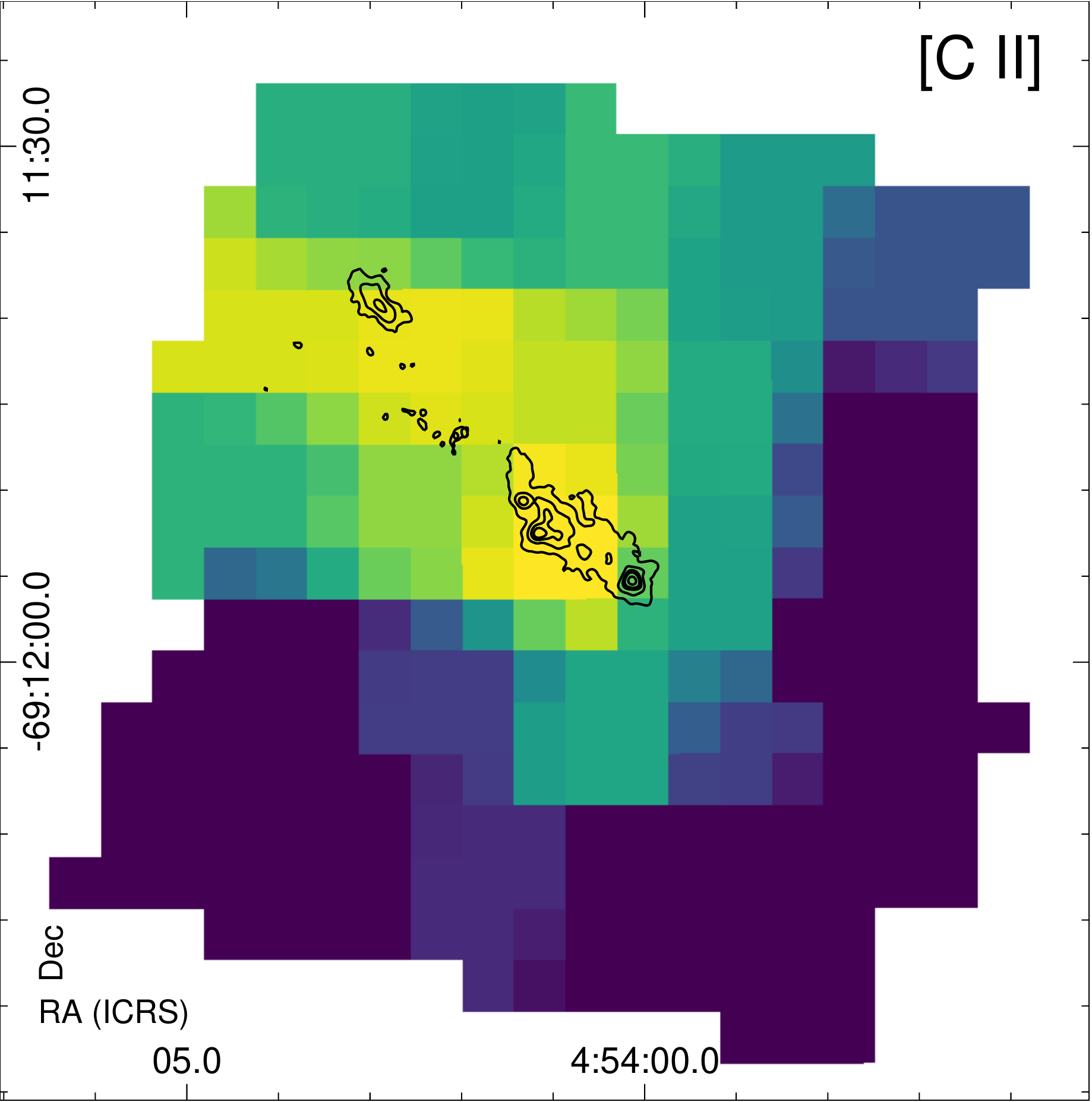}
\includegraphics[width=0.329\textwidth]{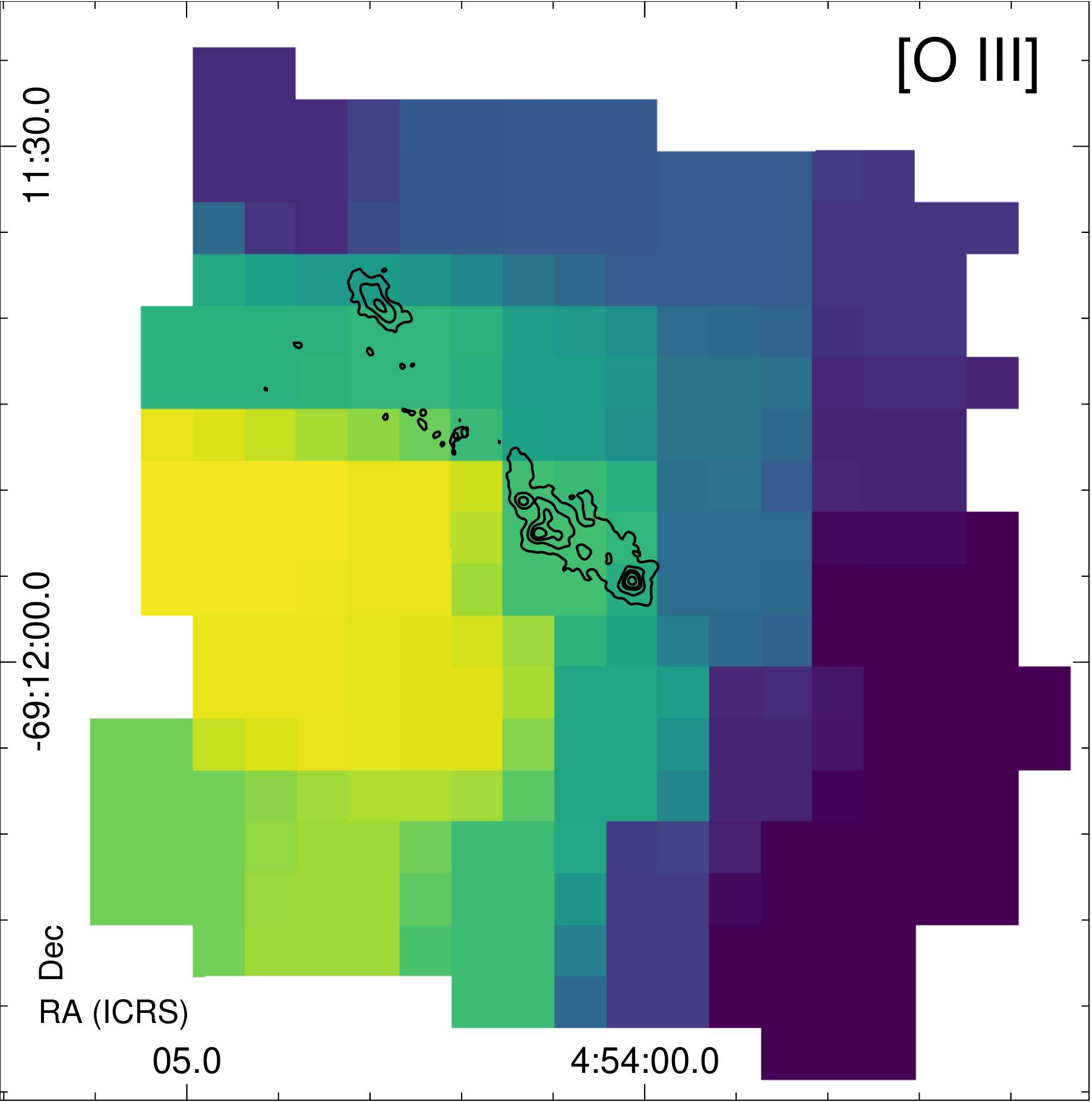}
\includegraphics[width=0.329\textwidth]{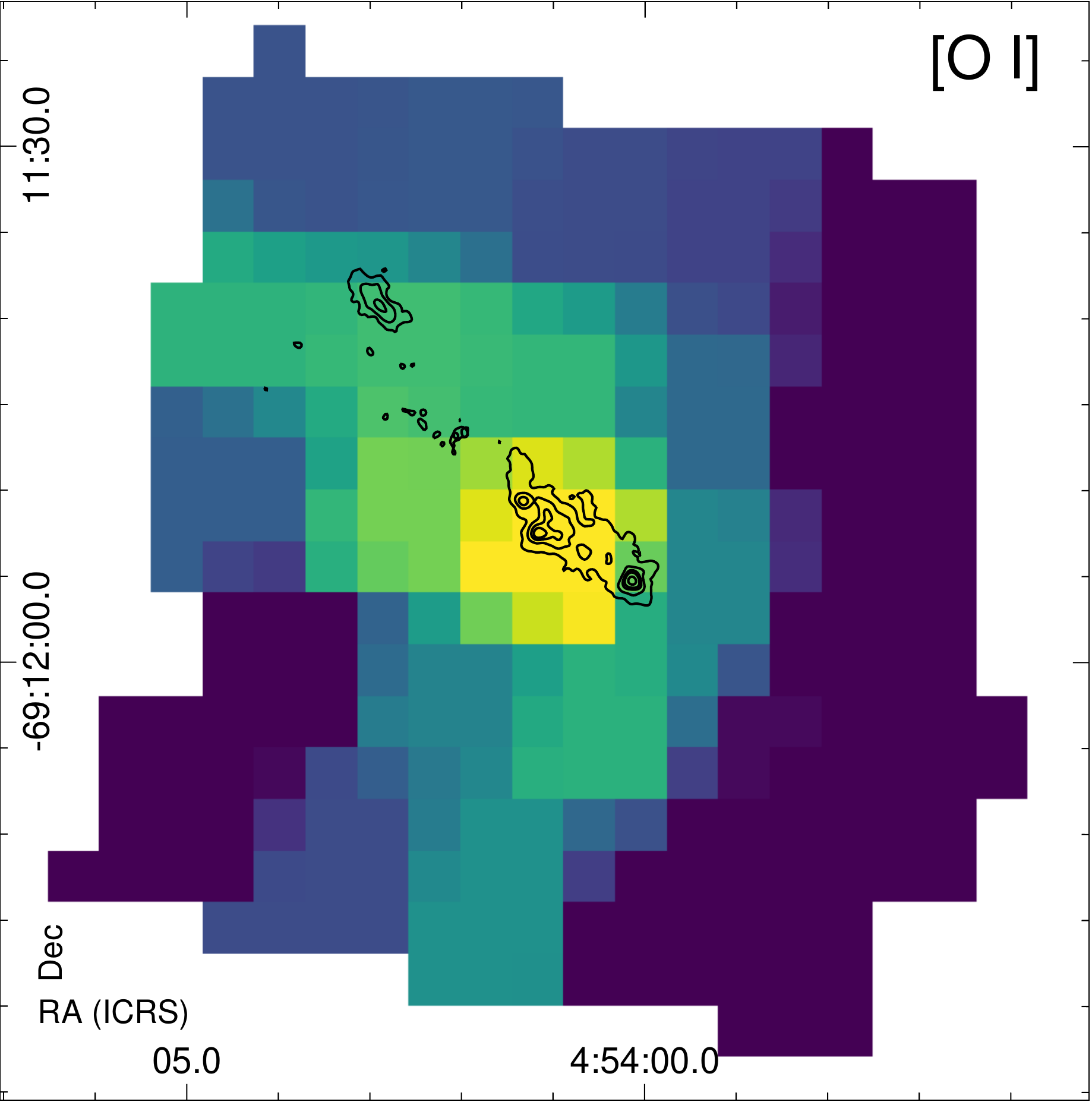}
\caption{The {\it Herschel} PACS line emission maps for [C {\sc ii}] 157.74 $\mu$m (left), [O {\sc iii}] 88.36 $\mu$m (center), and [O {\sc i}] 63.18 $\mu$m (right; \citealt{oliveira2019}). The 1.2 mm contours are overlaid in black with contour levels the same as in Figure~\ref{f:N83B6cont}. \label{f:herspec}} 
\end{figure*}

\begin{figure*}[hb!]
\centering
\includegraphics[width=0.495\textwidth]{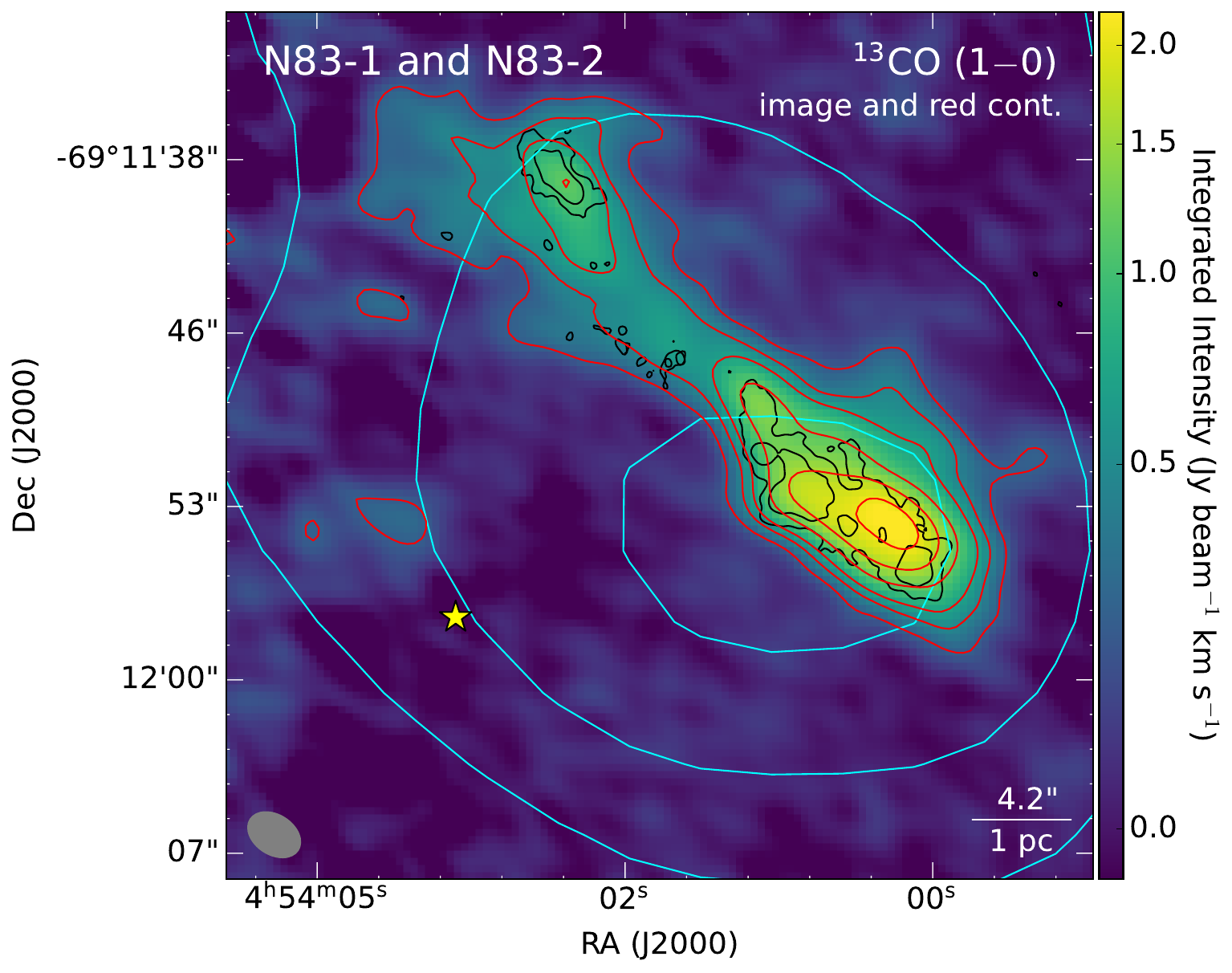} 
\includegraphics[width=0.495\textwidth]{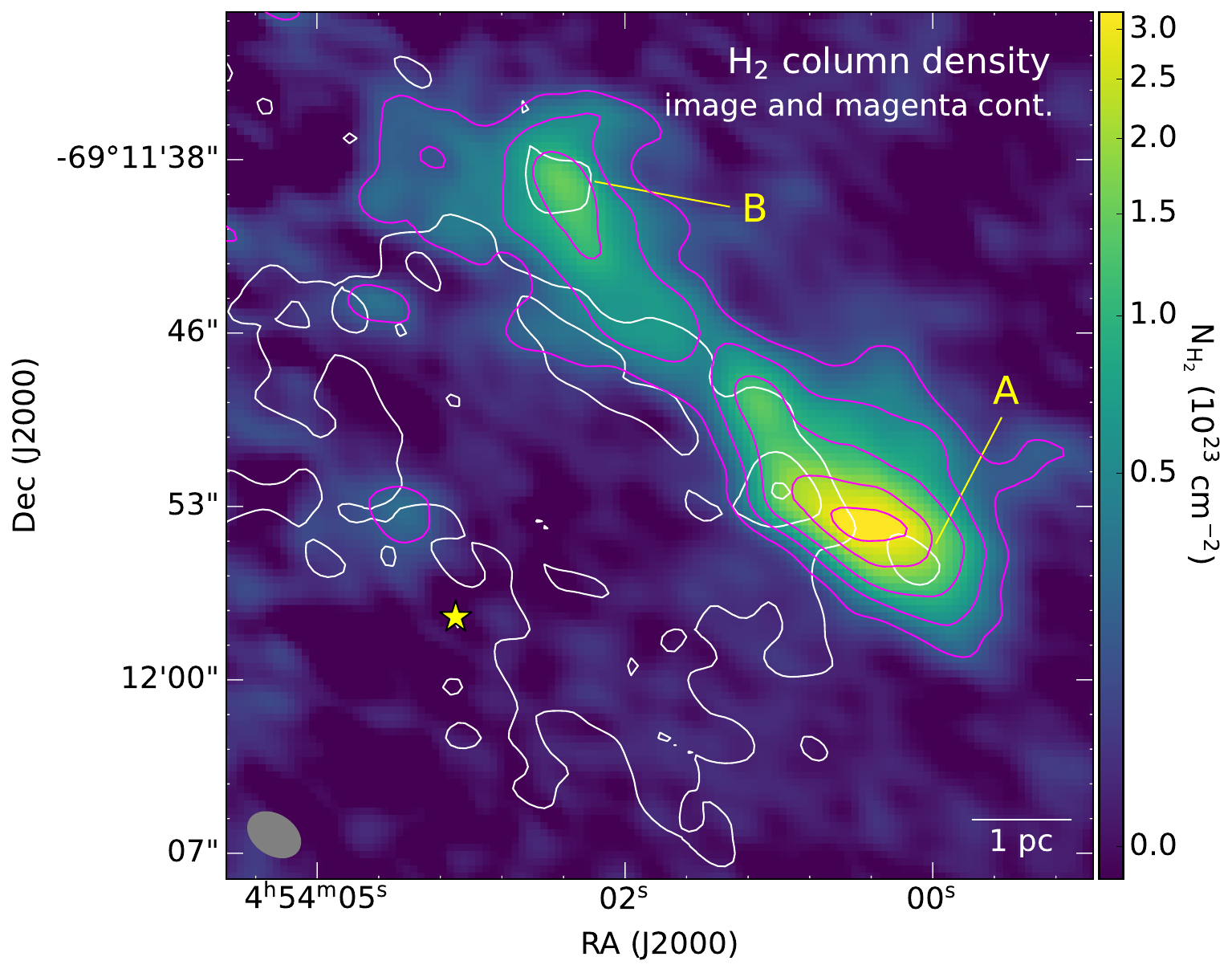}
\caption{{\it Left panel}: The $^{13}$CO (1--0) integrated intensity image of the same area in N\,83 as in Figure~\ref{f:N83B3B6cont}. The $^{13}$CO contours are shown in red with contour levels corresponding to (10, 20, 30, 50, 70, 90)\% of the $^{13}$CO integrated intensity peak of 2.3 Jy~beam$^{-1}$~km~s$^{-1}$.  The cyan contours correspond to the C$^{18}$O (2--1) emission with contour levels of (30, 60, 90)\% of the  C$^{18}$O integrated intensity peak of 6.55 Jy~beam$^{-1}$~km~s$^{-1}$.  The 1.2 mm continuum contours are shown in black and correspond to (3, 10)$\sigma_1$ and (3, 10)$\sigma_2$ for the ALMA fields N\,83--1 and N\,83--2, respectively, where $\sigma_1$ and $\sigma_2$ are the rms noise level in the corresponding image (see Figure~\ref{f:N83B3B6cont} and Table~\ref{t:imagesinfo}).  The yellow star is the same as in Figure~\ref{f:N83B3B6cont}.  {\it Right panel}: The H$_2$ column density, $N({\rm H_2})$, map constructed based on the  $^{13}$CO (1--0) data (see Section~\ref{s:NH2}).  The $N({\rm H_2})$ contours are shown in red with contour levels of (0.2, 0.5, 1, 2, 3) $\times$ 10$^{23}$ cm$^{-2}$.  The yellow star and white contours are the same as in Figure~\ref{f:N83B3B6cont}. The positions of hot cores N\,83A--mm\,A and N\,83A--mm\,B are indicated.  In both panels, the size of the ALMA synthesized beam of the $^{13}$CO (1--0) observations ($2\rlap.{''}36\times1\rlap.{''}64$)  is shown in the lower left corner of the image.  For the C$^{18}$O observations, the beam is $29\rlap.{''}6\times29\rlap.{''}6$. \label{f:N83B3B6_13CO}}
\end{figure*}

\begin{figure*}
\centering
\includegraphics[width=0.7\textwidth]{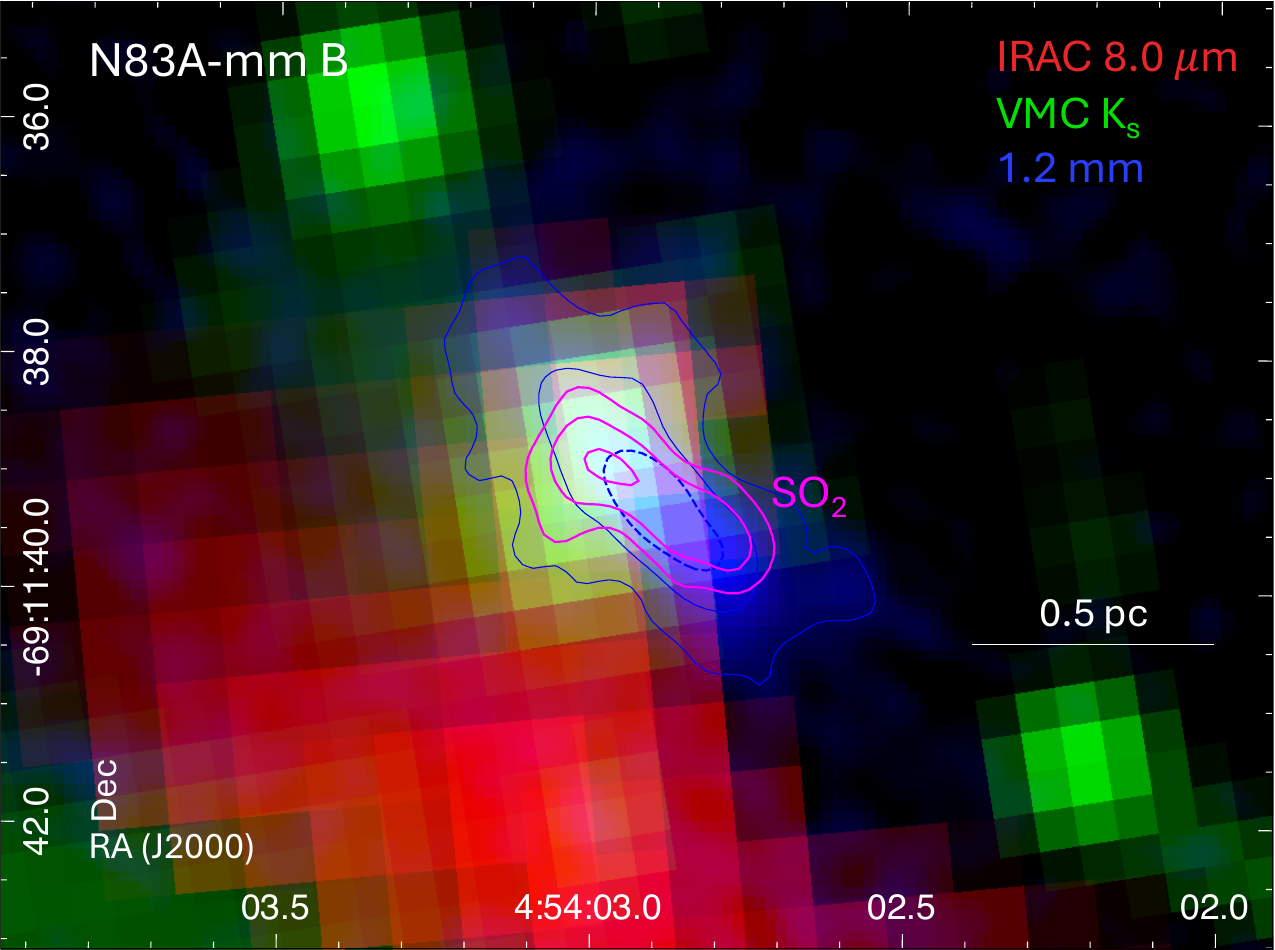}
\caption{Three-color mosaic of N\,83A--mm\,B and its surrounding combining the {\it Spitzer}/IRAC 8.0 $\mu$m (red), VMS $K_{s}$ (green), and ALMA 1.2 mm continuum (blue; blue contours as in Figure~\ref{f:N83B6cont}) images.  The magenta contours corresponds to the \ce{SO2} 5$_{2,4}$--4$_{1,3}$ ($E_{\rm U}$ = 23.6~K) emission with contour levels of (20, 50, 90)\% of the \ce{SO2} emission peak of 108 mJy beam$^{-1}$ km s$^{-1}$. \label{f:Bmultiple}}
\end{figure*}

\begin{figure*}
\centering
\includegraphics[width=0.495\textwidth]{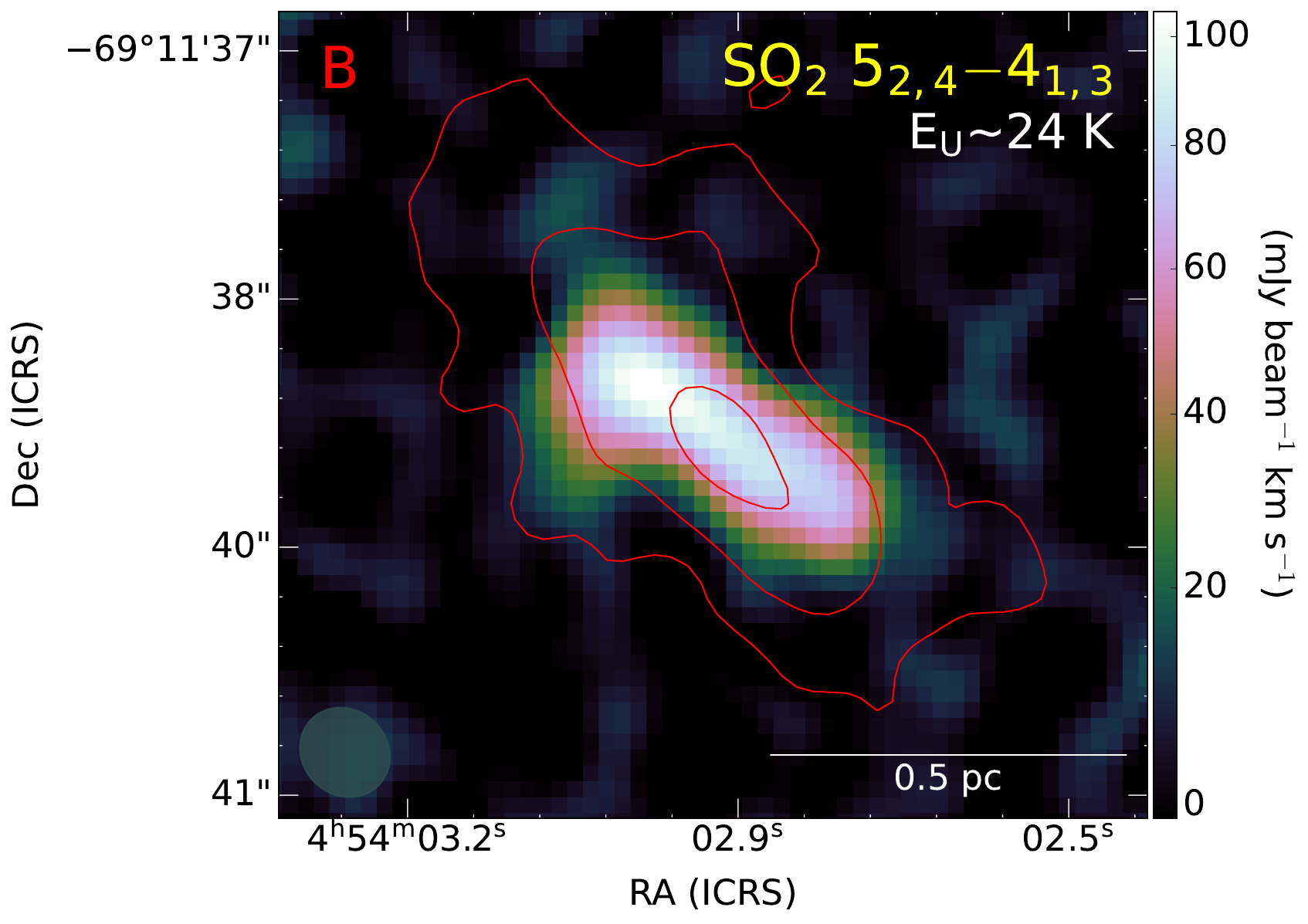}
\includegraphics[width=0.495\textwidth]{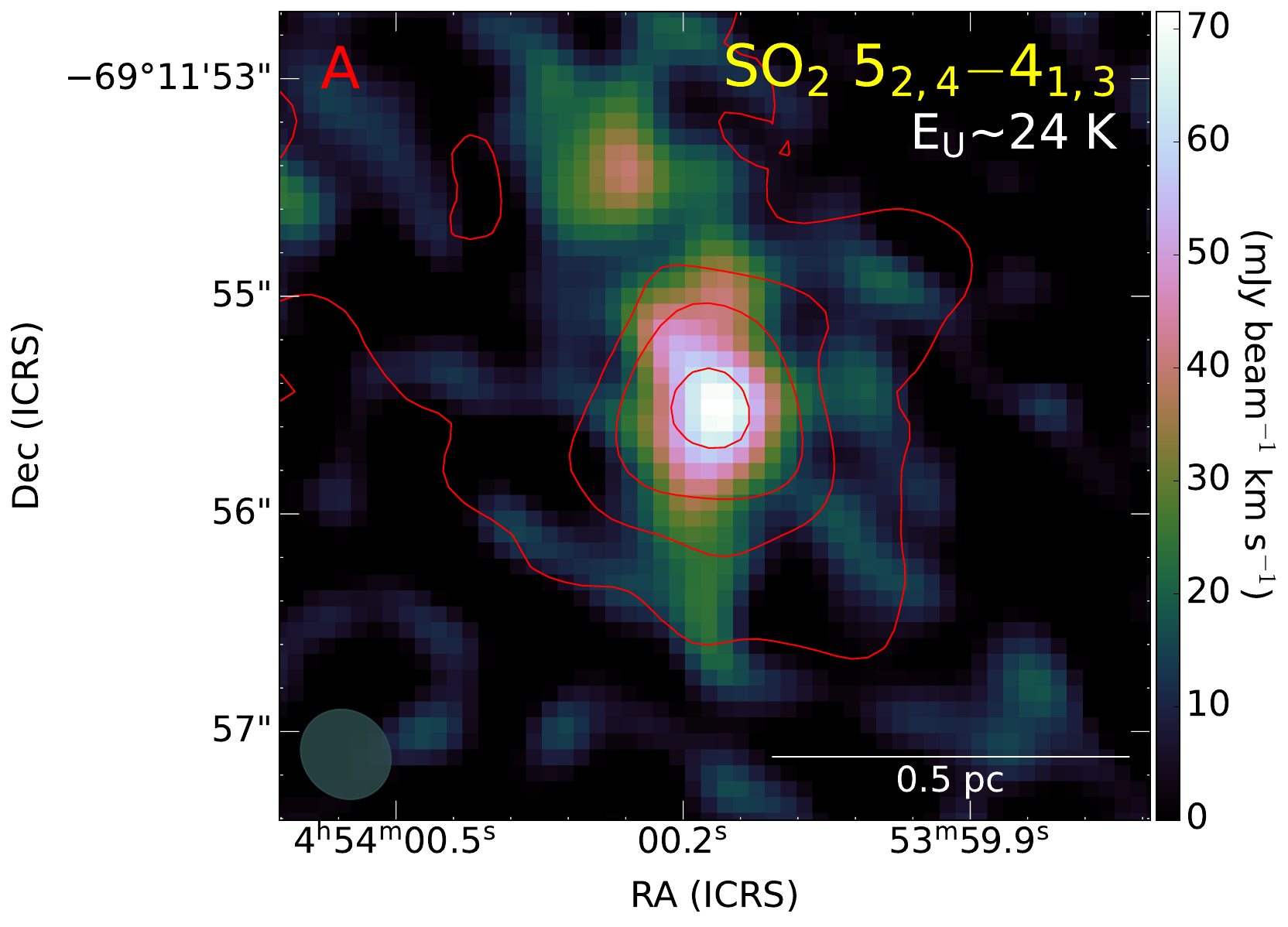}
\includegraphics[width=0.495\textwidth]{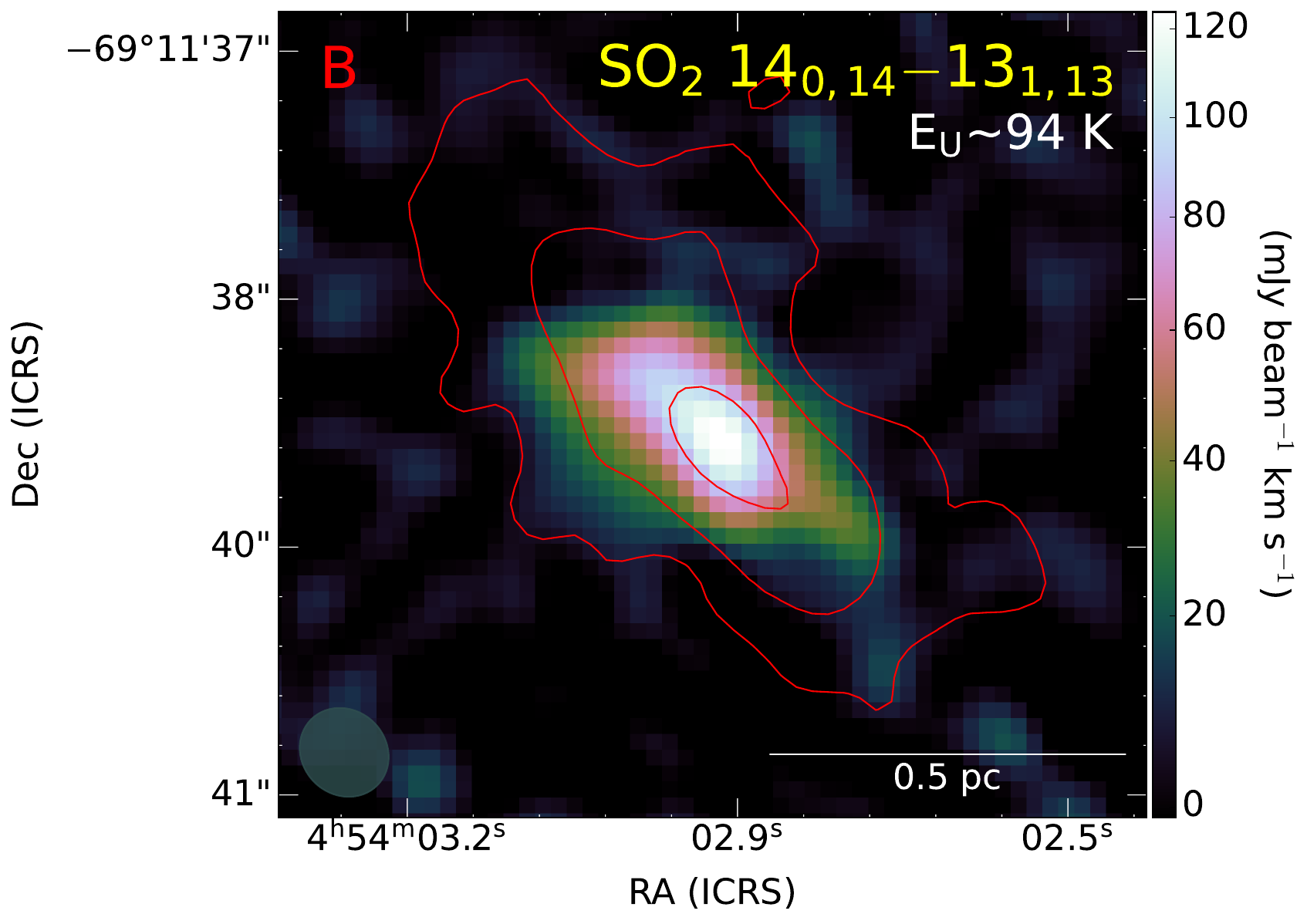}
\includegraphics[width=0.495\textwidth]{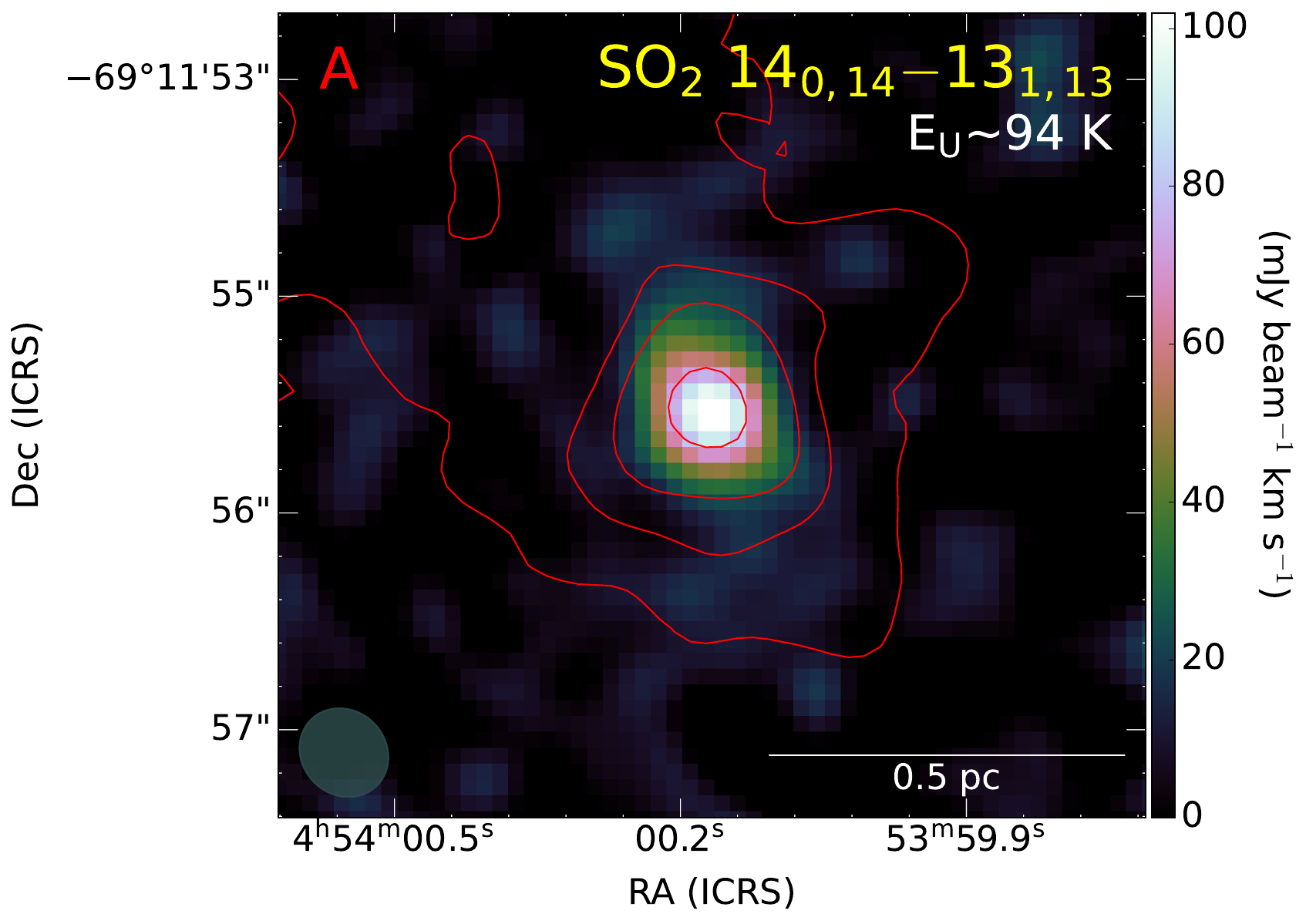}
\caption{The \ce{SO2} 5$_{2,4}$--4$_{1,3}$ (upper panel;  $E_{\rm U}$ = 23.6~K) and  \ce{SO2} 14$_{0,14}$--13$_{1,13}$ (lower panel;  $E_{\rm U}$ = 93.9~K) integrated intensity images of hot cores N\,83A--mm\,B (left) and A (right). The 1.2 mm  continuum contours are overlaid with contour levels of (3, 10, 40)\% and (3, 10, 20, 60)\% of the corresponding continuum intensity peak (see Figure~\ref{f:N83B6cont} caption) for sources B and A, respectively. \label{f:so2int}}
\end{figure*}

\begin{figure*}
\centering
\includegraphics[width=\textwidth]{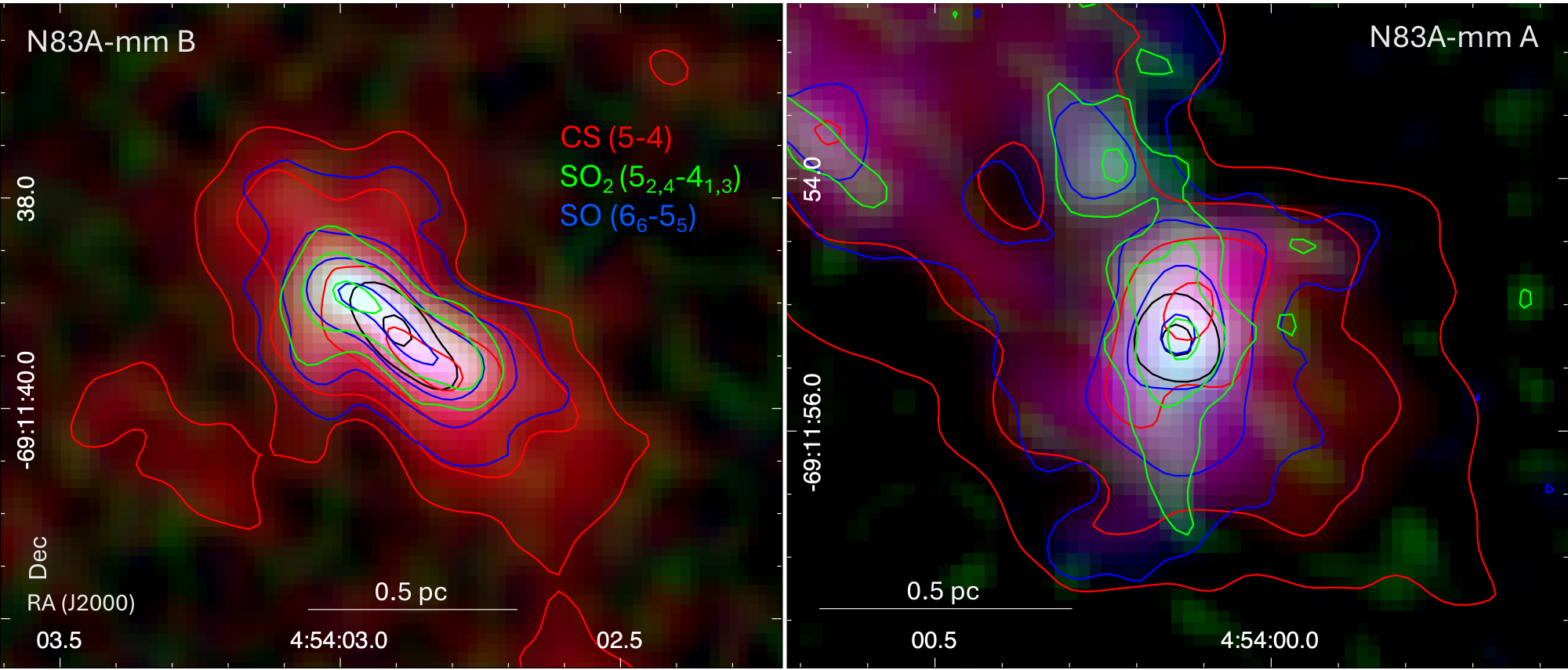}
\caption{Three-color mosaics combining the \ce{CS} 5-4 (red; $E_{\rm U}$ = 35.3~K), \ce{SO2} 5$_{2,4}$--4$_{1,3}$ (green;  23.6~K), and \ce{SO} 6$_6$--5$_5$ (blue; 56.5~K) integrated intensity images of hot cores N\,83A--mm\,B (left) and N\,83A--mm\,A (right).  The \ce{CS}, \ce{SO2}, and \ce{SO} contours are overlaid in the corresponding colors.  For hot cores N\,83A--mm (B, A), the \ce{CS} contour levels are (5, 20, 50, 90)\% of the \ce{CS} integrated intensity peaks of  (0.70, 1.08) Jy~beam$^{-1}$~km~s$^{-1}$.    The \ce{SO2} contour levels are (20, 50, 90)\% and (25, 50, 90)\% for sources B and A, respectively, of the corresponding \ce{SO2} integrated intensity peaks of  0.11 and 0.07 Jy~beam$^{-1}$~km~s$^{-1}$. The \ce{SO} contour levels are (5, 20, 50, 90)\% of the \ce{SO} integrated intensity peaks of  (0.48, 0.50) Jy~beam$^{-1}$~km~s$^{-1}$ for B and A, respectively. The 1.2 mm continuum contours are shown in black with contour levels of (50, 90)\% of the peak continuum intensity of (2.86, 4.97) mJy~beam$^{-1}$ for (B, A). \label{f:alma3color}}
\end{figure*}

\begin{figure*}
\centering
\includegraphics[width=0.328\textwidth]{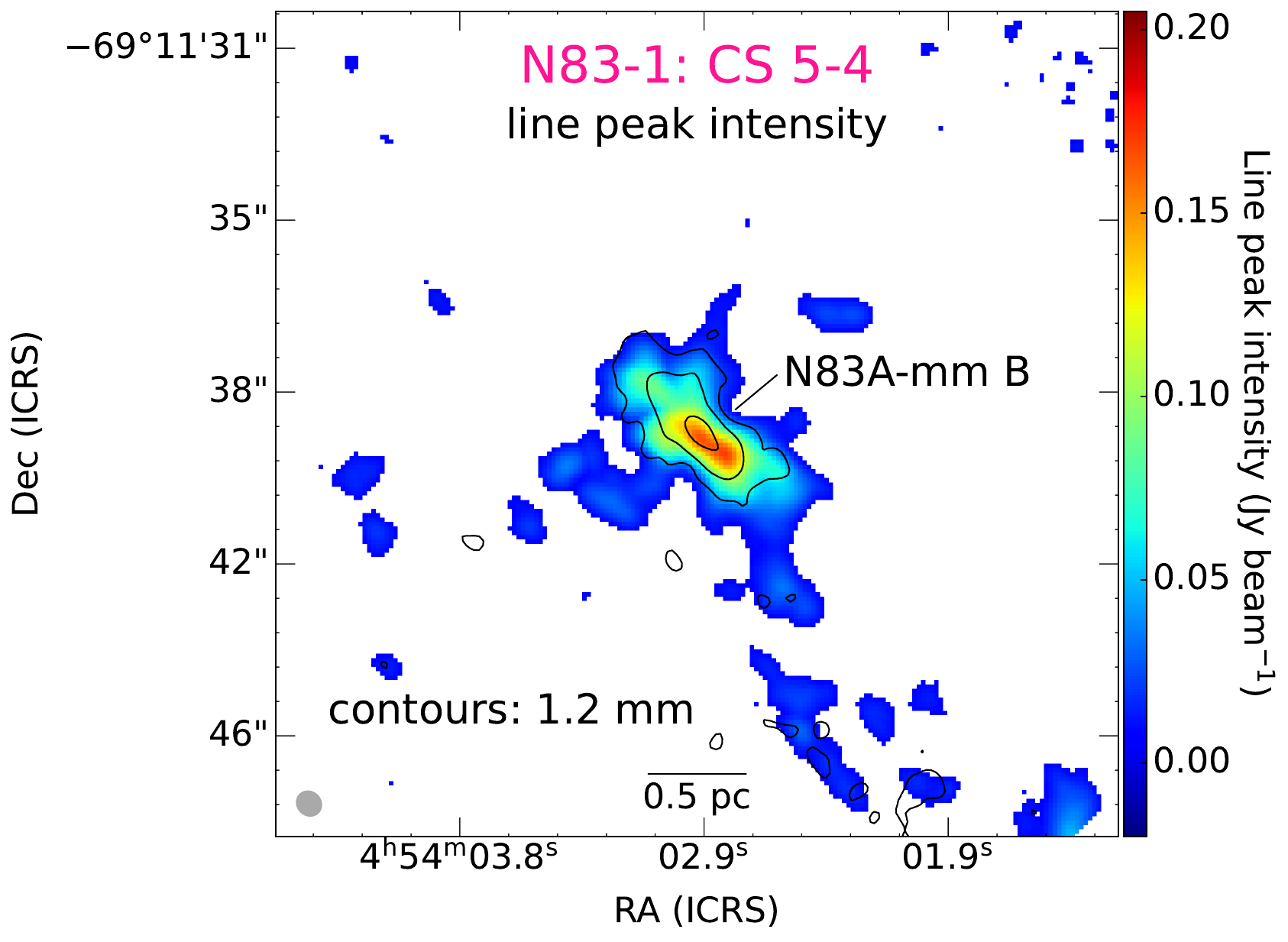}
\includegraphics[width=0.318\textwidth]{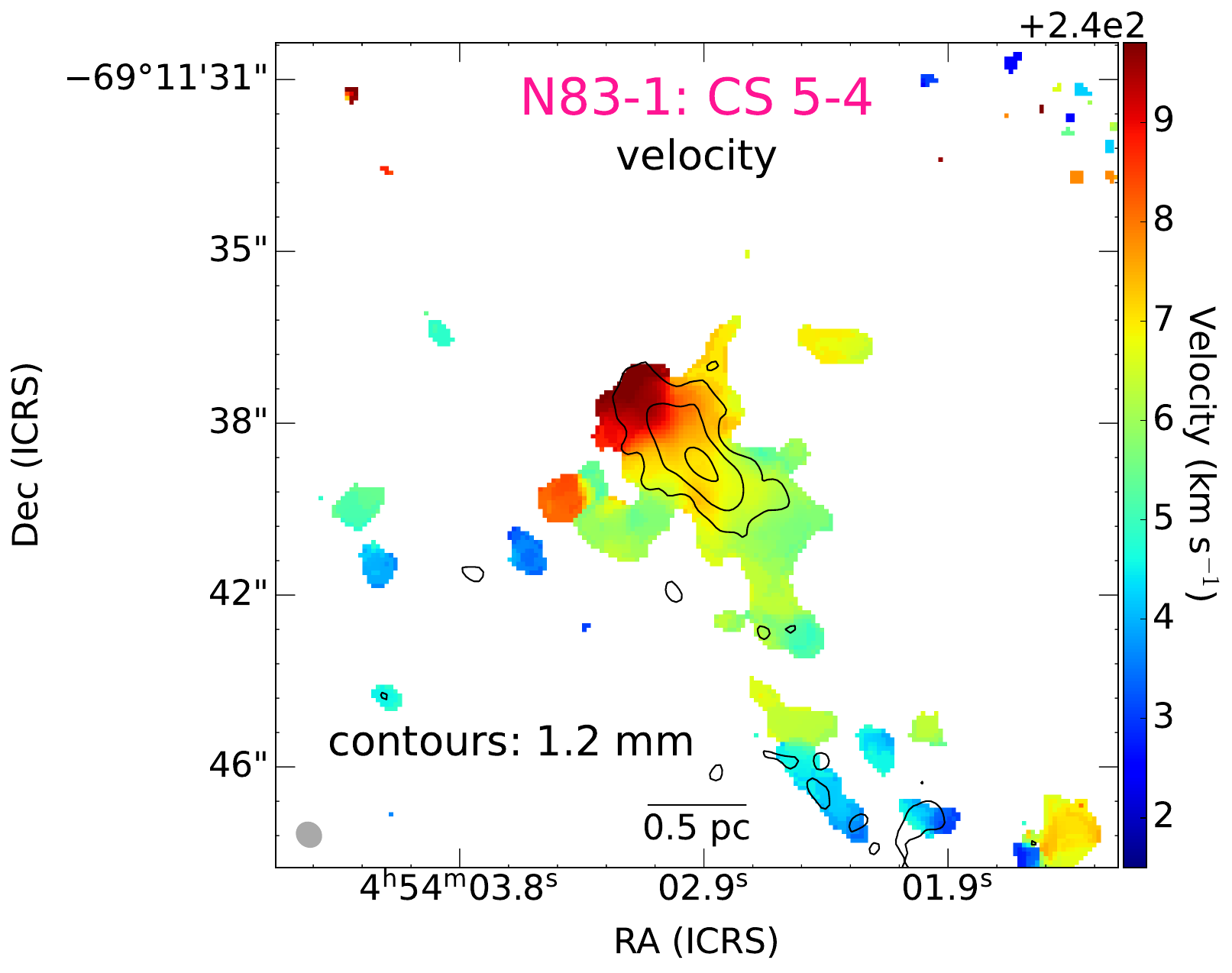}
\includegraphics[width=0.328\textwidth]{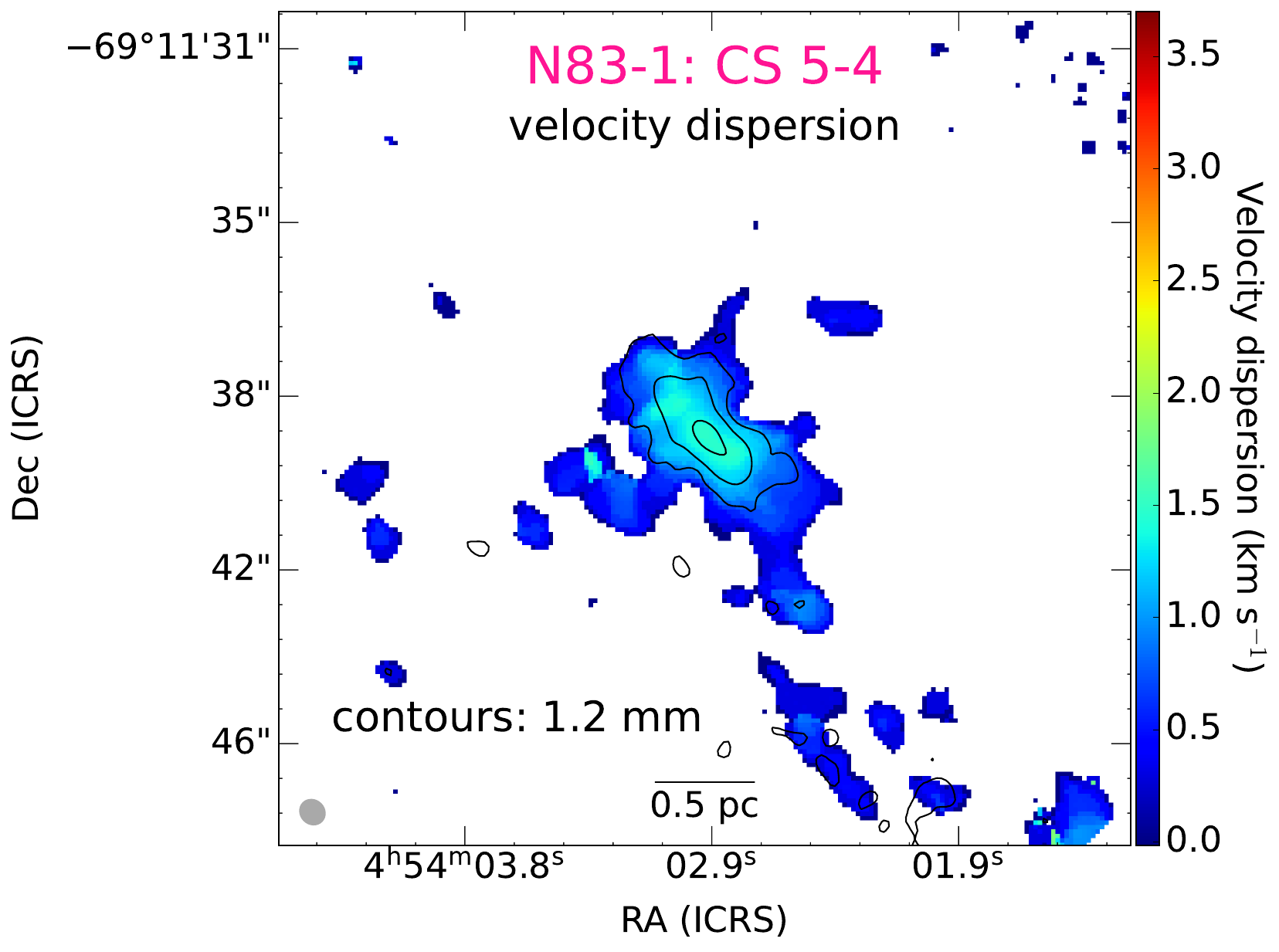}
\includegraphics[width=0.328\textwidth]{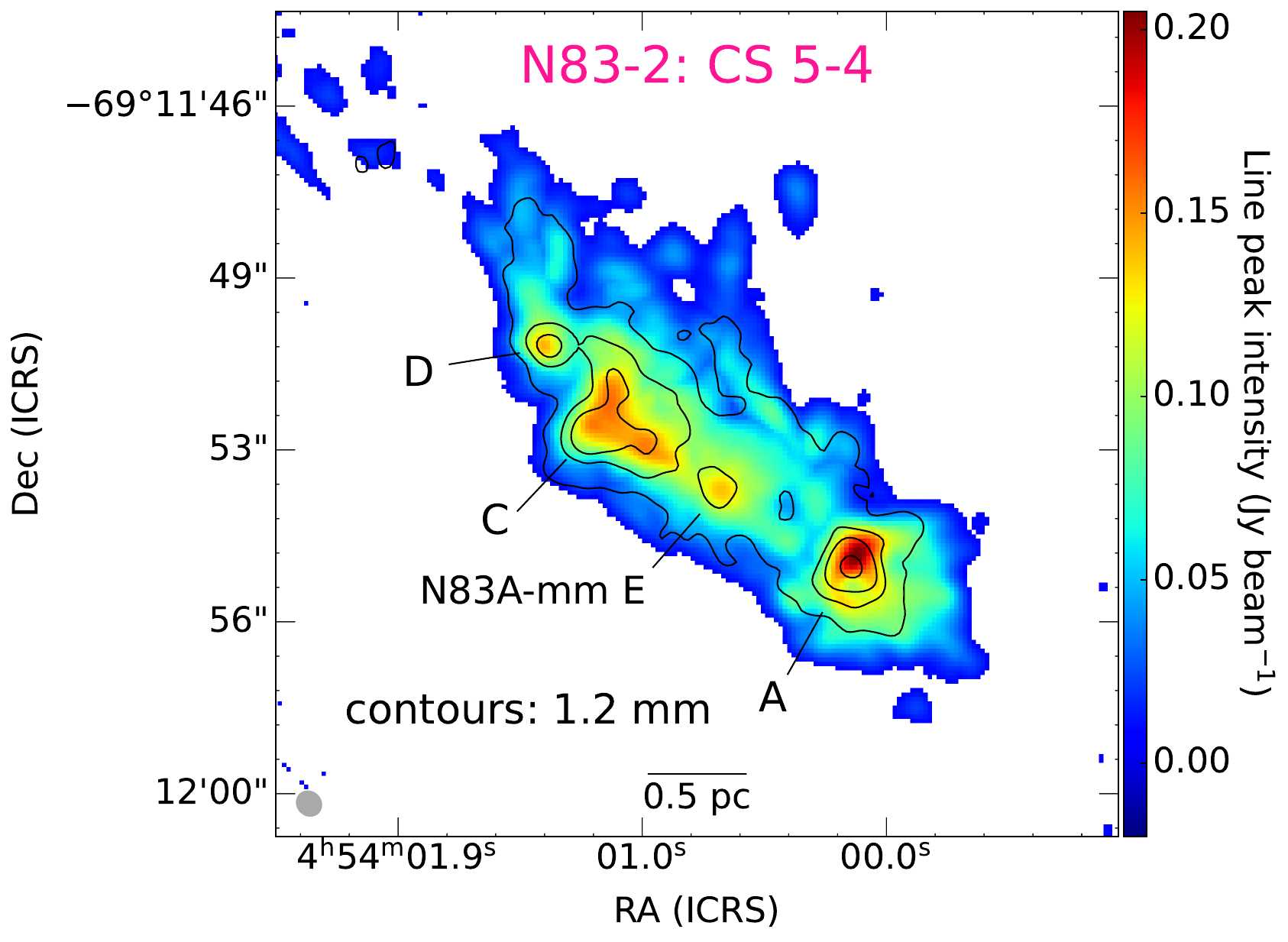}
\includegraphics[width=0.328\textwidth]{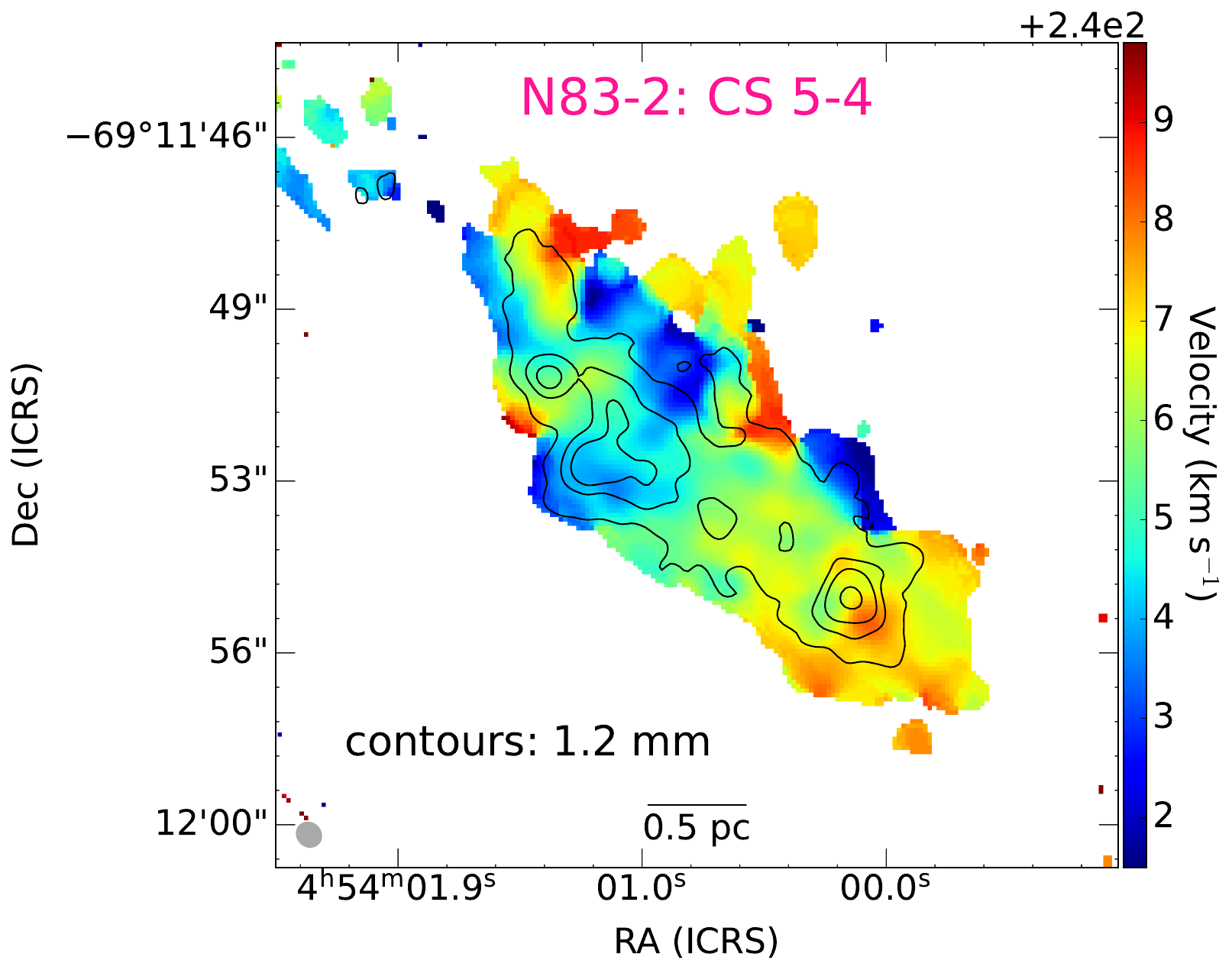}
\includegraphics[width=0.328\textwidth]{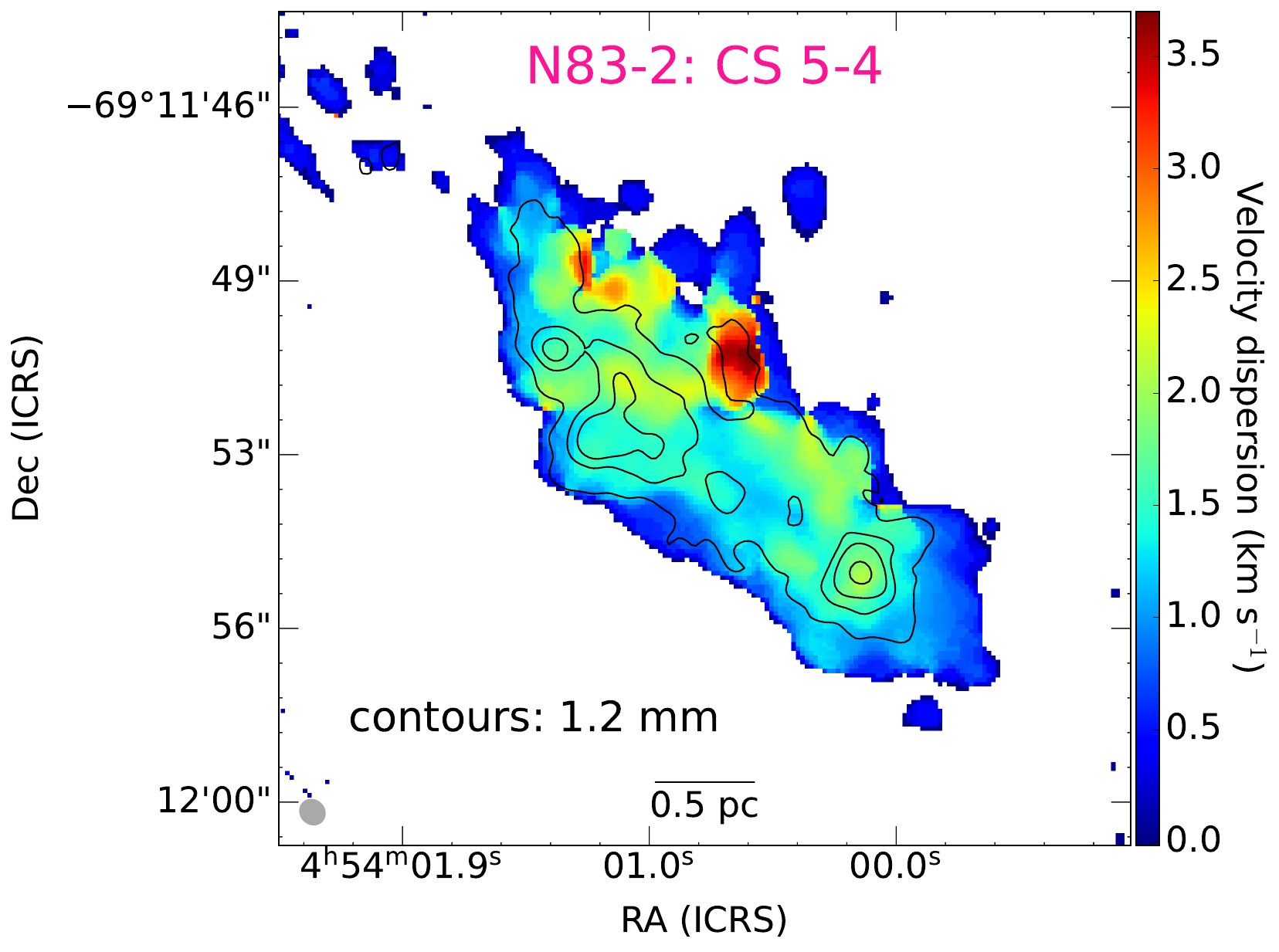}\caption{The CS 5--4 emission line peak intensity (moment 8; left), velocity (moment 1; center), and velocity dispersion (moment 2; right panel) images of the ALMA field N\,83--1 (top) and N\,83--2 (bottom).  For each parameter, we use the same color scale for both fields. All pixels with intensities below 5$\sigma$ are masked. \label{f:kinematicsCS} } 
\end{figure*}

\begin{figure*}
\centering
\includegraphics[width=0.328\textwidth]{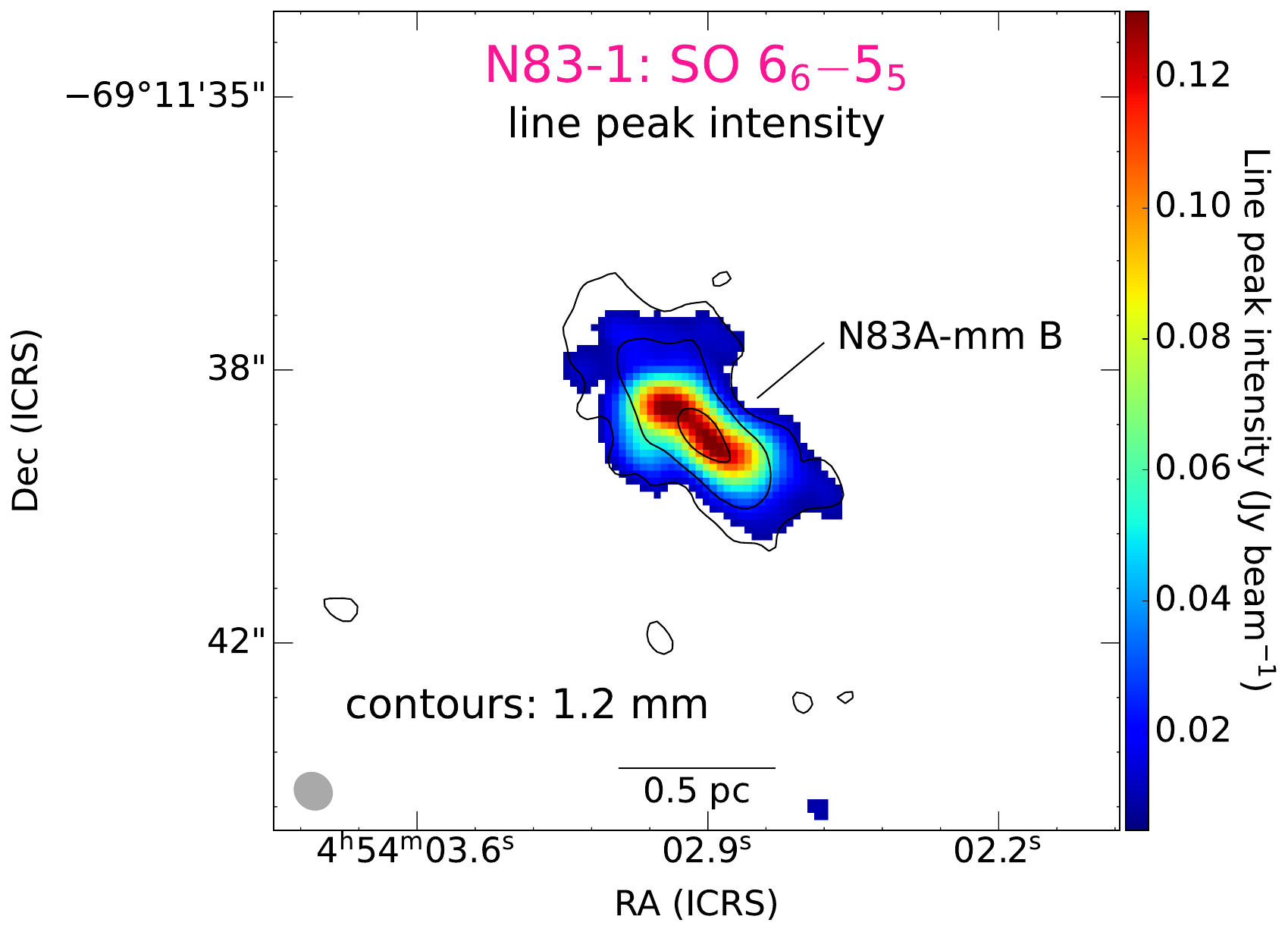}
\includegraphics[width=0.318\textwidth]{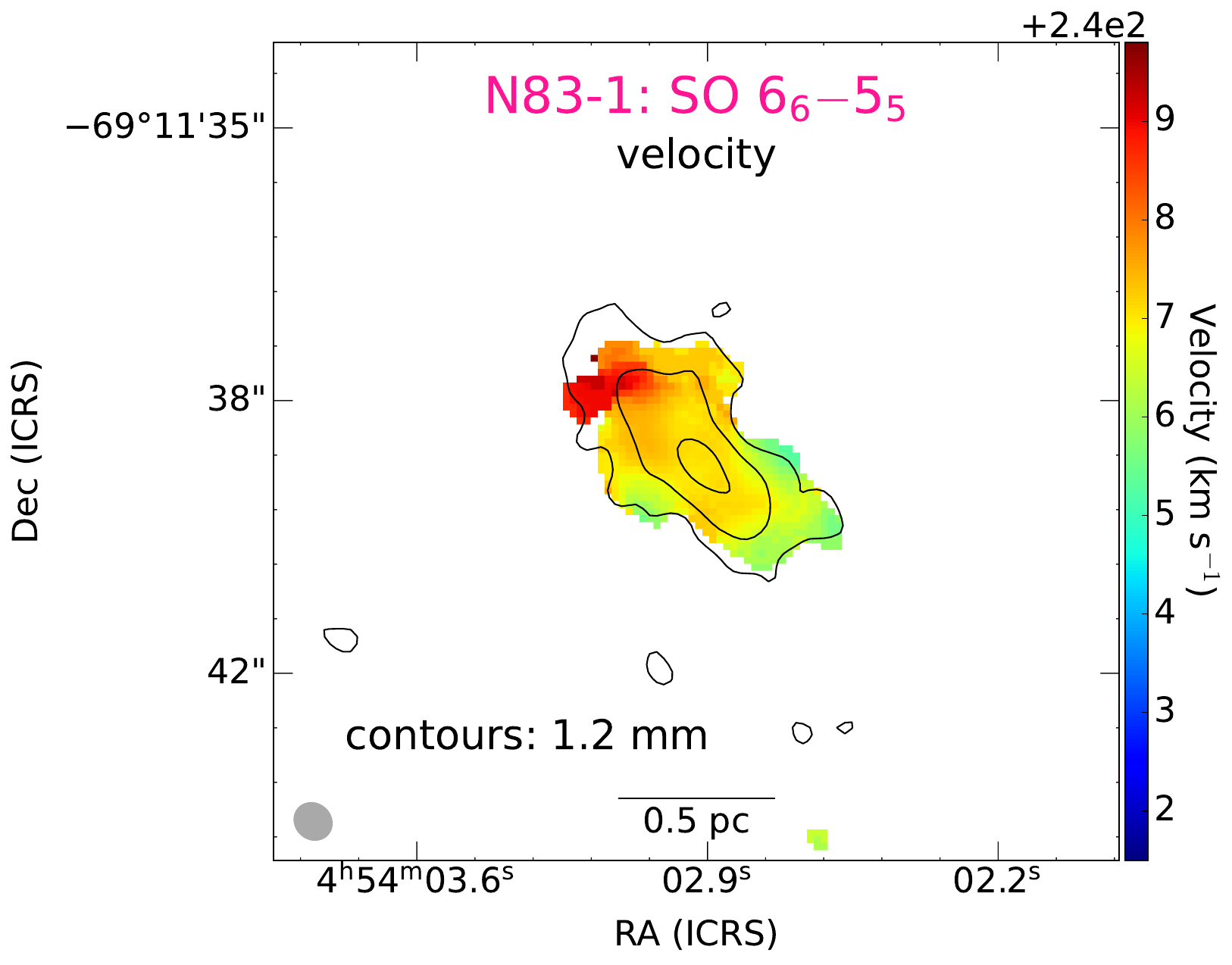}
\includegraphics[width=0.328\textwidth]{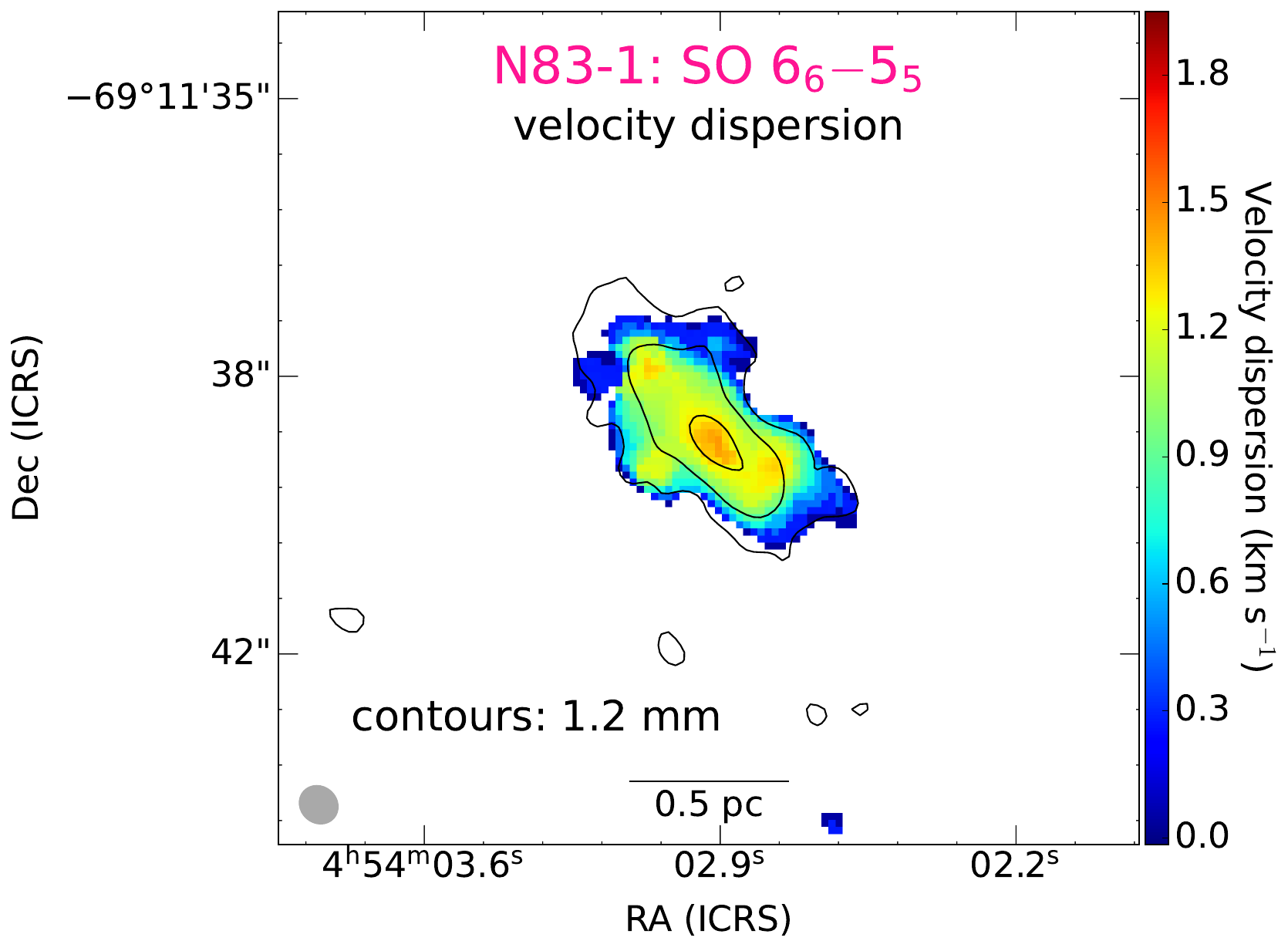}
\includegraphics[width=0.328\textwidth]{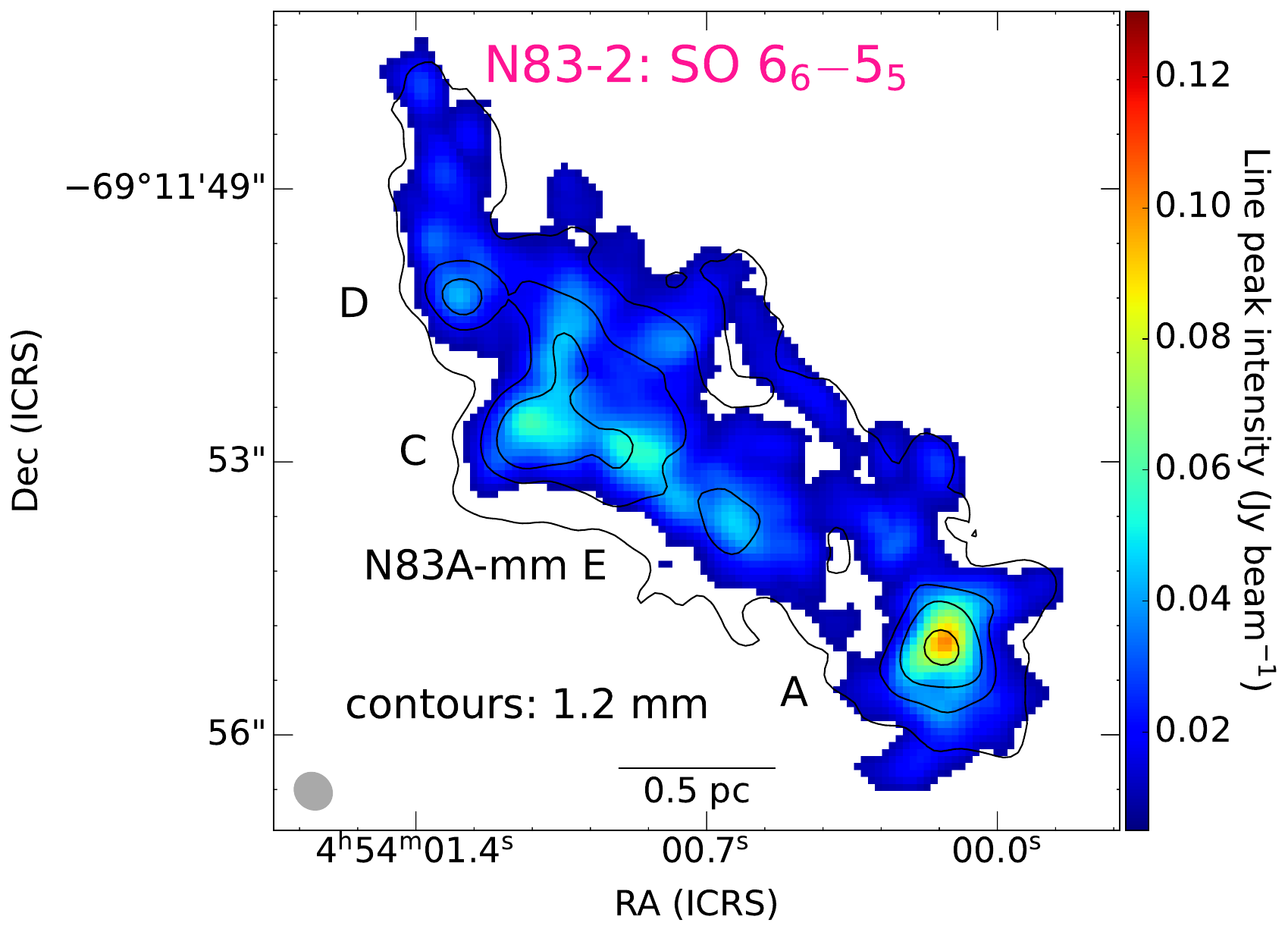}
\includegraphics[width=0.318\textwidth]{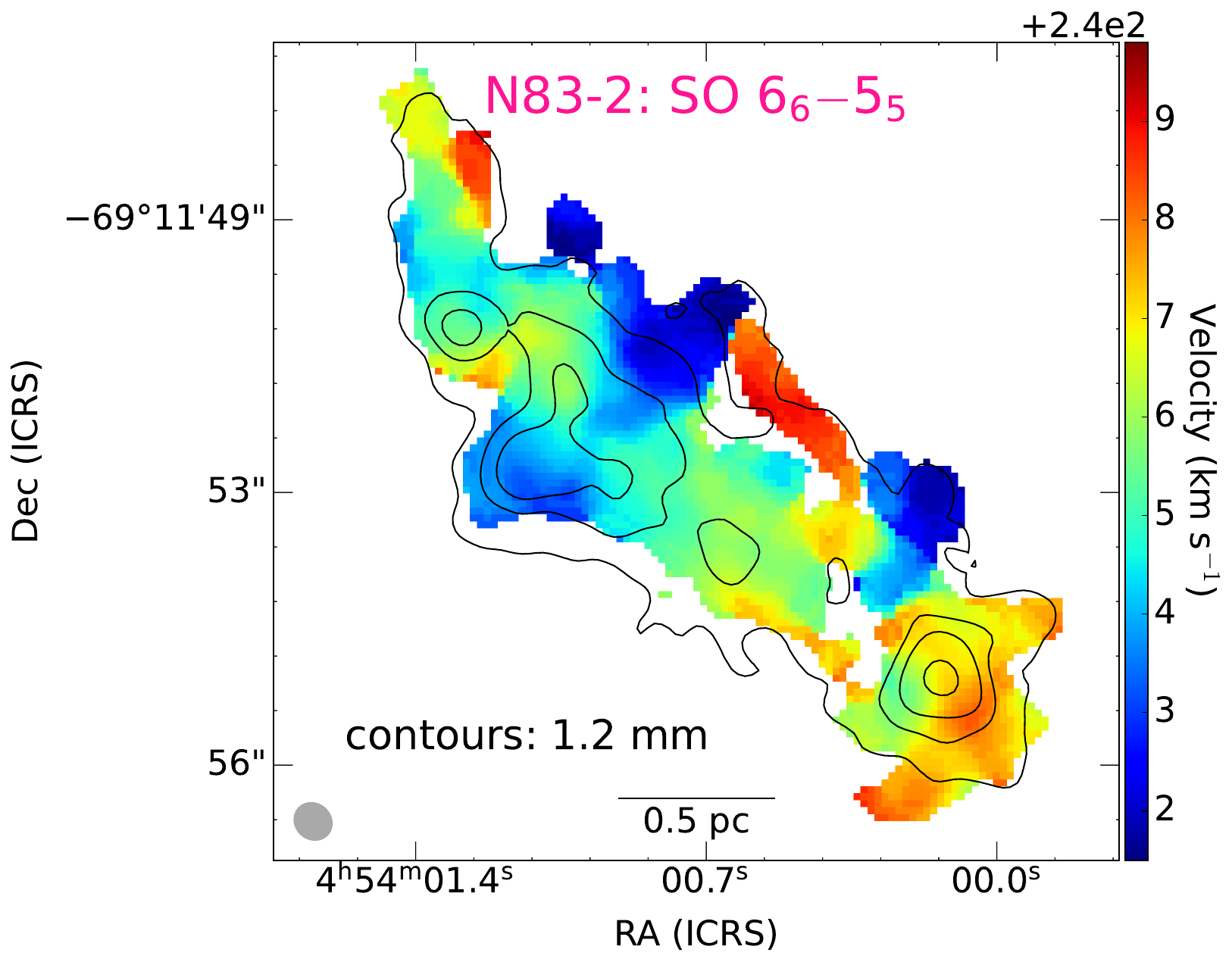}
\includegraphics[width=0.328\textwidth]{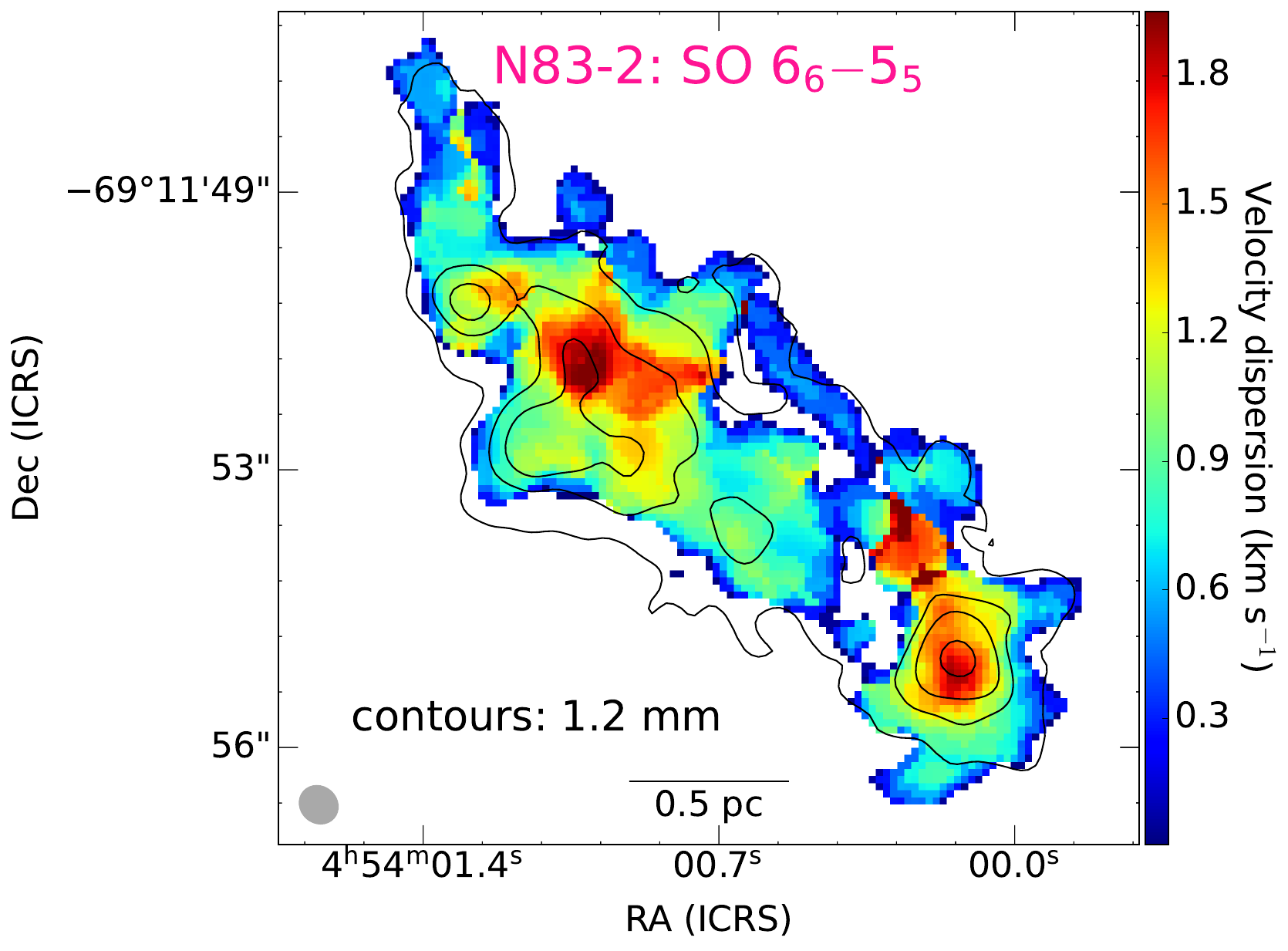}
\caption{The same as Figure~\ref{f:kinematicsCS}, but for SO $6_{6}-5_{5}$.
 \label{f:kinematicsSO}}
\end{figure*}

\begin{figure*}
\centering
\includegraphics[width=0.328\textwidth]{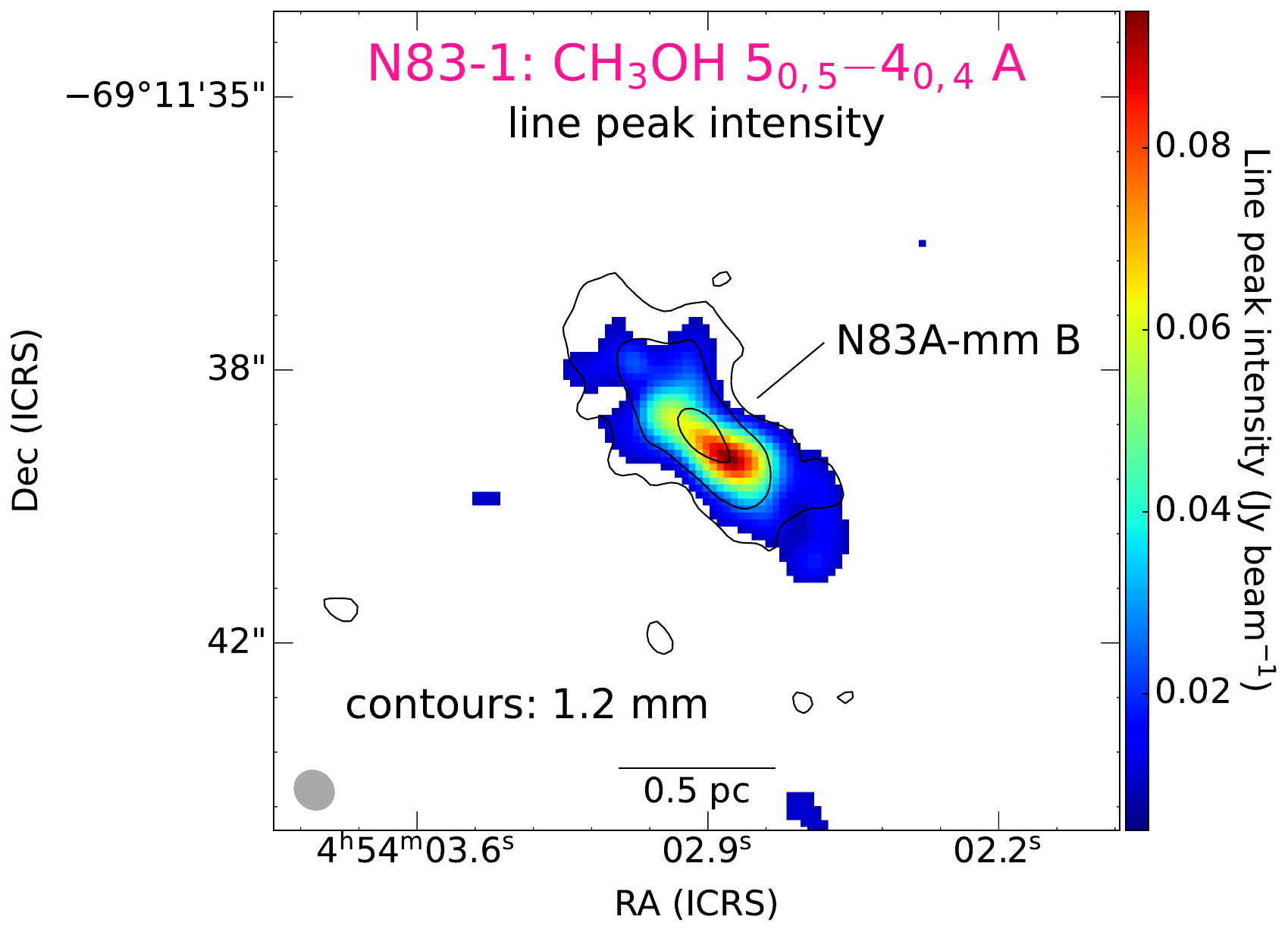}
\includegraphics[width=0.318\textwidth]{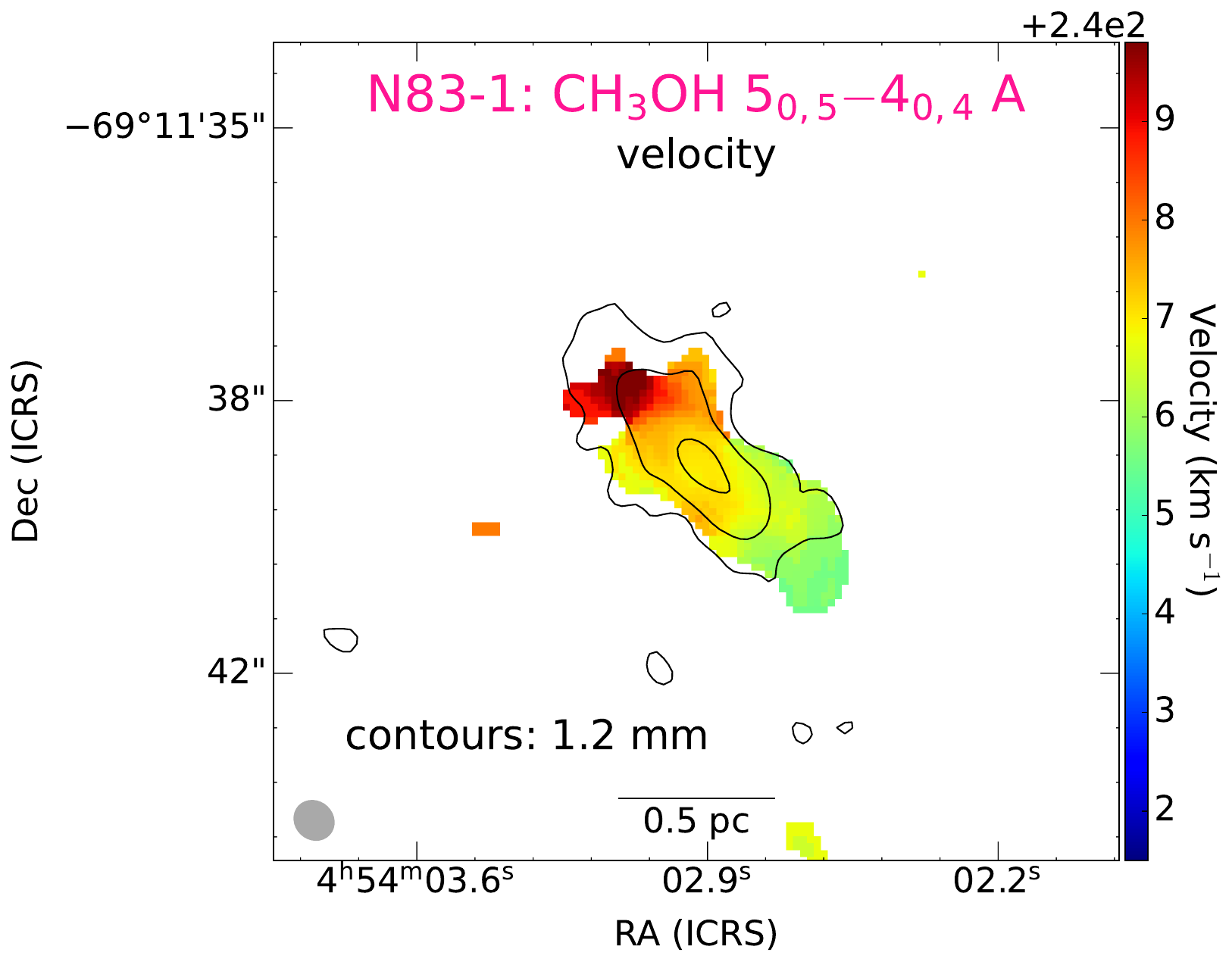}
\includegraphics[width=0.328\textwidth]{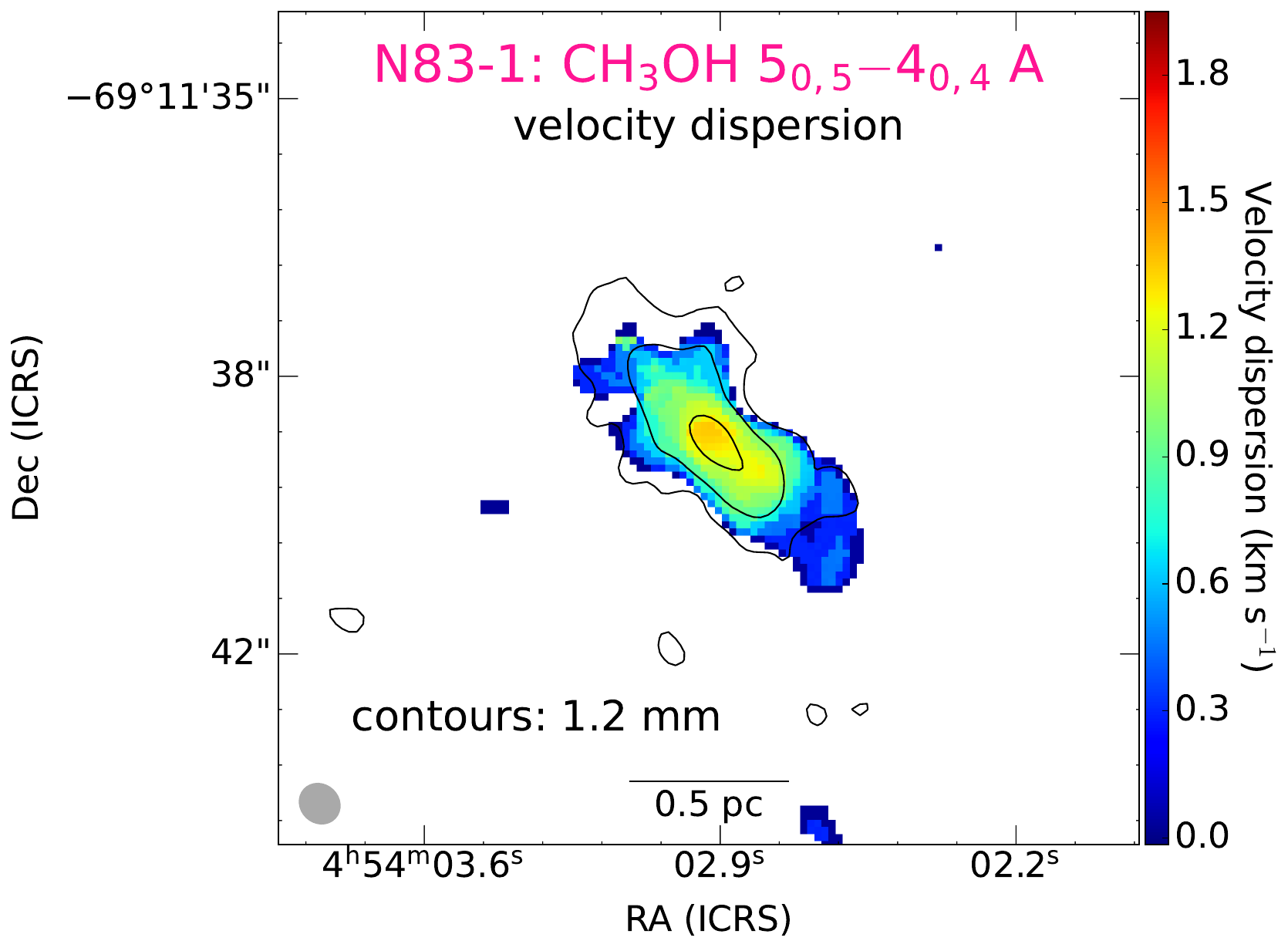}
\includegraphics[width=0.328\textwidth]{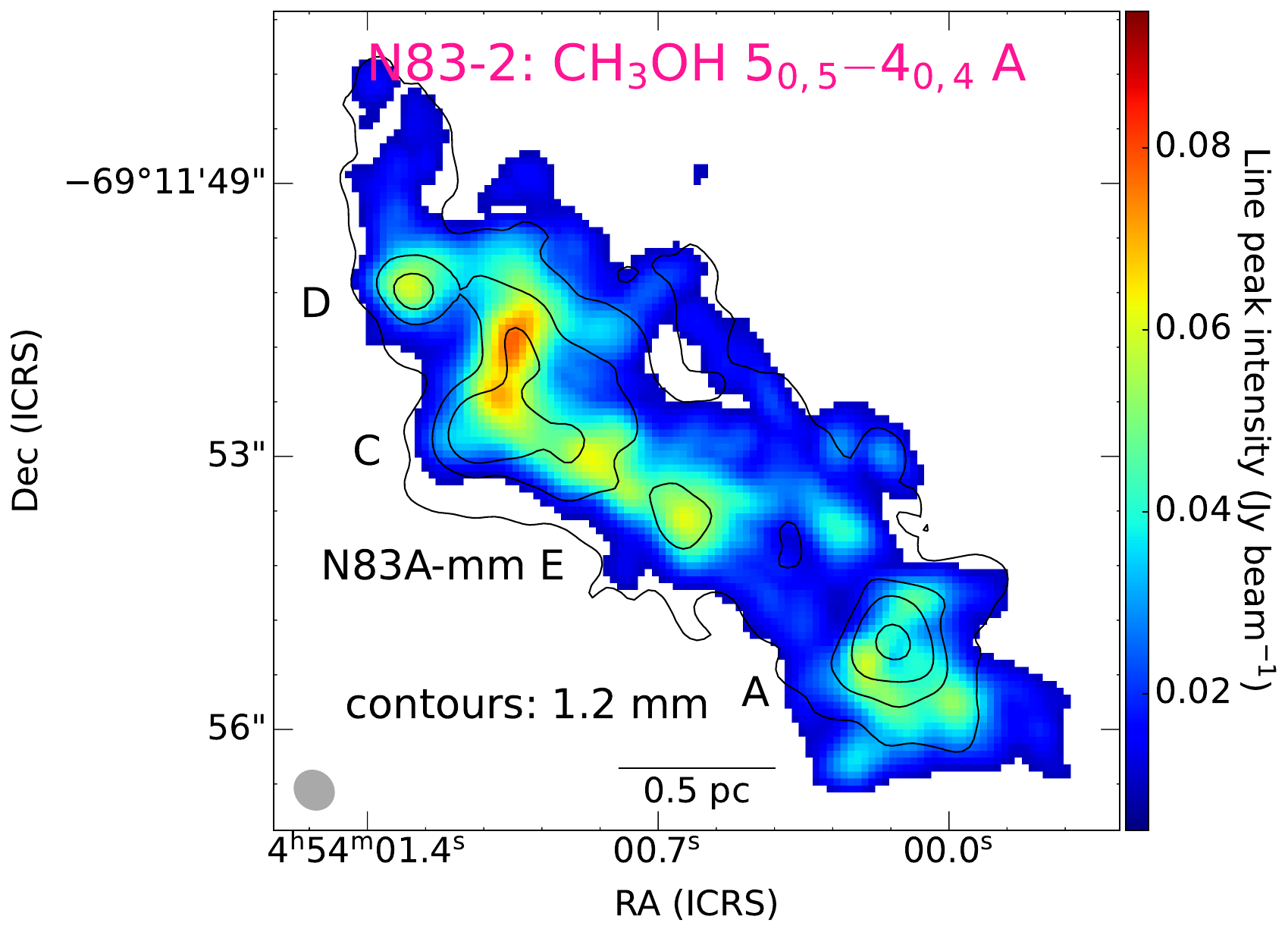}
\includegraphics[width=0.318\textwidth]{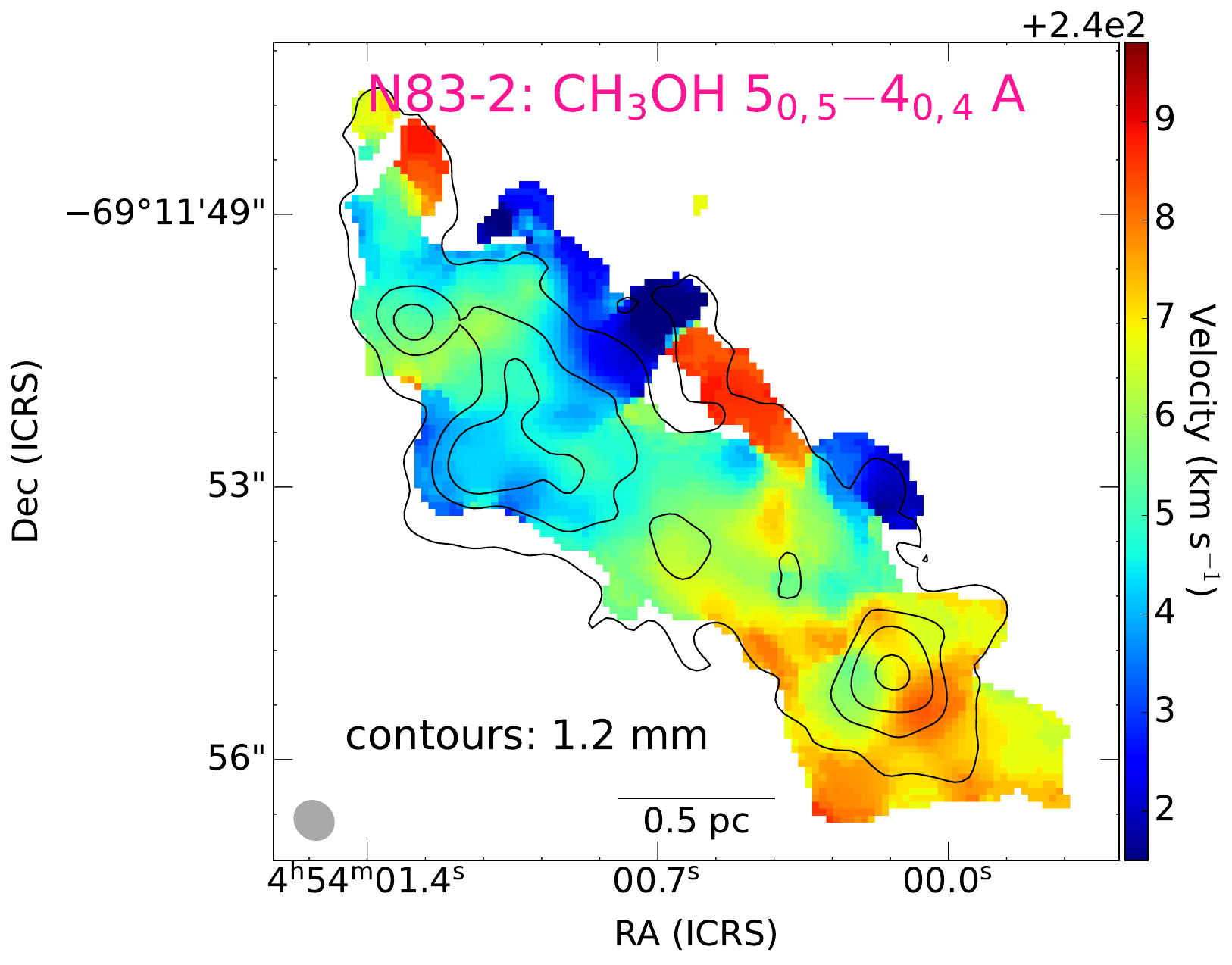}
\includegraphics[width=0.328\textwidth]{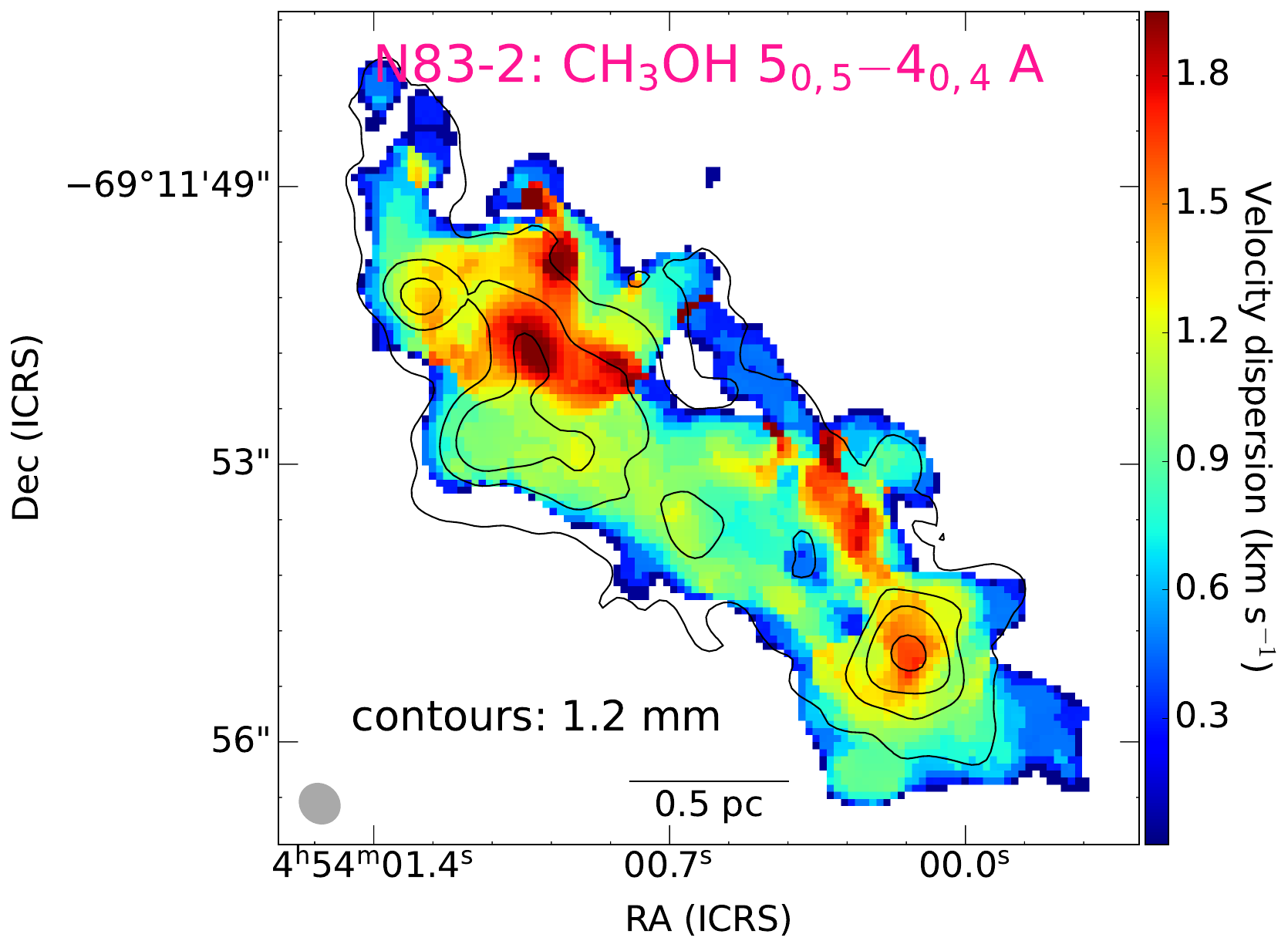}
\caption{The same as Figure~\ref{f:kinematicsCS}, but for \ce{CH3OH} (the 5$_{0, 5}$--4$_{0, 4}$ A transition). \label{f:kinematicsCH3OH} }
\end{figure*}

\begin{figure*}
\centering
\includegraphics[width=0.9\textwidth]{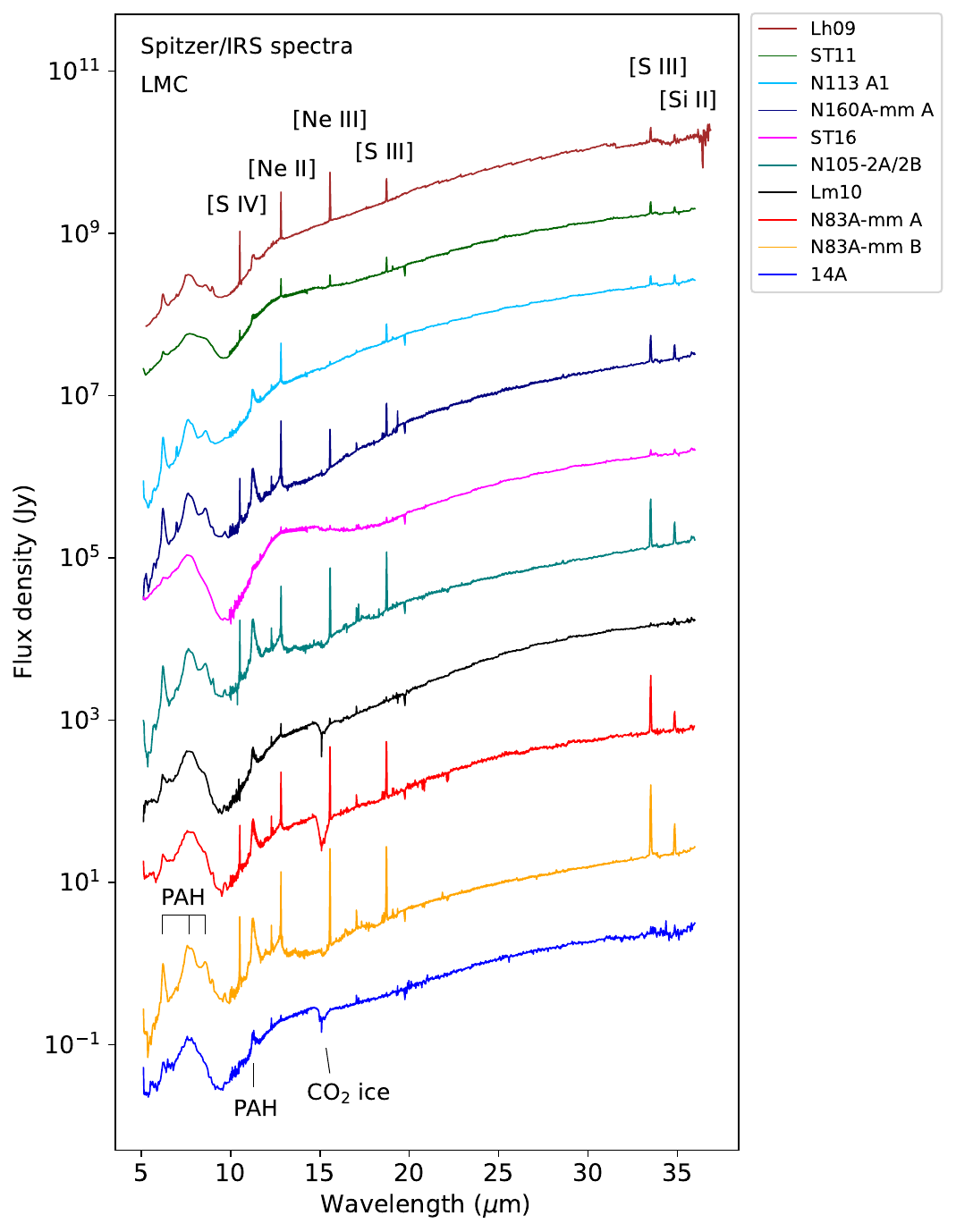}
\caption{The {\it Spitzer}/IRS spectra of YSOs associated with hot cores in the LMC, color-coded as indicated in the legend. The CO$_2$ ice and PAH features, and fine-structure lines are labeled. The spectra have been scaled by the following multiplicative factors for clarity (from bottom to top): 1, 8, 180, 1900, $2\times10^4$, $1.1\times10^5$, $10^6$, $7.5\times10^6$, $4\times10^7$, $1.6\times10^8$. \label{f:SpitzerIRSseparate}}
\end{figure*}

In Figures~\ref{f:N83AspitzerFOV}--\ref{f:SpitzerIRSseparate}, we provide the ancillary images and spectra supporting the discussion in Section~\ref{s:discussion}. Below, we provide additional information on Figures~\ref{f:N83B3B6cont} and \ref{f:herspec}. 

The ALMA 3 mm continuum image with the resolution of $\sim$2$''$ shown in Figure~\ref{f:N83B3B6cont} reveals the compact and extended components in the field of view covering both N\,83--1 and N\,83--2 ALMA fields.  Both the thermal emission from dust and the free-free emission from ionized gas likely contribute to the 3 mm continuum emission. The extended 3 mm continuum component detected in N\,83A follows the interior of the {\it Spitzer} bubble and has a similar distribution to other ionized gas tracers between the ionized star and the rim of the bubble where N\,83--1 and N\,83--2 are located.  Both hot cores (N\,83A--mm A and B) and N\,83A--mm C are associated with the 3 mm continuum peaks. Source C is associated with the brightest of the three 3 mm continuum sources (and it is the brightest source in the image) with the peak intensity of 1.85 mJy~beam$^{-1}$, followed by source B (1.07 mJy~beam$^{-1}$), and A (0.93 mJy~beam$^{-1}$). 

In Figure~\ref{f:herspec}, we show the [C\,{\sc II}] 157.74 $\mu$m,  [O\,{\sc I}]  63.18 $\mu$m, and [O\,{\sc III}] 88.36 $\mu$m line emission maps from \citet{oliveira2019} obtained with the  {\it Herschel} Photodetector Array Camera and Spectrometer (PACS).  The [O\,{\sc III}] emission traces the highly-ionized gas within the {\it Spitzer} bubble, while both the [O\,{\sc I}]  and [C\,{\sc II}]  emission originates predominantly from the PDRs (i.e., with a negligible contribution from shock excitation; \citealt{oliveira2019}) on spatial scales probed by {\it Herschel}/PACS (the nominal resolution of 4$\rlap.{''}$5/$\sim$1.1 pc at 63 $\mu$m and 11$\rlap.{''}$4/$\sim$2.7 pc at 158 $\mu$m).  The [C\,{\sc II}]  emission has a similar distribution to the 8.0 $\mu$m emission with the emission peaks coincident with the ALMA fields N\,83--1  and N\,83--2 hosting hot cores N\,83A--mm B and A, respectively. The peak of the [O\,{\sc I}] emission is associated with N\,83--2, implying that the gas in this field is denser than in N\,83--1 since the critical density of the [O\,{\sc I}] line is higher than that of the [C\,{\sc II}] line. This conclusion is supported by the ALMA $^{13}$CO (1--0) and C$^{18}$O (2--1) observations (see Section~\ref{s:multiwav} and Figure~\ref{f:N83B3B6_13CO}).

\end{document}